\documentclass[12pt]{article}
\usepackage[utf8]{inputenc}
\usepackage[left=2cm,right=2cm,top=2cm,bottom=2cm]{geometry}
\usepackage{lmodern}
\usepackage{soul}
\usepackage{makecell} 
\usepackage{lipsum}
\usepackage{graphicx}
\usepackage[dvipsnames]{xcolor}
\usepackage{array}
\usepackage{dirtytalk}
\usepackage{adjustbox}
\usepackage{setspace} 
\usepackage{hhline} 
\usepackage{rotating}
\usepackage{float} 
\usepackage{array} 
\usepackage{eurosym} 
\usepackage{arydshln} 
\usepackage{longtable} 
\usepackage{csquotes} 
\usepackage{nccmath} 
\usepackage{bm} 
\usepackage{booktabs} 
\usepackage{xcolor} 
\usepackage{multirow} 
\usepackage{bm} 
\usepackage{physics} 
\usepackage{systeme} 
\usepackage{mathtools} 
\usepackage{amsmath, amsthm, amssymb, amsfonts} 
\usepackage{graphicx} 
\usepackage{amsmath}
\usepackage{amssymb}

\usepackage{graphicx}
\usepackage{subcaption}

\usepackage{pgfplots} 
\pgfplotsset{compat=1.16} 
\usepackage{tikz} 
\usetikzlibrary{intersections,calc} 
\usepackage{caption} 
\usepackage{soul} 
\usepackage{sidecap} 
\usepackage[space]{grffile} 
\usepackage{subcaption} 
\usepackage{makebox} 
\usepackage[short]{optidef} 
\usepackage{abraces} 
\usepackage{enumitem} 
\usepackage{comment} 
\usepackage{cancel} 
\usepackage{accents} 
\usepackage{stackrel} 
\usepackage{dashrule} 
\usepackage{scalerel} 
\usepackage{centernot} 
\usepackage{pifont} 
\usepackage{pdflscape}
\usepackage{chngcntr}
\usepackage[flushleft]{threeparttable}
\usepackage[round]{natbib}
\usepackage{hyperref,etoolbox}
\usepackage{etoc} 
\hypersetup{
    colorlinks=true,
    linkcolor=black,
    filecolor=magenta,
    citecolor=black, 
    urlcolor=black,
    }
\makeatletter
\def\@footnotecolor{red}
\define@key{Hyp}{footnotecolor}{%
 \HyColor@HyperrefColor{#1}\@footnotecolor%
}
\patchcmd{\@footnotemark}{\hyper@linkstart{link}}{\hyper@linkstart{footnote}}{}{}
\makeatother
\hypersetup{footnotecolor=black}

\usepackage[ruled,vlined]{algorithm2e}
\SetKwInput{KwInput}{Input}                
\SetKwInput{KwOutput}{Output}

\usetikzlibrary{arrows,decorations.markings}
\usetikzlibrary{decorations.pathreplacing}
\tikzset{myarr/.style={decoration={markings, mark=between positions 0 and 1 step 4mm with {\arrow{stealth}},},postaction=decorate}}
\tikzset{myarrrev/.style={decoration={markings, mark=between positions 0 and 1 step 4mm with {\arrowreversed{stealth}},},postaction=decorate}}
\DeclareCaptionLabelFormat{subfig}{\figurename #1~\thefigure.\alph{subfigure}:}

\newcolumntype{L}[1]{>{\raggedright\let\newline\\\arraybackslash\hspace{0pt}}m{#1}}
\newcolumntype{C}[1]{>{\centering\let\newline\\\arraybackslash\hspace{0pt}}m{#1}}
\newcolumntype{R}[1]{>{\raggedleft\let\newline\\\arraybackslash\hspace{0pt}}m{#1}}

\theoremstyle{definition}

\theoremstyle{definition}

\theoremstyle{definition}

\theoremstyle{definition}

\theoremstyle{definition}

\theoremstyle{theorem}

\theoremstyle{definition}

\theoremstyle{definition}

\definecolor{blue}{RGB}{0,114,178}
\definecolor{red}{RGB}{204,51,17}
\definecolor{yellow}{RGB}{240,228,66}
\definecolor{green}{RGB}{0,158,115}

\tikzset{
solid node/.style={circle,draw,inner sep=1.5,fill=black},
hollow node/.style={circle,draw,inner sep=1.5}
}

\usepackage{xr}
\title{\Large{Large datasets for the euro area and its member countries and the dynamic effects of the common monetary policy}}
\author{\normalsize Matteo Barigozzi$^*$  
\hspace{2cm} 
  Claudio Lissona$^*$%
  \hspace{2cm} 
Lorenzo Tonni$^{\dagger}$%
  }
 \vspace{5pt}
\date{\small This version: \today}

\begin{document}
\setstretch{1.5}
\maketitle

\begin{abstract}
We introduce EA-MD-QD, a new publicly available dataset comprising 1136 macroeconomic time series for the euro area (EA) and its ten largest member countries observed at monthly or quarterly frequency. Since January 2024, EA-MD-QD has been updated monthly and continuously revised, providing a valuable resource for policy analysis in the EA. Using EA-MD-QD, we study country-specific impulse responses to an EA-wide monetary policy shock. Results reveal moderate yet significant cross-country heterogeneity, with differences between so-called core countries, such as France and Germany, and peripheral countries, such as Italy and Spain, in their price and interest rate responses, together with meaningful differences in real activity, while stock price responses are relatively homogeneous.
Evidence points to homeownership and saving behavior as potential drivers of the observed cross-country differences.


\end{abstract}

\renewcommand{\thefootnote}{$\ast$} 
\footnotetext{Department of Economics, University of Bologna, Italy.}

\renewcommand{\thefootnote}{$\dagger$} 

\footnotetext{Department of Economics, Management and Quantitative Methods, University of Milan, Italy.\\ 
 All authors gratefully acknowledge financial support from the Italian Ministry of Education, University and Research (PRIN 2020, Grant  2020N9YFFE\_003). This paper uses data from the Eurosystem Household Finance and Consumption Survey. The results published and the related observations and analysis may not correspond to results or analysis of the data producers.} 

\renewcommand{\thefootnote}{\arabic{footnote}}
\newpage

\addtocontents{toc}{\protect\etocsettocdepth{-1}} 
\section{Introduction}
\label{sec::intro}

%
%
%
%
%
%
%
%
%
%

The recent and ongoing \enquote{data revolution} has enabled researchers to collect and process large amounts of information, thereby broadening the scope of macroeconomic research. The availability of high-dimensional datasets—where the number of series, $N$, can be comparable to or even exceed the time dimension, $T$—allows researchers to exploit the rich information contained in large sets of macroeconomic and financial indicators. This, in turn, facilitates addressing relevant policy questions using econometric methods specifically designed to handle, and in some cases benefit from, such large datasets.

In this paper, we present and document a new large-scale dataset for macroeconomic analysis, denoted as EA-MD-QD. The dataset comprises time series for both the euro area (EA) as a whole and its ten largest member countries, providing an accessible and comprehensive resource for policy analysis of EA-wide outcomes. It also enables comparisons across data vintages and empirical studies.
Overall, EA-MD-QD includes $1136$ time series, at either monthly or quarterly frequency depending on data availability. The first available vintage spans the period January 2000-January 2024. After that date, the data have been updated every month. The countries covered are Austria, Belgium, France, Germany, Greece, Ireland, Italy, the Netherlands, Portugal, and Spain. All the series are retrieved from institutional sources, namely Eurostat, ECB, OECD and FRED. The EA-MD-QD is made publicly available online and updated every month and comes jointly with Matlab and Python codes for data transformation, missing-value imputation, and the treatment of outliers and the COVID period.\footnote{\url{https://zenodo.org/doi/10.5281/zenodo.10514667}}

Although the cost of data collection has substantially decreased in recent years, several sunk costs still hinder data availability and, consequently, the reproducibility of economic research. First, the manual updating of time series becomes increasingly burdensome as the dimensionality of the dataset grows, highlighting the need for systematic procedures to facilitate data acquisition. Second, once collected, data are often not immediately suitable for analysis due to issues such as seasonality, non-stationarity, and missing values, which require additional pre-processing. In this work, we aim to reduce, if not entirely eliminate, these costs by providing a dataset that is periodically updated and accompanied by codes enabling users to generate, within seconds, a research-ready dataset. 

Since data are updated on a monthly basis, real-time monthly vintages are available. Specifically, updates occur at the end of each calendar month, either on the last working day or on the day when all updates scheduled for that month by the data providers have been released. Making vintages publicly available serves two main purposes. First, it enables users to account for the effects of data revisions across different periods, which have been shown to matter in many macroeconomic applications \citep[see, e.g.,][]{orphanides2002unreliability}. Second, it fosters the reproducibility of research findings, as users can identify and retrieve the exact data vintage employed in a given analysis.

Building on the seminal contribution of \citet{sw1996}, several large-scale datasets have been developed to foster macroeconomic research. Among the most prominent examples are FRED-MD and FRED-QD by \citet{fredMD,fredQD}, two publicly available collections of monthly and quarterly macroeconomic and financial time series for the United States. Similarly, \citet{stevanovic} constructed a large macroeconomic dataset for Canada.

To the best of our knowledge, this is the first attempt to provide a comprehensive dataset for macroeconomic research that encompasses both the EA as a whole and its largest member countries. The closest reference to our work using EA data is the real-time database developed by \citet{giannone2012area}, which relies on the information contained in the ECB's Monthly Bulletins, i.e., the rawest information set available to policymakers on the data of the ECB governing council. Our dataset differs from theirs along several dimensions.
First, the relatively small number of vintages of our data prevents a fully real-time analysis.
Second, both the timing of data collection and the information content of our dataset do not correspond to any specific policy meeting or information set available to policymakers. Instead, we provide a snapshot of macroeconomic conditions in the EA and its largest member countries at a given point in time. Hence, the purpose and scope of the two datasets are different.
Finally, unlike \citet{giannone2012area}, our dataset also covers the largest EA countries, with the aim of offering comparable information at both the national and EA aggregate levels. Moreover, we provide practitioners with all the codes needed to make the dataset readily usable for empirical macroeconomic research.

As a second contribution, we employ EA-MD-QD to study the Impulse Response Functions (IRF) of a common monetary policy shock in the EA and, particularly, across individual EA member countries. The availability of country-level data enables policymakers to account for cross-country heterogeneity, which may ultimately influence both the effectiveness and the transmission of the common monetary policy. In this context, understanding whether, and to what extent, heterogeneous dynamics arise across countries is essential for designing and implementing more effective EA-level policies.

To this end, we employ the recently proposed Common Component Vector Autoregression (CC-VAR) methodology by \citet{forni2020common}, which relies on the natural assumption of an underlying factor structure in the data. In a first step, standard Principal Component Analysis (PCA) is employed to extract the common factors and compute the common components of all variables. In a second step, a VAR model is estimated on some selected common components of interest along with other observables. Intuitively, this allows us to study IRFs to common shocks, as the EA monetary policy one, without any contamination from idiosyncratic dynamics. More precisely, provided we include in the VAR step at least as many variables as common factors, we can fully identify the space spanned by the common shocks, by focussing on their common components only. This implies that the IRFs to common shocks for the variables considered in the VAR step do not change, even if some variables are substituted by others. 
Such an invariance property motivates our preference for the CC-VAR over more classical alternatives such as a standard VAR or a FAVAR \citep{bernanke2005measuring}.

In our baseline specification we identify an EA-wide common monetary policy shock using monthly data and the Instrumental Variables (IV) approach proposed by \citet{stock2012disentangling} and \citet{mertens2013dynamic}, in combination with the High-Frequency Identification (HFI) strategy of \citet{gertler2015monetary}. We test several instruments from the pool provided by \citet{altavilla2019measuring} and select the 1-year overnight index swap (OIS), which is associated with the highest first-stage $F$-statistic.

We find four key results. First, looking at the share of variance explained by the common factors, we uncover non-negligible cross-country  heterogeneity, particularly in unemployment and interest rate dynamics, whereas industrial production, stock prices, and prices display a higher degree of homogeneity. These findings provide a preliminary assessment of the degree of comovement among EA countries across different economic dimensions.

Second, the IRFs are consistent with standard economic theory. In particular, we observe a decline in the IRFs of prices in all EA economies following a contractionary monetary policy shock--although with different magnitudes, while at the EA level
their magnitude aligns with other results in the literature 
\citep{jarocinski2020deconstructing}.

Third, we find evidence of a moderate yet meaningful heterogeneity in impulse responses across countries for several key variables. In particular, our estimates point to a core-periphery pattern in price and interest rate dynamics. For prices, core countries’ responses are on average aligned with--or even more pronounced than--the EA aggregate, whereas peripheral countries appear less affected. Symmetrically, interest rate responses in peripheral countries display stronger medium-term effects on average. No clear pattern emerges for real indicators, though some countries--such as Italy and Greece--show notable deviations from the aggregate. By contrast, stock prices exhibit a relatively homogeneous behavior across countries. On average,  Germany’s responses closely mirror the EA aggregate, while Greece shows the most pronounced deviations.

Finally, to explore potential drivers of this heterogeneity, we examine the correlations between the peak country-level IRFs and selected structural characteristics related to labor markets, households, firms, and national economic structures \citep[][]{corsetti2022one}. While no causal inference is implied, the evidence suggests that monetary policy transmission is strongly related to several economic channels--including homeownership rates and savings behavior--that either dampen or amplify the effects of common shocks across countries.

Over the years, a large body of literature has studied the effects of common monetary policy in the EA \citep[]{jarocinski2020deconstructing, andrade2021delphic}. In this paper, we also analyse potential asymmetries in the transmission mechanism of monetary policy shocks across EA member countries. In particular, by employing a novel large-dimensional dataset, more recent data vintages, and a novel econometric framework, we extend and update the previous studies by \citet{barigozzi2014euro}, \citet{georgiadis2015examining}, \citet{burriel2018uncovering}, \citet{corsetti2022one}, \citet{mandler2022heterogeneity}, \citet{gefang2025does}. 

Overall, we find moderate heterogeity in the transmission of monetary policy across EA countries, in line with \citet{gefang2025does}. 
Whenever present, the heterogeneity highlighted by our results broadly aligns with \citet{barigozzi2014euro}, who document similar cross-country differences in prices and unemployment rates between so-called core and peripheral countries. Unlike our results, \citet{corsetti2022one} report price increases for several countries in response to monetary tightening. Nevertheless, despite this difference in sign, the relative cross-country patterns we identify are broadly consistent with their findings. Conversely, \citet{mandler2022heterogeneity} obtain an inverted cross-country ranking in price responses compared with our results.
Along similar lines, \citet{altavilla2016financial} study the effect of Outright Monetary Transactions on four major EA countries and find a similar core–periphery pattern in the responses of national interest rates.
Relatedly, \citet{sciacovelli2025monetary} and \citet{pica2021housing} focus on cross-country asymmetries in the consumption response to monetary policy and show that these differences are largely explained by housing market characteristics, such as the share of adjustable-rate mortgages, liquidity-constrained households, and homeownership rates.

Section~\ref{sec::data} describes the EA-MD-QD dataset. Section~\ref{sbsec::bench} introduces the baseline specification adopted for the empirical application. Section~\ref{sbsec::factors} discusses the factor analysis, while Section~\ref{sbsec::comovements} examines the comovements across variables and countries explained by the factors. Section~\ref{sec::method} describes the estimation and identification of the EA and country-specific IRFs to a common monetary policy shock. Section~\ref{sec::results} reports the estimated IRFs, as well as the analysis of cross-country heterogeneity and its potential drivers. Section~\ref{conclusion} concludes. In a Supplementary Appendix we provide a detailed description of the data, a step-by-step guide to data preparation and estimation of comovements,  and additional empirical results, based on different data and identification approaches, showing the robustness of our results.

\section{The EA-MD-QD dataset}
\label{sec::data}

\subsection{Data description}
\label{sbsec::datades}
The dataset is constructed according to three guiding principles: (i) \textit{accessibility}: all series are sourced from public, institutional databases; (ii) \textit{timeliness}: the dataset is fully web-scraped, enabling monthly updates of all series in the panel according to their respective release calendars; and (iii) \textit{coverage}: it includes variables representing the most relevant sources for modern macro-financial analysis. Importantly, the monthly updates provide practitioners with a new vintage of the dataset each month, incorporating all data releases and revisions to previously published series, if any. All monthly vintages are available from the initial release of EA-MD-QD in January 2024.\\
\indent \textbf{Euro area}. The dataset for the EA  comprises $N = 118$ series, of which $N_M = 47$ are monthly and $N_Q = 71$ are quarterly. All series span from 2000:M1 (2000:Q1) to the most recent available observation. The selection of variables follows established datasets for the US \citep{fredMD,fredQD} and covers: (1) National Accounts, (2) Labor Market Indicators, (3) Credit Aggregates, (4) Labor Costs, (5) Exchange Rates, (6) Financial Markets, (7) Industrial Production and Turnover, (8) Prices, (9) Confidence Indicators, and (10) Monetary Aggregates. Among the 118 series, 104 are sourced from Eurostat, 10 from the ECB Data Warehouse, 3 from the OECD Statistical Database, and 1 from the FRED portal.

\indent \textbf{Countries}. In addition to the EA dataset, we provide datasets for the ten largest EA countries: Austria, Belgium, France, Germany, Greece, Ireland, Italy, the Netherlands, Portugal, and Spain. Each country-specific dataset is constructed according to the same principles used for the EA data, with a few exceptions. For instance, variables related to Monetary Aggregates are not included at the country level, as they cannot be straightforwardly attributed to individual countries. Other groups of variables (e.g., Credit Aggregates or Industrial Production and Turnover) are smaller for some countries, such as Ireland, due to data unavailability. The country-specific variables are sourced from the same providers as those of the EA dataset.

Table \ref{tab::dataEA0} provides a summary of the series included in the dataset, along with a brief description, their frequency (column F), and provider (column P). 
For each series, the last columns of Table \ref{tab::dataEA0} indicate the countries for which the series is available. A checkmark denotes availability, while a dash indicates that the series is not available for that country. A more detailed description of the series, with additional information, is in Table A2 in Appendix A. 

Table \ref{tab::nseries} provides a detailed breakdown of the numerosity of series by country, organized according to the ten macro-categories included in the dataset. Broadly, each category is similarly represented across countries, and the frequency of individual series is generally uniform across countries. Specifically, for both the EA and each individual country, approximately 60\% of the series are quarterly, while the remaining 40\% are monthly. There are two notable exceptions: indicators of Industrial Production and Turnover are unavailable for Ireland, and Producer Prices are missing for both Ireland and Portugal. 

 Due to their mixed-frequency nature, the dataset is inherently unbalanced, with quarterly series recorded in the first month of each corresponding quarter and missing values in the remaining months.
The provided codes allow users to handle this unbalanced structure. Practitioners can choose to: (i) retain the mixed-frequency format; (ii) subset the dataset to include only monthly or only quarterly variables; or (iii) aggregate the monthly data to the quarterly frequency.
In the latter case, monthly series are aggregated to the quarterly level by summing the values of monthly \textit{flow} variables and taking the mean of monthly \textit{stock} variables.\footnote{Alternatively, one could take the value from the last month of the reference quarter as the quarterly observation. However, this approach would overlook intra-quarter dynamics that may be informative. For instance, if 2020:Q2 were represented solely by June 2020 data, much of the COVID-19 shock observed in April 2020 would be disregarded.}

\begin{table}[htbp]
\centering
\caption{Data Description by country}
\begin{footnotesize}
\begin{threeparttable}
\resizebox{6.4in}{!}{%
\begin{tabular}{  c | c | l | c | c || c | c | c | c | c | c | c | c | c | c | c | c | c | c | c |}
\hline
\hline
\textbf{N} & \textbf{ID} & \hspace{130pt}\textbf{Series} & \textbf{F} & \textbf{P} & \textbf{EA} & \textbf{AT} & \textbf{BE} & \textbf{DE} & \textbf{EL} & \textbf{ES} & \textbf{FR} & \textbf{IE} & \textbf{IT} & \textbf{NL} & \textbf{PT}\\ 
\hline
\hline
\multicolumn{16}{c}{(1) \textbf{National Accounts}}\\
\hline
1  & GDP    & Real Gross Domestic Product                                 & Q & EUR & \checkmark & \checkmark & \checkmark & \checkmark & \checkmark & \checkmark & \checkmark & \checkmark & \checkmark & \checkmark & \checkmark \\ 
2  & EXPGS  & Real Export Goods and services                              & Q & EUR & \checkmark & \checkmark & \checkmark & \checkmark & \checkmark & \checkmark & \checkmark & \checkmark & \checkmark & \checkmark & \checkmark \\ 
3  & IMPGS  & Real Import Goods and services                              & Q & EUR & \checkmark & \checkmark & \checkmark & \checkmark & \checkmark & \checkmark & \checkmark & \checkmark & \checkmark & \checkmark & \checkmark \\ 
4  & GFCE   & Real Government Final consumption expenditure               & Q & EUR & \checkmark & \checkmark & \checkmark & \checkmark & \checkmark & \checkmark & \checkmark & \checkmark & \checkmark & \checkmark & \checkmark \\ 
5  & HFCE   & Real Households consumption expenditure                     & Q & EUR & \checkmark & \checkmark & \checkmark & \checkmark & \checkmark & \checkmark & \checkmark & \checkmark & \checkmark & \checkmark & \checkmark \\ 
6  & CONSD  & Real Households consumption expenditure: Durable Goods      & Q & EUR & \checkmark & \checkmark & \checkmark & \checkmark & \checkmark & \checkmark & \checkmark & \checkmark & \checkmark & \checkmark & \checkmark \\ 
7  & CONSSD & Real Households consumption expenditure: Semi-Durable Goods & Q & EUR & - & \checkmark & - & \checkmark & - & - & \checkmark & \checkmark & \checkmark & \checkmark & - \\
8  & CONSSV & Real Households consumption expenditure: Services           & Q & EUR & - & \checkmark & - & \checkmark & - & - & \checkmark & \checkmark & \checkmark & \checkmark & - \\ 
9  & CONSND & Real Households consumption expenditure: Non-Durable Goods  & Q & EUR & \checkmark & \checkmark & - & \checkmark & - & - & \checkmark & \checkmark & \checkmark & \checkmark & - \\ 
10 & GCF    & Real Gross capital formation                                & Q & EUR & \checkmark & \checkmark & \checkmark & \checkmark & \checkmark & \checkmark & \checkmark & \checkmark & \checkmark & \checkmark & \checkmark \\ 
11 & GCFC   & Real Gross fixed capital formation                          & Q & EUR & \checkmark & \checkmark & \checkmark & \checkmark & \checkmark & \checkmark & \checkmark & \checkmark & \checkmark & \checkmark & \checkmark \\ 
12 & GFACON & Real Gross Fixed Capital Formation: Construction            & Q & EUR & \checkmark & \checkmark & - & \checkmark & \checkmark & \checkmark & \checkmark & \checkmark & \checkmark & \checkmark & \checkmark \\ 
13 & GFAMG  & Real Gross Fixed Capital Formation: Machinery and Equipment & Q & EUR & \checkmark & \checkmark & - & \checkmark & \checkmark & \checkmark & \checkmark & \checkmark & \checkmark & \checkmark & \checkmark \\
14 & AHRDI  & Adjusted Households Real Disposable Income                  & Q & EUR & \checkmark & - & - & - & - & - & - & - & - & - & - \\ 
15 & AHFCE  & Actual Final Consumption Expenditure of Households          & Q & EUR & \checkmark & - & - & - & - & - & - & - & - & - & - \\ 
16 & GNFCPS & Gross Profit Share of Non-Financial Corporations            & Q & EUR & \checkmark & \checkmark & \checkmark & \checkmark & \checkmark & \checkmark & \checkmark & \checkmark & \checkmark & \checkmark & \checkmark \\ 
17 & GNFCIR & Gross Investment Share of Non-Financial Corporations        & Q & EUR & \checkmark & \checkmark & \checkmark & \checkmark & \checkmark & \checkmark & \checkmark & \checkmark & \checkmark & \checkmark & \checkmark \\ 
18 & GHIR   & Gross Investment Rate of Households 					      & Q & EUR & \checkmark & \checkmark & - & \checkmark & - & \checkmark & \checkmark & \checkmark & \checkmark & \checkmark & \checkmark \\ 
19 & GHSR   & Gross Households Savings Rate 						      & Q & EUR & \checkmark & \checkmark & \checkmark & \checkmark & - & \checkmark & \checkmark & \checkmark & \checkmark & \checkmark & \checkmark \\ 
\hline
\multicolumn{16}{c}{(2) \textbf{Labor Market Indicators}}\\
\hline
20 & TEMP   & Total Employment (domestic concept)                                  & Q & EUR & \checkmark & \checkmark & \checkmark & \checkmark & \checkmark & \checkmark & \checkmark & \checkmark & \checkmark & \checkmark & \checkmark \\ 
21 & EMP    & Employees (domestic concept)                                         & Q & EUR & \checkmark & \checkmark & \checkmark & \checkmark & \checkmark & \checkmark & \checkmark & \checkmark & \checkmark & \checkmark & \checkmark \\ 
22 & SEMP   & Self Employment (domestic concept)                                   & Q & EUR & \checkmark & \checkmark & \checkmark & \checkmark & \checkmark & \checkmark & \checkmark & \checkmark & \checkmark & \checkmark & \checkmark \\ 
23 & THOURS & Hours Worked: Total                                                  & Q & EUR & \checkmark & \checkmark & - & \checkmark & \checkmark & \checkmark & \checkmark & \checkmark & \checkmark & \checkmark & \checkmark \\
24 & EMPAG  & Quarterly Employment: Agriculture, Forestry, Fishing                 & Q & EUR & \checkmark & \checkmark & \checkmark & \checkmark & \checkmark & \checkmark & \checkmark & \checkmark & \checkmark & \checkmark & \checkmark \\ 
25 & EMPIN  & Quarterly Employment: Industry                                       & Q & EUR & \checkmark & \checkmark & \checkmark & \checkmark & \checkmark & \checkmark & \checkmark & \checkmark & \checkmark & \checkmark & \checkmark \\ 
26 & EMPMN  & Quarterly Employment: Manufacturing                                  & Q & EUR & \checkmark & \checkmark & \checkmark & \checkmark & \checkmark & \checkmark & \checkmark & \checkmark & \checkmark & \checkmark & \checkmark \\ 
27 & EMPCON & Quarterly Employment: Construction                                   & Q & EUR & \checkmark & \checkmark & \checkmark & \checkmark & \checkmark & \checkmark & \checkmark & \checkmark & \checkmark & \checkmark & \checkmark \\ 
28 & EMPRT  & Quarterly Employment: Wholesale/Retail trade, transport, food        & Q & EUR & \checkmark & \checkmark & \checkmark & \checkmark & \checkmark & \checkmark & \checkmark & \checkmark & \checkmark & \checkmark & \checkmark \\ 
29 & EMPIT  & Quarterly Employment: Information and Communication                  & Q & EUR & \checkmark & \checkmark & \checkmark & \checkmark & \checkmark & \checkmark & \checkmark & \checkmark & \checkmark & \checkmark & \checkmark \\ 
30 & EMPFC  & Quarterly Employment: Financial and Insurance activities             & Q & EUR & \checkmark & \checkmark & \checkmark & \checkmark & \checkmark & \checkmark & \checkmark & \checkmark & \checkmark & \checkmark & \checkmark \\ 
31 & EMPRE  & Quarterly Employment: Real Estate                                    & Q & EUR & \checkmark & \checkmark & \checkmark & \checkmark & \checkmark & \checkmark & \checkmark & \checkmark & \checkmark & \checkmark & \checkmark \\ 
32 & EMPPR  & Quarterly Employment: Professional, Scientific, Technical activities & Q & EUR & \checkmark & \checkmark & \checkmark & \checkmark & \checkmark & \checkmark & \checkmark & \checkmark & \checkmark & \checkmark & \checkmark \\ 
33 & EMPPA  & Quarterly Employment: PA, education, health ad social services       & Q & EUR & \checkmark & \checkmark & \checkmark & \checkmark & \checkmark & \checkmark & \checkmark & \checkmark & \checkmark & \checkmark & \checkmark \\ 
34 & EMPENT & Quarterly Employment: Arts and recreational activities               & Q & EUR & \checkmark & \checkmark & \checkmark & \checkmark & \checkmark & \checkmark & \checkmark & \checkmark & \checkmark & \checkmark & \checkmark \\ 
35 & UNETOT & Unemployment: Total (\% active population)                           & M & EUR & \checkmark & \checkmark & \checkmark & \checkmark & \checkmark & \checkmark & \checkmark & \checkmark & \checkmark & \checkmark & \checkmark \\ 
36 & UNEO25 & Unemployment: Over 25 years (\% active population)                   & M & EUR & \checkmark & \checkmark & \checkmark & \checkmark & \checkmark & \checkmark & \checkmark & \checkmark & \checkmark & \checkmark & \checkmark \\
37 & UNEU25 & Unemployment: Under 25 years (\% active population)                  & M & EUR & \checkmark & \checkmark & \checkmark & \checkmark & \checkmark & \checkmark & \checkmark & \checkmark & \checkmark & \checkmark & \checkmark \\
38 & RPRP   & Real Labour Productivity (person)                                    & Q & EUR & \checkmark & \checkmark & \checkmark & \checkmark & \checkmark & \checkmark & \checkmark & \checkmark & \checkmark & \checkmark & \checkmark \\ 
39 & WS     & Wages and salaries                                                   & Q & EUR & \checkmark & \checkmark & \checkmark & \checkmark & \checkmark & \checkmark & \checkmark & \checkmark & \checkmark & \checkmark & \checkmark \\ 
40 & ESC    & Employers' Social Contributions                                      & Q & EUR & \checkmark & \checkmark & \checkmark & \checkmark & \checkmark & \checkmark & \checkmark & \checkmark & \checkmark & \checkmark & \checkmark \\ 
\hline
\multicolumn{16}{c}{(3) \textbf{Credit Aggregates}}\\
\hline
41 & TASS.SDB   & Total Economy - Assets: Short-Term Debt Securities          & Q & EUR & \checkmark & \checkmark & \checkmark & \checkmark & \checkmark & \checkmark & \checkmark & \checkmark & \checkmark & \checkmark & \checkmark \\ 
42 & TASS.LDB   & Total Economy - Assets: Long-Term Debt Securities           & Q & EUR & \checkmark & \checkmark & \checkmark & \checkmark & \checkmark & \checkmark & \checkmark & \checkmark & \checkmark & \checkmark & \checkmark \\ 
43 & TASS.SLN   & Total Economy - Assets: Short-Term Loans                    & Q & EUR & \checkmark & \checkmark & \checkmark & \checkmark & \checkmark & \checkmark & \checkmark & \checkmark & \checkmark & \checkmark & \checkmark \\ 
44 & TASS.LLN   & Total Economy - Assets: Long-Term Loans                     & Q & EUR & \checkmark & \checkmark & \checkmark & \checkmark & \checkmark & \checkmark & \checkmark & \checkmark & \checkmark & \checkmark & \checkmark \\ 
45 & TLB.SDB    & Total Economy - Liabilities: Short-Term Debt Securities     & Q & EUR & \checkmark & \checkmark & \checkmark & \checkmark & - & \checkmark & \checkmark & \checkmark & \checkmark & \checkmark & \checkmark \\ 
46 & TLB.LDB    & Total Economy - Liabilities: Long-Term Debt Securities      & Q & EUR & \checkmark & \checkmark & \checkmark & \checkmark & \checkmark & \checkmark & \checkmark & \checkmark & \checkmark & \checkmark & \checkmark \\ 
47 & TLB.SLN    & Total Economy - Liabilities: Short-Term Loans               & Q & EUR & \checkmark & \checkmark & \checkmark & \checkmark & \checkmark & \checkmark & \checkmark & \checkmark & \checkmark & \checkmark & \checkmark \\ 
48 & TLB.LLN    & Total Economy - Liabilities: Long-Term Loans                & Q & EUR & \checkmark & \checkmark & \checkmark & \checkmark & \checkmark & \checkmark & \checkmark & \checkmark & \checkmark & \checkmark & \checkmark \\ 
49 & NFCASS     & Non-Financial Corporations: Total Financial Assets          & Q & EUR & \checkmark & \checkmark & \checkmark & \checkmark & \checkmark & \checkmark & \checkmark & \checkmark & \checkmark & \checkmark & \checkmark \\ 
50 & NFCASS.SLN & Non-Financial Corporations - Assets: Short-Term Loans       & Q & EUR & \checkmark & \checkmark & \checkmark & \checkmark & - & \checkmark & \checkmark & \checkmark & \checkmark & \checkmark & \checkmark \\ 
51 & NFCASS.LLN & Non-Financial Corporations - Assets: Long-Term Loans        & Q & EUR & \checkmark & \checkmark & \checkmark & \checkmark & - & \checkmark & \checkmark & \checkmark & \checkmark & \checkmark & \checkmark \\ 
52 & NFCLB      & Non-Financial Corporations: Total Financial Liabilities     & Q & EUR & \checkmark & \checkmark & \checkmark & \checkmark & \checkmark & \checkmark & \checkmark & \checkmark & \checkmark & \checkmark & \checkmark \\ 
53 & NFCLB.SLN  & Non-Financial Corporations - Liabilities - Short-Term Loans & Q & EUR & \checkmark & \checkmark & \checkmark & \checkmark & \checkmark & \checkmark & \checkmark & - & \checkmark & \checkmark & \checkmark \\ 
54 & NFCLB.LLN  & Non-Financial Corporations - Liabilities - Long-Term Loans  & Q & EUR & \checkmark & \checkmark & \checkmark & \checkmark & \checkmark & \checkmark & \checkmark & - & \checkmark & \checkmark & \checkmark \\ 
55 & GGASS      & General Government: Total Financial Assets                  & Q & EUR & \checkmark & \checkmark & \checkmark & \checkmark & \checkmark & \checkmark & \checkmark & \checkmark & \checkmark & \checkmark & \checkmark \\ 
56 & GGASS.SLN  & General Government - Assets: Short-Term Loans               & Q & EUR & \checkmark & \checkmark & - & - & - & - & - & - & - & \checkmark & - \\ 
57 & GGASS.LLN  & General Government - Assets: Short-Term Loans               & Q & EUR & \checkmark & \checkmark & - & \checkmark & - & - & \checkmark & - & - & \checkmark & \checkmark \\ 
58 & GGLB       & General Government: Total Financial Liabilities             & Q & EUR & \checkmark & \checkmark & \checkmark & \checkmark & \checkmark & \checkmark & \checkmark & \checkmark & \checkmark & \checkmark & \checkmark \\ 
59 & GGLB.SLN   & General Government - Liabilities: Short-Term Loans          & Q & EUR & \checkmark & \checkmark & \checkmark & \checkmark & - & \checkmark & \checkmark & \checkmark & \checkmark & \checkmark & \checkmark \\ 
60 & GGLB.LLN   & General Government - Liabilities: Long-Term Loans           & Q & EUR & \checkmark & \checkmark & \checkmark & \checkmark & - & \checkmark & \checkmark & \checkmark & \checkmark & \checkmark & \checkmark \\ 
61 & HHASS      & Households: Total Financial Assets                          & Q & EUR & \checkmark & \checkmark & \checkmark & \checkmark & \checkmark & \checkmark & \checkmark & \checkmark & \checkmark & \checkmark & \checkmark \\ 
62 & HHASS.SLN  & Households - Assets: Short-Term Loans                       & Q & EUR & \checkmark & - & - & - & - & - & \checkmark & - & - & - & - \\ 
63 & HHASS.LLN  & Households - Assets: Long-Term Loans                        & Q & EUR & \checkmark & - & - & - & - & - & \checkmark & - & - & \checkmark & \checkmark \\
64 & HHLB       & Households: Total Financial Liabilities                     & Q & EUR & \checkmark & \checkmark & \checkmark & \checkmark & \checkmark & \checkmark & \checkmark & \checkmark & \checkmark & \checkmark & \checkmark \\
65 & HHLB.SLN   & Households - Liabilities: Short-Term Loans                  & Q & EUR & \checkmark & \checkmark & \checkmark & \checkmark & \checkmark & \checkmark & \checkmark & - & \checkmark & \checkmark & \checkmark \\
66 & HHLB.LLN   & Households - Liabilities: Long-Term Loans                   & Q & EUR & \checkmark & \checkmark & \checkmark & \checkmark & \checkmark & \checkmark & \checkmark & - & \checkmark & \checkmark & \checkmark \\
\hline
\multicolumn{16}{c}{(4) \textbf{Labor Costs}}\\
\hline
67 & ULCIN  & Nominal Unit Labor Costs: Industry                                       & Q & EUR & \checkmark & \checkmark & \checkmark & \checkmark & \checkmark & \checkmark & \checkmark & \checkmark & \checkmark & \checkmark & \checkmark \\ 
68 & ULCMQ  & Nominal Unit Labor Costs: Mining and Quarrying                           & Q & EUR & \checkmark & \checkmark & \checkmark & - & \checkmark & - & \checkmark & \checkmark & - & \checkmark & \checkmark \\ 
69 & ULCMN  & Nominal Unit Labor Costs: Manufacturing                                  & Q & EUR & \checkmark & \checkmark & \checkmark & \checkmark & \checkmark & \checkmark & \checkmark & \checkmark & \checkmark & \checkmark & \checkmark \\ 
70 & ULCCON & Nominal Unit Labor Costs: Construction                                   & Q & EUR & \checkmark & \checkmark & \checkmark & \checkmark & \checkmark & \checkmark & \checkmark & - & \checkmark & \checkmark & \checkmark \\ 
71 & ULCRT  & Nominal Unit Labor Costs: Wholesale/Retail Trade, Transport, Food, IT    & Q & EUR & \checkmark & \checkmark & \checkmark & \checkmark & \checkmark & \checkmark & \checkmark & \checkmark & \checkmark & \checkmark & \checkmark \\ 
72 & ULCFC  & Nominal Unit Labor Costs: Financial Activities                           & Q & EUR & \checkmark & \checkmark & \checkmark & \checkmark & \checkmark & \checkmark & \checkmark & \checkmark & \checkmark & \checkmark & \checkmark \\ 
73 & ULCRE  & Nominal Unit Labor Costs: Real Estate                                    & Q & EUR & \checkmark & \checkmark & \checkmark & \checkmark & \checkmark & \checkmark & \checkmark & - & \checkmark & \checkmark & \checkmark \\ 
74 & ULCPR  & Nominal Unit Labor Costs: Professional, Scientific, Technical activities & Q & EUR & \checkmark & \checkmark & \checkmark & \checkmark & \checkmark & \checkmark & \checkmark & \checkmark & \checkmark & \checkmark & \checkmark \\ 
\hline 
\multicolumn{16}{c}{(5) \textbf{Financial Markets}}\\
\hline
75 & REER42 & Real Exchange Rate (42 main industrial countries) & M & EUR  & \checkmark & \checkmark & \checkmark & \checkmark & \checkmark & \checkmark & \checkmark & \checkmark & \checkmark & \checkmark & \checkmark \\
76 & ERUS   & Exchange Rate (US dollar)                         & M & EUR  & \checkmark & - & - & - & - & - & - & - & - & - & - \\ 
77 & SHIX  & Stock Price  Index                                     & M & OECD & \checkmark & \checkmark & \checkmark & \checkmark & \checkmark & \checkmark & \checkmark & \checkmark & \checkmark & \checkmark & \checkmark \\   
\hline 
\multicolumn{16}{c}{(6) \textbf{Interest Rates}}\\
\hline
79 & IRT3M & 3-Months Interest Rates                  & M & EUR & \checkmark & - & - & - & - & - & - & - & - & - & - \\ 
79 & IRT6M & 6-Months Interest Rates                  & M & EUR & \checkmark & - & - & - & - & - & - & - & - & - & - \\ 
80 & LTIRT & Long-Term Interest Rates (EMU Criterion) & M & EUR & \checkmark & \checkmark & \checkmark & \checkmark & \checkmark & \checkmark & \checkmark & \checkmark & \checkmark & \checkmark & \checkmark \\ 
\hline
\hline
\end{tabular}%
}
\end{threeparttable}
\end{footnotesize}
\label{tab::dataEA0}
\end{table}

\newpage

\begin{table}[htbp]
\ContinuedFloat
\centering
\caption{Data Description by country}
\begin{footnotesize}
\begin{threeparttable}
\resizebox{6.4in}{!}{%
\begin{tabular}{  c | c | l | c | c || c | c | c | c | c | c | c | c | c | c | c | c | c | c | c |}
\hline
\hline
\textbf{N} & \textbf{ID} & \hspace{130pt}\textbf{Series} & \textbf{F} & \textbf{P} & \textbf{EA} & \textbf{AT} & \textbf{BE} & \textbf{DE} & \textbf{EL} & \textbf{ES} & \textbf{FR} & \textbf{IE} & \textbf{IT} & \textbf{NL} & \textbf{PT}\\ 
\hline
\hline
\multicolumn{16}{c}{(7) \textbf{Industrial Production and Turnover}}\\
\hline
81 & IPMN     & Industrial Production Index: Manufacturing              & M & EUR & \checkmark & \checkmark & \checkmark & \checkmark & \checkmark & \checkmark & \checkmark & - & \checkmark & \checkmark & \checkmark \\ 
82 & IPCAG    & Industrial Production Index: Capital Goods              & M & EUR & \checkmark & \checkmark & \checkmark & \checkmark & \checkmark & \checkmark & \checkmark & - & \checkmark & \checkmark & \checkmark \\ 
83 & IPCOG    & Industrial Production Index: Consumer Goods             & M & EUR & \checkmark & \checkmark & \checkmark & \checkmark & \checkmark & \checkmark & \checkmark & - & \checkmark & - & - \\ 
84 & IPDCOG   & Industrial Production Index: Durable Consumer Goods     & M & EUR & \checkmark & \checkmark & \checkmark & \checkmark & \checkmark & \checkmark & \checkmark & - & \checkmark & \checkmark & \checkmark \\ 
85 & IPNDCOG  & Industrial Production Index: Non Durable Consumer Goods & M & EUR & \checkmark & \checkmark & \checkmark & \checkmark & \checkmark & \checkmark & \checkmark & - & \checkmark & \checkmark & \checkmark \\ 
86 & IPING    & Industrial Production Index: Intermediate Goods 		& M & EUR & \checkmark & \checkmark & \checkmark & \checkmark & \checkmark & \checkmark & \checkmark & - & \checkmark & \checkmark & \checkmark \\ 
87 & IPNRG    & Industrial Production Index: Energy                     & M & EUR & \checkmark & \checkmark & \checkmark & \checkmark & \checkmark & \checkmark & \checkmark & - & \checkmark & \checkmark & \checkmark \\ 
\hline
88 & TRNMN    & Turnover Index: Manufacturing              & M & EUR & \checkmark & \checkmark & \checkmark & \checkmark & \checkmark & \checkmark & \checkmark & - & - & \checkmark & \checkmark \\ 
89 & TRNCAG   & Turnover Index: Capital Goods              & M & EUR & \checkmark & \checkmark & \checkmark & \checkmark & \checkmark & \checkmark & \checkmark & - & \checkmark & \checkmark & \checkmark \\ 
90 & TRNCOG   & Turnover Index: Consumer Goods             & M & EUR & \checkmark & \checkmark & \checkmark & \checkmark & \checkmark & \checkmark & \checkmark & - & \checkmark & \checkmark & \checkmark \\ 
91 & TRNDCOG  & Turnover Index: Durable Consumer Goods     & M & EUR & \checkmark & \checkmark & \checkmark & \checkmark & \checkmark & \checkmark & \checkmark & - & \checkmark & \checkmark & \checkmark \\ 
92 & TRNNDCOG & Turnover Index: Non Durable Consumer Goods & M & EUR & \checkmark & \checkmark & \checkmark & \checkmark & \checkmark & \checkmark & \checkmark & - & \checkmark & \checkmark & \checkmark \\ 
93 & TRNING   & Turnover Index: Intermediate Goods         & M & EUR & \checkmark & \checkmark & \checkmark & \checkmark & \checkmark & \checkmark & \checkmark & - & \checkmark & \checkmark & \checkmark \\ 
94 & TRNNRG   & Turnover Index: Energy                     & M & EUR & \checkmark & \checkmark & \checkmark & \checkmark & \checkmark & \checkmark & \checkmark & - & \checkmark & - & \checkmark \\ 
\hline
95 & CAREG & Passenger's Cars Registrations  & M & ECB  & \checkmark & - & - & - & - & - & - & - & - & - & - \\ 
\hline
\multicolumn{16}{c}{(8) \textbf{Prices}}\\
\hline 
96  & PPICAG   & Producer Price Index: Capital Goods                               & M & EUR  & \checkmark & \checkmark & \checkmark & \checkmark & \checkmark & \checkmark & \checkmark & - & \checkmark & \checkmark & - \\ 
97  & PPICOG   & Producer Price Index: Consumer Goods                              & M & EUR  & \checkmark & \checkmark & \checkmark & \checkmark & \checkmark & \checkmark & \checkmark & - & \checkmark & \checkmark & - \\ 
98  & PPIDCOG  & Producer Price Index: Durable Consumer Goods                      & M & EUR  & \checkmark & \checkmark & \checkmark & \checkmark & \checkmark & \checkmark & \checkmark & - & \checkmark & \checkmark & - \\ 
99  & PPINDCOG & Producer Price Index: Non Durable Consumer Goods  				   & M & EUR  & \checkmark & \checkmark & \checkmark & \checkmark & \checkmark & \checkmark & \checkmark & - & \checkmark & \checkmark & - \\ 
100  & PPIING   & Producer Price Index: Intermediate Goods                         & M & EUR  & \checkmark & \checkmark & \checkmark & \checkmark & \checkmark & \checkmark & \checkmark & - & \checkmark & \checkmark & - \\ 
101  & PPINRG   & Producer Price Index: Energy 								       & M & EUR  & \checkmark & \checkmark & \checkmark & \checkmark & \checkmark & \checkmark & \checkmark & - & \checkmark & \checkmark & - \\ 
102 & HICPOV   & Harmonized Index of Consumer Prices: Overall Index                & M & ECB  & \checkmark & \checkmark & \checkmark & \checkmark & \checkmark & \checkmark & \checkmark & \checkmark & \checkmark & \checkmark & \checkmark \\ 
103 & HICPNEF  & Harmonized Index of Consumer Prices: All Items, no Energy{\&}Food & M & ECB  & \checkmark & \checkmark & \checkmark & \checkmark & \checkmark & \checkmark & \checkmark & \checkmark & \checkmark & \checkmark & \checkmark \\ 
104 & HICPG    & Harmonized Index of Consumer Prices: Goods 	                   & M & ECB  & \checkmark & \checkmark & \checkmark & \checkmark & \checkmark & \checkmark & \checkmark & \checkmark & \checkmark & \checkmark & \checkmark \\ 
105 & HICPIN   & Harmonized Index of Consumer Prices: Industrial Goods             & M & ECB  & \checkmark & \checkmark & \checkmark & \checkmark & \checkmark & \checkmark & \checkmark & \checkmark & \checkmark & \checkmark & \checkmark \\ 
106 & HICPSV   & Harmonized Index of Consumer Prices: Services                     & M & ECB  & \checkmark & \checkmark & \checkmark & \checkmark & \checkmark & \checkmark & \checkmark & \checkmark & \checkmark & \checkmark & \checkmark \\ 
107 & HICPNG   & Harmonized Index of Consumer Prices: Energy                       & M & EUR  & \checkmark & \checkmark & \checkmark & \checkmark & \checkmark & \checkmark & \checkmark & \checkmark & \checkmark & \checkmark & \checkmark \\ 
108 & DFGDP    & Real Gross Domestic Product Deflator                              & Q & EUR  & \checkmark & \checkmark & \checkmark & \checkmark & \checkmark & \checkmark & \checkmark & \checkmark & \checkmark & \checkmark & \checkmark \\ 
109 & HPRC     & Residential Property Prices (BIS)                                 & Q & FRED & \checkmark & \checkmark & \checkmark & \checkmark & - & \checkmark & \checkmark & \checkmark & \checkmark & \checkmark & - \\ 
\hline
\multicolumn{16}{c}{(9) \textbf{Confidence Indicators}}\\
\hline
110 & ICONFIX  & Industrial Confidence Indicator   & M & EUR  & \checkmark & \checkmark & \checkmark & \checkmark & \checkmark & \checkmark & \checkmark & \checkmark & \checkmark & \checkmark & \checkmark \\  
111 & CCONFIX  & Consumer Confidence Index         & M & EUR  & \checkmark & \checkmark & \checkmark & \checkmark & \checkmark & \checkmark & \checkmark & \checkmark & \checkmark & \checkmark & \checkmark \\  
112 & ESENTIX  & Economic Sentiment Indicator      & M & EUR  & \checkmark & \checkmark & \checkmark & \checkmark & \checkmark & \checkmark & \checkmark & \checkmark & \checkmark & \checkmark & \checkmark \\  
113 & KCONFIX  & Construction Sentiment Indicator  & M & EUR  & \checkmark & \checkmark & \checkmark & \checkmark & \checkmark & \checkmark & \checkmark & \checkmark & \checkmark & \checkmark & \checkmark \\  
114 & RTCONFIX & Retail Confidence Indicator       & M & EUR  & \checkmark & \checkmark & \checkmark & \checkmark & \checkmark & \checkmark & \checkmark & \checkmark & \checkmark & \checkmark & \checkmark \\  
115 & SCONFIX  & Services Confidence Indicator     & M & EUR  & \checkmark & \checkmark & \checkmark & \checkmark & \checkmark & \checkmark & \checkmark & \checkmark & \checkmark & \checkmark & \checkmark \\  
116 & BCI      & Business Confidence Index         & M & OECD & \checkmark & \checkmark & \checkmark & \checkmark & \checkmark & \checkmark & \checkmark & \checkmark & \checkmark & \checkmark & \checkmark \\  
117 & CCI      & Consumer Confidence Index         & M & OECD & \checkmark & \checkmark & \checkmark & \checkmark & \checkmark & \checkmark & \checkmark & \checkmark & \checkmark & \checkmark & \checkmark \\  
\hline
\multicolumn{16}{c}{(10) \textbf{Monetary Aggregates}}\\
\hline
118 & CURR & Money Stock: Currency & M & ECB & \checkmark & - & - & - & - & - & - & - & - & - & - \\ 
119 & M1   & Money Stock: M1       & M & ECB & \checkmark & - & - & - & - & - & - & - & - & - & - \\  
120 & M2   & Money Stock: M2       & M & ECB & \checkmark & - & - & - & - & - & - & - & - & - & - \\  
\hline
\hline
\end{tabular}%
}
\end{threeparttable}
\end{footnotesize}
\end{table}

\begin{table}[htbp]
\caption{Number of series: EA and individual countries}
\centering
\scriptsize
\setlength{\tabcolsep}{6pt}
\resizebox{6.2in}{!}{
\begin{tabular}{l| c|  c|  c|  c|  c|  c|  c|  c|  c|  c|  c|| c}
\hline
\hline
& \textbf{EA} & \textbf{AT} & \textbf{BE} & \textbf{DE} & \textbf{EL} & \textbf{ES} & \textbf{FR} & \textbf{IE} & \textbf{IT} & \textbf{NL} & \textbf{PT} & \textbf{TOTAL} \\
\hline
\hline
\scriptsize{(1) National Accounts} & 17 & 17 & 11 & 17 & 12 & 14 & 17 & 17 & 17 & 17 & 14 & 161\\
\scriptsize{(2) Labor Market Indicators} &21 & 21 & 20 & 21 & 21 & 21 & 21 & 21 & 21 & 21 & 21 & 229\\
\scriptsize{(3) Credit Aggregates} & 26 & 24 & 22 & 23 & 17 & 22 & 25 & 18 & 22 & 25 & 24 & 248\\
\scriptsize{(4) Labor Costs} & 8 & 8 & 8 & 7 & 8 & 7 & 8 & 6 & 7 & 8 & 8 & 85\\
\scriptsize{(5) Financial Markets} & 3 & 2 & 2 & 2 & 2 & 2 & 2 & 2 & 2 & 2 & 2 & 23\\
\scriptsize{(6) Interest Rates} & 3 & 1 & 1 & 1 & 1 & 1 & 1 & 1 & 1 & 1 & 1 & 12\\
\scriptsize{(7) Production and Turnover} & 14 & 14 & 14 & 14 & 14 & 14 & 14 & 0 & 13 & 12 & 13 & 136\\
\scriptsize{(8) Prices} & 14 & 14 & 14 & 14 & 13 & 14 & 14 & 8 & 14 & 14 & 7 & 140\\
\scriptsize{(9) Confidence Indicators} & 8 & 8 & 8 & 8 & 8 & 8 & 8 & 8 & 8 & 8 & 8 & 88\\
\scriptsize{(10) Monetary Aggregates} & 3 & 0 & 0 & 0 & 0 & 0 & 0 & 0 & 0 & 0 & 0 & 3\\
\hline
\hline
\scriptsize{\textbf{(1)-(10) TOTAL}} 
& 118 & 109 & 100 & 107 & 96 & 103 & 110 & 81 & 105 & 108 & 99 & 1136\\
\hline
\hline
\end{tabular}
}
\label{tab::nseries}
\end{table}

Besides its mixed frequency nature, the dataset is unbalanced also because of both data availability and ragged edges arising from the asynchronous release of different series. Regarding data availability, while most series begin at or before 2000:Q1 (2000:M1), some start a few periods later (see Appendix A for details). As for release timing, the series are updated at different intervals across categories: some variables (e.g., National Accounts) are published roughly one to two months after the end of the reference quarter, while others (e.g., Credit Aggregates) may be released up to four months later. The companion codes allow practitioners to impute these missing values by means of two different procedures described in Section \ref{sbsec::outs}.

Finally, while most series are already seasonally adjusted at the source, some are only available in raw form. For these, we retrieve the unadjusted series and apply seasonal adjustment \textit{ex post}. In particular, we employ standard seasonal filtering methods \citep{findley1998new} for monthly variables, while quarterly variables are adjusted using a simple dummy-variable approach.\footnote{It is well known that filtering techniques can suffer from end-of-sample issues, requiring many observations to obtain reliable estimates of the trend component. Given the limited time span at the quarterly frequency, end-of-period estimates for seasonally adjusted quarterly variables obtained with these filters are therefore unreliable.} Only Credit Aggregates and Producer Prices are not seasonally adjusted at the source, accounting on average for about 30\% of the total series. Starting from the October 2025 release, we also provide users with raw data, where these series remain unadjusted.
\subsection{Transformations to stationarity}
\label{sbsec::transform}

\indent Being composed of variables that capture a wide range of economic and financial aggregates, the EA-MD-QD includes both stationary and non-stationary series. Although recent advances in econometric methods allow for the presence of non-stationarity in large-dimensional settings \citep{barigozzi_large-dimensional_2021}, many empirical applications still require stationary data. For this reason, we provide users with the raw data together with transformation codes designed to achieve stationarity.


We consider three possible sets of transformations. First, \textit{statistical} transformations which remove any $I(1)$ or $I(2)$ dynamics according to standard unit root tests \citep{dickey1979distribution,phillips1988testing}. Overall, both across variable categories and countries, only a small number of series exhibit $I(2)$ dynamics. Most of these belong to the group of Credit Aggregates, distributed relatively evenly across countries, with a few notable exceptions related to Unit Labor Costs and Labor Market Indicators. Conversely, the majority of the series in the panel are $I(1)$. A smaller subset of variables is instead $I(0)$, primarily concentrated among confidence indicators and, for some countries, national accounts variables. The transformations are applied uniformly across countries, with only minor deviations reflecting country-specific idiosyncrasies that generate dynamics differing from those of the corresponding EA series. We refer to Table A2 in Appendix A for details.

Second, we consider the statistical set of transformations described above with the exception of interest rates which are kept in levels. This choice is in line with the literature on EA monetary policy transmission, which is also the focus of the second part of the paper \citep{corsetti2022one}. Third, we consider the statistical set of transformations described above with the exception of interest rates and unemployment rates in order to keep all rates in levels.

The companion codes allow the users either to keep the data in levels or to apply either of the three sets of transformations described above. 


\subsection{Treatment of missing values}
\label{sbsec::outs}
As discussed above, regardless of the dataset frequency, missing values remain due to ragged edges arising from data availability, the asynchronous timing of data releases, as well as from removals of outliers. Moreover, the transformations applied to the series, as described in Section \ref{sbsec::transform}, introduce additional missing values because of observation losses due to differencing. While the untreated data are always provided, the companion codes to the dataset allow practitioners to choose between two different strategies for imputing missing values.

Specifically, missing values can be imputed  using either the Expectation Maximization (EM) algorithm proposed by \cite{stock2002macroeconomic} and also employed by \cite{fredMD}, or the one by \cite{banbura2014maximum}.
The former represents the most straightforward approach commonly used in the literature for imputing both outliers and missing values, though not necessarily the most sophisticated. The latter is particularly suitable when the data exhibit a factor structure with autocorrelated factors. 
\subsection{Treatment of outliers and COVID}
\label{sbsec::outs2}
Following \cite{fredQD}, an observation is considered an outlier if it deviates from the sample median by more than ten interquartile ranges. Once outliers are detected and removed, we can impute the corresponding missing values by either of the methods described in Section \ref{sbsec::outs}. This applies to the entire sample except the Covid period.

Unlike \enquote{standard} outliers, the COVID shock is pervasive, affecting most series in the panel--particularly those representing the real economy--with substantial heterogeneity even within these series. Moreover, it is unclear how much of the COVID shock should be attributed to economic versus exogenous forces \citep{ng2021modeling}. In light of these considerations, treating the COVID period as composed of sporadic outliers would result in the loss of potentially important economic information relevant for understanding the current economic stance. 

In the companion codes, we provide users with two alternatives related to the treatment of the COVID period. In the first option, data for variables representing the real side of the economy, as, e.g., Industrial Production, are treated as missing in 2020-2021 and imputed via the Kalman smoother using information from financial and nominal variables, as, e.g., the Stock Price Index and Prices. 
Alternatively, users can retain the transformed data without any specific treatment for COVID.

\section{Baseline specification}
\label{sbsec::bench}
As discussed above, the user has many possible choices about the data to be analysed and their treatment. Clearly, the specific choices depend on the aim of the empirical analysis. In the rest of the paper, we exploit the EA-MD-QD database to study the transmission of EA monetary policy across countries, with the goal of highlighting the advantages the database offers for structural macroeconomic analysis. To this end, hereafter, we adopt the following baseline specification. 
\begin{enumerate}
\item We consider only monthly data. Results for quarterly or monthly data plus GDP, i.e., mixed frequency data, are in Appendices I and J, respectively.

\item All variables are differenced according to the statistical transformations, except for  interest rates, which are kept in levels. Results based on the two other sets of transformations discussed in Section \ref{sbsec::transform} are in Appendices F and G.

\item Missing values are imputed according to the method by \citet{stock2002macroeconomic} as discussed in Section \ref{sbsec::outs}. Results obtained using the imputation method by \citet{banbura2014maximum} are virtually identical and thus omitted. 

\item Outliers are detected as in \citet{fredQD} as discussed in Section \ref{sbsec::outs2}, with the exception of COVID period, which is not treated as an outlier, but it is explicitly accounted for in the empirical analysis. Results based on a sample ending in 2019:M12, thus excluding COVID, are in Appendix D.

 \end{enumerate}
 
Specifically, we construct a balanced panel $\mathbf{x} = \{x_{it}, i=1,\ldots,N ; t=1,\ldots,T\}$ consisting of 47 EA-wide monthly variables and 200 national series, all transformed as described above, for a total of $N = 247$ time series, and covering the period 2002:M1-2023:M10, corresponding to $T=263$ monthly observations. 

These choices are based on a series of practical considerations. First, regarding the chosen sample, we begin the analysis in 2002:M1, following the literature on the identification of EA monetary policy shocks via instrumental variables (IV) \citep{altavilla2019measuring, andrade2021delphic}, since liquidity in the Overnight Index Swap (OIS) market was limited before then, hindering identification. The sample ends in 2023:M10, which is the latest date for which monetary policy surprises from \cite{altavilla2019measuring}—used to identify the shocks—are currently available. 

Second, when the goal is extracting common factors, as in our empirical analysis, it is well documented that adding more series to the dataset does not necessarily help in recovering the factors \citep{boivin2006more}. Indeed, one should retain only those series that are most likely to be driven by the same common shocks. Not surprisingly these are the most aggregated series. Hence our choice of appending  to the EA monthly dataset the following national variables: Industrial Production Indexes (IPMN, IPCAG, IPCOG, IPDCOG, IPNDCOG, IPING, IPNRG), Harmonized Indexes of Consumer Prices (HICPOV, HICPNEF, HICPG, HICPIN, HICPSV, HICPNG), Producer Price Indexes (PPICAG, PPICOG, PPINDCOG, PPIDCOG, PPIING, PPINRG), 10-years Interest Rates (LTIRT), Stock Price Indexes (SHIX), and Unemployment Rates (UNETOT).\footnote{Indeed, when including all the monthly variables in the dataset, the variance of the idiosyncratic components of the most relevant variables included in our analysis grows by 20\% on average.}

Third, although for consistent estimation of the factors we do not require any constraint between the cross-sectional dimension $N$ and the sample size $T$, still it is advisable to have a panel where these quantities have a comparable value. As pointed out by \citet{onatski2010determining} if $N$ is much larger than $T$ it is harder to recover consistently the number of factors by studying the behavior of sample eigenvalues.
Since the considered time span is relatively short, we prefer working with just a subset of the whole EA-MD-QD data.

A step-by-step description of the procedure adopted in this section and Sections \ref{sbsec::factors} and \ref{sbsec::comovements} to prepare and analyze the data through a factor model is given in Appendix B.


\section{Factor analysis}
\label{sbsec::factors}

We assume that the $N$-dimensional vector of observed data at time $t$, $\mathbf x_t$, follows a factor model: 
\begin{equation}
\mathbf x_t\ =\ \bm\mu+ \bm\Lambda \mathbf f_t+\bm \xi_t\ =\ \bm\chi_t+\bm\xi_t \quad\quad t=1,\ldots, T,\label{eq:FM}
\end{equation}
where $\bm\Lambda$ is the $N\times r$ vector of loadings associated to the $r$-dimensional vector of zero-mean common factors $\mathbf{f}_t$, $\bm\xi_t$ is a $N\times 1$ vector of zero-mean idiosyncratic components, and $\bm\mu$ is a $N \times 1$ vector of constants, hence $\mathbb E[\mathbf x_t]=\bm \mu$. 

Both the factors and the idiosyncratic components are allowed to be serially correlated. Moreover, when $N$ is large, it is reasonable to allow the idiosyncratic components to be also  (weakly) cross-sectionally correlated, and, in this case, we say that \eqref{eq:FM} is an approximate factor model. We refer to \citet{bai2003inferential} for the formal assumptions.

The loadings and the factors in \eqref{eq:FM} are not identified, since the factor model can equivalently be expressed with $\bm\Lambda \mathbf H$ as the loadings matrix and $\mathbf H^{-1}\mathbf f_t$ as the factors vector, for some invertible $r\times r$ matrix $\mathbf H$. To identify the factors, additional assumptions would be needed, but since in this paper our interest is only in estimating the common component $\bm\chi_t=\bm\mu+ \bm\Lambda \mathbf f_t$, we do not explore this path further. Indeed, $\bm\chi_t$ is always identified once we determine the number of factors $r$, so that we can disentangle it from the idiosyncratic component.


The factors and loadings are estimated via the classical PCA approach. So we estimate the loadings, denoted as $\widehat{\bm\Lambda}$, as $\sqrt N$ times the $r$ normalized eigenvectors corresponding to the $r$-largest eigenvalues of the sample covariance matrix of the standardized data $\mathbf x_t$, i.e., of $T^{\,-1}\sum_{t=1}^T \widehat{\bm\Omega}^{-1/2}(\mathbf x_t-\widehat{\bm\mu})(\mathbf x_t-\widehat{\bm\mu})'\widehat{\bm\Omega}^{-1/2}$, where
$\widehat{\bm\mu}$ and $\widehat{\bm\Omega}$ are, respectively,  the $N$ dimensional vector of sample means and a diagonal $N\times N$ matrix with entries the $N$ sample variances of each element of $\mathbf x_t$. Then, the factors, $\widehat{\mathbf f}_t$, are obtained by projecting the estimated loadings onto the data, i.e., $\widehat{\mathbf f}_t=(\widehat{\bm\Lambda}'\widehat{\bm\Lambda})^{-1}\widehat{\bm\Lambda}'\widehat{\bm\Omega}^{-1/2}(\mathbf x_t-\widehat{\bm\mu})= N^{-1}\widehat{\bm\Lambda}'\widehat{\bm\Omega}^{-1/2}(\mathbf x_t-\widehat{\bm\mu})$ (due to the normalization of the eigenvectors). The estimated common components are then de-standardized and de-centered by multiplying them by the standard deviation of the original series and adding the corresponding mean. The resulting common component $N$-dimensional vector is denoted as $\widehat{\bm\chi}_t=\widehat{\bm\mu}+\widehat{\bm\Omega}^{1/2}\widehat{\bm\Lambda}\widehat{\mathbf f}_t$.
From the results in \cite{stock2002forecasting} and \citet{bai2003inferential}, it immediately follows that $\widehat{\bm\chi}_t$ is a consistent estimator of $\bm\chi_t$, as $N,T\to\infty$.  This in practice shows the necessity of working with a high-dimensional panel in order to consistently disentangle the common components capturing all main comovements from the idiosyncratic ones.

In Table \ref{tab:numf} we report the number of common factors obtained by employing the following standard methods: (i) the log-information criterion (IC2) of \cite{bai2002determining}, implemented also when (ii) tuning the penalty as suggested by \cite{alessi2010improved}, (iii) the eigenvalue-ratio criterion by \cite{ahn2013eigenvalue}, and (iv) the test by \cite{onatski2010determining}.\footnote{For those methods requiring it, we set the maximum number of factors to $r_{\max}=15$.}  Hereafter, we set $r= 6$.


\begin{table}[t!]
\begin{center}
\caption{Estimated number of factors} \label{tab:numf}
\footnotesize
\begin{tabular}{l | c }
\hline
\hline
Method & {Number of factors $r$}\\
\hline
\cite{bai2002determining}			&	8	\\
\cite{alessi2010improved}			&	6	\\
\cite{onatski2010determining}		&	2	\\
\cite{ahn2013eigenvalue}			&	2	\\
\hline
\hline
\end{tabular}
\end{center}
\end{table}

\section{Comovements across EA countries}
\label{sbsec::comovements}
For a given variable, the share of total variance explained by the common component offers a straightforward measure of cross-country comovement. Specifically, for a given country and variable, a high proportion of variance explained by the common component indicates that the variable is primarily driven by EA-wide common factors rather than by country-specific idiosyncratic dynamics.
To provide an intuition of the explanatory power of the common factors, Table \ref{tab::comovements} reports the share of variance explained by the common component for selected key variables. This analysis offers a preliminary assessment of the degree of synchronization across EA countries. Indeed, if country dynamics were perfectly aligned, we would expect relatively similar levels of comovement across countries for each variable.  Confidence intervals for the explained variance are obtained using 1000 replications of the bootstrap procedure by \citet{barigozzi2018simultaneous} and described in Appendix B.

\begin{table}[t!]
  \centering \scriptsize
  \caption{Share of Explained Variance by the common factors}
  \resizebox{\textwidth}{!}{  
    \begin{tabular}{c|C{.2\textwidth}|C{.2\textwidth}|C{.2\textwidth}|C{.2\textwidth}|C{.2\textwidth}}
\hline
\hline
    & & & & & \\[-10pt]
    \multirow{2}{*}{Country} & IP: Manufacturing & HICP: Overall & 10-years Interest Rate & Stock Price Index & Unemployment Rate \\
    &  (IPMN) &  (HICPOV) & (LTIRT)  & (SHIX) & (UNETOT) \\
    & & & & & \\[-10pt]
\hline
\hline
    & & & & & \\[-10pt]
    EA    & 0.88  & 0.82  & 0.93  & 0.90  & 0.64 \\
          & (0.83-0.91) & (0.77-0.86) & (0.87-0.95) & (0.86-0.92) & (0.53-0.68) \\
    AT    & 0.63  & 0.68  & 0.93  & 0.78  & 0.10 \\
          & (0.52-0.70) & (0.61-0.73) & (0.86-0.94) & (0.73-0.80) & (0.08-0.15) \\
    BE    & 0.18  & 0.50  & 0.94  & 0.81  & 0.11 \\
          & (0.12-0.26) & (0.43-0.58) & (0.88-0.95) & (0.77-0.84) & (0.08-0.17) \\
    DE    & 0.72  & 0.58  & 0.91  & 0.81  & 0.14 \\
          & (0.63-0.78) & (0.51-0.65) & (0.82-0.91) & (0.77-0.84) & (0.11-0.28) \\
    EL    & 0.15  & 0.42  & 0.13  & 0.61  & 0.12 \\
          & (0.11-0.22) & (0.34-0.49) & (0.12-0.39) & (0.55-0.65) & (0.10-0.21) \\
    ES    & 0.81  & 0.64  & 0.79  & 0.77  & 0.55 \\
          & (0.72-0.86) & (0.56-0.70) & (0.74-0.86) & (0.72-0.79) & (0.49-0.64) \\
    FR    & 0.86  & 0.65  & 0.94  & 0.88  & 0.17 \\
          & (0.79-0.89) & (0.60-0.73) & (0.86-0.94) & (0.84-0.90) & (0.12-0.22) \\
    IE    & -     & 0.57  & 0.65  & 0.71  & 0.29 \\
          & (-) & (0.51-0.67) & (0.59-0.77) & (0.66-0.75) & (0.22-0.40) \\
    IT    & 0.77  & 0.52  & 0.81  & 0.85  & 0.41 \\
          & (0.69-0.82) & (0.46-0.61) & (0.77-0.86) & (0.81-0.86) & (0.32-0.48) \\
    NL    & 0.29  & 0.37  & 0.92  & 0.83  & 0.27 \\
          & (0.21-0.36) & (0.30-0.46) & (0.84-0.92) & (0.78-0.85) & (0.20-0.34) \\
    PT    & 0.52  & 0.47  & 0.41  & 0.67  & 0.32 \\
          & (0.39-0.61) & (0.39-0.55) & (0.35-0.61) & (0.62-0.70) & (0.26-0.40) \\
    \hline
    \hline
    \end{tabular}%
    }
\begin{tabular}{p{.98\textwidth}}
\scriptsize {\sc Notes:} \rm Each entry in the table corresponds to the share of variability within each variable (in the columns) for each country (in the rows) explained by the common component, $\widehat{\chi}_{i,t}$. Numbers in parentheses indicate the lower and upper bounds of the 68\% confidence interval, computed using the bootstrap procedure described in Appendix B.
\end{tabular}
  \label{tab::comovements}
\end{table}

Comovement across variables is relatively high at the EA level. At the country level, however, some heterogeneity emerges across indicators. For industrial production, Belgium, Greece, the Netherlands, and, to a lesser extent, Portugal, deviate from the high commonality observed at the EA level.
Hence, larger industrial economies are more closely aligned with the EA, whereas smaller economies exhibit a higher degree of idiosyncratic variation in production. 
In contrast, prices display a relatively more homogeneous pattern across countries, but the average share of variance explained is lower than for industrial production. Indeed, nominal variables have been shown to comove less than real variables in standard large macroeconomic datasets \citep{ahn2025common,lissona2025heterogeneous}. 
The degree of commonality for interest rates exhibits a clear core-periphery pattern: Northern European countries display a high level of commonality, whereas interest rate dynamics in Southern countries appear more driven by country-specific factors. However, even within this group, the extent of idiosyncrasy varies considerably: Italy and Spain comove more strongly with the EA, while Greece is almost entirely idiosyncratic. Stock prices show a relatively high and homogeneous degree of commonality across countries, with the exceptions of Greece and Portugal. 

The variable exhibiting the highest degree of heterogeneity is the unemployment rate. Recall that this variable is taken in first differences under our baseline specification. As such its degree of comovement is not very large, but still there are considerable differences across countries and these are unaffected by the chosen transformation.\footnote{If the unemployment rate were taken in levels, then its  high-persistence would result in an anomalously large explained variance of its common component. The choice of taking first differences is precisely made to avoid such ``overfitting'' phenomenon.}


Overall, these results provide preliminary insights into the degree of heterogeneity across variables and EA countries. While no structural claims can be made at this stage, they motivate a deeper analysis to determine whether this heterogeneity persists conditional on a monetary policy shock (see  Section \ref{sbsec::correlations}). 

Finally, we acknowledge that a more rigorous analysis of comovements across EA variables would require explicitly accounting for lag and lead relationships both across variables and countries, as in \cite{d2016nowcasting} and \cite{cascaldi2024back}. However, given the large dimensionality of our data--both within and across countries--such an approach would require modifications of the employed methodology which are beyond the scope of this paper.

\section{Estimation and identification of IRFs}
\label{sec::method}

\subsection{Common Component VAR}
\label{sbsec::ccvar}
It is widely acknowledged that by using large datasets we can retrieve structural shocks via factor analysis \citep[see, e.g., the theoretical and empirical findings by][]{GiannoneReichlin2006,forni2009opening,Forni2014}. 
In this spirit, here we adopt the CC-VAR by \cite{forni2020common} in order to estimate and identify the IRFs to the EA monetary policy shock. This approach simply consists in fitting a VAR on a vector $\mathbf{Y}_t$ of $n$ endogenous variables with $n \ll N$, and containing $n^*$ estimated common components of selected variables, $\widehat{\bm\chi}_t$, with  $r\le n^*\le n$, along with any additional observable of interest. 

{The key feature of the CC-VAR is that, when the VAR includes as many common components as there are latent factors, that is, $n^*=r$, the space spanned by the structural shocks driving the $N$ common components is the same as the space spanned by the reduced-form VAR innovations. Intuitively, this is because all common components arise from the same underlying factor structure and are therefore driven by the same shocks. As a result, replacing one common component with another in the VAR leaves the impulse response functions of the remaining variables unchanged.}

Our specification of the vector $\mathbf Y_t$ is given in Table \ref{tab::Yt}. First, we include the observed EA 2-years Interest Rate $R_t$, which is also our policy rate (see Section \ref{sbsbsec::idIRF} for its motivation). Then, we include the common component for five key EA variables: the Industrial Production growth in manufacturing (IPMN), the Overall Harmonized Consumer Price Index inflation (HICPOV), the 10-years Interest Rate (LTIRT), the Stock Price Index growth (SHIX), and the monthly change in the Unemployment Rate (UNETOT). Finally, we include, one at a time, the common components of various national variables of interest, for which we aim to study the IRF. In particular, we consider a total of 49 different national variables resulting in 49 possible choices for $\mathbf Y_t$.\footnote{The five key variables (IPMN, HICPOV, LTIRT, SHIX, UNETOT) for the ten countries covered by the EA-MD-QD, and recalling that IPMN for Ireland is unavailable.} For any of those choices, $\mathbf Y_t$ has always dimension $n=7$ and  $n^* =n-1=6$ so that $n^*=r$, and, as a consequence, the IRFs for the first six variables in $\mathbf Y_t$ are unchanged in all 49 VARs  (see the results in Section \ref{ssec::EAITFSbase}).



\begin{table}[t!]
\centering\footnotesize
\caption{Components of $\mathbf Y_t$ in the CC-VAR}\label{tab::Yt}
\begin{tabular}{l l l }
\hline
\hline
Notation &ID& name\\
\hline\\[-6pt]
$R_t$&-&EA 2-years Interest Rate\\[2pt]
$\widehat\chi_{{\rm IPMN\, EA},t}$& IPMN &EA IP: Manufacturing (common component)\\[2pt]
$\widehat\chi_{{\rm HICPOV\, EA},t}$&HICPOV& EA HICP: Overall (common component)\\[2pt]
$\widehat\chi_{{\rm LTIRT\, EA},t}$&LTIRT&  EA 10-years Interest Rate (common component)\\[2pt]
$\widehat\chi_{{\rm SHIX\, EA},t}$ &SHIX& EA Stock Price Index (common component)\\[2pt]
$\widehat\chi_{{\rm UNETOT\, EA},t}$&UNETOT & EA Unemployment Rate (common component)\\[2pt]
$\widehat\chi_{{\rm nat.,}t} $ &-& National variable (common component)\\[2pt]
\hline
\hline
\end{tabular}
\end{table}

Specifically, for any choice of $\mathbf Y_t$, we estimate the following reduced-form VAR:
\begin{equation}\label{reduced_form}
    \mathbf Y_t = \mathbf c + \sum_{i=1}^{p} \mathbf B_i \mathbf Y_{t-i} + \sigma_t \mathbf u_t, \quad t=1,\ldots, T,
\end{equation}
where $\mathbf c$ is a $n \times 1$ vector of reduced-form constants, $\mathbf B_i$, $i=1,\ldots p$, are $n \times n$ matrices of reduced-form coefficients and $\mathbf u_t$ is the $n \times 1$ zero-mean vector of reduced-form errors, with zero mean and covariance matrix $\mathbf \Sigma$.  Throughout, we choose $p=8$ lags in the VAR. Estimation of the VAR in \eqref{reduced_form} gives the estimated coefficients $\widehat{\mathbf B}_i$ and by VAR inversion we obtain the estimated reduced form IRFs.
As $N,T\to\infty$, the estimated IRFs are consistent with rate $\min(\sqrt T,\sqrt N)$ \citep{forni2020common}. This result holds under the standard approximate factor model assumptions: the common factors are pervasive in the cross-section, while the idiosyncratic components are only weakly correlated \citep[see, e.g.,][]{bai2003inferential}.


The scaling factor $\sigma_t$ is included to address the significant fluctuations observed during the Covid period by adjusting the model residual volatility, as suggested by \cite{lenza2022estimate}. In particular, $\sigma_t$ takes a value of 1 for all periods preceding the Covid onset period, while from 2020:M3 until the end of the sample, at $T=$ 2023:M10, $\sigma_t$ is estimated in each period by maximum likelihood.\footnote{Our approach differs slightly from the approach of \cite{lenza2022estimate} as their approach is based on US data. 
Nevertheless, the variations introduced do not significantly affect our results. Moreover, adjusting the IRFs for Covid introduces slight deviations from this invariance due to changes in the volatility parameter $\sigma_t$, leading to unwanted dispersion in the EA IRFs. To address this and preserve the invariance property of the CC-VAR, we first estimate $\sigma_t$ for each specification differing only in the last variable. We then take the median of the estimated $\sigma_t$ vectors across specifications to obtain an average volatility, which is subsequently used to compute the national IRFs as described.}


{Given our task of identifying the common EA monetary policy shock and its effects on country-specific variables, the CC-VAR  seems to be a more suitable choice than the FAVAR \citep{bernanke2005measuring}. Indeed, while augmenting a classical VAR with latent factors allows the FAVAR to recover the space spanned by structural shocks, this space is also contaminated by idiosyncratic dynamics. Hence, we cannot guarantee the invariance of the IRFs which characterizes the CC-VAR. This is because in the FAVAR we use the observed variables and not their common components.\footnote{Note that if we augmented the CC-VAR with factors, as for example in the original application by \citet{forni2020common}, the invariance of the IRFs would obviously still be preserved, and we could think of it as a FAVAR for common components.}
Moreover, because the factors are not identified (see Section \ref{sbsec::factors}), including them directly in the VAR might complicate identification. By contrast, the CC-VAR uses common components which are always identified through their associated observed variables, allowing to identify the structural shocks by means of any of the standard structural VAR methods.  Admittedly, the IRFs estimated via a FAVAR have a rate of convergence $\min(\sqrt T,N)$  \citep{bai2006confidence}, which is faster than the rate $\min(\sqrt T,\sqrt N)$ derived for the CC-VAR. Nevertheless, given that in our application $N=241$ and $T=263$, we do not expect large differences in the estimation uncertainty of the two methods.}

Finally, a classical VAR is also likely to be inferior to the CC-VAR, since not only lacks the invariance property of IRFs, but it also does not guarantee that we can recover the space of the structural shocks \citep{alessi2011non}.

\subsection{Identification of the monetary policy shock}
\label{sbsbsec::idIRF}
The structural counterpart of (\ref{reduced_form}) is:
\begin{equation}\label{structural_form}
    \mathbf A_0 \mathbf Y_t = \mathbf c^{*} + \sum_{i=1}^{p} \mathbf A_i \mathbf Y_{t-i} + \sigma_t \bm \varepsilon_t,\quad t=1,\ldots, T,
\end{equation}
where the reduced-form errors $\mathbf u_t$ are related to the structural errors $\bm \varepsilon_t$ through the following relationship:
\begin{equation}
    \mathbf u_t = \mathbf A_0^{-1} \bm \varepsilon_t,\quad t=1,\ldots, T.
\end{equation}
Hereafter, let $\mathbf S \equiv \mathbf A_0^{-1}$ for simplicity of notation. 

As is well known in the VAR literature, the matrix $\mathbf S$ is unobserved and we need an identification strategy to estimate it and give an economic interpretation to the elements of $\bm{\varepsilon}_t$. For our application, focused on the monetary policy shock only, it is sufficient to identify the associated elements in the column of the matrix $\mathbf S$. We denote such column as $\mathbf s$ and the corresponding monetary policy shock as $\varepsilon_t^p$, while all other structural shocks are collected into the vector $\bm\varepsilon_t^q$. The entry of $\mathbf s$ corresponding to $\varepsilon_t^p$, which is the contemporaneous impact of the monetary policy shock on the policy rate, is denoted as $s^p$, while all other entries are collected into the vector $\mathbf s^q$.
Once the EA-wide monetary policy shock is identified, the corresponding IRFs are then computed from the, truncated, VMA($\infty$) representation of the estimated structural VAR defined in (\ref{structural_form}).

To identify $\mathbf s$ and $\varepsilon_t^p$, we employ the high-frequency Proxy-SVAR method from \citet{gertler2015monetary}.\footnote{In Appendix H we consider also sign restrictions as an alternative identification strategy.}
This approach requires jointly specifying two types of variables: a policy indicator and a monetary policy instrument. 
The policy indicator, denoted as $R_t$, is a variable capturing the central bank's monetary policy stance; it  is included in $\mathbf Y_t$ and thus enters directly into the CC-VAR model.
The monetary policy instrument,  denoted as $Z_t$, must satisfy two properties. First, it must be correlated with the monetary policy shock:
\begin{equation}\label{relevance}
    \text E[Z_t \varepsilon_t^p] >0.
\end{equation}
Second, it must be exogenous to all other shocks, collected in the vector $ \bm \varepsilon_t^q$, i.e.,
\begin{equation}\label{exogeneity}
 \text   E[Z_t \bm \varepsilon_t^q]=\mathbf 0.
\end{equation}

For a given choice of $R_t$ and $Z_t$, after estimating the reduced-form VAR in (\ref{reduced_form}), and its vector of estimated reduced-form residuals  $\widehat{\mathbf u}_t $, we identify $\mathbf s$ using a two-step procedure. First, we project the residual of  the policy indicator  $R_t$, which we denote as $\widehat{u}_t^p$, on the instrument $Z_t$. That is, we estimate the linear regression:
\begin{equation}
\label{stage_1}
    \widehat{u}_t^{\,p} = \alpha + \beta Z_t + \zeta_t,\quad t=1,\ldots, T.
\end{equation}
Then, letting $\widehat\alpha$ and $\widehat\beta$ be the OLS estimates of $\alpha$ and $\beta$ and letting $\widehat{\widehat{u}}_t^p=\widehat{\alpha}+\widehat{\beta} Z_t$, 
we estimate the linear regression,
\begin{equation}
\label{stage_2}
    \widehat{\mathbf u}^q_t = \bm\gamma\, \widehat{\widehat{u}}_t^p + \bm \delta_t,\quad t=1,\ldots, T,
\end{equation}
where $\widehat{\mathbf u}^q_t $ contains  all other reduced-form residuals.
The OLS estimate of ${\bm\gamma}$, denoted as $\widehat{\bm\gamma}$, is then an estimate of ${\mathbf s^q}/{s^{p}} $.
This identifies $\mathbf{s}$ up to a scaling factor. To fully identify $\mathbf{s}$, we normalize the impact of the monetary policy shock on the interest rate to one, i.e., we set $s^p=1$, so that $\widehat{\bm\gamma}=\widehat{\mathbf s}^q$. This results in an identified monetary policy shock that increases the policy rate $R_t$ by one percentage point (100 basis points) at impact.

As in \citet{gertler2015monetary}, we test different combinations of instrument-policy indicators. Specifically, our selection of policy indicator $R_t$ candidates includes the EA 1-, 2-, and 3-year interest rates downloaded from the Eurostat database, and our choice for the pool of instrument candidates is drawn from the work of \citet{altavilla2019measuring}.\footnote{To convert high-frequency data into monthly frequency, daily observations are cumulated over the last 30 days and then averaged over the corresponding month, following \citet{gertler2015monetary}. This procedure substantially mitigates the temporal aggregation bias \citep{kilian2024construct}.}
We then select the pair $(R_t,Z_t)$ that exhibits the highest $F$-statistic related to Equation \ref{stage_1}. The $F$-statistic is used to test the null hypothesis of instrument irrelevance (low $F$-statistic) against the alternative hypothesis of instrument relevance (high $F$-statistic), with a threshold of 10 commonly used to distinguish between strong and weak instruments \citep{stock2002survey}. 
Based on this test we choose $R_t$ as the EA 2-years interest rate, while for $Z_t$ we choose the 1-year OIS. This choice gives a standard $F$-statistic equal to 16.6 and a robust $F$-statistic, computed following \citet{olea2013robust}, equal to 11.3.\footnote{We obtained similar results using other instrument-policy variable combinations. Specifically, we also tested the 1-year OIS together with the 1-year interest rate ($F$-statistic 13.9), the 1-year OIS together with the 3-year interest rate ($F$-statistic 12.0) and the 2-year OIS together with the 2-year interest rate ($F$-statistic 12.8). All these specifications yielded similar results.}

The OIS price is measured immediately before and after both the ECB's press statement and press conference. The policy surprise measure is derived by summing these two price differences. The rationale is that, before the ECB policy announcements, the swap price incorporates market expectations regarding the future path of the OIS. Therefore, any adjustment in the price immediately after the policy announcements is interpreted as a recalibration of market expectations in response to an unforeseen monetary policy surprise. This satisfies the correlation requirement in \eqref{relevance}. Regarding the exogeneity condition in \eqref{exogeneity}, we assume that no other significant macroeconomic shock occurs in the time span between the press statement and the press conference. Additionally, we distinguish conventional monetary policy shocks from information shocks \citep{jarocinski2020deconstructing, andrade2021delphic} by retaining only those high-frequency observations for the OIS price that are associated with a simultaneous movement of opposite sign in the swap price on the Euro Stoxx 50 index. 

Finally, to quantify uncertainty around the estimated IRFs, we employ a standard Wild bootstrap \citep{gonccalves2004bootstrapping} with $1000$ bootstrap repetitions, also accounting for the well-known bias due to the estimation of the VAR in finite samples \citep{kilian1998small}.

\section{The dynamic effects of the EA monetary policy}\label{sec::results}

\subsection{EA IRFs}
\label{ssec::EAITFSbase}
\begin{figure}[h!]
\centering \footnotesize \sc \smallskip
\setlength{\tabcolsep}{.005\textwidth}
\caption{EA IRFs}
\begin{tabular}{ccc}
\scriptsize 2-years Interest Rate ($R_t$) & \scriptsize IP: Manufacturing (IPMN) & \scriptsize HICP: Overall (HICPOV) \\[3pt]
\includegraphics[trim= .5cm 8cm .5cm 8.5cm, clip, width=0.32\textwidth]{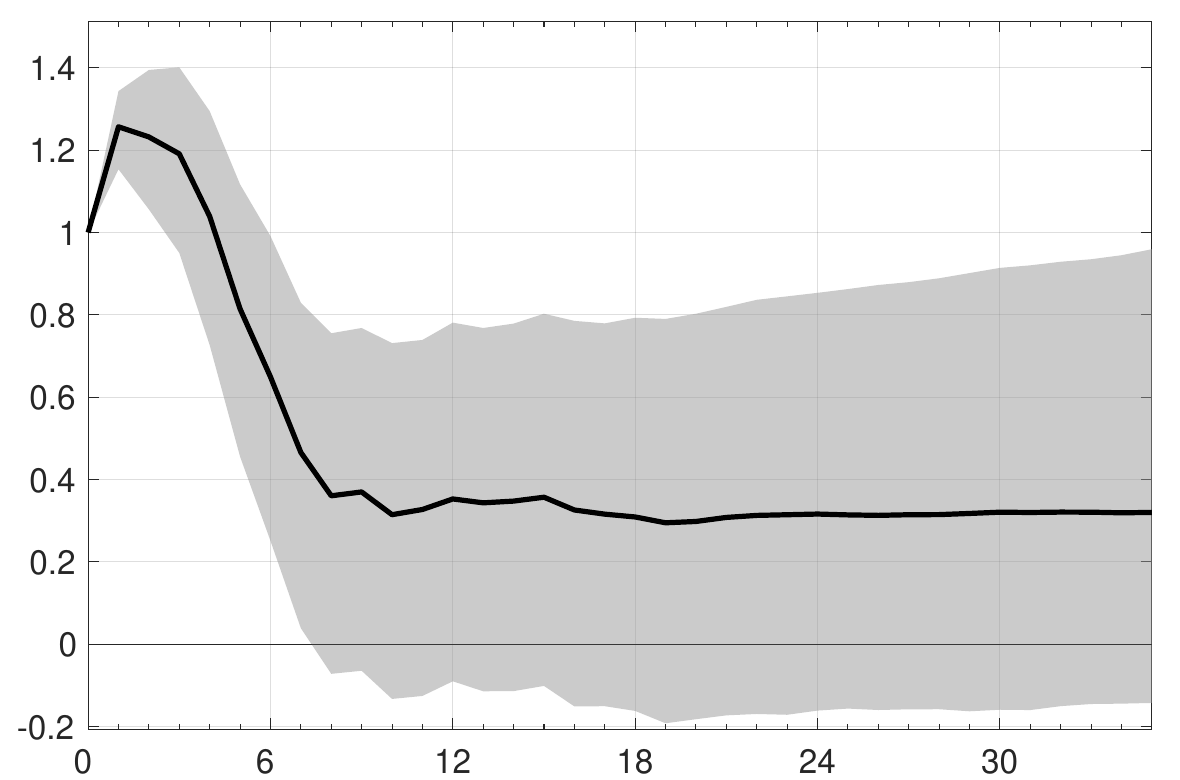} &
\includegraphics[trim= .5cm 8cm .5cm 8.5cm, clip, width=0.32\textwidth]{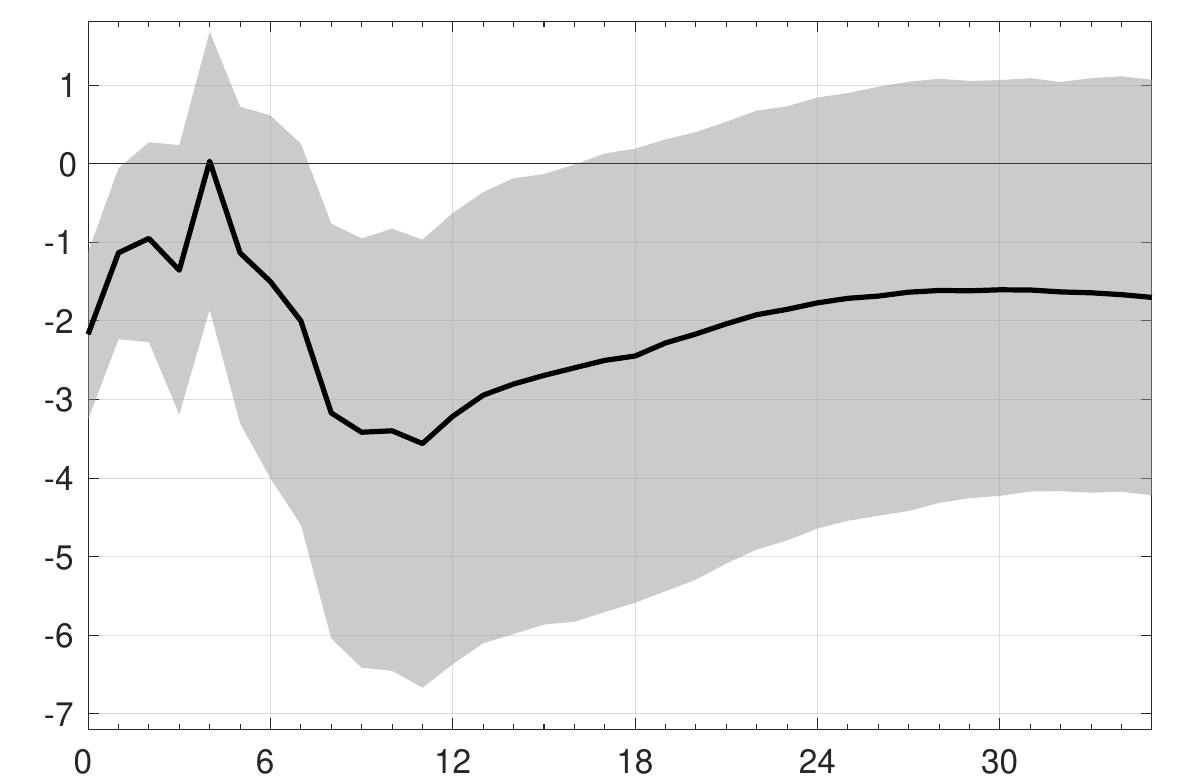} &
\includegraphics[trim= .5cm 8cm .5cm 8.5cm, clip, width=0.32\textwidth]{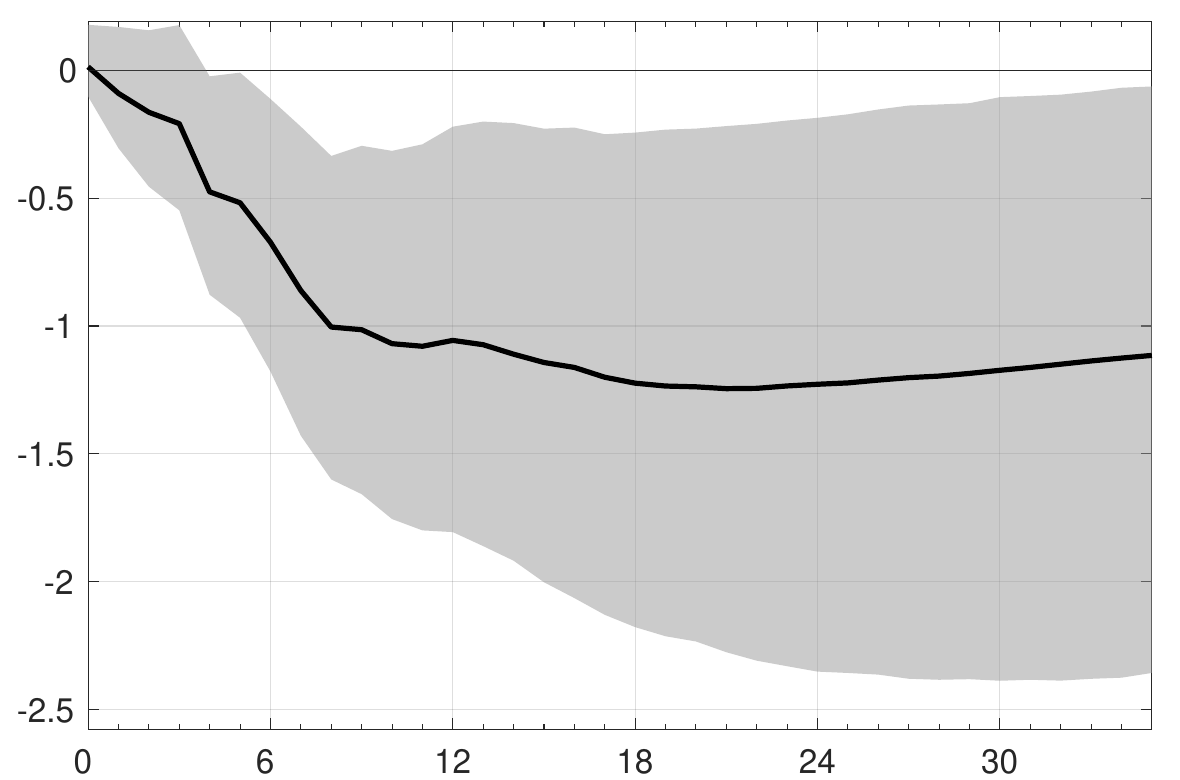} \\[5pt]
\scriptsize Stock Price Index (SHIX) &\scriptsize 10-years Interest Rate (LTIRT) &\scriptsize Unemployment Rate (UNETOT) \\[3pt]
\includegraphics[trim= .5cm 8cm .5cm 8.5cm, clip, width=0.32\textwidth]{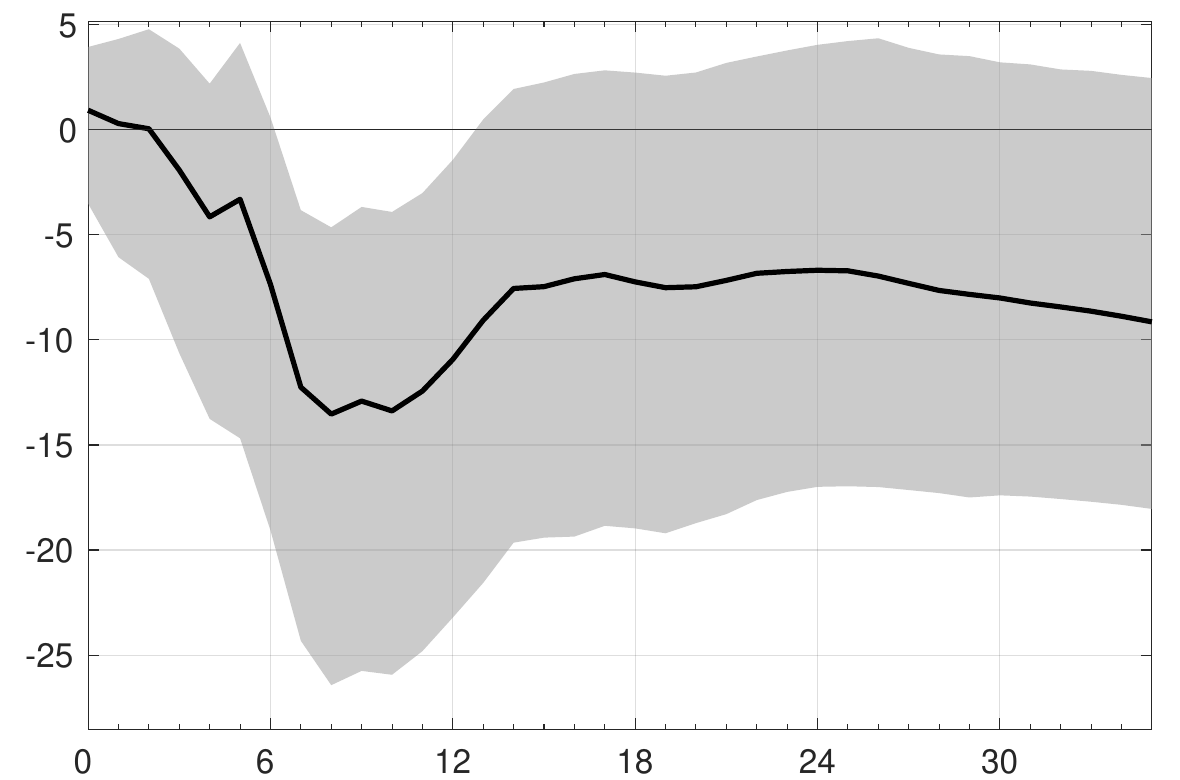} &
\includegraphics[trim= .5cm 8cm .5cm 8.5cm, clip, width=0.32\textwidth]{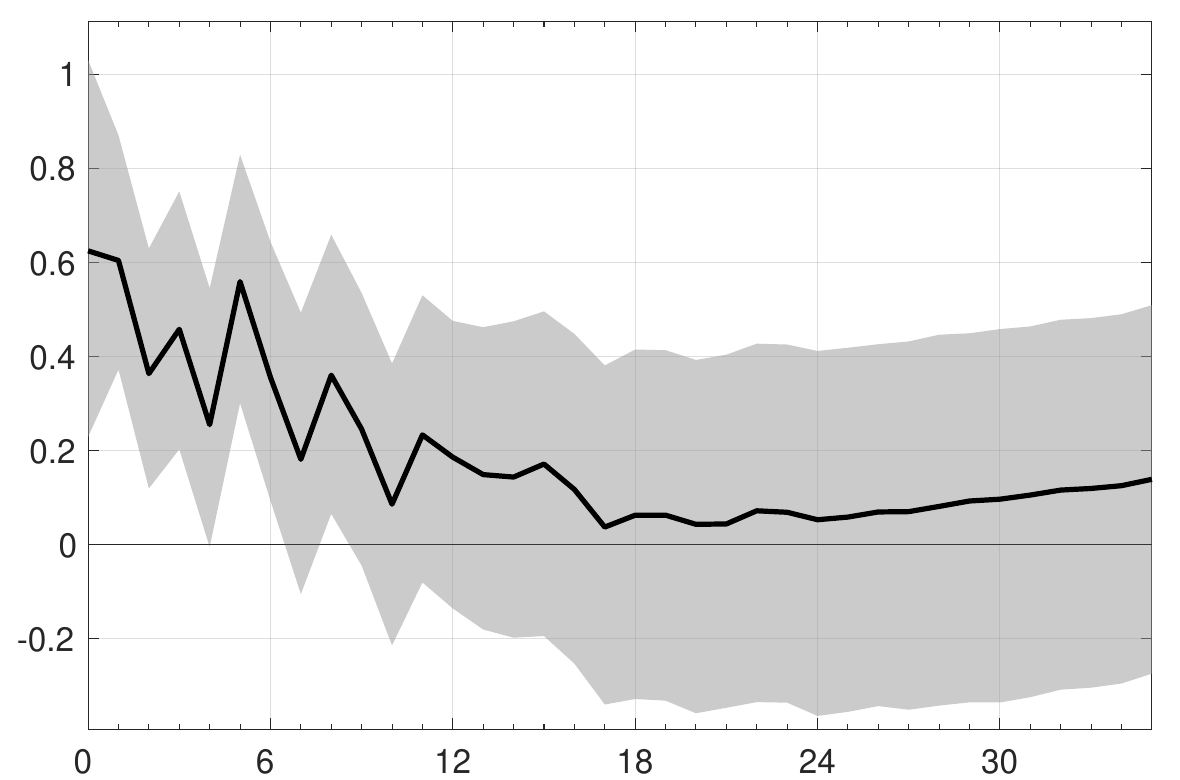} &
\includegraphics[trim= .5cm 8cm .5cm 8.5cm, clip, width=0.32\textwidth]{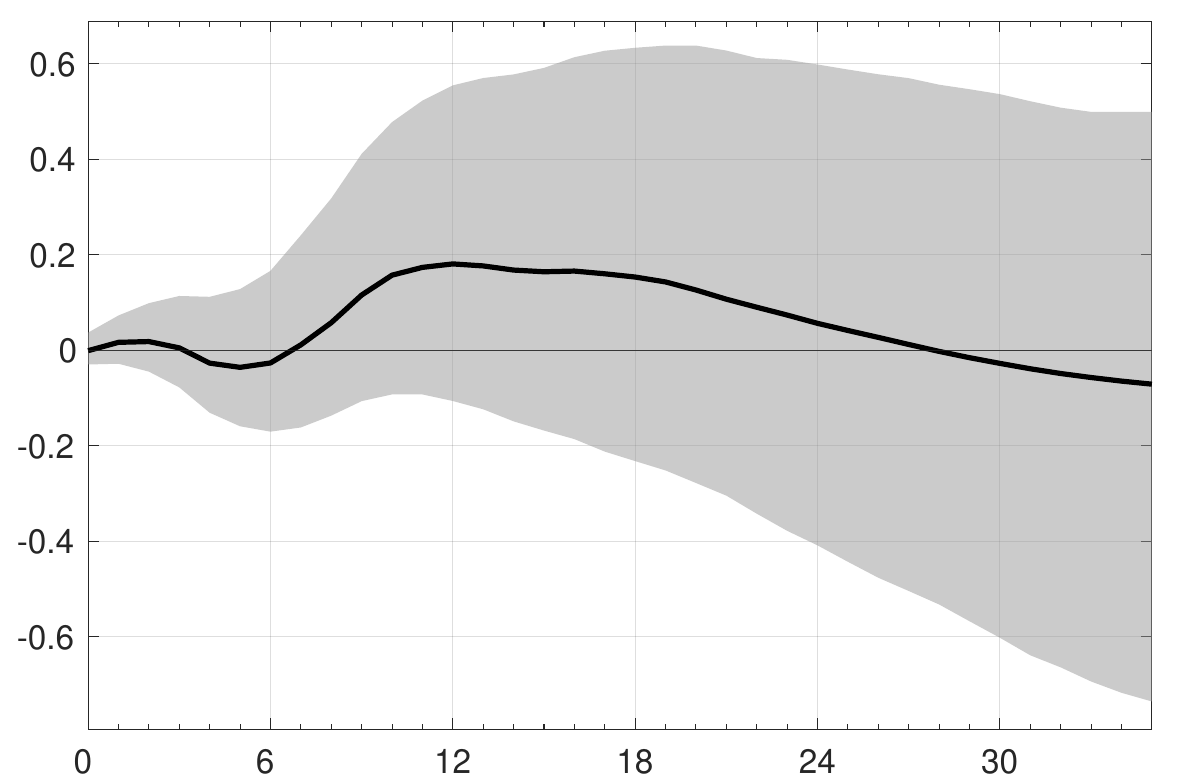} \\
\end{tabular}
\begin{tabular}{p{.98\textwidth}} \scriptsize Notes: \rm Each sub-figure plots the impulse response of one EA variable to a 100bps contractionary monetary policy shock. The black solid line is the point estimate in our baseline setting, while the gray shaded area is the corresponding 68\% confidence interval. For IPMN, HICPOV and SHIX, which enter the VAR in log-differences, and UNETOT, which enters in first-differences, the IRFs are cumulated. For $R_t$ and LTIRT, which enter the VAR in levels, the IRFs are shown in levels.
\end{tabular}
\label{fig::EAirfs}
\end{figure}

Figure \ref{fig::EAirfs} presents the estimated IRFs of the EA variables included in the CC-SVAR in \eqref{reduced_form} in response to a 100 basis-point (bps) shock to the 2-years interest rate, along with their one-standard-deviation, i.e., 68\%, confidence intervals.\footnote{The choice of this confidence level is standard in works on the euro area  \citep{barigozzi2014euro,andrade2021delphic,corsetti2022one} and motivated by the relative short sample used for estimation. See also \citet{forni2020common} for a similar study on U.S. data.}
 As already highlighted in Section \ref{sbsec::ccvar}, we notice that the same EA IRFs are obtained for any of the 49 VARs considered, each with a different national common component included. 
When comparing these results with classical VAR or FAVAR models (see Figures E1 and E2 in Appendix E), we see that 
neither of those approaches gives EA IRFs which are invariant with respect to the use of different national variables. Indeed, both for the VAR and the FAVAR we obtain 49 different IRFs for each of the six EA variables. This is what motivates our choice of the CC-VAR.

A contractionary monetary policy shock leads to an increase in both short-term and long-term interest rates, with the latter rising to roughly half the magnitude of the former at impact. This pattern is expected, as monetary policy shocks typically have a stronger effect on shorter maturities than on the upper end of the yield curve \citep{altavilla2019measuring}. As anticipated, industrial production, prices, and the stock price in the EA decline. In contrast, the unemployment rate rises with a lag of a few months, although the response is not statistically significant. 

The magnitudes of the responses of prices and the stock price are comparable to those reported by \citet{jarocinski2020deconstructing} (after rescaling the interest rate increase to 100 basis points therein), but contrast with \citet{corsetti2022one}, who document a muted reaction of prices. This difference likely reflects our explicit control for so-called \textit{information shocks} when constructing the instruments, consistently with the findings of \citet{andrade2021delphic} and \citet{jarocinski2020deconstructing}. The pointwise response of the unemployment rate is similar in magnitude to the one reported by \citet{corsetti2022one}.



These results, and those in the next sections, are coherent with those obtained using alternative sets of transformations (Appendices F and G), 
monthly or quarterly data and sign restrictions (Appendices H and I), as well as when augmenting the monthly data considered so far with quarterly EA and national GDPs (Appendix J).


\subsection{Country-level IRFs}
\begin{figure}[htbp]
\centering \footnotesize \sc
\setlength{\tabcolsep}{.005\textwidth}
\caption{Country-level IRFs}
\begin{tabular}{ccc}

\scriptsize IP: Manufacturing (IPMN) & \scriptsize HICP: Overall (HICPOV) &\scriptsize  Stock Price Index (SHIX) \\

\includegraphics[trim= .5cm 8cm .5cm 8.5cm, clip, width=0.325\textwidth]{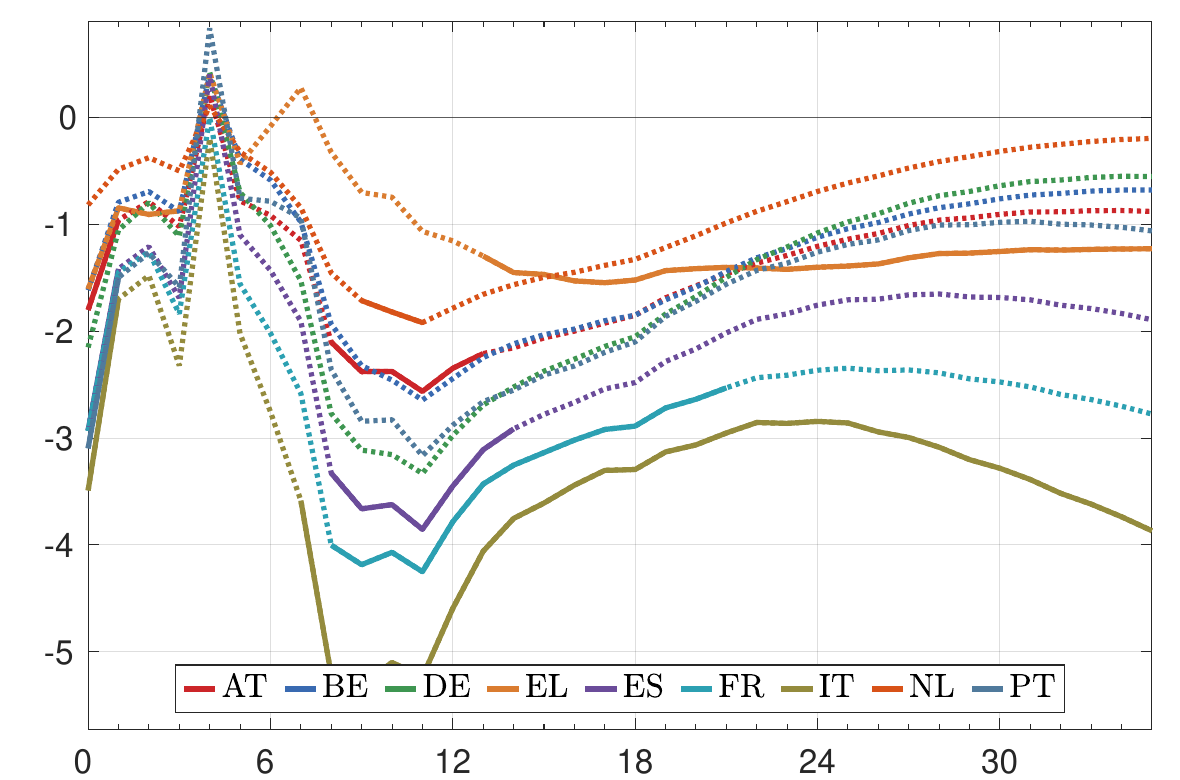} &
\includegraphics[trim= .5cm 8cm .5cm 8.5cm, clip, width=0.325\textwidth]{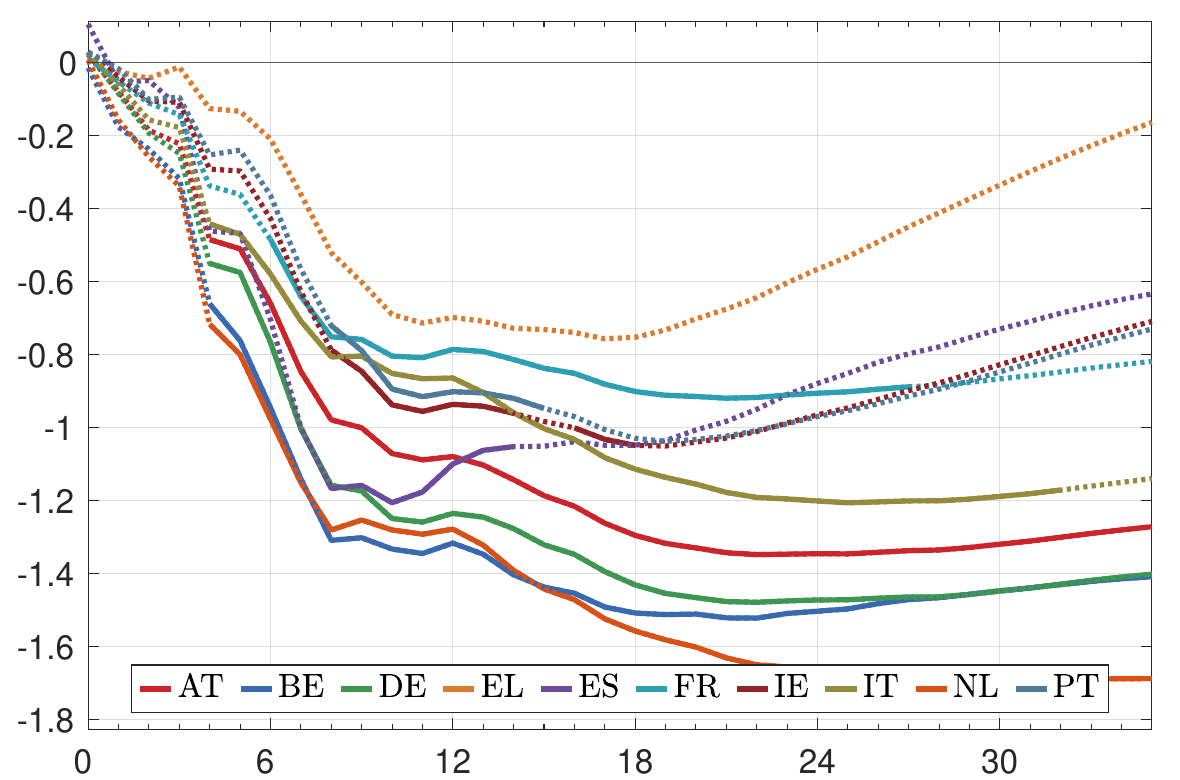} &
\includegraphics[trim= .5cm 8cm .5cm 8.5cm, clip, width=0.325\textwidth]{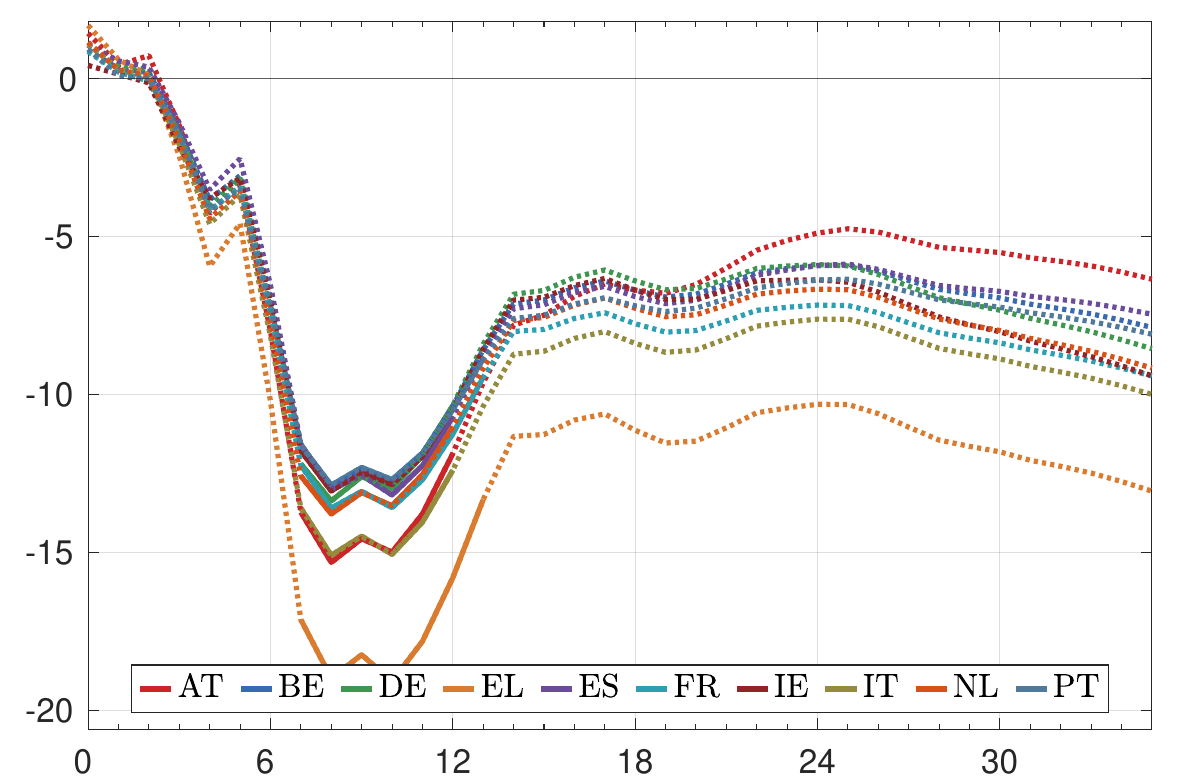} \\[5pt]
\end{tabular}

\begin{tabular}{cc}
\scriptsize 10-years Interest Rate (LTIRT) &\scriptsize  Unemployment Rate (UNETOT) \\
\includegraphics[trim= .5cm 8cm .5cm 8.5cm, clip, width=0.325\textwidth]{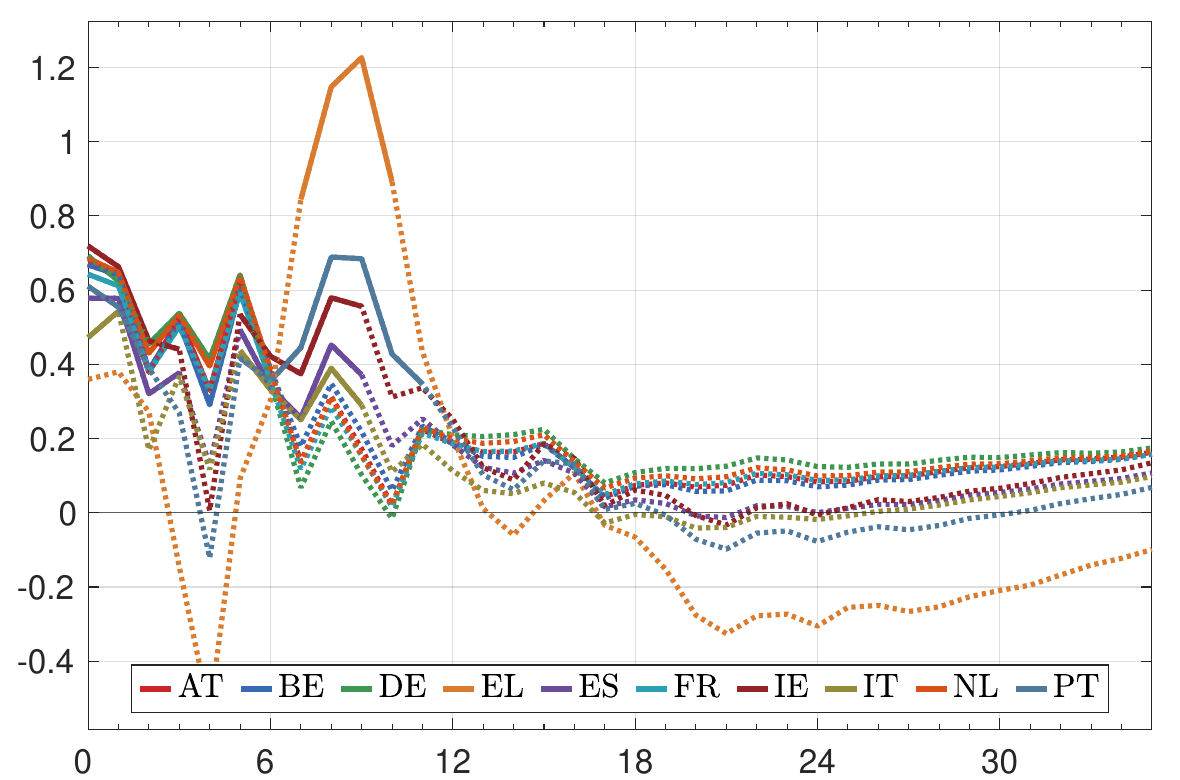} 
& \includegraphics[trim= .5cm 8cm .5cm 8.5cm, clip, width=0.325\textwidth]{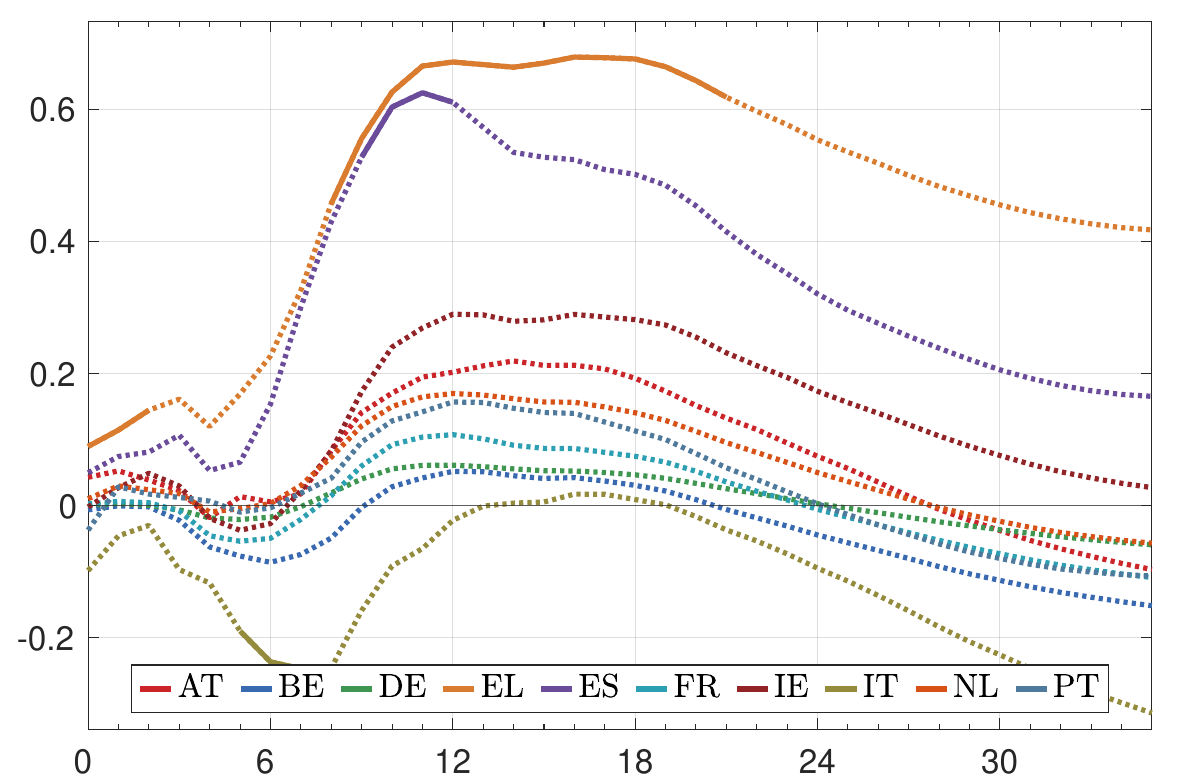}
\end{tabular}
\begin{tabular}{p{.98\textwidth}} \scriptsize Notes: \rm Each sub-figure plots the impulse responses, for all countries, of one variable to a 100bps contractionary monetary policy shock. Within each sub-figure, at each horizon $h=0,\ldots,36$ the country-level impulse responses are denoted with a solid line if the IRF is statistically significant at the 68\% level at that horizon, and with a dotted line otherwise. For IPMN, HICPOV and SHIX, which enter the VAR in log-differences, and UNETOT, which enters in first-differences, the IRFs are cumulated. For $R_t$ and LTIRT, which enter the VAR in levels, the IRFs are shown in levels.
\end{tabular}
\label{fig::national_irf_monthly}
\end{figure}

Figure \ref{fig::national_irf_monthly} presents the IRFs for the national variables. At each horizon, responses that are significant at the one-standard-deviation, i.e., 68\%, confidence level are shown with solid lines, while non-significant responses are depicted with dotted lines. Individual responses with their confidence intervals can be found in Appendix C.

As with the EA IRFs, the direction of the IRFs for each country is consistent with standard economic theory. Moreover, the IRFs exhibit broadly similar dynamics across countries, with the main differences appearing in the magnitude of the response.
For industrial production, the peak decline in the IRFs ranges from approximately 6\% for Italy to around 2\% for Greece, which stands out both in terms of magnitude and the shape of its response relative to other countries. These results highlight how the largest manufacturing economies in the EA are more strongly affected by the adverse effects of a contractionary shock.
Price responses are particularly homogeneous across countries, with a slightly more muted reaction in Greece. The maximum difference between the IRFs is approximately 1.5 percentage points, indicating only minor misalignments across countries.

Similar patterns are observed for stock prices, although Greece exhibits a particularly strong response to the contractionary shock, with its stock price declining by roughly 20\%.

On the interest rate side, Greece again stands out with an exceptionally strong response, peaking at around 120 bps. More generally, interest rates increase more in other peripheral countries (i.e., Portugal, Ireland, Italy, and Spain) compared to core countries. This core-periphery distinction is less clear for the unemployment rate. While Greece and Spain experience notably higher responses, Italy shows a counterintuitive, though mostly non-significant, effect. Overall, most unemployment responses are borderline significant at best.

These results suggest only moderate heterogeneity across countries, which is mostly concentrated in a few selected cases. However, this analysis alone does not allow us to fully assess this claim, as these observed differences may be entirely obscured by estimation uncertainty.

\subsection{Assessing cross-country heterogeneity}
\label{sbsec::het}
To quantify heterogeneity across EA countries, for each variable of interest we compute the difference between the national IRFs and the corresponding EA IRF. 
 This procedure allows for a direct comparison across individual countries. Specifically, at each horizon, a positive (negative) value indicates that the national response lies above (below) the corresponding EA response. Importantly, this relative positioning does not necessarily imply a stronger or weaker response in absolute terms, as it depends on the sign and magnitude of both IRFs. For instance, a positive difference may arise either when both IRFs are positive and the national response is larger, or when both are negative and the national response is smaller. By repeating this procedure for each bootstrap repetition, we can also quantify the uncertainty around these point estimates. 

Figure \ref{fig::diffIRF} presents these differences along with one-standard-deviation, i.e., 68\%, confidence intervals. For industrial production, France and Italy stand out from the other countries, as they are the only ones exhibiting a stronger downward response relative to the EA aggregate. This pattern differs for Germany, which, despite being one of the main manufacturing economies in the EA like Italy and France, shows no substantial deviation from the EA aggregate response. 
\begin{figure}[ht!]
\centering \footnotesize \sc
\caption{Difference between country-level and EA IRFs}
\label{fig::diffIRF}
\setlength{\tabcolsep}{.005\textwidth}

\begin{tabular}{ccccc}
\multicolumn{5}{c}{\normalfont\footnotesize\textbf{Panel A.\ IP: Manufacturing (IPMN)}}\\[4pt]
\scriptsize AT & \scriptsize BE & \scriptsize DE & \scriptsize FR & \scriptsize NL \\[2pt]
\includegraphics[trim=5cm 12cm 5cm 12.5cm, clip, width=0.19\textwidth]{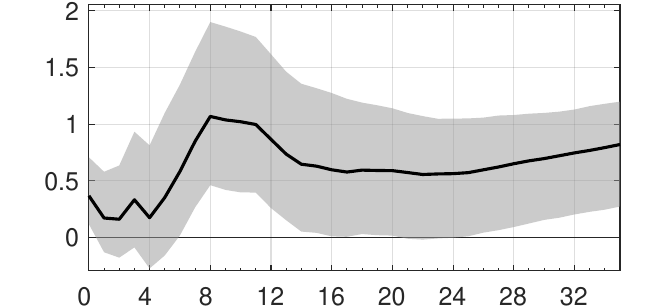} &
\includegraphics[trim=5cm 12cm 5cm 12.5cm, clip, width=0.19\textwidth]{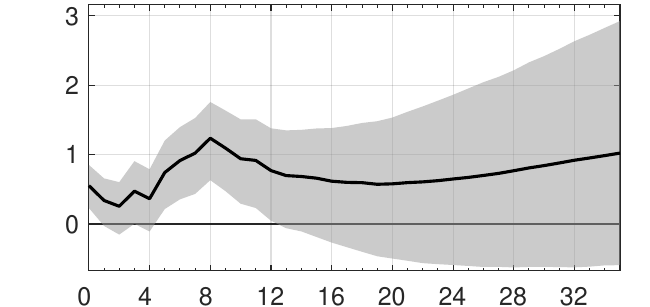} &
\includegraphics[trim=5cm 12cm 5cm 12.5cm, clip, width=0.19\textwidth]{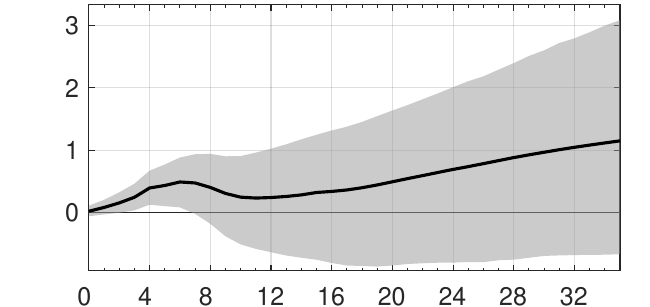} &
\includegraphics[trim=5cm 12cm 5cm 12.5cm, clip, width=0.19\textwidth]{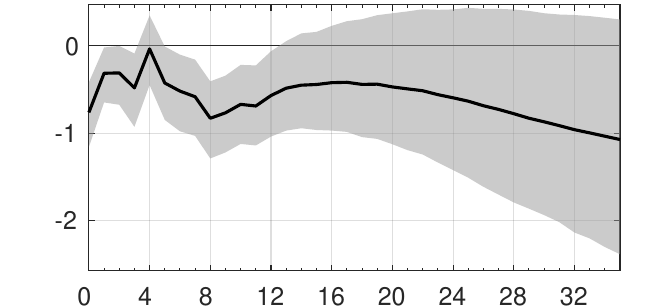} &
\includegraphics[trim=5cm 12cm 5cm 12.5cm, clip, width=0.19\textwidth]{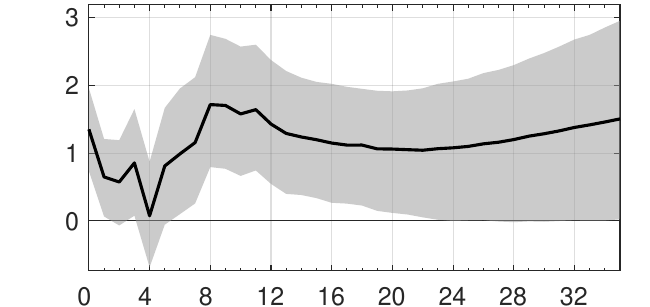} \\
\scriptsize EL & \scriptsize ES &  \scriptsize IT & \scriptsize PT \\[2pt]
\includegraphics[trim=5cm 12cm 5cm 12.5cm, clip, width=0.19\textwidth]{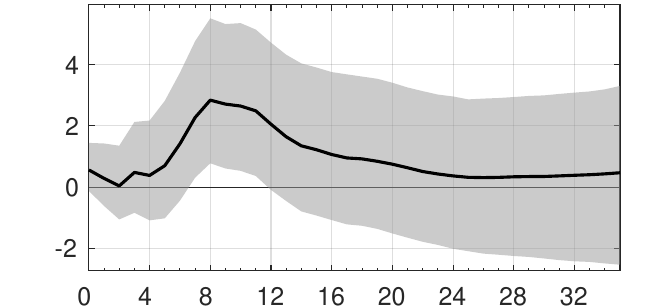} &
\includegraphics[trim=5cm 12cm 5cm 12.5cm, clip, width=0.19\textwidth]{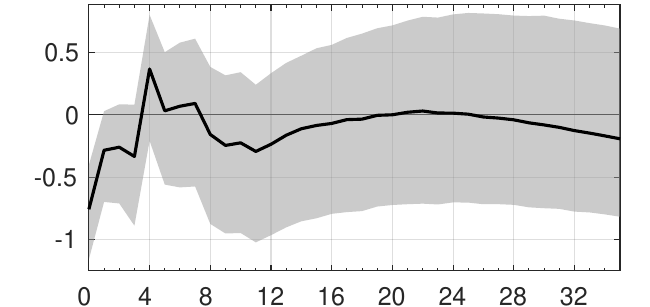} &
\includegraphics[trim=5cm 12cm 5cm 12.5cm, clip, width=0.19\textwidth]{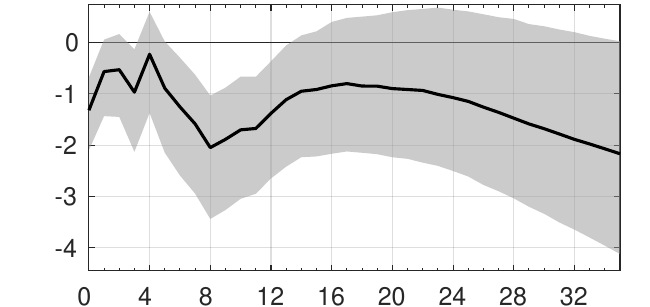} &
\includegraphics[trim=5cm 12cm 5cm 12.5cm, clip, width=0.19\textwidth]{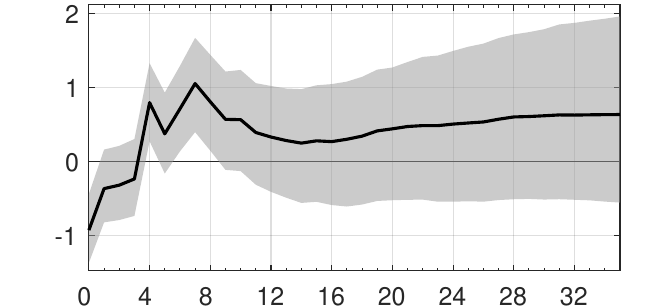} \\
\end{tabular}

\vspace{10pt}

\begin{tabular}{ccccc}
\multicolumn{5}{c}{\normalfont\footnotesize\textbf{Panel B.\ HICP: Overall (HICPOV)}}\\[4pt]
\scriptsize AT & \scriptsize BE & \scriptsize DE & \scriptsize FR & \scriptsize NL \\[2pt]
\includegraphics[trim=5cm 12cm 5cm 12.5cm, clip, width=0.19\textwidth]{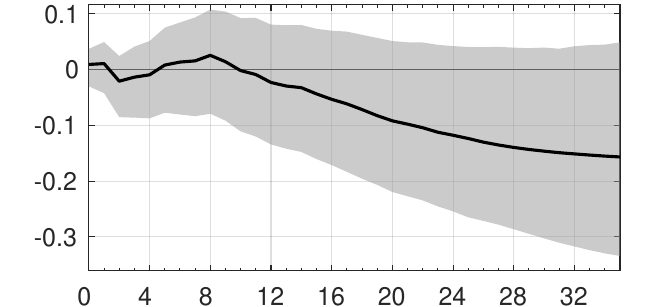} &
\includegraphics[trim=5cm 12cm 5cm 12.5cm, clip, width=0.19\textwidth]{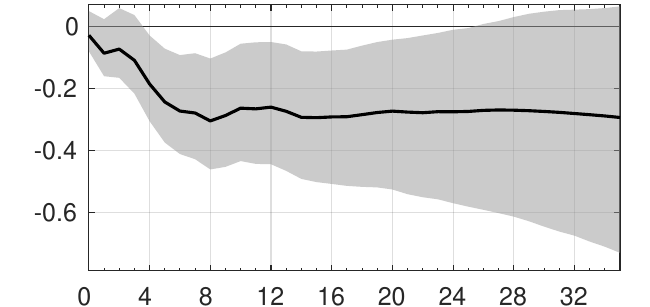} &
\includegraphics[trim=5cm 12cm 5cm 12.5cm, clip, width=0.19\textwidth]{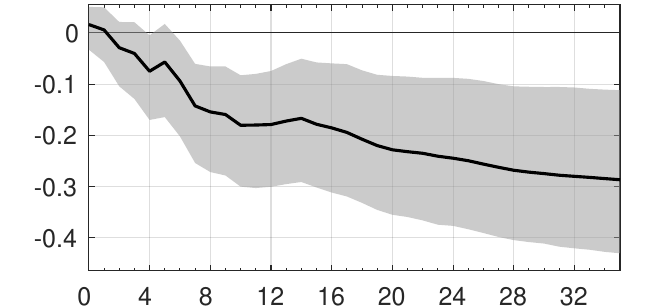} &
\includegraphics[trim=5cm 12cm 5cm 12.5cm, clip, width=0.19\textwidth]{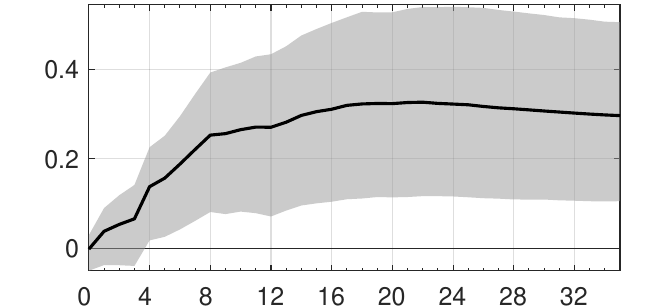} &
\includegraphics[trim=5cm 12cm 5cm 12.5cm, clip, width=0.19\textwidth]{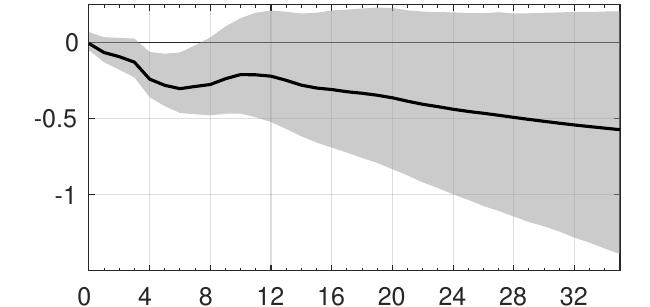} \\
\scriptsize EL & \scriptsize ES & \scriptsize IE & \scriptsize IT & \scriptsize PT \\[2pt]
\includegraphics[trim=5cm 12cm 5cm 12.5cm, clip, width=0.19\textwidth]{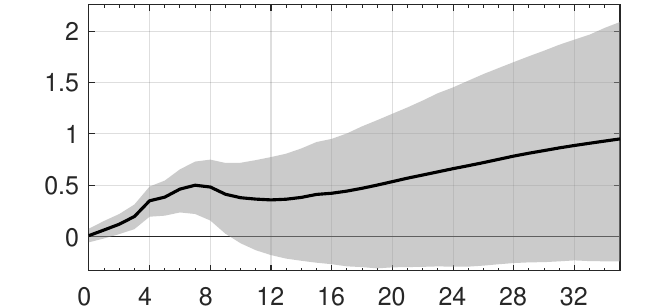} &
\includegraphics[trim=5cm 12cm 5cm 12.5cm, clip, width=0.19\textwidth]{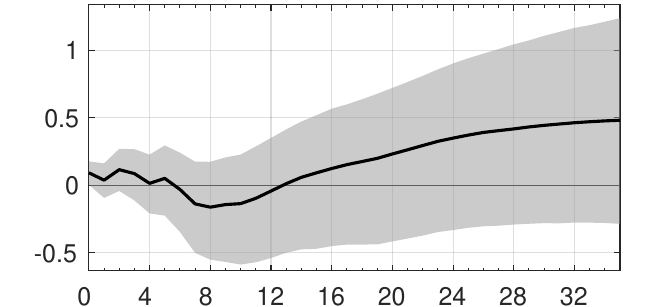} &
\includegraphics[trim=5cm 12cm 5cm 12.5cm, clip, width=0.19\textwidth]{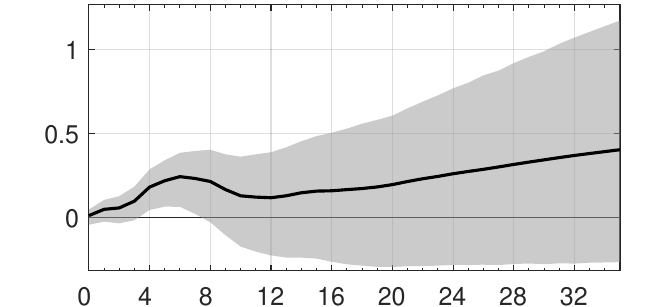} &
\includegraphics[trim=5cm 12cm 5cm 12.5cm, clip, width=0.19\textwidth]{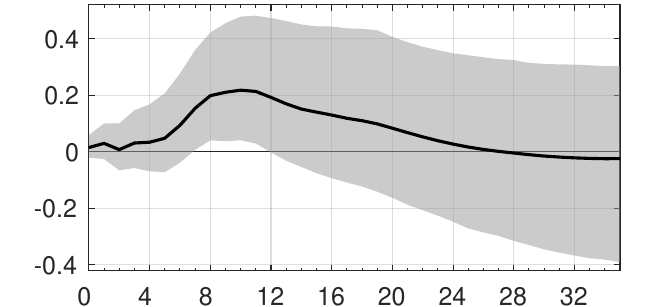} &
\includegraphics[trim=5cm 12cm 5cm 12.5cm, clip, width=0.19\textwidth]{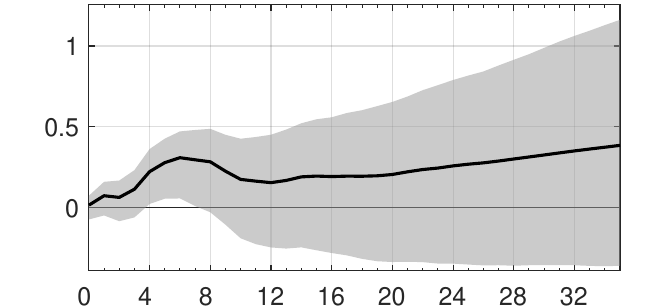} \\
\end{tabular}
\end{figure}

\begin{figure}[ht!]
\ContinuedFloat
\centering \footnotesize \sc
\caption{Difference between country-level and EA IRFs (continued)}
\setlength{\tabcolsep}{.005\textwidth}

\begin{tabular}{ccccc}
\multicolumn{5}{c}{\normalfont\footnotesize\textbf{Panel C.\ 10-years Interest Rate (LTIRT)}}\\[4pt]
\scriptsize AT & \scriptsize BE & \scriptsize DE & \scriptsize FR & \scriptsize NL \\[2pt]
\includegraphics[trim=5cm 12cm 5cm 12.5cm, clip, width=0.19\textwidth]{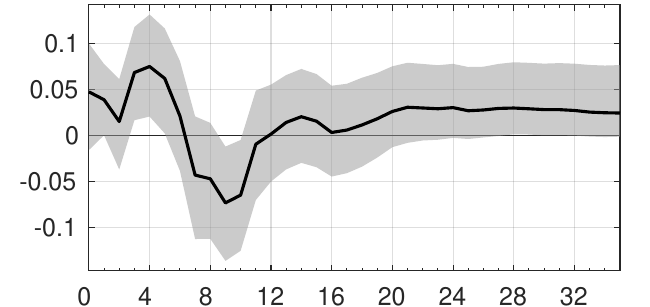} &
\includegraphics[trim=5cm 12cm 5cm 12.5cm, clip, width=0.19\textwidth]{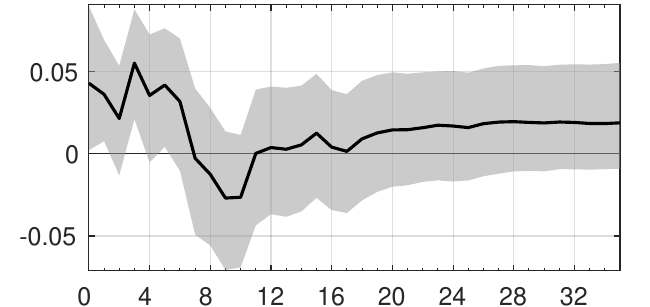} &
\includegraphics[trim=5cm 12cm 5cm 12.5cm, clip, width=0.19\textwidth]{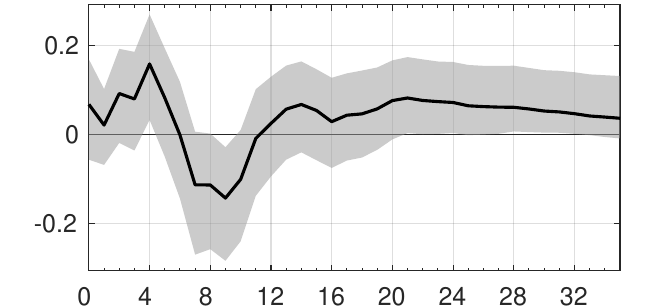} &
\includegraphics[trim=5cm 12cm 5cm 12.5cm, clip, width=0.19\textwidth]{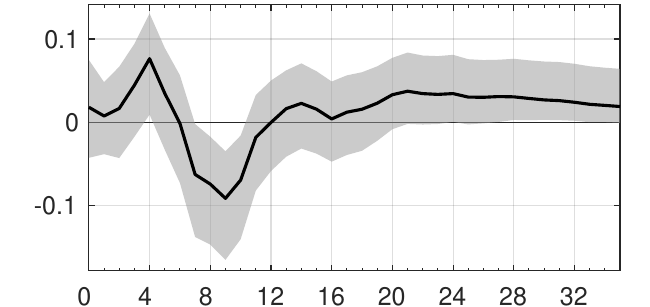} &
\includegraphics[trim=5cm 12cm 5cm 12.5cm, clip, width=0.19\textwidth]{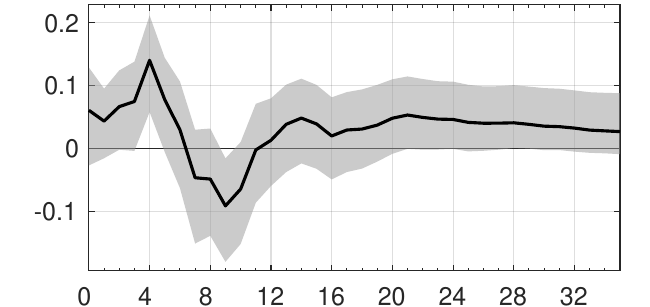} \\
\scriptsize EL & \scriptsize ES & \scriptsize IE & \scriptsize IT & \scriptsize PT \\[2pt]
\includegraphics[trim=5cm 12cm 5cm 12.5cm, clip, width=0.19\textwidth]{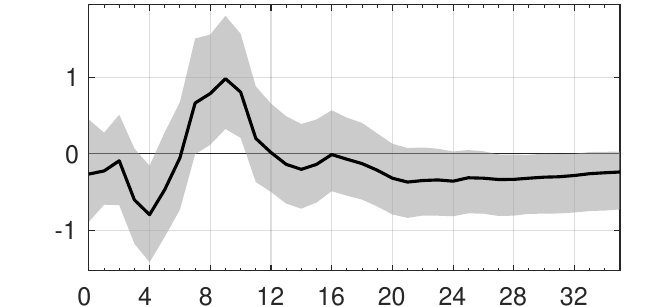} &
\includegraphics[trim=5cm 12cm 5cm 12.5cm, clip, width=0.19\textwidth]{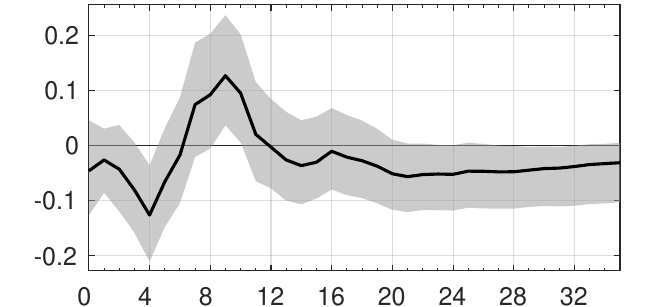} &
\includegraphics[trim=5cm 12cm 5cm 12.5cm, clip, width=0.19\textwidth]{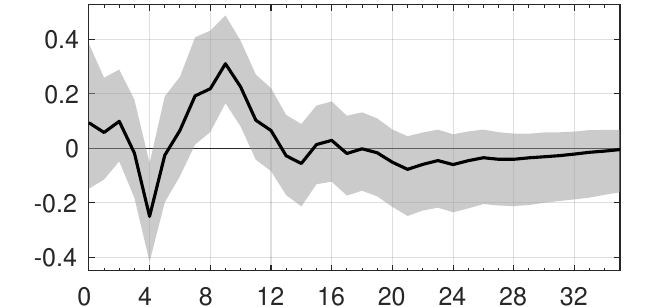} &
\includegraphics[trim=5cm 12cm 5cm 12.5cm, clip, width=0.19\textwidth]{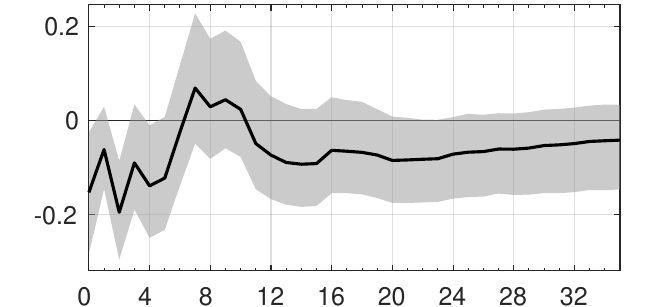} &
\includegraphics[trim=5cm 12cm 5cm 12.5cm, clip, width=0.19\textwidth]{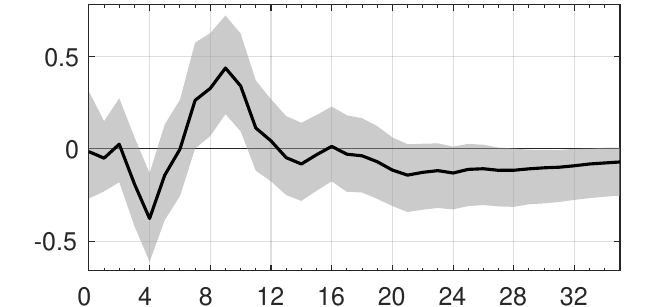} \\
\end{tabular}

\vspace{10pt}

\begin{tabular}{ccccc}
\multicolumn{5}{c}{\normalfont\footnotesize\textbf{Panel D.\ Stock Price Index (SHIX)}}\\[4pt]
\scriptsize AT & \scriptsize BE & \scriptsize DE & \scriptsize FR & \scriptsize NL \\[2pt]
\includegraphics[trim=5cm 12cm 5cm 12.5cm, clip, width=0.19\textwidth]{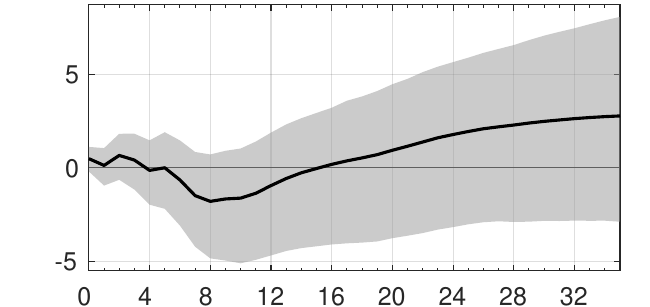} &
\includegraphics[trim=5cm 12cm 5cm 12.5cm, clip, width=0.19\textwidth]{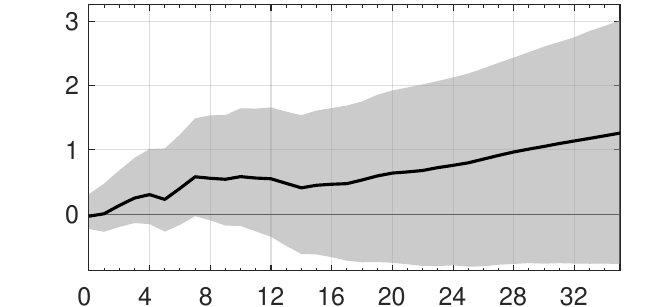} &
\includegraphics[trim=5cm 12cm 5cm 12.5cm, clip, width=0.19\textwidth]{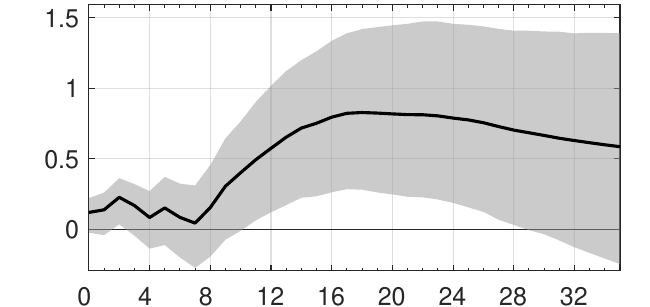} &
\includegraphics[trim=5cm 12cm 5cm 12.5cm, clip, width=0.19\textwidth]{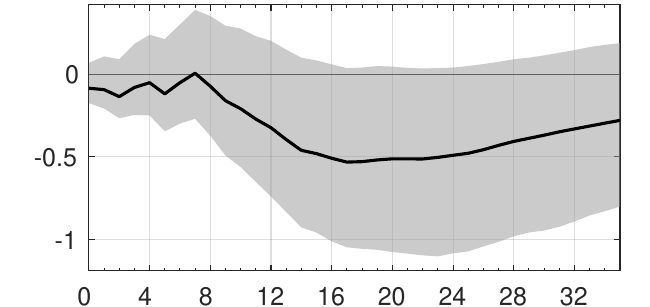} &
\includegraphics[trim=5cm 12cm 5cm 12.5cm, clip, width=0.19\textwidth]{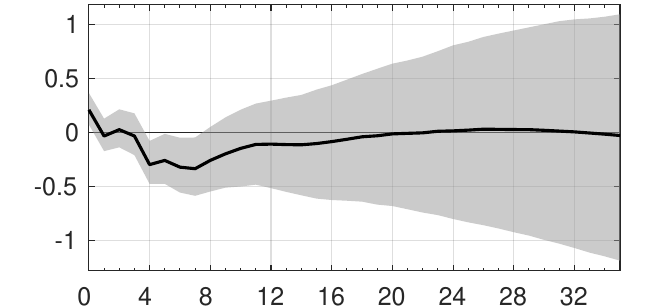} \\
\scriptsize EL & \scriptsize ES & \scriptsize IE & \scriptsize IT & \scriptsize PT \\[2pt]
\includegraphics[trim=5cm 12cm 5cm 12.5cm, clip, width=0.19\textwidth]{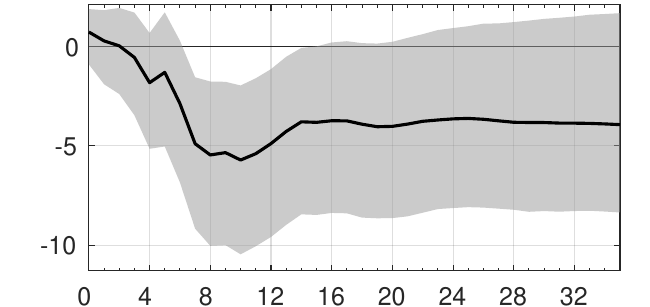} &
\includegraphics[trim=5cm 12cm 5cm 12.5cm, clip, width=0.19\textwidth]{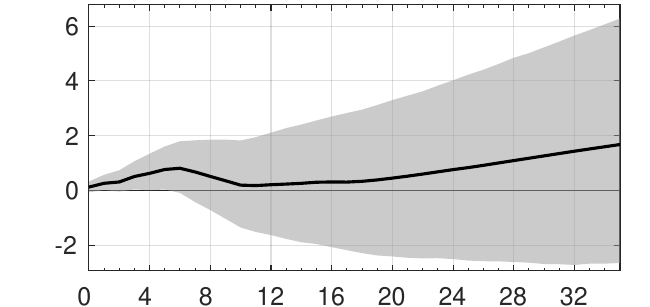} &
\includegraphics[trim=5cm 12cm 5cm 12.5cm, clip, width=0.19\textwidth]{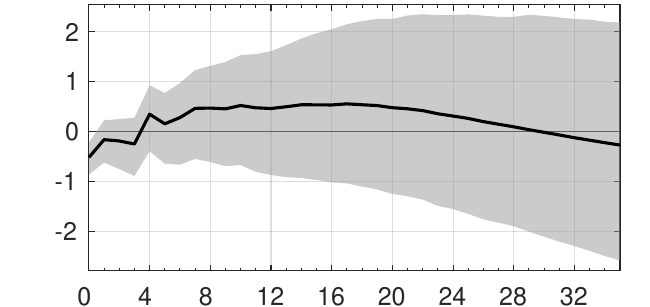} &
\includegraphics[trim=5cm 12cm 5cm 12.5cm, clip, width=0.19\textwidth]{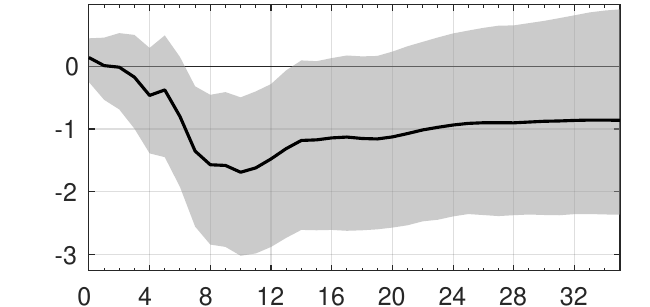} &
\includegraphics[trim=5cm 12cm 5cm 12.5cm, clip, width=0.19\textwidth]{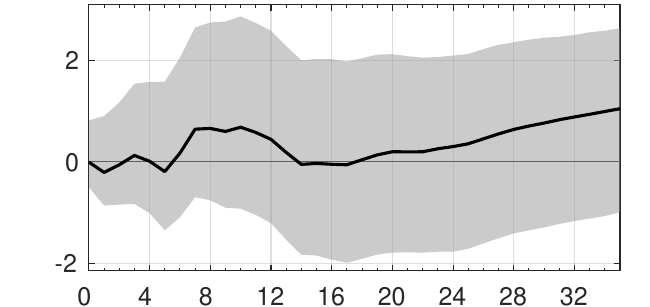} \\
\end{tabular}

\vspace{10pt}

\begin{tabular}{ccccc}
\multicolumn{5}{c}{\normalfont\footnotesize\textbf{Panel E.\ Unemployment Rate (UNETOT)}}\\[4pt]
\scriptsize AT & \scriptsize BE & \scriptsize DE & \scriptsize FR & \scriptsize NL \\[2pt]
\includegraphics[trim=5cm 12cm 5cm 12.5cm, clip, width=0.19\textwidth]{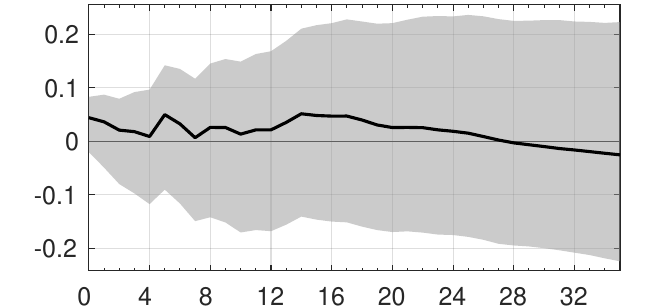} &
\includegraphics[trim=5cm 12cm 5cm 12.5cm, clip, width=0.19\textwidth]{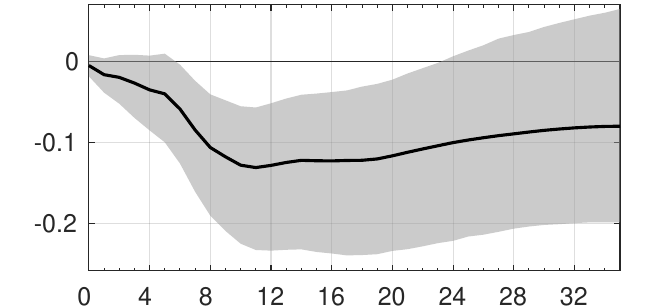} &
\includegraphics[trim=5cm 12cm 5cm 12.5cm, clip, width=0.19\textwidth]{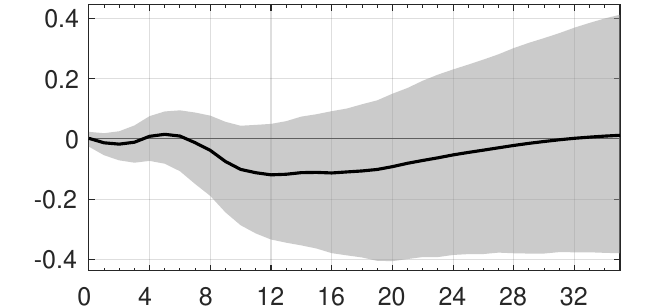} &
\includegraphics[trim=5cm 12cm 5cm 12.5cm, clip, width=0.19\textwidth]{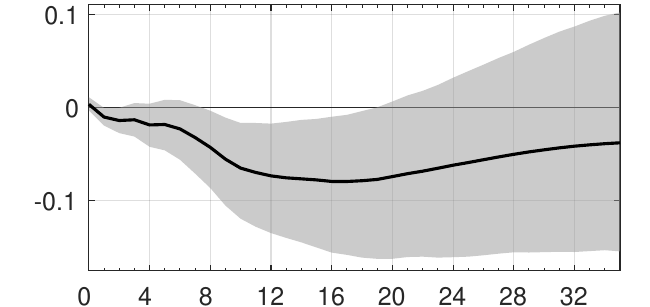} &
\includegraphics[trim=5cm 12cm 5cm 12.5cm, clip, width=0.19\textwidth]{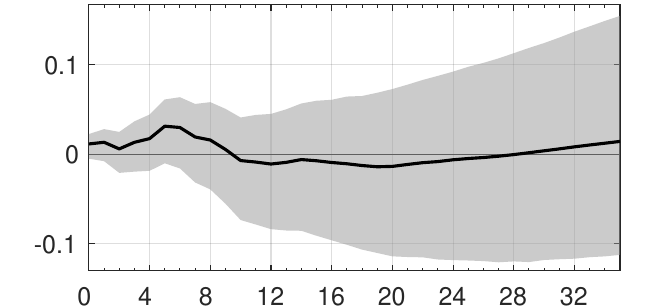} \\
\scriptsize EL & \scriptsize ES & \scriptsize IE & \scriptsize IT & \scriptsize PT \\[2pt]
\includegraphics[trim=5cm 12cm 5cm 12.5cm, clip, width=0.19\textwidth]{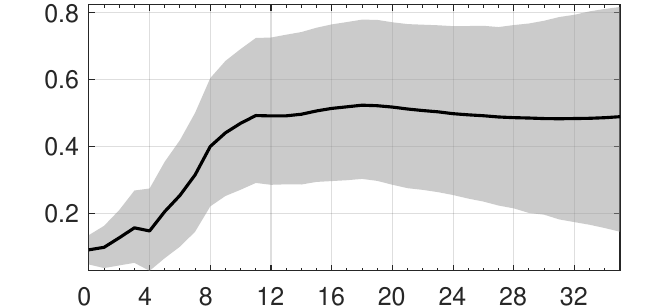} &
\includegraphics[trim=5cm 12cm 5cm 12.5cm, clip, width=0.19\textwidth]{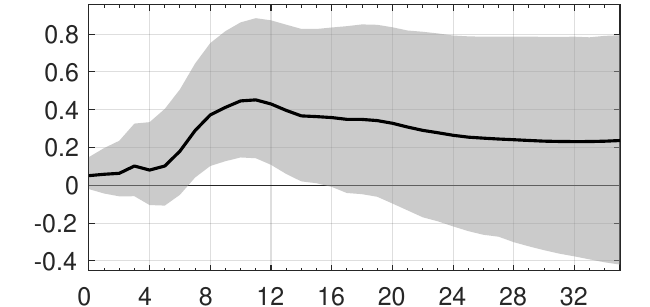} &
\includegraphics[trim=5cm 12cm 5cm 12.5cm, clip, width=0.19\textwidth]{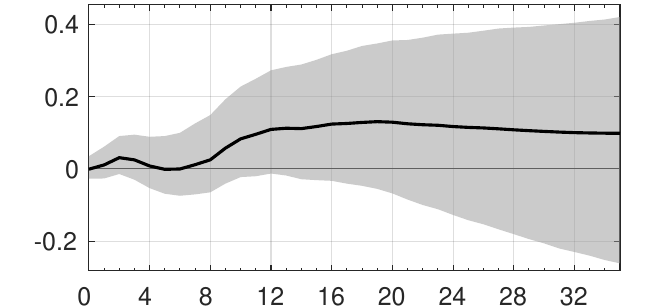} &
\includegraphics[trim=5cm 12cm 5cm 12.5cm, clip, width=0.19\textwidth]{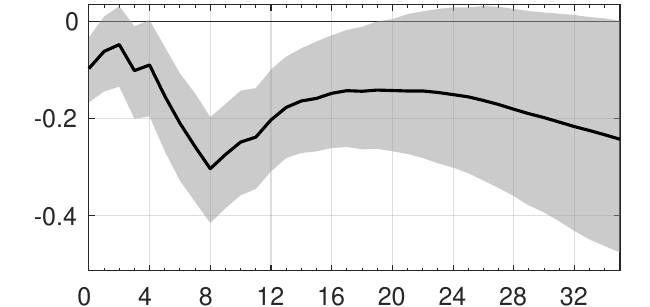} &
\includegraphics[trim=5cm 12cm 5cm 12.5cm, clip, width=0.19\textwidth]{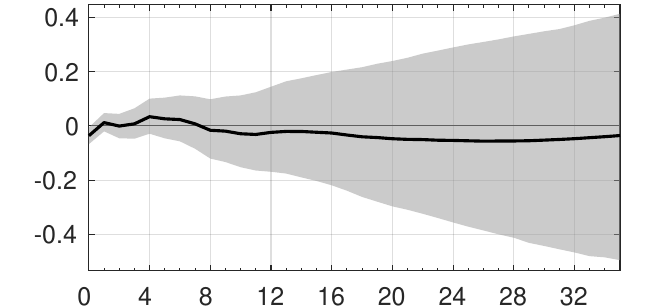} \\
\end{tabular}

\vspace{8pt}
\resizebox{6.9in}{!}{
\begin{tabular}{p{1\textwidth}} \scriptsize Notes: \rm Each sub-figure plots the difference between the country-level IRF and the corresponding EA counterpart for the variable indicated in each panel, one country per sub-figure. From top to bottom, the panels report: 
(A) IP: Manufacturing (IPMN); 
(B) HICP: Overall (HICPOV); 
(C) the 10-years Interest Rate (LTIRT); 
(D) the Stock Price Index (SHIX); 
(E) the Unemployment Rate (UNETOT). IPMN is unavailable for Ireland. The black solid line is the point estimate in our baseline setting, while the gray shaded area is the corresponding 68\% confidence interval. The IRFs are cumulated for IPMN, HICPOV, and SHIX, which enter the model in log-differences, and for UNETOT, which enters in first differences, while they are reported in levels for LTIRT. The scale in the vertical axis differs across countries.
\end{tabular}
}
\end{figure}

All other countries display weaker responses compared with the EA, particularly core economies such as Austria, Belgium, and the Netherlands. 

In contrast, for prices, some differences emerge between core and peripheral countries. Core countries are either closely aligned with the EA aggregate (e.g., Austria) or exhibit a stronger response (e.g., Germany), whereas peripheral countries generally show a more muted reaction, with the exception of Spain, where the difference is not statistically significant. This pattern seems to suggests that the transmission of monetary policy to prices differs between core and peripheral countries \citep{barigozzi2014euro}.

Stock prices present a relatively homogeneous picture, with only Germany, Italy, and Greece showing statistically significant deviations from the EA aggregate. However, the magnitude of these differences varies markedly: Germany and Italy deviate by at most approximately 2\%, whereas Greece exhibits a substantially stronger response, with a difference exceeding 5\% compared to the EA counterpart.

An interesting pattern emerges for interest rates. First, for nearly all countries, the peak differences relative to the EA occur between 6 and 8 months, coinciding with the horizon at which most responses are statistically significant. Second, there is a hint of a core-periphery pattern: responses in peripheral countries tend to lie above the EA aggregate at medium horizons, whereas those of core countries peak below the EA at the same horizons. Notably, the magnitude of the response for Greece is particularly strong.
On the labor market side, the most pronounced differences are observed for Greece and Spain, whose responses exceed the EA aggregate, and for Italy and France, whose responses are comparatively lower. This pattern is consistent with the one observed by \citet{barigozzi2014euro}, despite the different samples considered.

Overall, our results indicate the presence of moderate, yet non-negligible, heterogeneity across EA countries. The estimates point to a core-periphery pattern in price and interest rate dynamics. In terms of prices, core economies react broadly in line with---or somewhat more strongly than---the EA aggregate, whereas the responses of peripheral countries are more muted. Conversely, interest rates in peripheral countries tend to display comparatively stronger effects over the medium term. No systematic pattern emerges for real variables, although some countries, such as Italy and Greece, deviate more visibly from the aggregate dynamics. In contrast, stock prices behave in a largely uniform manner. Among all members, Germany’s reactions most closely match the EA aggregate, while Greece exhibits the largest departures.

These results do not signal major concerns for the conduct and transmission of monetary policy in the EA. However, they highlight the importance of closely monitoring within-country dynamics to further smooth policy transmission and mitigate adverse effects on domestic production, labor market outcomes, and--most importantly--financing costs, particularly in light of the recent post-Covid surge in public spending.

\subsection{Cross-country characteristics and IRF dynamics} \label{sbsec::correlations}
To investigate cross-country differences in responses to monetary policy shocks, we follow the methodology outlined in \cite{corsetti2022one}. For each country, we compile data on selected institutional characteristics, including--but not limited to--labor market features, the financial conditions of firms and households, wage and price dynamics, and other country-specific factors. Whenever possible, these variables are averaged over the sample period; otherwise, we use data up to the most recent available period. We then compute the correlation between each explanatory variable and the peak response of the country-specific IRFs. The results of this analysis are reported in Table \ref{tab::CRanalysis}.

We begin by examining labor market indicators, including wage flexibility and job security. For wages, we consider three levels of wage adjustment frequency: more than once a year, once a year, and less than once a year. We find no statistically significant relationship between downward wage rigidity and heterogeneous responses across countries. Nevertheless, the signs of the correlations suggest that higher wage rigidity is associated with a stronger response in real activity and a more muted response in prices, consistent with a flatter aggregate demand curve \citep{christiano2005nominal}. Results for the employment protection index are weaker: as expected, higher employment protection corresponds to a smaller increase in unemployment, but its effect on other variables is relatively muted.

Turning to firm characteristics, lower price stickiness is associated with a stronger transmission of monetary policy to prices, along with a more muted response in real activity \citep[see, e.g.][]{alvarez2016real}. Regarding the financial structure of firms, the results suggest that a higher leverage ratio of non-financial corporations relative to GDP is linked to a weaker response in real activity and stock prices, but a stronger response in prices. Similar findings are reported by \cite{dedola2005monetary}, as higher leverage proxies for borrowing capacity and, consequently, firm performance.

For households, we focus on their housing situation and consumption behavior. Our results indicate that homeownership per se provides little information on cross-country heterogeneity. In contrast, the method of financing housing appears more relevant. Countries with a higher share of homeowners with a mortgage experience a stronger impact on prices (and a weaker impact on stock prices), whereas the opposite holds for homeowners without a mortgage. This finding is consistent with recent literature, which, building on rational inattention models, shows that households with mortgages are more attentive to central bank communication and interest rate decisions, thereby enhancing monetary policy transmission \citep{ahn2024effects}. We do not find any meaningful relationship between the type of mortgages (i.e., fixed versus floating) and country-specific responses. However, in countries where households have higher loan-to-value ratios on their mortgages, the impact on stock prices is smaller, further reinforcing the mechanism described in \cite{dedola2005monetary}.
Unlike \cite{corsetti2022one}, we do not find any statistically significant results related to households’ consumption habits. 

\begin{table}[htbp]
\centering
\caption{Cross-country characteristics and IRFs dynamics: correlation analysis}
\setlength{\tabcolsep}{.03\textwidth}
\resizebox{6.5in}{!}{ 
\begin{tabular}{l|l|ccccc}
\hline
\hline
& & & & & & \\[-12pt]
\textbf{ID} & \textbf{Variable} & \textbf{IPMN}  & \textbf{HICPOV}  & \textbf{LTIRT} & \textbf{SHIX}  & \textbf{UNETOT} \\
 & & & & & & \\[-12pt]
\hline
& & & & & & \\[-10pt]
1 & Employment Protection & -0.03 & -0.20 & -0.03 & -0.10 & -0.32 \\
  & &\footnotesize (0.38) &\footnotesize (0.35) &\footnotesize (0.35) &\footnotesize (0.35) &\footnotesize (0.34) \\

& & & & & & \\[-10pt]
         
2 & Wage Adj. Frequency: more than once a year & 0.28  & -0.28 & 0.31  & 0.33  & -0.46 \\
  & &\footnotesize (0.43) &\footnotesize (0.39) &\footnotesize (0.39) &\footnotesize (0.39) &\footnotesize (0.36) \\

& & & & & & \\[-10pt]
       
3 & Wage Adj. Frequency: once a year & 0.55  & -0.08 & 0.32  & -0.06 & 0.35 \\
  & &\footnotesize (0.37) &\footnotesize (0.41) &\footnotesize (0.39) &\footnotesize (0.41) &\footnotesize (0.38) \\
 
& & & & & & \\[-10pt]
       
4 & Wage Adj. Frequency: less than once a year & -0.61 & 0.01  & -0.54 & -0.35 & -0.46 \\
  & &\footnotesize (0.35) &\footnotesize (0.41) &\footnotesize (0.34) &\footnotesize (0.38) &\footnotesize (0.36) \\

& & & & & & \\[-10pt]
          
5 & Price Flexibility & 0.17  & -0.39 & -0.21 & 0.65  & -0.02 \\
  & &\footnotesize (0.44) &\footnotesize (0.41) &\footnotesize (0.44) &\footnotesize (0.34) &\footnotesize (0.45) \\

& & & & & & \\[-10pt] 
     
6 & NFC Leverage & 0.22  & -0.38 & -0.33 & 0.62* & -0.21 \\
  & &\footnotesize (0.37) &\footnotesize (0.33) &\footnotesize (0.33) &\footnotesize (0.28) &\footnotesize (0.35) \\

& & & & & & \\[-10pt]
                  
7 & Homeownership & -0.06 & 0.37  & 0.14  & -0.07 & 0.44 \\  
  & &\footnotesize (0.38) &\footnotesize (0.33) &\footnotesize (0.35) &\footnotesize (0.35) &\footnotesize (0.32) \\

& & & & & & \\[-10pt]

8 & Homeownership with mortgage & 0.35  & -0.66** & -0.33 & 0.60* & -0.26 \\
  & &\footnotesize (0.35) &\footnotesize (0.27) &\footnotesize (0.33) &\footnotesize (0.28) &\footnotesize (0.34) \\

& & & & & & \\[-10pt]
       
9 & Homeownership without mortgage & -0.33 & 0.76** & 0.35  & -0.54 & 0.47 \\
  & &\footnotesize (0.36) &\footnotesize (0.23) &\footnotesize (0.33) &\footnotesize (0.30) &\footnotesize (0.31) \\

& & & & & & \\[-10pt]
             
10 & Fixed-Rate Mortgages & -0.07 & -0.59* & -0.34 & 0.37  & -0.63* \\
   & &\footnotesize (0.38) &\footnotesize (0.28) &\footnotesize (0.33) &\footnotesize (0.33) &\footnotesize (0.28) \\

& & & & & & \\[-10pt]
          
11 & Loan to Value & -0.17 & -0.27 & -0.37 & 0.61* & -0.20 \\
   & &\footnotesize (0.40) &\footnotesize (0.36) &\footnotesize (0.35) &\footnotesize (0.30) &\footnotesize (0.37) \\

& & & & & & \\[-10pt]
          
12 & Share of HtM & 0.33  & 0.41  & 0.39  & 0.06  & 0.38 \\
   & &\footnotesize (0.36) &\footnotesize (0.32) &\footnotesize (0.33) &\footnotesize (0.35) &\footnotesize (0.33) \\

& & & & & & \\[-10pt]
       
13 & Share of WHtM & 0.26  & 0.53  & 0.50  & -0.10 & 0.53 \\
   & &\footnotesize (0.37) &\footnotesize (0.30) &\footnotesize (0.31) &\footnotesize (0.35) &\footnotesize (0.30) \\

& & & & & & \\[-10pt]
          
14 & Saving Rate & -0.08 & -0.78*** & -0.65** & 0.63* & -0.44 \\
   & &\footnotesize (0.38) &\footnotesize (0.22) &\footnotesize (0.27) &\footnotesize (0.28) &\footnotesize (0.32) \\

& & & & & & \\[-10pt]

15 & Explained Variance & -0.82*** & 0.12  & -0.78*** & 0.47  & 0.25 \\
   & &\footnotesize (0.22) &\footnotesize (0.35) &\footnotesize (0.22) &\footnotesize (0.31) &\footnotesize (0.34) \\
\hline
\hline
\end{tabular}%
} 
\resizebox{6.9in}{!}{ 
\begin{tabular}{p{1\textwidth}} \scriptsize {\sc Notes}: \rm Each entry of the table corresponds to the correlation between each indicator (in the row) and the peak of the impulse response for a specific variable (in the columns). The variables considered are: Industrial Production: Manufacturing (IPMN), HICP: Overall (HICPOV), Stock Price Index (SHIX), 10-years Interest Rates (LTIRT) and Unemployment Rate (UNETOT). Standard errors in parentheses, asterisks denote statistical significance at 99\% (***), 95\% (**), and 90\% (*). The first row reports the Employment Protection Index (OECD). Rows 2–4 show the frequency of wage adjustments by firms \citep{branten2018nominal}, while row 5 reports the frequency with which firms adjust output prices \citep{gautier2024new}.
Row 6 shows the leverage of non-financial corporations relative to GDP (Eurostat). Row 7 reports the homeownership rate as a share of the population (Eurostat). Rows 8 and 9 indicate, respectively, the shares of residential properties with and without an outstanding mortgage (Eurostat).
Row 10 reports the share of fixed-rate mortgages \citep{de2025long}. Row 11 shows the average household loan-to-value ratio from the Eurosystem Household Finance and Consumption Survey (HFCS).
Rows 12 and 13 display, respectively, the shares of \enquote{wealthy hand-to-mouth} and \enquote{poor hand-to-mouth} households, constructed as in \citet{slacalek2020household} based on HFCS data.
Row 14 reports the household saving rate as a percentage of GDP (OECD). Finally, row 15 corresponds to the share of variance explained by the common factors for each variable in the columns, as reported in Table~\ref{tab::comovements}. 
\end{tabular}
}
\label{tab::CRanalysis}
\end{table}

From a broader perspective, higher saving rates are associated with a stronger impact on prices. In contrast, long-term interest rates exhibit a less pronounced increase, while the negative effects on stock prices are mitigated. 

Finally, we examine how the degree of cross-sectional comovement for each variable, as quantified in Section \ref{sbsec::comovements}, influences dynamic responses across countries. We find that countries whose industrial production is more closely aligned with the rest of the panel are notably more affected by the contractionary shock. This finding is consistent with Table \ref{tab::comovements}, which shows the largest comovements in industrial production among the main manufacturing countries in the EA. In contrast, the impact on long-term interest rates is smaller for countries that are more aligned with the panel. This reflects the fact that countries with more idiosyncratic dynamics experience the strongest transmission of monetary policy to sovereign yield movements.

Overall, the results highlight several interesting channels through which a common monetary policy is transmitted across countries. However, these findings should be interpreted with caution, as they are based on variability across a relatively small cross-section of countries and should be regarded primarily as descriptive.

\section{Conclusions}\label{conclusion}


We introduce EA-MD-QD, a new high-dimensional dataset for the euro area and its ten largest member countries. The dataset, including all vintages, is publicly available and continuously updated, providing a valuable resource for macroeconomic analysis. Using a Common Component SVAR, we study the effects of a common monetary policy shock at both the EA and country levels and assess cross-country heterogeneity in transmission.


Our results reveal moderate but meaningful heterogeneity across countries. Price and interest rate responses display a core-periphery pattern, while real variables show less systematic variation and stock prices react more uniformly. An exploratory analysis further suggests that differences in factors such as homeownership and saving behavior may help explain the observed heterogeneity, highlighting the role of domestic economic characteristics in shaping the transmission of common monetary policy shocks.




\bibliographystyle{abbrvnat} 
\bibliography{refsdataEA.bib}
\clearpage

\begin{titlepage}
\centering
{\Large Large datasets for the euro area and its member countries and the dynamic effects of the common monetary policy\\
Supplementary Appendix\par}
\vspace{1.5cm}
    {\normalsize Matteo Barigozzi  
\hspace{2cm} 
  Claudio Lissona%
  \hspace{2cm} 
Lorenzo Tonni \par}
\end{titlepage}

\tableofcontents
\addtocontents{toc}{\protect\etocsettocdepth{2}} 
\appendix
\setstretch{1.2}

\counterwithin{figure}{section}
\counterwithin{table}{section}
\renewcommand\thefigure{\thesection\arabic{figure}}
\renewcommand\thetable{\thesection\arabic{table}}

\newpage
\section{Data description}
\label{app::data}
\noindent This appendix provides a detailed description of the dataset for the EA and its ten largest member countries. All series are retrieved from institutional sources. For each series, we provide a brief description, the data source, the measurement unit, the seasonal adjustment procedure, and the transformation (if any) applied. Table~\ref{tab::gloss} presents a glossary to facilitate the interpretation of the data description reported in Table~\ref{tab::dataEA}.

The first eight columns of Table~\ref{tab::dataEA} contain several identifiers for each series, namely: a numeric indicator (N), an alphabetical identifier (ID), a short description (Series), the unit of measure (Unit), any information about seasonal adjustment (SA), the frequency (F), the data provider (P), and the class (C) to which each variable belongs (Real, Nominal, or Financial). The remaining columns list all countries included in the dataset, starting with the EA aggregate, followed by individual countries in alphabetical order. Each entry in the country columns denotes the transformation applied to the corresponding series for that country. These \textit{statistical transformations} ensure stationarity for each series, and are based on standard unit root tests \citep{dickey1979distribution, phillips1988testing}. Two alternative sets of transformations are described in Section 2.2, and require minor changes to the transformations described in Table \ref{tab::dataEA}. Throughout the table, when a series is unavailable for a given country, the corresponding entry is marked with a \enquote{$\text{-}$}.

The dataset for the EA and individual countries is constructed with the objective of ensuring consistency across countries in terms of time coverage, number of indicators, and treatment of variables. Due to data limitations, full harmonization is not always possible. Specifically, the main differences across countries are as follows:
\begin{enumerate}
\item Labor Market Indicators are seasonally and calendar adjusted for all countries, except for France, Portugal, and Greece, where they are only seasonally adjusted.
\item Credit Aggregates are available from 2000:Q1 for all countries, except for Ireland, for which they start in 2002:Q1.
\item Labor Costs are available from 2000:Q1 for all countries, except for the EA, where they are available from 2001:Q1.
\item Turnover Indicators are available from 2000:M1 for all countries, except for France, where data start in 2002:M1.
\item The Harmonized Index of Consumer Prices (HICP) is available for all countries from 2000:M1, except for Spain (2000:M12), Ireland (2002:M1), and Belgium (2001:M1).
\end{enumerate}

\begin{table}[H]
\caption{Glossary}
\centering \scriptsize
\label{tab::gloss}
\begin{tabular}{ l | l }
\hline
\hline
\multicolumn{1}{c|}{\textbf{Source}} & \multicolumn{1}{c}{\textbf{Unit}} \\ 
\hline
\hline
EUR = Eurostat & CLV15 = Chain-linked volumes (2015=100) \\
OECD = Organization for Economic Co-operation and Development & $\text{CP}$ = Current Prices (Million\euro) \\
ECB = European Central Bank & 1000p = Thousands of persons  \\
FRED = Federal Reserve Economic Data &  1000U = Thousands of Units\\
& I\textit{xx} = Index, 20\textit{xx} = 100\\
\hline
\end{tabular}%
\\[10pt]
\setlength\tabcolsep{2.42em}
\begin{tabular}{ l | l | l }
\hline
\hline
\multicolumn{1}{c|}{\textbf{Seasonal Adjustment}} & \multicolumn{1}{c|}{\textbf{Frequency}}  & \multicolumn{1}{c}{\textbf{Transformation}} \\ 
\hline
\hline
NSA = No Seasonal Adjustment & Q = Quarterly & 1 = $100\times\log(x_t)$ \\
SA = Seasonal Adjustment & M = Monthly & 2 = $100\times \Delta\log(x_t)$\\
SCA = Seasonal and Calendar Adjustment & & 3 = $100\times\Delta^2\log(x_t)$ \\
MSA = Manual adjustment & & 4 =  $x_t$ (No Transformation) \\
& & 5 = $\Delta x_t$\\
\hline
\end{tabular}%
\end{table}

\newgeometry{left=1.5cm,right=1.5cm,top=1in,bottom=1in}
\begin{table}[ht!]
\caption{Data Description and Transformation by country}
\begin{footnotesize}
\begin{threeparttable}
\resizebox{7in}{!}{%
\begin{tabular}{  c | c | l | c | c | c | c | c || c | c | c | c | c | c | c | c | c | c | c | c | c | c | c |}
\hline
\hline
\textbf{N} & \textbf{ID} & \hspace{130pt}\textbf{Series} & \textbf{Unit} & \textbf{SA} & \textbf{F} & \textbf{P} & \textbf{C} & \textbf{EA} & \textbf{AT} & \textbf{BE} & \textbf{DE} & \textbf{EL} & \textbf{ES} & \textbf{FR} & \textbf{IE} & \textbf{IT} & \textbf{NL} & \textbf{PT}\\ 
\hline
\hline
\multicolumn{19}{c}{(1) \textbf{National Accounts}}\\
\hline
1  & GDP    & Real Gross Domestic Product                                 & CLV15 & SCA & Q & EUR & R & 2 & 2 & 2 & 2 & 2 & 2 & 2 & 2 & 2 & 2 & 2 \\ 
2  & EXPGS  & Real Export Goods and services                              & CLV15 & SCA & Q & EUR & R & 2 & 2 & 2 & 2 & 2 & 2 & 2 & 2 & 2 & 2 & 2 \\ 
3  & IMPGS  & Real Import Goods and services                              & CLV15 & SCA & Q & EUR & R & 2 & 2 & 2 & 2 & 2 & 2 & 2 & 2 & 2 & 2 & 2 \\ 
4  & GFCE   & Real Government Final consumption expenditure               & CLV15 & SCA & Q & EUR & R & 2 & 2 & 2 & 2 & 2 & 2 & 2 & 2 & 2 & 2 & 2 \\ 
5  & HFCE   & Real Households consumption expenditure                     & CLV15 & SCA & Q & EUR & R & 2 & 2 & 2 & 2 & 2 & 2 & 2 & 2 & 2 & 2 & 2 \\ 
6  & CONSD  & Real Households consumption expenditure: Durable Goods      & CLV15 & SCA & Q & EUR & R & 2 & 2 & 2 & 2 & 2 & 2 & 2 & 2 & 2 & 2 & 2 \\ 
7  & CONSSD & Real Households consumption expenditure: Semi-Durable Goods & CLV15 & SCA & Q & EUR & R & - & 2 & - & 2 & - & - & 2 & 2 & 2 & 2 & - \\
8  & CONSSV & Real Households consumption expenditure: Services           & CLV15 & SCA & Q & EUR & R & - & 2 & - & 2 & - & - & 2 & 2 & 2 & 2 & - \\ 
9  & CONSND & Real Households consumption expenditure: Non-Durable Goods  & CLV15 & SCA & Q & EUR & R & 2 & 2 & - & 2 & - & - & 2 & 2 & 2 & 2 & - \\ 
10 & GCF    & Real Gross capital formation                                & CLV15 & SCA & Q & EUR & R & 2 & 2 & 2 & 2 & 2 & 2 & 2 & 2 & 2 & 2 & 2 \\ 
11 & GCFC   & Real Gross fixed capital formation                          & CLV15 & SCA & Q & EUR & R & 2 & 2 & 2 & 2 & 2 & 2 & 2 & 2 & 2 & 2 & 2 \\ 
12 & GFACON & Real Gross Fixed Capital Formation: Construction            & CLV15 & SCA & Q & EUR & R & 2 & 2 & - & 2 & 2 & 3 & 2 & 2 & 2 & 2 & 2 \\ 
13 & GFAMG  & Real Gross Fixed Capital Formation: Machinery and Equipment & CLV15 & SCA & Q & EUR & R & 2 & 2 & - & 2 & 2 & 2 & 2 & 2 & 2 & 2 & 2 \\
14 & AHRDI  & Adjusted Households Real Disposable Income                  & \%chg & SCA & Q & EUR & R & 4 & - & - & - & - & - & - & - & - & - & - \\ 
15 & AHFCE  & Actual Final Consumption Expenditure of Households          & \%chg & SCA & Q & EUR & R & 4 & - & - & - & - & - & - & - & - & - & - \\ 
16 & GNFCPS & Gross Profit Share of Non-Financial Corporations            & \%    & SCA & Q & EUR & R & 5 & 5 & 5 & 5 & 5 & 5 & 5 & 5 & 5 & 5 & 5 \\ 
17 & GNFCIR & Gross Investment Share of Non-Financial Corporations        & \%    & SCA & Q & EUR & R & 5 & 5 & 5 & 5 & 5 & 5 & 5 & 5 & 5 & 5 & 5 \\ 
18 & GHIR   & Gross Investment Rate of Households 							    & \%    & SCA & Q & EUR & R & 5 & 5 & - & 5 & - & 5 & 5 & 5 & 5 & 5 & 5 \\ 
19 & GHSR   & Gross Households Savings Rate 								    & \%    & SCA & Q & EUR & R & 5 & 5 & 5 & 5 & - & 5 & 5 & 5 & 5 & 5 & 5 \\ 
\hline
\multicolumn{19}{c}{(2) \textbf{Labor Market Indicators}}\\
\hline
20 & TEMP   & Total Employment (domestic concept)                                  & 1000p & SCA & Q & EUR & R & 2 & 2 & 2 & 2 & 2 & 2 & 2 & 2 & 2 & 2 & 2 \\ 
21 & EMP    & Employees (domestic concept)                                         & 1000p & SCA & Q & EUR & R & 2 & 2 & 2 & 2 & 2 & 2 & 2 & 2 & 2 & 3 & 2 \\ 
22 & SEMP   & Self Employment (domestic concept)                                   & 1000p & SCA & Q & EUR & R & 2 & 2 & 3 & 3 & 2 & 2 & 3 & 2 & 2 & 2 & 2 \\ 
23 & THOURS & Hours Worked: Total                                                  & I15   & SCA & Q & EUR & R & 2 & 2 & - & 2 & 2 & 2 & 2 & 2 & 2 & 2 & 2 \\
24 & EMPAG  & Quarterly Employment: Agriculture, Forestry, Fishing                 & 1000p & SCA & Q & EUR & R & 2 & 2 & 2 & 2 & 2 & 2 & 2 & 2 & 2 & 2 & 2 \\ 
25 & EMPIN  & Quarterly Employment: Industry                                       & 1000p & SCA & Q & EUR & R & 2 & 2 & 2 & 2 & 2 & 2 & 2 & 2 & 2 & 3 & 2 \\ 
26 & EMPMN  & Quarterly Employment: Manufacturing                                  & 1000p & SCA & Q & EUR & R & 2 & 2 & 2 & 2 & 2 & 2 & 2 & 2 & 2 & 3 & 2 \\ 
27 & EMPCON & Quarterly Employment: Construction                                   & 1000p & SCA & Q & EUR & R & 2 & 2 & 3 & 2 & 2 & 2 & 3 & 2 & 2 & 2 & 2 \\ 
28 & EMPRT  & Quarterly Employment: Wholesale/Retail trade, transport, food        & 1000p & SCA & Q & EUR & R & 2 & 2 & 2 & 2 & 2 & 2 & 2 & 2 & 2 & 2 & 2 \\ 
29 & EMPIT  & Quarterly Employment: Information and Communication                  & 1000p & SCA & Q & EUR & R & 2 & 2 & 2 & 2 & 2 & 2 & 2 & 2 & 2 & 2 & 2 \\ 
30 & EMPFC  & Quarterly Employment: Financial and Insurance activities             & 1000p & SCA & Q & EUR & R & 2 & 2 & 2 & 2 & 2 & 2 & 2 & 2 & 2 & 2 & 2 \\ 
31 & EMPRE  & Quarterly Employment: Real Estate                                    & 1000p & SCA & Q & EUR & R & 2 & 2 & 2 & 2 & 2 & 2 & 2 & 2 & 2 & 2 & 2 \\ 
32 & EMPPR  & Quarterly Employment: Professional, Scientific, Technical activities & 1000p & SCA & Q & EUR & R & 2 & 2 & 2 & 2 & 2 & 2 & 2 & 2 & 2 & 2 & 2 \\ 
33 & EMPPA  & Quarterly Employment: PA, education, health ad social services       & 1000p & SCA & Q & EUR & R & 2 & 2 & 2 & 2 & 2 & 2 & 2 & 2 & 2 & 2 & 2 \\ 
34 & EMPENT & Quarterly Employment: Arts and recreational activities               & 1000p & SCA & Q & EUR & R & 2 & 2 & 2 & 2 & 2 & 2 & 2 & 2 & 2 & 2 & 2 \\ 
35 & UNETOT & Unemployment: Total (\% active population)                           & \%    & SA  & M & EUR & R & 4 & 4 & 4 & 4 & 4 & 4 & 4 & 4 & 4 & 4 & 4 \\ 
36 & UNEO25 & Unemployment: Over 25 years (\% active population)                   & \%    & SA  & M & EUR & R & 4 & 4 & 4 & 4 & 4 & 4 & 4 & 4 & 4 & 4 & 4 \\
37 & UNEU25 & Unemployment: Under 25 years (\% active population)                  & \%    & SA  & M & EUR & R & 4 & 4 & 4 & 4 & 4 & 4 & 4 & 4 & 4 & 4 & 4 \\
38 & RPRP   & Real Labour Productivity (person)                                    & I15   & SCA & Q & EUR & R & 2 & 2 & 2 & 2 & 2 & 2 & 2 & 2 & 2 & 2 & 2 \\ 
39 & WS     & Wages and salaries                                                   & CP    & SCA & Q & EUR & N & 2 & 2 & 2 & 2 & 2 & 2 & 2 & 2 & 2 & 2 & 2 \\ 
40 & ESC    & Employers' Social Contributions                                      & CP    & SCA & Q & EUR & N & 2 & 2 & 2 & 2 & 2 & 2 & 2 & 2 & 2 & 2 & 2 \\ 
\hline
\multicolumn{19}{c}{(3) \textbf{Credit Aggregates}}\\
\hline
41 & TASS.SDB   & Total Economy - Assets: Short-Term Debt Securities          & MLN\euro & MSA & Q & EUR & F & 2 & 2 & 2 & 2 & 2 & 2 & 2 & 2 & 2 & 2 & 2 \\ 
42 & TASS.LDB   & Total Economy - Assets: Long-Term Debt Securities           & MLN\euro & MSA & Q & EUR & F & 2 & 2 & 2 & 2 & 2 & 2 & 2 & 2 & 2 & 2 & 2 \\ 
43 & TASS.SLN   & Total Economy - Assets: Short-Term Loans                    & MLN\euro & MSA & Q & EUR & F & 2 & 2 & 2 & 2 & 2 & 3 & 2 & 2 & 2 & 2 & 3 \\ 
44 & TASS.LLN   & Total Economy - Assets: Long-Term Loans                     & MLN\euro & MSA & Q & EUR & F & 2 & 2 & 2 & 2 & 2 & 3 & 2 & 2 & 3 & 2 & 2 \\ 
45 & TLB.SDB    & Total Economy - Liabilities: Short-Term Debt Securities     & MLN\euro & MSA & Q & EUR & F & 2 & 2 & 2 & 2 & - & 2 & 2 & 2 & 2 & 2 & 2 \\ 
46 & TLB.LDB    & Total Economy - Liabilities: Long-Term Debt Securities      & MLN\euro & MSA & Q & EUR & F & 2 & 2 & 2 & 2 & 2 & 2 & 2 & 2 & 2 & 2 & 2 \\ 
47 & TLB.SLN    & Total Economy - Liabilities: Short-Term Loans               & MLN\euro & MSA & Q & EUR & F & 2 & 2 & 2 & 2 & 2 & 2 & 2 & 2 & 2 & 2 & 2 \\ 
48 & TLB.LLN    & Total Economy - Liabilities: Long-Term Loans                & MLN\euro & MSA & Q & EUR & F & 2 & 2 & 2 & 2 & 2 & 3 & 2 & 2 & 3 & 2 & 3 \\ 
49 & NFCASS     & Non-Financial Corporations: Total Financial Assets          & MLN\euro & MSA & Q & EUR & F & 2 & 2 & 2 & 2 & 2 & 2 & 2 & 2 & 2 & 2 & 2 \\ 
50 & NFCASS.SLN & Non-Financial Corporations - Assets: Short-Term Loans       & MLN\euro & MSA & Q & EUR & F & 2 & 2 & 2 & 2 & - & 2 & 2 & 2 & 2 & 2 & 2 \\ 
51 & NFCASS.LLN & Non-Financial Corporations - Assets: Long-Term Loans        & MLN\euro & MSA & Q & EUR & F & 2 & 2 & 2 & 2 & - & 2 & 2 & 2 & 2 & 2 & 2 \\ 
52 & NFCLB      & Non-Financial Corporations: Total Financial Liabilities     & MLN\euro & MSA & Q & EUR & F & 2 & 2 & 2 & 2 & 2 & 2 & 2 & 2 & 2 & 2 & 2 \\ 
53 & NFCLB.SLN  & Non-Financial Corporations - Liabilities - Short-Term Loans & MLN\euro & MSA & Q & EUR & F & 2 & 2 & 2 & 2 & 2 & 2 & 2 & - & 2 & 2 & 2 \\ 
54 & NFCLB.LLN  & Non-Financial Corporations - Liabilities - Long-Term Loans  & MLN\euro & MSA & Q & EUR & F & 2 & 2 & 2 & 2 & 3 & 2 & 2 & - & 3 & 2 & 3 \\ 
55 & GGASS      & General Government: Total Financial Assets                  & MLN\euro & MSA & Q & EUR & F & 2 & 2 & 2 & 2 & 2 & 2 & 2 & 2 & 2 & 2 & 2 \\ 
56 & GGASS.SLN  & General Government - Assets: Short-Term Loans               & MLN\euro & MSA & Q & EUR & F & 2 & 2 & - & - & - & - & - & - & - & 2 & - \\ 
57 & GGASS.LLN  & General Government - Assets: Short-Term Loans               & MLN\euro & MSA & Q & EUR & F & 2 & 2 & - & 2 & - & - & 2 & - & - & 2 & 2 \\ 
58 & GGLB       & General Government: Total Financial Liabilities             & MLN\euro & MSA & Q & EUR & F & 2 & 2 & 2 & 2 & 2 & 2 & 2 & 2 & 2 & 2 & 2 \\ 
59 & GGLB.SLN   & General Government - Liabilities: Short-Term Loans          & MLN\euro & MSA & Q & EUR & F & 2 & 2 & 2 & 2 & - & 2 & 2 & 2 & 2 & 2 & 2 \\ 
60 & GGLB.LLN   & General Government - Liabilities: Long-Term Loans           & MLN\euro & MSA & Q & EUR & F & 2 & 2 & 2 & 2 & - & 2 & 2 & 2 & 2 & 2 & 3 \\ 
61 & HHASS      & Households: Total Financial Assets                          & MLN\euro & MSA & Q & EUR & F & 2 & 2 & 2 & 2 & 2 & 2 & 2 & 2 & 2 & 2 & 2 \\ 
62 & HHASS.SLN  & Households - Assets: Short-Term Loans                       & MLN\euro & MSA & Q & EUR & F & 2 & - & - & - & - & - & 2 & - & - & - & - \\ 
63 & HHASS.LLN  & Households - Assets: Long-Term Loans                        & MLN\euro & MSA & Q & EUR & F & 2 & - & - & - & - & - & 2 & - & - & 2 & 2 \\
64 & HHLB       & Households: Total Financial Liabilities                     & MLN\euro & MSA & Q & EUR & F & 3 & 2 & 2 & 3 & 3 & 3 & 3 & 2 & 3 & 2 & 3 \\
65 & HHLB.SLN   & Households - Liabilities: Short-Term Loans                  & MLN\euro & MSA & Q & EUR & F & 2 & 2 & 2 & 2 & 3 & 2 & 2 & - & 2 & 2 & 2 \\
66 & HHLB.LLN   & Households - Liabilities: Long-Term Loans                   & MLN\euro & MSA & Q & EUR & F & 3 & 2 & 2 & 3 & 3 & 3 & 3 & - & 3 & 3 & 3 \\
\hline
\multicolumn{19}{c}{(4) \textbf{Labor Costs}}\\
\hline
67 & ULCIN  & Nominal Unit Labor Costs: Industry                                       & I16 & SCA & Q & EUR & N & 2 & 2 & 2 & 2 & 2 & 2 & 2 & 2 & 2 & 2 & 2 \\ 
68 & ULCMQ  & Nominal Unit Labor Costs: Mining and Quarrying                           & I16 & SCA & Q & EUR & N & 2 & 2 & 2 & - & 2 & - & 2 & 2 & - & 2 & 2 \\ 
69 & ULCMN  & Nominal Unit Labor Costs: Manufacturing                                  & I16 & SCA & Q & EUR & N & 2 & 2 & 2 & 2 & 2 & 2 & 2 & 2 & 2 & 2 & 2 \\ 
70 & ULCCON & Nominal Unit Labor Costs: Construction                                   & I16 & SCA & Q & EUR & N & 2 & 2 & 2 & 2 & 2 & 2 & 2 & - & 2 & 2 & 2 \\ 
71 & ULCRT  & Nominal Unit Labor Costs: Wholesale/Retail Trade, Transport, Food, IT    & I16 & SCA & Q & EUR & N & 2 & 2 & 3 & 2 & 2 & 2 & 2 & 2 & 2 & 2 & 2 \\ 
72 & ULCFC  & Nominal Unit Labor Costs: Financial Activities                           & I16 & SCA & Q & EUR & N & 2 & 2 & 2 & 2 & 2 & 2 & 2 & 2 & 2 & 2 & 2 \\ 
73 & ULCRE  & Nominal Unit Labor Costs: Real Estate                                    & I16 & SCA & Q & EUR & N & 2 & 2 & 2 & 2 & 2 & 2 & 2 & - & 2 & 2 & 2 \\ 
74 & ULCPR  & Nominal Unit Labor Costs: Professional, Scientific, Technical activities & I16 & SCA & Q & EUR & N & 2 & 2 & 3 & 2 & 2 & 2 & 2 & 2 & 2 & 2 & 2 \\ 
\hline 
\multicolumn{19}{c}{(5) \textbf{Financial Markets}}\\
\hline
75 & REER42 & Real Exchange Rate (42 main industrial countries) & I10 & NSA & M & EUR & F & 2 & 2 & 2 & 2 & 2 & 2 & 2 & 2 & 2 & 2 & 2 \\
76 & ERUS   & Exchange Rate (US dollar) & I10 & NSA & M & EUR & F & 2 & - & - & - & - & - & - & - & - & - & - \\  
77 & SHIX  & Stock Price Index                  & I10   & SA  & M & OECD & F & 2 & 2 & 2 & 2 & 2 & 2 & 2 & 2 & 2 & 2 & 2 \\  
\hline 
\multicolumn{19}{c}{(6) \textbf{Interest Rates}}\\
\hline
78 & IRT3M & 3-Months Interest Rates                  & \% & NSA & M & EUR & F & 1 & - & - & - & - & - & - & - & - & - & - \\ 
79 & IRT6M & 6-Months Interest Rates                  & \% & NSA & M & EUR & F & 1 & - & - & - & - & - & - & - & - & - & - \\ 
80 & LTIRT & Long-Term Interest Rates (EMU Criterion) & \% & NSA & M & EUR & F & 1 & 1 & 1 & 1 & 1 & 1 & 1 & 1 & 1 & 1 & 1 \\ 
\hline
\hline
\end{tabular}%
}
\end{threeparttable}
\end{footnotesize}
\label{tab::dataEA}
\end{table}

\newpage

\begin{table}[ht!]
\ContinuedFloat
\caption{Data Description and Transformation by country}
\begin{footnotesize}
\begin{threeparttable}
\resizebox{7in}{!}{%
\begin{tabular}{  c | c | l | c | c | c | c | c || c | c | c | c | c | c | c | c | c | c | c | c | c | c | c |}
\hline
\hline
\textbf{N} & \textbf{ID} & \hspace{130pt}\textbf{Series} & \textbf{Unit} & \textbf{SA} & \textbf{F} & \textbf{P} & \textbf{C} & \textbf{EA} & \textbf{AT} & \textbf{BE} & \textbf{DE} & \textbf{EL} & \textbf{ES} & \textbf{FR} & \textbf{IE} & \textbf{IT} & \textbf{NL} & \textbf{PT}\\ 
\hline
\hline
\multicolumn{19}{c}{(7) \textbf{Industrial Production and Turnover}}\\
\hline
81 & IPMN     & Industrial Production Index: Manufacturing              & I21 & SCA & M & EUR & R & 2 & 2 & 2 & 2 & 2 & 2 & 2 & - & 2 & 2 & 2 \\ 
82 & IPCAG    & Industrial Production Index: Capital Goods              & I21 & SCA & M & EUR & R & 2 & 2 & 2 & 2 & 2 & 2 & 2 & - & 2 & 2 & 2 \\ 
83 & IPCOG    & Industrial Production Index: Consumer Goods             & I21 & SCA & M & EUR & R & 2 & 2 & 2 & 2 & 2 & 2 & 2 & - & 2 & - & - \\ 
84 & IPDCOG   & Industrial Production Index: Durable Consumer Goods     & I21 & SCA & M & EUR & R & 2 & 2 & 2 & 2 & 2 & 2 & 2 & - & 2 & 2 & 2 \\ 
85 & IPNDCOG  & Industrial Production Index: Non Durable Consumer Goods & I21 & SCA & M & EUR & R & 2 & 2 & 2 & 2 & 2 & 2 & 2 & - & 2 & 2 & 2 \\ 
86 & IPING    & Industrial Production Index: Intermediate Goods 		& I21 & SCA & M & EUR & R & 2 & 2 & 2 & 2 & 2 & 2 & 2 & - & 2 & 2 & 2 \\ 
87 & IPNRG    & Industrial Production Index: Energy                     & I21 & SCA & M & EUR & R & 2 & 2 & 2 & 2 & 2 & 2 & 2 & - & 2 & 2 & 2 \\ 
\hline
88 & TRNMN    & Turnover Index: Manufacturing              & I21 & SCA & M & EUR & R & 2 & 2 & 2 & 2 & 2 & 2 & 2 & - & - & 2 & 2 \\ 
89 & TRNCAG   & Turnover Index: Capital Goods              & I21 & SCA & M & EUR & R & 2 & 2 & 2 & 2 & 2 & 2 & 2 & - & 2 & 2 & 2 \\ 
90 & TRNCOG   & Turnover Index: Consumer Goods             & I15 & SCA & M & EUR & R & 2 & 2 & 2 & 2 & 2 & 2 & 2 & - & 2 & 2 & 2 \\ 
91 & TRNDCOG  & Turnover Index: Durable Consumer Goods     & I21 & SCA & M & EUR & R & 2 & 2 & 2 & 2 & 2 & 2 & 2 & - & 2 & 2 & 2 \\ 
92 & TRNNDCOG & Turnover Index: Non Durable Consumer Goods & I21 & SCA & M & EUR & R & 2 & 2 & 2 & 2 & 2 & 2 & 2 & - & 2 & 2 & 2 \\ 
93 & TRNING   & Turnover Index: Intermediate Goods         & I21 & SCA & M & EUR & R & 2 & 2 & 2 & 2 & 2 & 2 & 2 & - & 2 & 2 & 2 \\ 
94 & TRNNRG   & Turnover Index: Energy                     & I21 & SCA & M & EUR & R & 2 & 2 & 2 & 2 & 2 & 2 & 2 & - & 2 & - & 2 \\ 
\hline
95 & CAREG & Passenger's Cars Registrations  & 1000U & SCA & M & ECB  & R & 2 & - & - & - & - & - & - & - & - & - & - \\
\hline
\multicolumn{19}{c}{(8) \textbf{Prices}}\\
\hline 
96  & PPICAG   & Producer Price Index: Capital Goods                               & I21 & MSA & M & EUR & N & 2 & 2 & 2 & 2 & 2 & 2 & 2 & - & 2 & 2 & - \\ 
97  & PPICOG   & Producer Price Index: Consumer Goods                              & I21 & MSA & M & EUR & N & 2 & 2 & 2 & 2 & 2 & 2 & 2 & - & 2 & 2 & - \\ 
98  & PPIDCOG  & Producer Price Index: Durable Consumer Goods                      & I21 & MSA & M & EUR & N & 2 & 2 & 2 & 2 & 2 & 2 & 2 & - & 2 & 2 & - \\ 
99  & PPINDCOG & Producer Price Index: Non Durable Consumer Goods  				   & I21 & MSA & M & EUR & N & 2 & 2 & 2 & 2 & 2 & 2 & 2 & - & 2 & 2 & - \\ 
100  & PPIING   & Producer Price Index: Intermediate Goods                          & I21 & MSA & M & EUR & N & 2 & 2 & 2 & 2 & 2 & 2 & 2 & - & 2 & 2 & - \\ 
101  & PPINRG   & Producer Price Index: Energy 								       & I21 & MSA & M & EUR & N & 2 & 2 & 2 & 2 & 2 & 2 & 2 & - & 2 & 2 & - \\ 
102 & HICPOV   & Harmonized Index of Consumer Prices: Overall Index                & I10 & SCA & M & ECB & N & 2 & 2 & 2 & 2 & 2 & 2 & 2 & 2 & 2 & 2 & 2 \\ 
103 & HICPNEF  & Harmonized Index of Consumer Prices: All Items, no Energy{\&}Food & I10 & SCA & M & ECB & N & 2 & 2 & 2 & 2 & 2 & 2 & 2 & 2 & 2 & 2 & 2 \\ 
104 & HICPG    & Harmonized Index of Consumer Prices: Goods 	                   & I10 & SCA & M & ECB & N & 2 & 2 & 2 & 2 & 2 & 2 & 2 & 2 & 2 & 2 & 2 \\ 
105 & HICPIN   & Harmonized Index of Consumer Prices: Industrial Goods             & I10 & SCA & M & ECB & N & 2 & 2 & 2 & 2 & 2 & 2 & 2 & 2 & 2 & 2 & 2 \\ 
106 & HICPSV   & Harmonized Index of Consumer Prices: Services                     & I10 & SCA & M & ECB & N & 2 & 2 & 2 & 2 & 2 & 2 & 2 & 2 & 2 & 2 & 2 \\ 
107 & HICPNG   & Harmonized Index of Consumer Prices: Energy                       & I10 & MSA & M & EUR & N & 2 & 2 & 2 & 2 & 2 & 2 & 2 & 2 & 2 & 2 & 2 \\ 
108 & DFGDP    & Real Gross Domestic Product Deflator                              & I15 & SCA & Q & EUR & N & 2 & 2 & 2 & 2 & 2 & 2 & 2 & 2 & 2 & 2 & 2 \\ 
109 & HPRC     & Residential Property Prices (BIS)                                 & MLN\euro & SCA & Q & FRED & N & 2 & 2 & 2 & 2 & - & 2 & 2 & 2 & 2 & 2 & - \\ 
\hline
\multicolumn{19}{c}{(9) \textbf{Confidence Indicators}}\\
\hline
110 & ICONFIX  & Industrial Confidence Indicator   & Index 	  & SA & M & EUR  & C & 4 & 4 & 4 & 4 & 4 & 4 & 4 & 4 & 4 & 4 & 4 \\  
111 & CCONFIX  & Consumer Confidence Index         & Index 	  & SA & M & EUR  & C & 4 & 4 & 4 & 5 & 5 & 5 & 5 & 5 & 5 & 4 & 5 \\  
112 & ESENTIX  & Economic Sentiment Indicator      & Index 	  & SA & M & EUR  & C & 4 & 4 & 4 & 4 & 5 & 4 & 4 & 4 & 4 & 4 & 4 \\  
113 & KCONFIX  & Construction Sentiment Indicator  & Index 	  & SA & M & EUR  & C & 4 & 5 & 4 & 5 & 5 & 5 & 5 & 4 & 5 & 5 & 5 \\  
114 & RTCONFIX & Retail Confidence Indicator       & Index 	  & SA & M & EUR  & C & 4 & 4 & 4 & 4 & 4 & 4 & 4 & 4 & 4 & 4 & 4 \\  
115 & SCONFIX  & Services Confidence Indicator     & Index 	  & SA & M & EUR  & C & 4 & 4 & 4 & 4 & 4 & 5 & 4 & 4 & 4 & 4 & 4 \\  
116 & BCI      & Business Confidence Index         & 2010=100 & SA & M & OECD & C & 4 & 4 & 4 & 4 & 5 & 4 & 4 & 4 & 4 & 4 & 4 \\  
117 & CCI      & Consumer Confidence Index         & 2010=100 & SA & M & OECD & C & 4 & 4 & 4 & 5 & 5 & 5 & 5 & 5 & 5 & 4 & 5 \\  
\hline
\multicolumn{19}{c}{(10) \textbf{Monetary Aggregates}}\\
\hline
118 & CURR & Money Stock: Currency & CP & SCA & M & ECB & F & 2 & - & - & - & - & - & - & - & - & - & - \\ 
119 & M1   & Money Stock: M1       & CP & SCA & M & ECB & F & 2 & - & - & - & - & - & - & - & - & - & - \\  
120 & M2   & Money Stock: M2       & CP & SCA & M & ECB & F & 2 & - & - & - & - & - & - & - & - & - & - \\  
\hline
\hline
\end{tabular}%
}
\end{threeparttable}
\end{footnotesize}
\end{table}

\section{Pseudo-code for replicating Sections 2.2-2.4}
\label{app::pseudocode}
\begin{algorithm}[H]
\NoCaptionOfAlgo
\SetAlgoLined
\SetNlSty{texttt}{[}{]}
\SetAlgoNlRelativeSize{0}
\vspace{3pt}
\indent \nlset{1} \textbf{Data processing}\\[5pt]
\begin{footnotesize}
\KwInput{De-seasonalized EA-MD-QD data from 2000:M1 to 2025:M3 $\rightarrow$  $\mathbf{Y}$ of size $T_1\times N_1$.}
\vspace{5pt}
\textbf{Step 1:} Subset $\mathbf{Y}$ to monthly data $\rightarrow$ $\mathbf Z$ of size $T_1\times N_0$.\\[3pt]
\textbf{Step 2:} Apply to $\mathbf Z$ statistical transformations + interest rates in levels   $\rightarrow$ $\mathbf{Z}^{(0)}$ of size $T_0\times N_0$.\\[3pt]
\textbf{Step 3:} Impute missing values as in Stock and Watson (2002b):\\[4pt]
\begin{footnotesize}
  \textbf{3.1:} Standardize using univariate sample moments from available sample $\rightarrow$ 
  $\check{\mathbf{Z}}^{(0)}=\{\mathbf{Z}^{(0)}-\widehat{\boldsymbol{\mu}}_{Z}^{(0)}\}\oslash\widehat{\boldsymbol{\sigma}}_{Z}^{(0)}$;\\[3pt]
  \textbf{3.2:} Impute missing values with 0 (sample mean of $\check{\mathbf{Z}}^{(0)}$) $\rightarrow$ $\widetilde{\mathbf{Z}}^{(0)}$;\\[3pt]  
  \textbf{3.3:} Determine the number of factors  $\rightarrow$ $r$;\\[3pt]
  \textbf{3.4:} Estimate factor model by first $r$ PCs of $\widetilde{\mathbf{Z}}^{(0)}$ $\rightarrow$  $\widehat{\boldsymbol{\mathcal X}}^{(0)} =\widehat{\mathbf{F}}^{(0)} \widehat{\boldsymbol{\Lambda}}^{(0)'} =(\widehat{\boldsymbol{\chi}}_1^{(0)}\cdots\widehat{\boldsymbol \chi}_{T_0}^{(0)})'$ of size $T_0\times N_0$;\\[3pt]
\textbf{3.5:} For a given threshold $\epsilon$ and maximum number of iterations $I$ (e.g., $\epsilon = 10^{-6}$, $I = 1000$), let $i = 1$:\\[5pt]
 \While{$i<I$}{
 \vspace{2pt}
  \textbf{3.5.1:} Impute missing values with $\widehat{\boldsymbol{\mathcal X}}^{(i-1)}$ and de-standardize 
  $\rightarrow$ $\mathbf{Z}^{(i)}=\widehat{\boldsymbol{\mathcal X}}^{(i-1)}\odot\widehat{\boldsymbol{\sigma}}_{Z}^{(i-1)}+\widehat{\boldsymbol{\mu}}_{Z}^{(i-1)}$;\\[3pt]
  \textbf{3.5.2:} Repeat steps 3.1--3.4 $\rightarrow$ $\widehat{\boldsymbol{\mathcal X}}^{(i)} = \widehat{\mathbf{F}}^{(i)}\widehat{\boldsymbol{\Lambda}}^{(i)'} =(\widehat{\boldsymbol{\chi}}_1^{(i)}\cdots\widehat{\boldsymbol \chi}_{T_0}^{(i)})'$ of size $T_0\times N_0$;\\[3pt]
  \textbf{3.5.3:} Impute missing values with $\widehat{\boldsymbol{\mathcal X}}^{(i)}$ and de-standardize $\rightarrow$ $\mathbf{Z}^{(i+1)}=\widehat{\boldsymbol{\mathcal X}}^{(i)}\odot \widehat{\boldsymbol{\sigma}}_{Z}^{(i)}+\widehat{\boldsymbol{\mu}}_{Z}^{(i)}$;\\[3pt]
    \textbf{3.5.4:} Compute $\operatorname{MSE} =\sum_{t=1}^{T_0} \Vert \widehat{\boldsymbol{\chi}}_t^{(i)}-\widehat{\boldsymbol{\chi}}_t^{(i-1)}\Vert^2/\sum_{t=1}^T \Vert \widehat{\boldsymbol{\chi}}_t^{(i-1)}\Vert^2$    
;\\[3pt]
  \textbf{if} $\operatorname{MSE} < \epsilon$; \quad $i^* = i+1$; \textbf{break};\\
  \textbf{else}\quad $i = i+1$.\\ 
  \textbf{end}
  }
 \end{footnotesize}
\KwOutput{$\mathbf{X}^{(0)} = \mathbf{Z}^{(i^*)}$ of size $T_0\times N_0$.}
\vspace{10pt}
\indent \nlset{2} \textbf{Estimation of the common components}\\[5pt]
\KwInput{Processed data from \texttt{[1]} from 2002:M1 to 2023:M10 $\rightarrow$ $\mathbf{X}^{(0)}$ of size $T\times N_0$.}
\vspace{5pt}
\textbf{Step 1:} Subset $\mathbf{X}^{(0)}$ to data for empirical analysis $\rightarrow$ $\mathbf{X}$ of size $T \times N$.\\[3pt]
\textbf{Step 2:} Standardize $\rightarrow$  $\widetilde{\mathbf{X}}= \{\mathbf X-\widehat{\boldsymbol \mu}_X\}\oslash \widehat{\boldsymbol\sigma}_X$.\\[3pt]
\textbf{Step 3:} Determine the number of factors $\rightarrow$ $r$.\\[3pt]
\textbf{Step 4:} Estimate factor model by first $r$ PCs of $\widetilde{\mathbf{X}}$ $\rightarrow$  $\widehat{\boldsymbol{\mathcal X}} = \widehat{\mathbf{F}} \widehat{\boldsymbol{\Lambda}}' =(\widehat{\boldsymbol{\chi}}_1\cdots\widehat{\boldsymbol \chi}_T)'$ of size $T\times N$.\\[3pt]
\textbf{Step 5:} Run the block bootstrap of Barigozzi et al. (2018):\\[3pt]
\begin{footnotesize}
\For{$b=1:B$}{
\textbf{5.1:} For the factors, resample rows of $\widehat{\mathbf{F}}$ $\rightarrow$ $\mathbf{F}^{(b)}$ of size $T\times r$;\\[4pt]
\textbf{5.3:} Generate bootstrap data $\rightarrow$ $\widetilde{\mathbf{X}}^{(b)} = {\mathbf{X}}+( \mathbf{F}^{(b)}-\widehat{ \mathbf{F}})\widehat{\boldsymbol{\Lambda}}'$ of size $T\times N$;\\[4pt]
\textbf{5.4:} Estimate factor model by first $r$ PCs of $\widetilde{\mathbf{X}}^{(b)}$ $\rightarrow$ 
$\widehat{\boldsymbol{\mathcal X}}^{(b)}=\widehat{\mathbf {F}}^{(b)}\widehat{\boldsymbol {\Lambda}}^{(b)} =(\widehat{\boldsymbol{\chi}}_1^{(b)}\cdots\widehat{\boldsymbol \chi}_T^{(b)})'$ of size $T\times N$.
}
\end{footnotesize}
\textbf{Step 6:} Compute the quantiles of $(\widehat{{\chi}}_{it}^{(1)}\cdots\widehat{{\chi}}_{it}^{(B)})'$ $\rightarrow$ $\widehat{q}_{it,\alpha/2}$, $\widehat{q}_{it,1-\alpha/2}$, $i=1,\ldots, N$, $t=1,\ldots,T$.\\[3pt]
\KwOutput{$\widehat{{\chi}}_{it}$ and $[\widehat{{\chi}}_{it}-\widehat{q}_{it,\alpha/2}, \widehat{{\chi}}_{it}+\widehat{q}_{t,1-\alpha/2} ]$,  $i=1,\ldots, N$, $t=1,\ldots,T$.}
\end{footnotesize}

\label{alg::methas}
\end{algorithm}

\newgeometry{left=2cm,right=2cm,top=2cm,bottom=2cm}

\section{Country-level IRFs: baseline specification}
\label{app::idcountries}
\begin{figure}[H]
\setlength{\tabcolsep}{.005\textwidth}
\centering \footnotesize \sc
\caption{Country-level IRFs}
\begin{tabular}{lccccc}
& \hspace{5pt}\scriptsize IPMN &  \hspace{3pt}\scriptsize HICPOV & 
 \hspace{2pt}\scriptsize SHIX &  \hspace{5pt}\scriptsize LTIRT &  \hspace{8pt}\scriptsize UNETOT \\[4pt]
\raisebox{1.5\height}{\rotatebox{90}{\scriptsize AT}} & \includegraphics[trim=5cm 11.8cm 5cm 12.4cm, clip, width=0.19\textwidth]{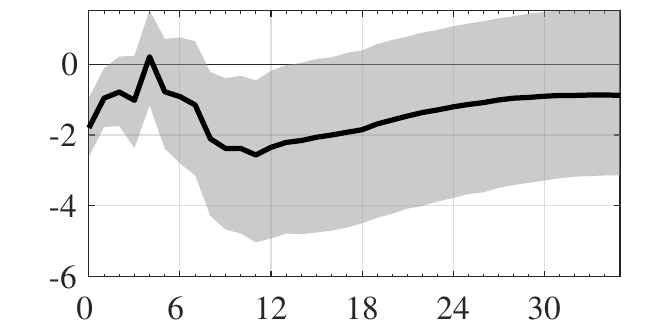} &
\includegraphics[trim=5cm 11.8cm 5cm 12.4cm, clip, width=0.19\textwidth]{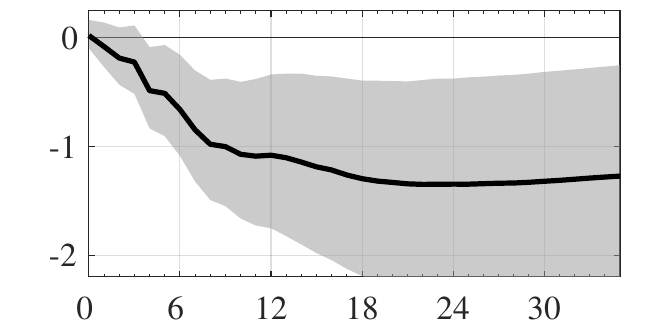} &
\includegraphics[trim=5cm 11.8cm 5cm 12.4cm, clip, width=0.19\textwidth]{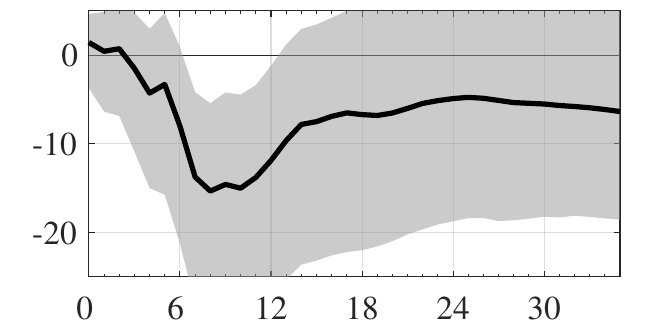} &
\includegraphics[trim=5cm 11.8cm 5cm 12.4cm, clip, width=0.19\textwidth]{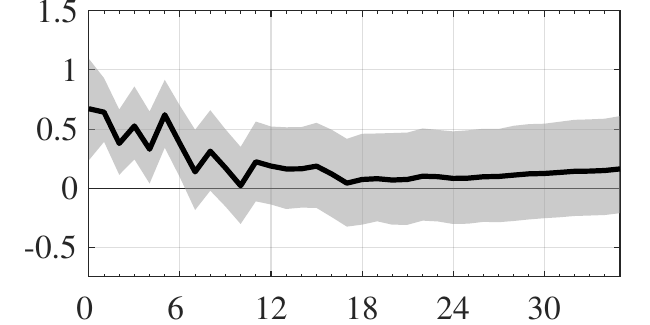} &
\includegraphics[trim=5cm 11.8cm 5cm 12.4cm, clip, width=0.19\textwidth]{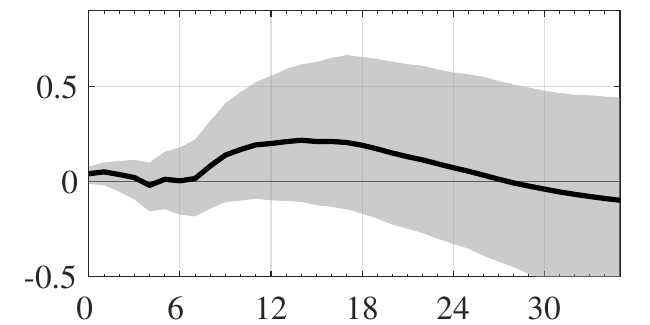} \\

\raisebox{1.5\height}{\rotatebox{90}{\scriptsize BE}} &
\includegraphics[trim=5cm 11.8cm 5cm 12.4cm, clip, width=0.19\textwidth]{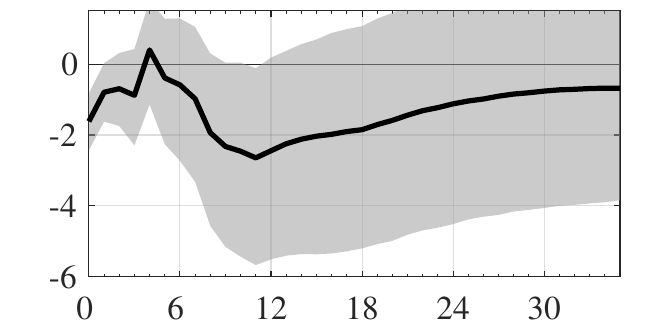} &
\includegraphics[trim=5cm 11.8cm 5cm 12.4cm, clip, width=0.19\textwidth]{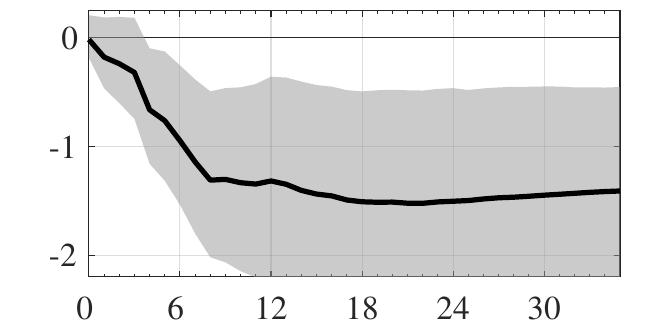} &
\includegraphics[trim=5cm 11.8cm 5cm 12.4cm, clip, width=0.19\textwidth]{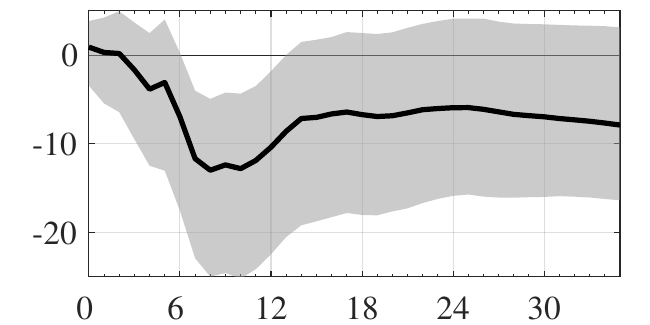} &
\includegraphics[trim=5cm 11.8cm 5cm 12.4cm, clip, width=0.19\textwidth]{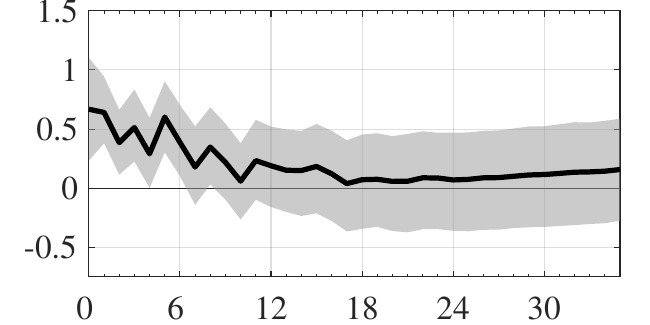} &
\includegraphics[trim=5cm 11.8cm 5cm 12.4cm, clip, width=0.19\textwidth]{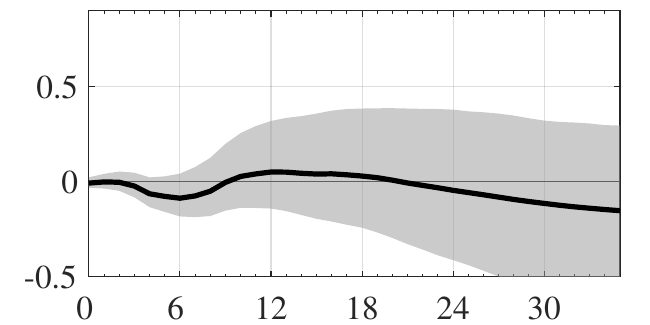} \\

\raisebox{1.5\height}{\rotatebox{90}{\scriptsize DE}} &
\includegraphics[trim=5cm 11.8cm 5cm 12.4cm, clip, width=0.19\textwidth]{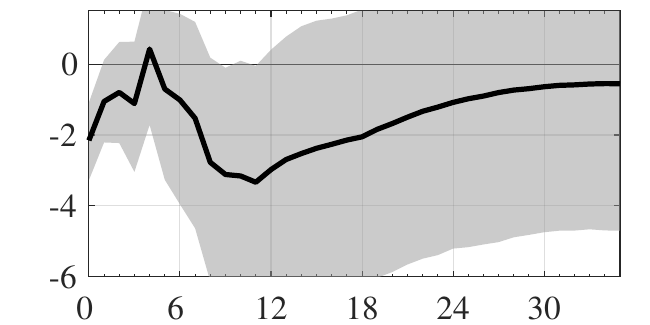} &
\includegraphics[trim=5cm 11.8cm 5cm 12.4cm, clip, width=0.19\textwidth]{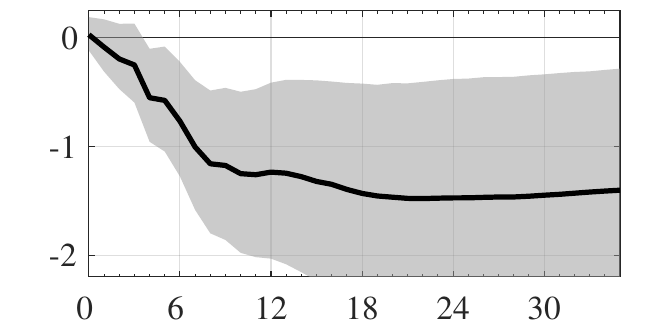} &
\includegraphics[trim=5cm 11.8cm 5cm 12.4cm, clip, width=0.19\textwidth]{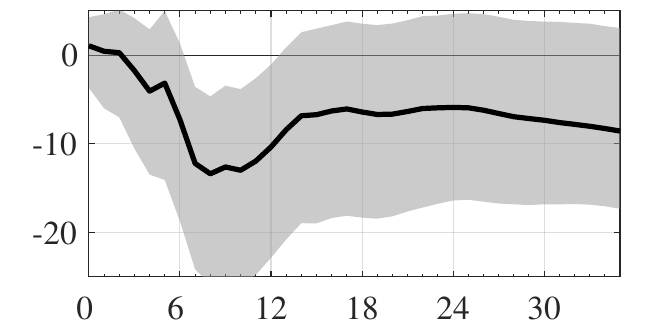} &
\includegraphics[trim=5cm 11.8cm 5cm 12.4cm, clip, width=0.19\textwidth]{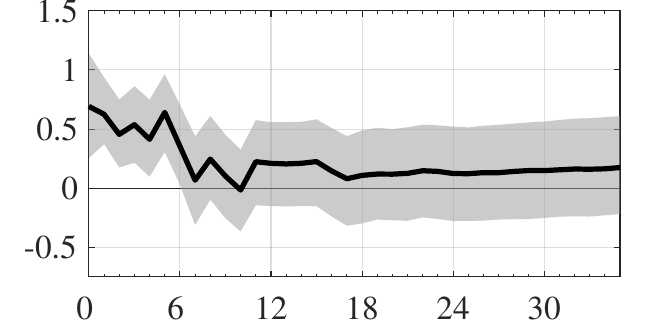} &
\includegraphics[trim=5cm 11.8cm 5cm 12.4cm, clip, width=0.19\textwidth]{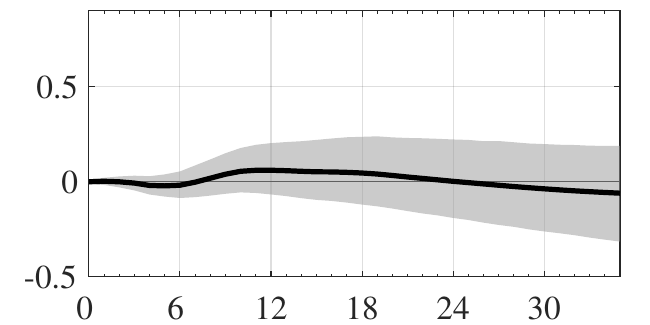} \\

\raisebox{1.5\height}{\rotatebox{90}{\scriptsize FR}} &
\includegraphics[trim=5cm 11.8cm 5cm 12.4cm, clip, width=0.19\textwidth]{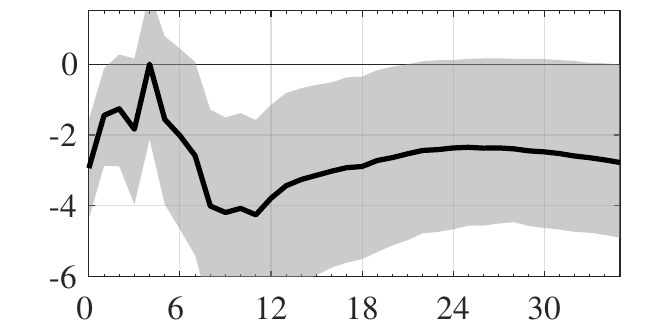} &
\includegraphics[trim=5cm 11.8cm 5cm 12.4cm, clip, width=0.19\textwidth]{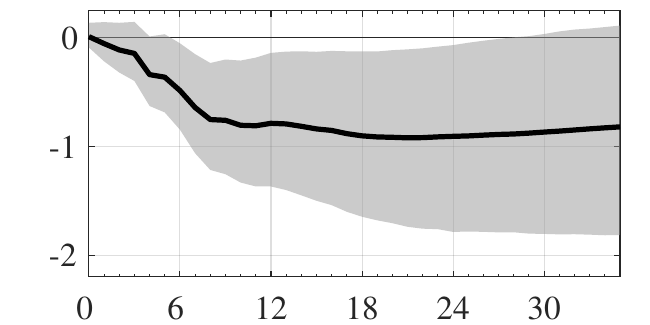} &
\includegraphics[trim=5cm 11.8cm 5cm 12.4cm, clip, width=0.19\textwidth]{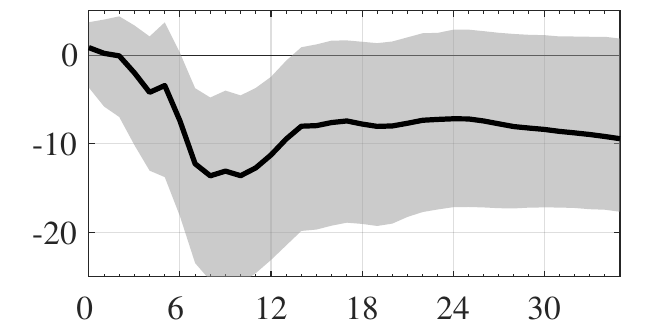} &
\includegraphics[trim=5cm 11.8cm 5cm 12.4cm, clip, width=0.19\textwidth]{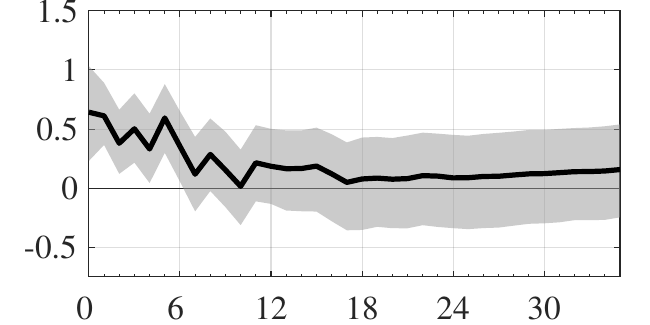} &
\includegraphics[trim=5cm 11.8cm 5cm 12.4cm, clip, width=0.19\textwidth]{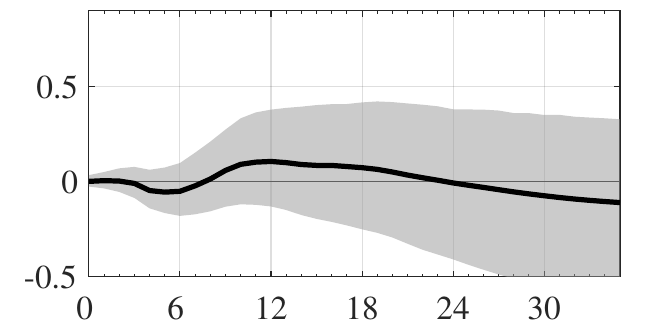} \\

\raisebox{1.5\height}{\rotatebox{90}{\scriptsize NL}} &
\includegraphics[trim=5cm 11.8cm 5cm 12.4cm, clip, width=0.19\textwidth]{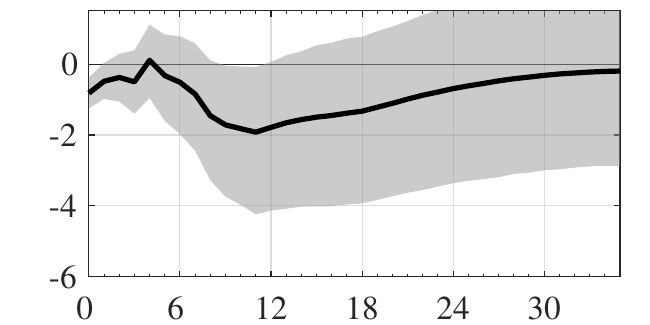} &
\includegraphics[trim=5cm 11.8cm 5cm 12.4cm, clip, width=0.19\textwidth]{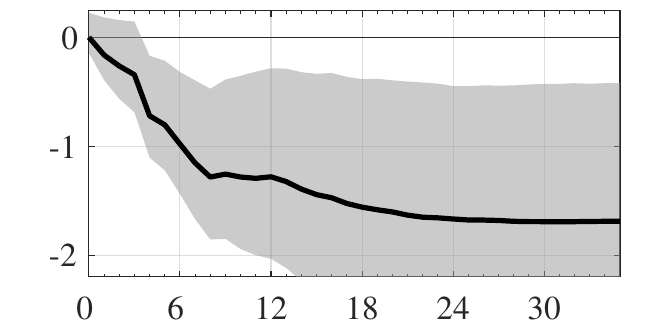} &
\includegraphics[trim=5cm 11.8cm 5cm 12.4cm, clip, width=0.19\textwidth]{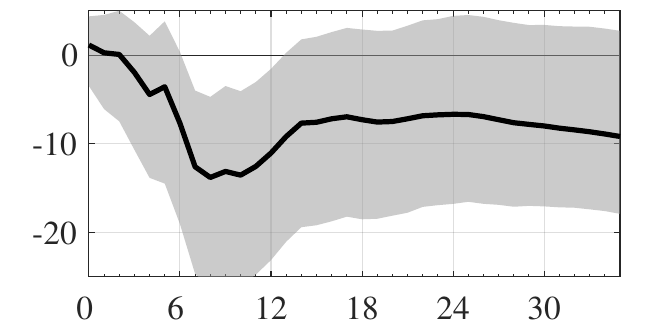} &
\includegraphics[trim=5cm 11.8cm 5cm 12.4cm, clip, width=0.19\textwidth]{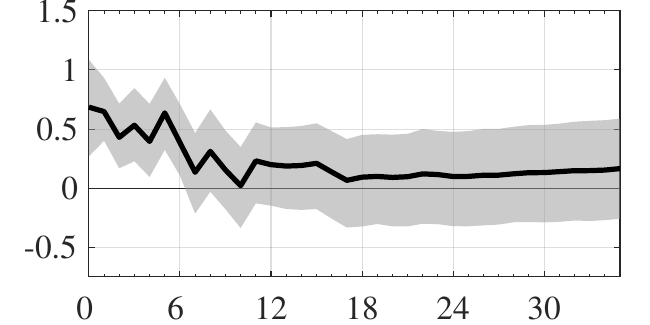} &
\includegraphics[trim=5cm 11.8cm 5cm 12.4cm, clip, width=0.19\textwidth]{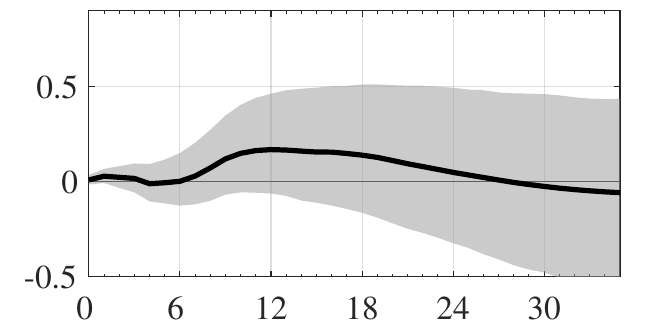} \\

\raisebox{1.7\height}{\rotatebox{90}{\scriptsize EL}} &
\includegraphics[trim=5cm 11.8cm 5cm 12.4cm, clip, width=0.19\textwidth]{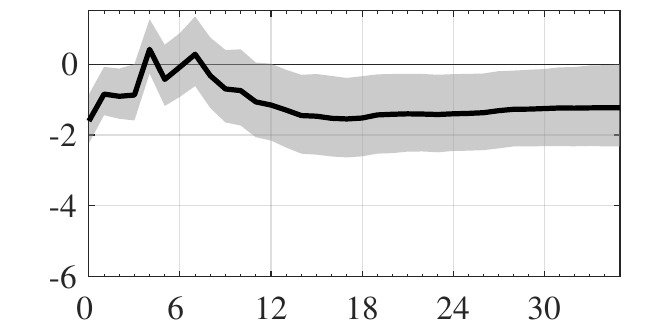} &
\includegraphics[trim=5cm 11.8cm 5cm 12.4cm, clip, width=0.19\textwidth]{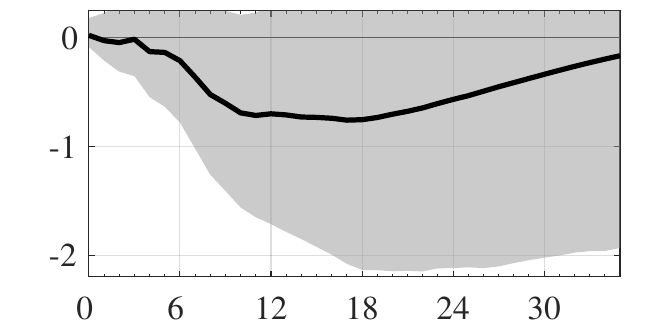} &
\includegraphics[trim=5cm 11.8cm 5cm 12.4cm, clip, width=0.19\textwidth]{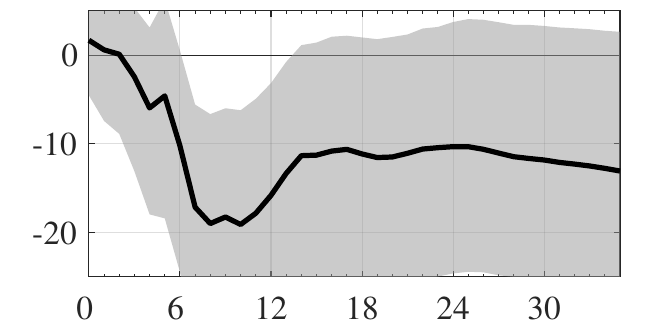} &
\includegraphics[trim=5cm 11.8cm 5cm 12.4cm, clip, width=0.19\textwidth]{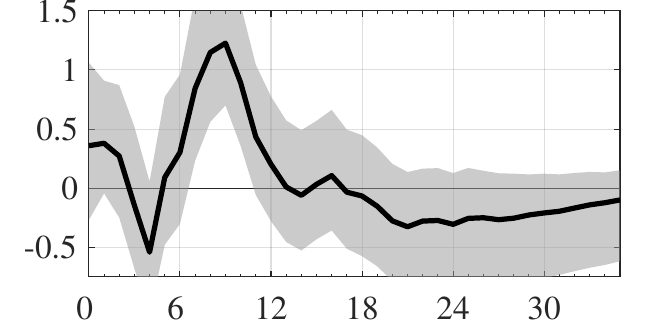} &
\includegraphics[trim=5cm 11.8cm 5cm 12.4cm, clip, width=0.19\textwidth]{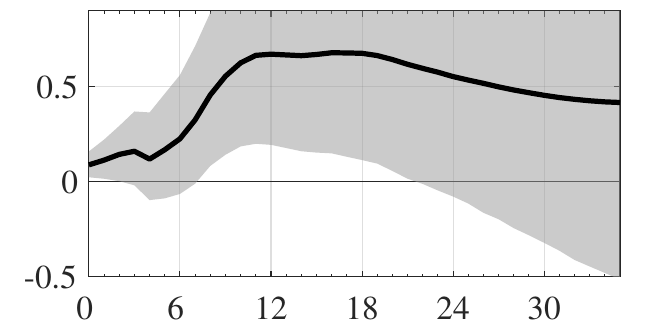} \\

\raisebox{1.7\height}{\rotatebox{90}{\scriptsize ES}} &
\includegraphics[trim=5cm 11.8cm 5cm 12.4cm, clip, width=0.19\textwidth]{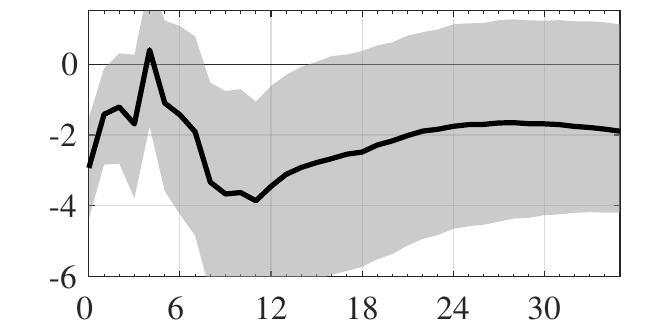} &
\includegraphics[trim=5cm 11.8cm 5cm 12.4cm, clip, width=0.19\textwidth]{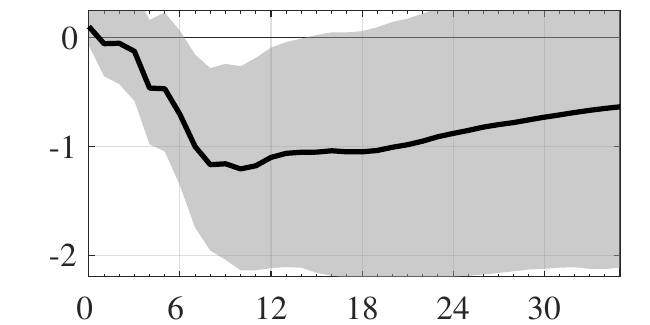} &
\includegraphics[trim=5cm 11.8cm 5cm 12.4cm, clip, width=0.19\textwidth]{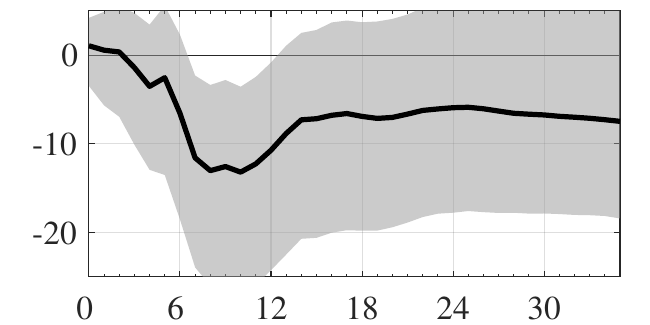} &
\includegraphics[trim=5cm 11.8cm 5cm 12.4cm, clip, width=0.19\textwidth]{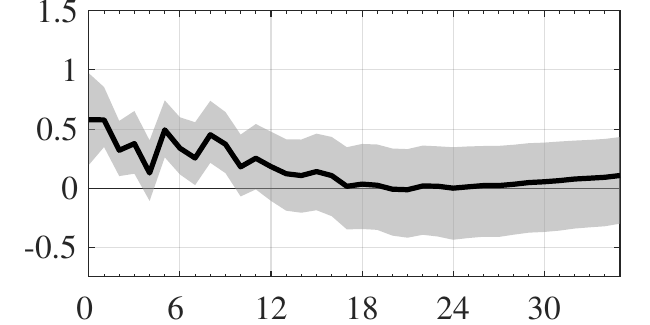} &
\includegraphics[trim=5cm 11.8cm 5cm 12.4cm, clip, width=0.19\textwidth]{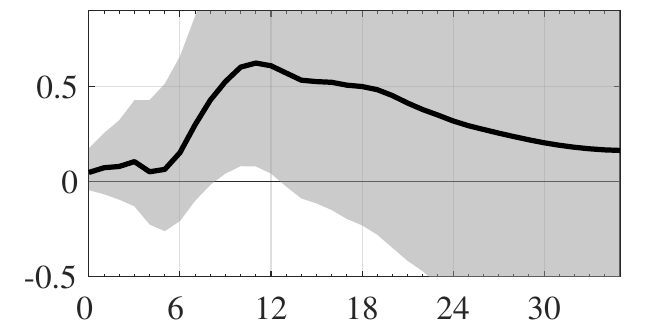} \\

\raisebox{2\height}{\rotatebox{90}{\scriptsize IE}} &
 &
\includegraphics[trim=5cm 11.8cm 5cm 12.4cm, clip, width=0.19\textwidth]{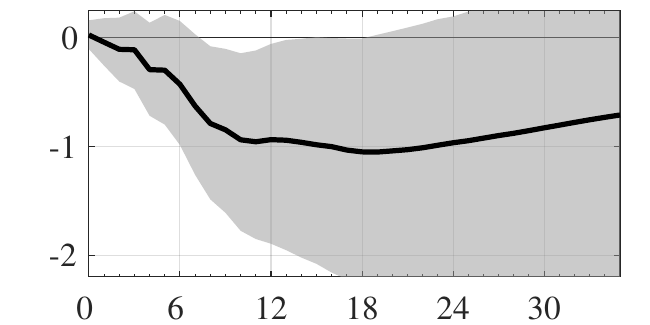} &
\includegraphics[trim=5cm 11.8cm 5cm 12.4cm, clip, width=0.19\textwidth]{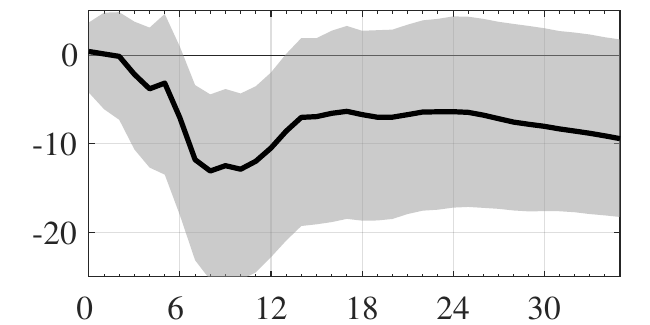} &
\includegraphics[trim=5cm 11.8cm 5cm 12.4cm, clip, width=0.19\textwidth]{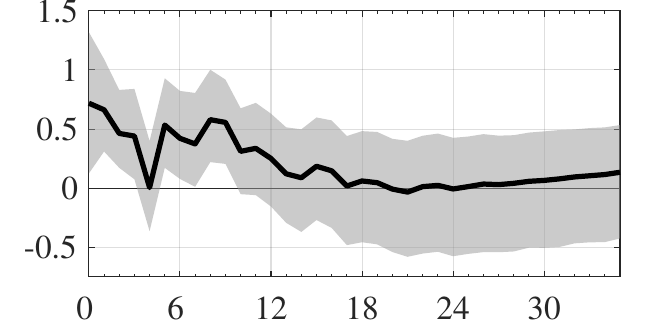} &
\includegraphics[trim=5cm 11.8cm 5cm 12.4cm, clip, width=0.19\textwidth]{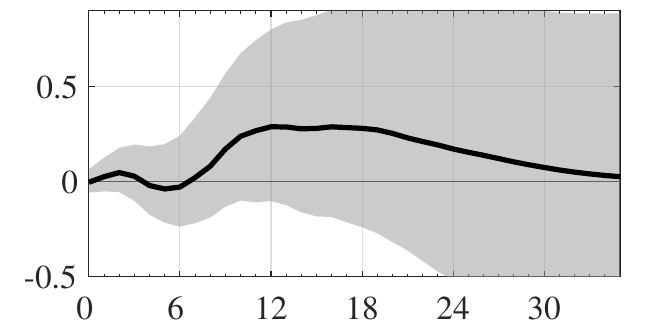} \\
                        
\raisebox{2\height}{\rotatebox{90}{\scriptsize IT}} &
\includegraphics[trim=5cm 11.8cm 5cm 12.4cm, clip, width=0.19\textwidth]{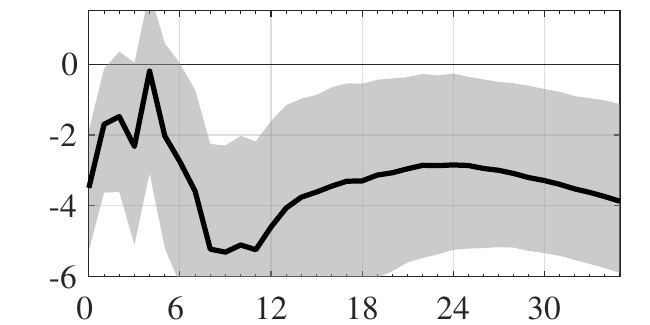} &
\includegraphics[trim=5cm 11.8cm 5cm 12.4cm, clip, width=0.19\textwidth]{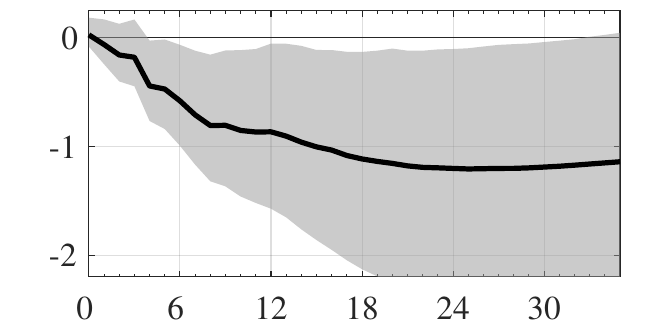} &
\includegraphics[trim=5cm 11.8cm 5cm 12.4cm, clip, width=0.19\textwidth]{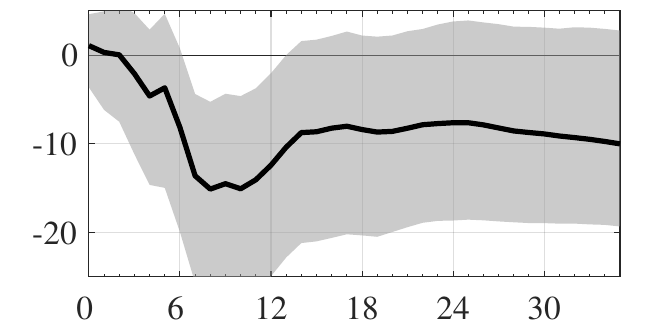} &
\includegraphics[trim=5cm 11.8cm 5cm 12.4cm, clip, width=0.19\textwidth]{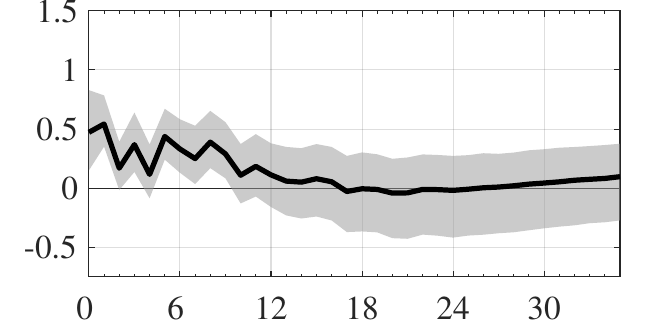} &
\includegraphics[trim=5cm 11.8cm 5cm 12.4cm, clip, width=0.19\textwidth]{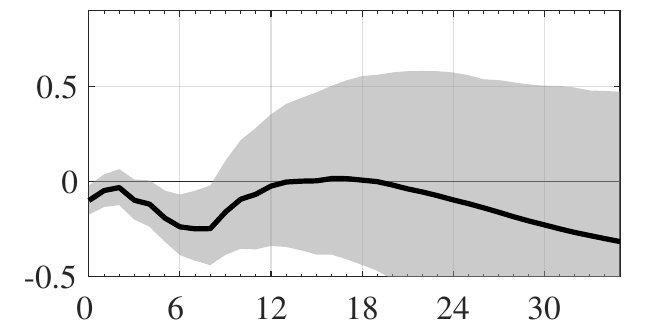} \\

\raisebox{1.5\height}{\rotatebox{90}{\scriptsize PT}} &
\includegraphics[trim=5cm 11.8cm 5cm 12.4cm, clip, width=0.19\textwidth]{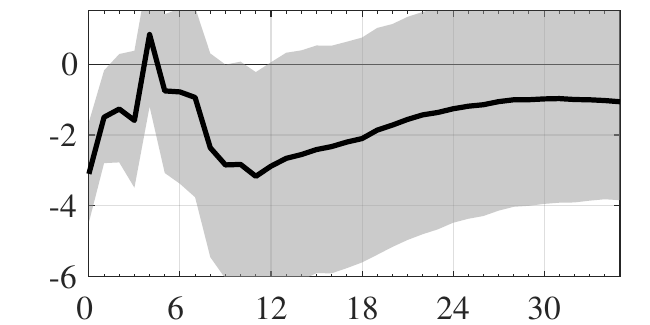} &
\includegraphics[trim=5cm 11.8cm 5cm 12.4cm, clip, width=0.19\textwidth]{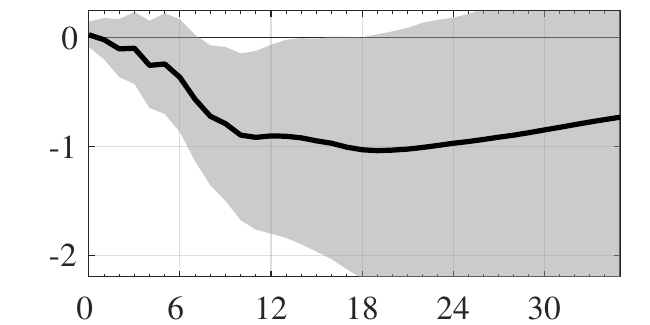} &
\includegraphics[trim=5cm 11.8cm 5cm 12.4cm, clip, width=0.19\textwidth]{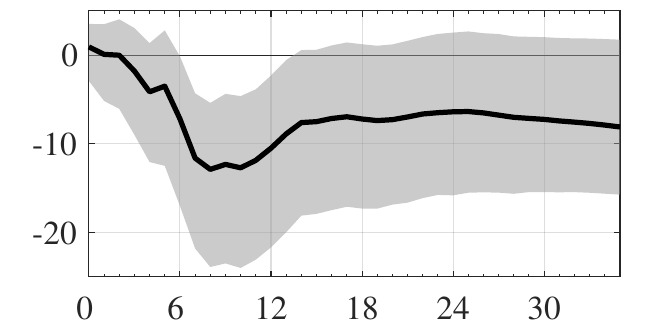} &
\includegraphics[trim=5cm 11.8cm 5cm 12.4cm, clip, width=0.19\textwidth]{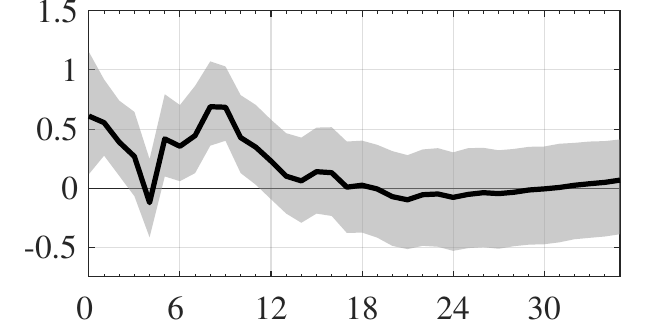} &
\includegraphics[trim=5cm 11.8cm 5cm 12.4cm, clip, width=0.19\textwidth]{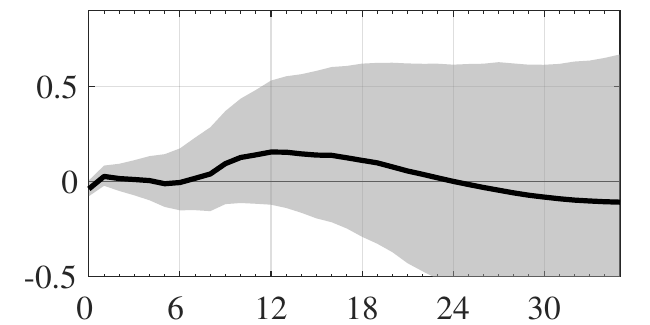} \\[5pt]
\end{tabular}
\resizebox{7.1in}{!}{ 
\begin{tabular}{p{1\textwidth}} \scriptsize Notes: \rm Each sub-figure plots the impulse response to a 100bps contractionary monetary policy shock for one variable and one country . Each column of the graph represents a variable, while each row represents a country. The variables considered are: Industrial Production: Manufacturing (IPMN), HICP: Overall (HICPOV), Stock Price Index (SHIX), 10-years Interest Rates (LTIRT) and Unemployment Rate (UNETOT). The black solid line is the point estimate in our baseline setting, while the gray shaded area is the corresponding 68\% confidence interval. The scale in the vertical axis has been fixed to be the same across countries, to allow for a one-to-one comparison with Figure 2 in the main text.
\end{tabular}
}
\label{fig::idcountries}
\end{figure}

\section{Comparison with pre-Covid estimates}
\label{app::preCOV}

\subsection{EA IRFs}
\begin{figure}[H]
\centering \footnotesize \sc \smallskip
\setlength{\tabcolsep}{.005\textwidth}
\caption{EA IRFs: full sample vs 2019:M12}
\begin{tabular}{ccc}
\scriptsize 2-years Interest Rate ($R_t$) &\scriptsize  IP: Manufacturing (IPMN) &\scriptsize  HICP: Overall (HICPOV)\\
\includegraphics[trim= .5cm 8cm .5cm 8.5cm, clip, width=0.32\textwidth]{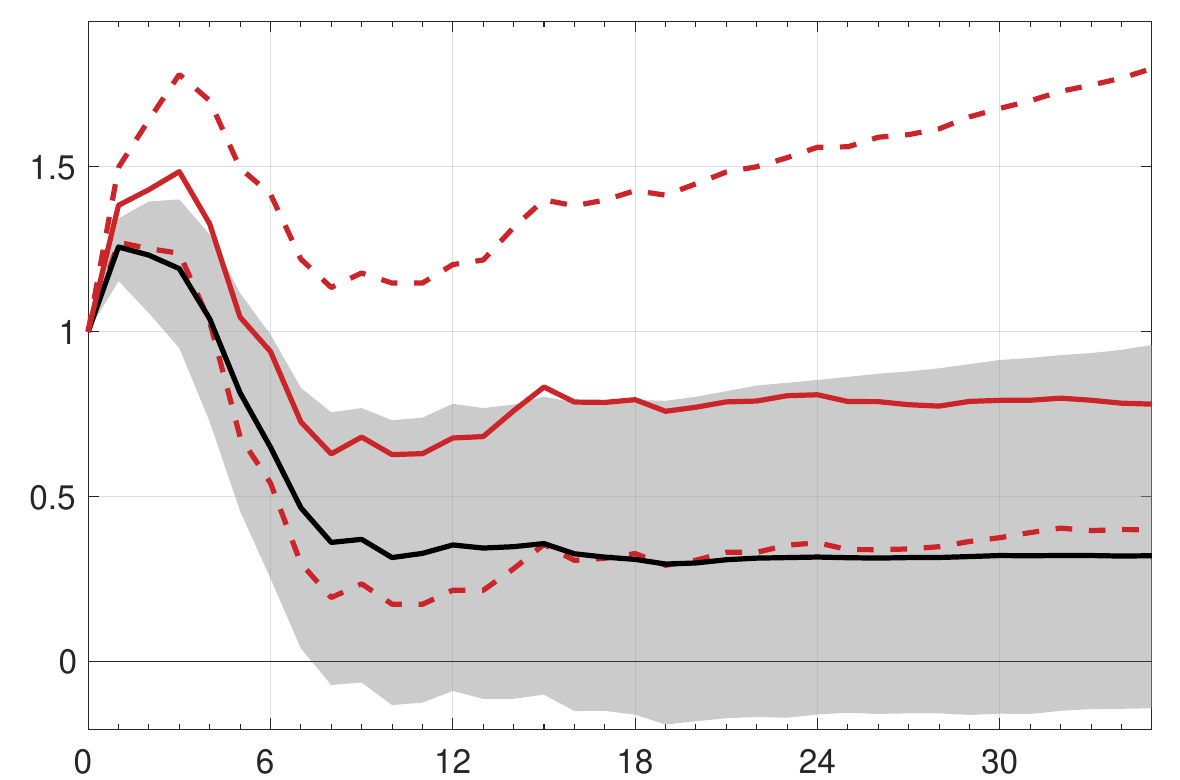} &
\includegraphics[trim= .5cm 8cm .5cm 8.5cm, clip, width=0.32\textwidth]{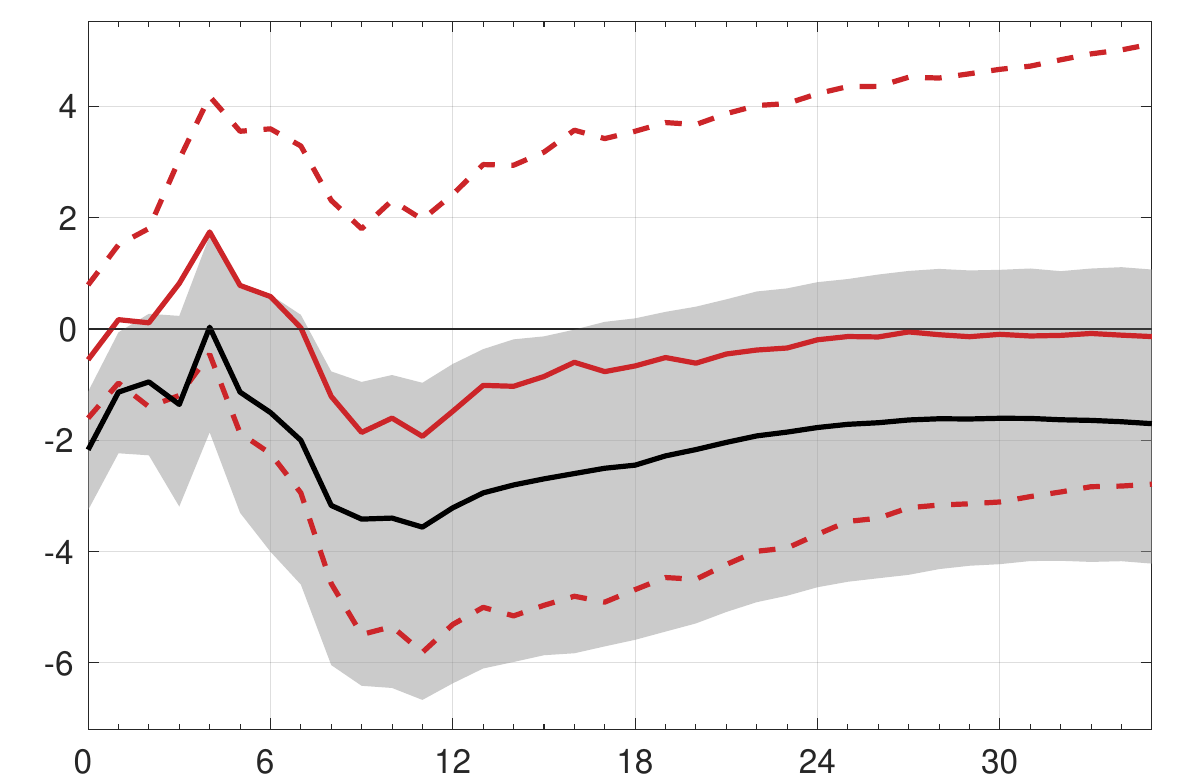} &
\includegraphics[trim= .5cm 8cm .5cm 8.5cm, clip, width=0.32\textwidth]{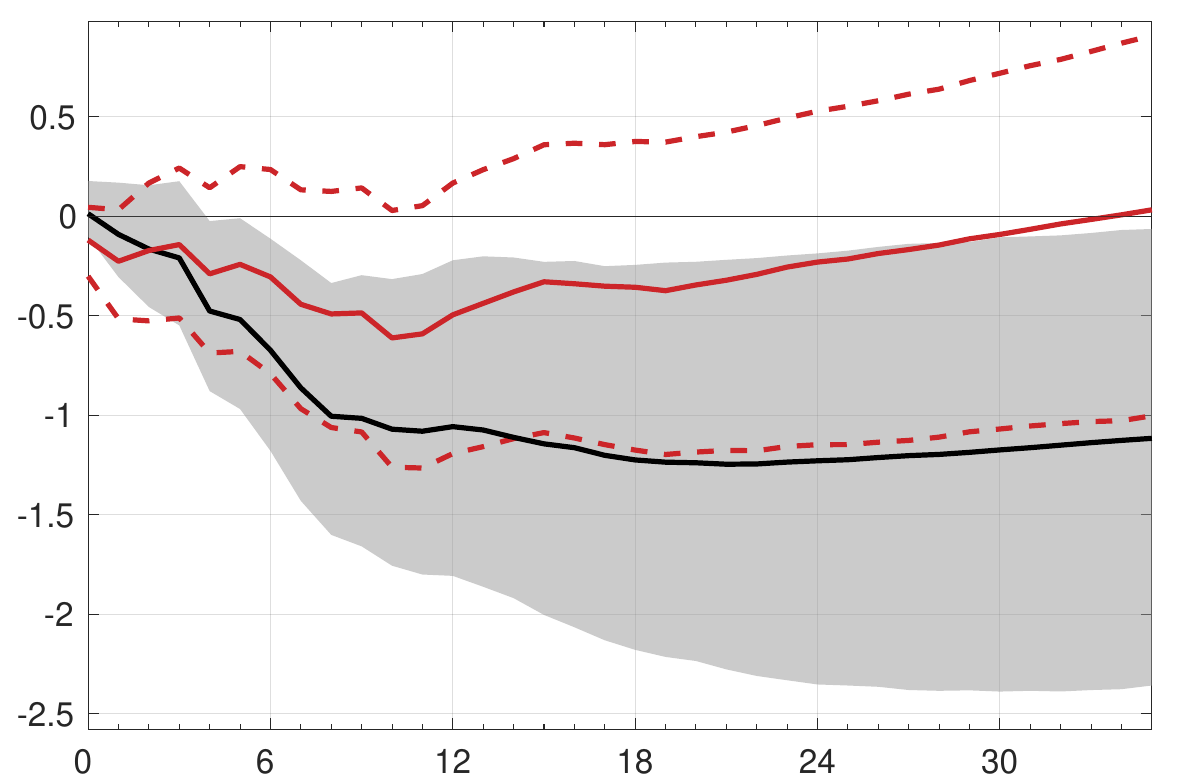} \\
\scriptsize 10-years Interest Rate (LTIRT) &\scriptsize Stock Price Index (SHIX) &\scriptsize Unemployment Rate (UNETOT) \\
\includegraphics[trim= .5cm 8cm .5cm 8.5cm, clip, width=0.32\textwidth]{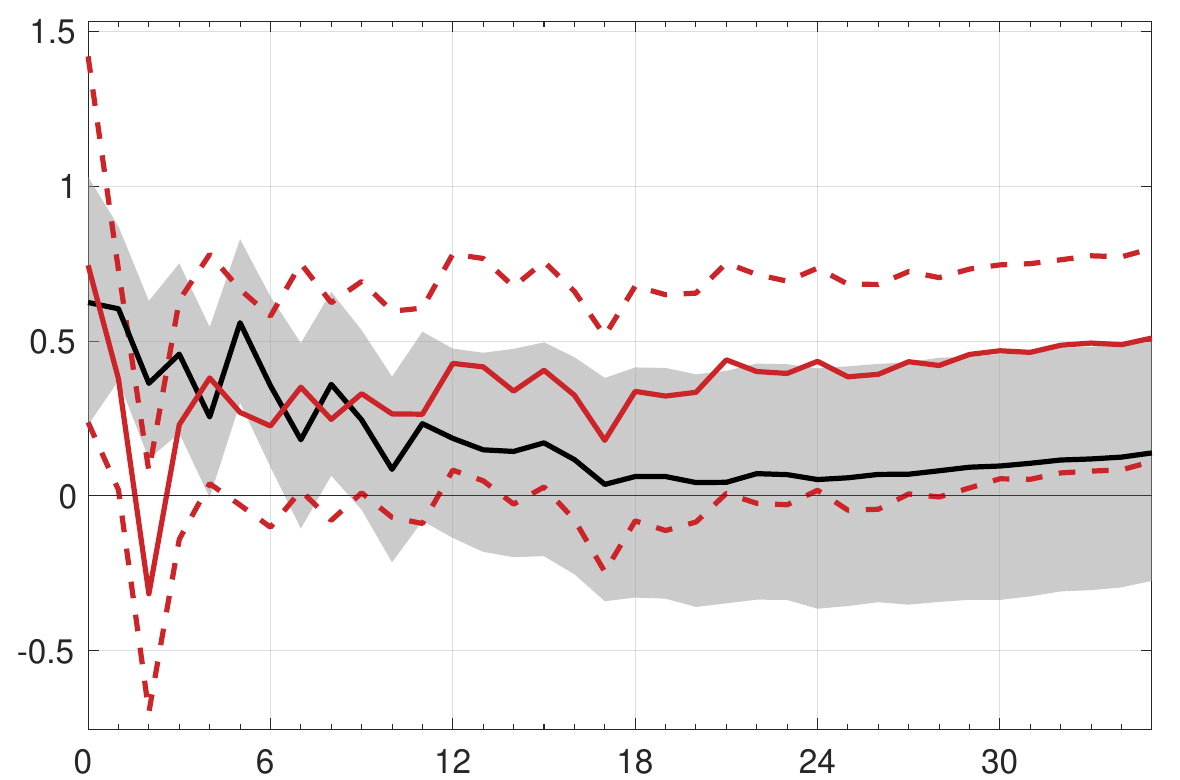} &
\includegraphics[trim= .5cm 8cm .5cm 8.5cm, clip, width=0.32\textwidth]{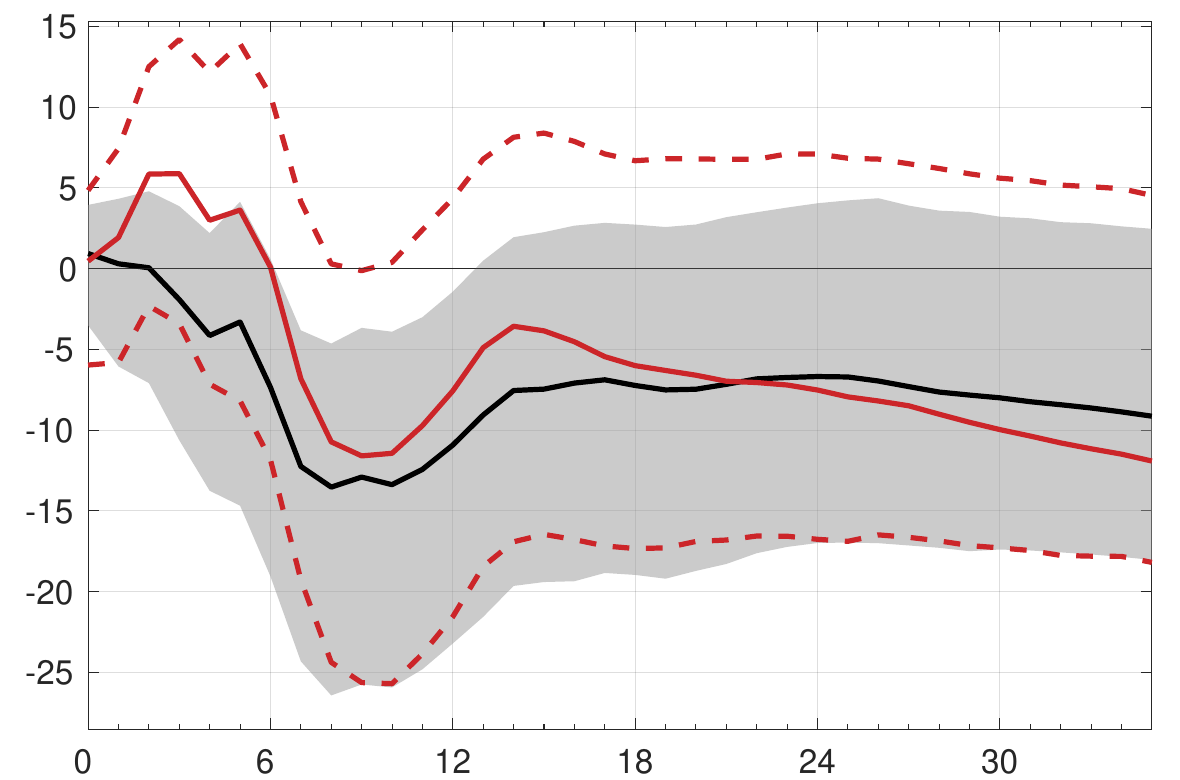} &
\includegraphics[trim= .5cm 8cm .5cm 8.5cm, clip, width=0.32\textwidth]{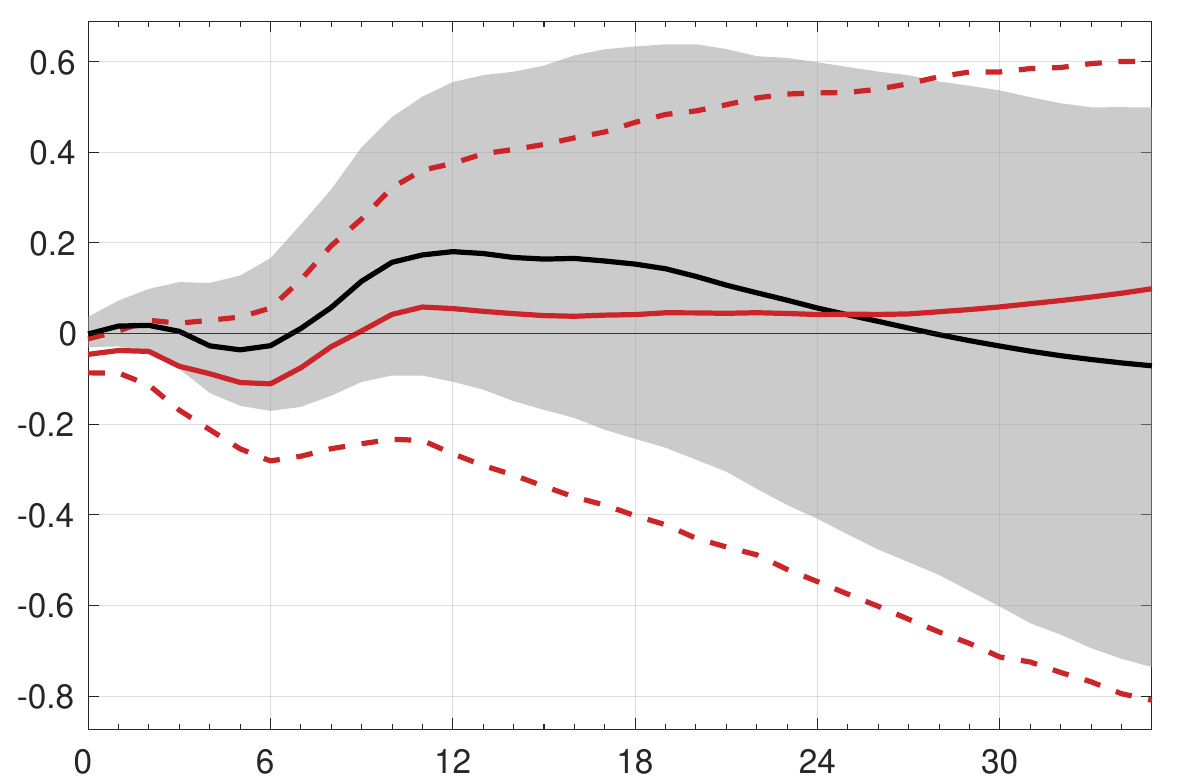} \\
\end{tabular}
\begin{tabular}{p{.98\textwidth}} \scriptsize Notes: \rm Each sub-figure plots the impulse response of one EA variable to a 100bps contractionary monetary policy shock. The black solid line is the point estimate in our baseline setting, while the red solid line is the point estimate obtained with data truncated in 2019:M12. The gray shaded area and red dotted lines are the corresponding 68\% confidence intervals.
\end{tabular}
\label{fig::EAirfs19}
\end{figure}

\subsection{Country-level IRFs}

\begin{figure}[H]
\centering \footnotesize \sc \smallskip
\setlength{\tabcolsep}{.005\textwidth}
\caption{Country-level IRFs:  statistical transformations}
\begin{tabular}{ccc}
\scriptsize IP: Manufacturing (IPMN)&\scriptsize HICP: Overall (HICPOV) &\scriptsize Stock Price Index (SHIX)  \\
\includegraphics[trim= .5cm 8cm .5cm 8.5cm, clip, width=0.32\textwidth]{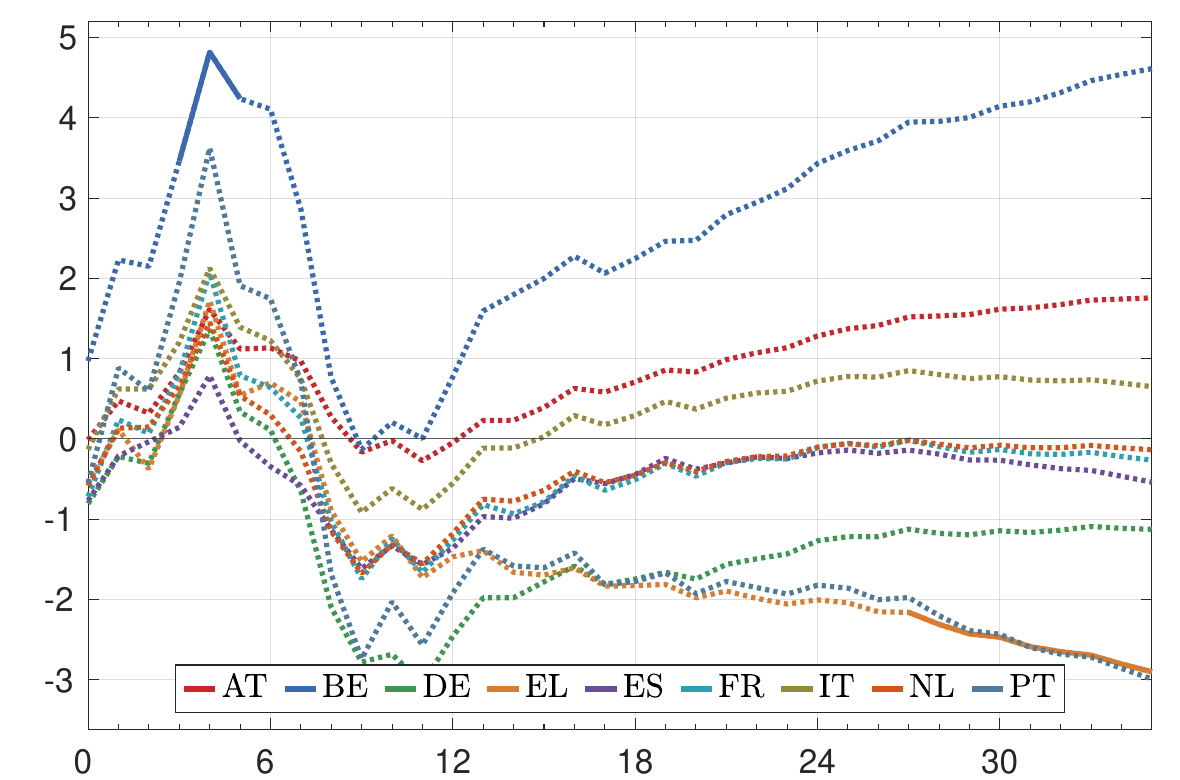} &
\includegraphics[trim= .5cm 8cm .5cm 8.5cm, clip, width=0.32\textwidth]{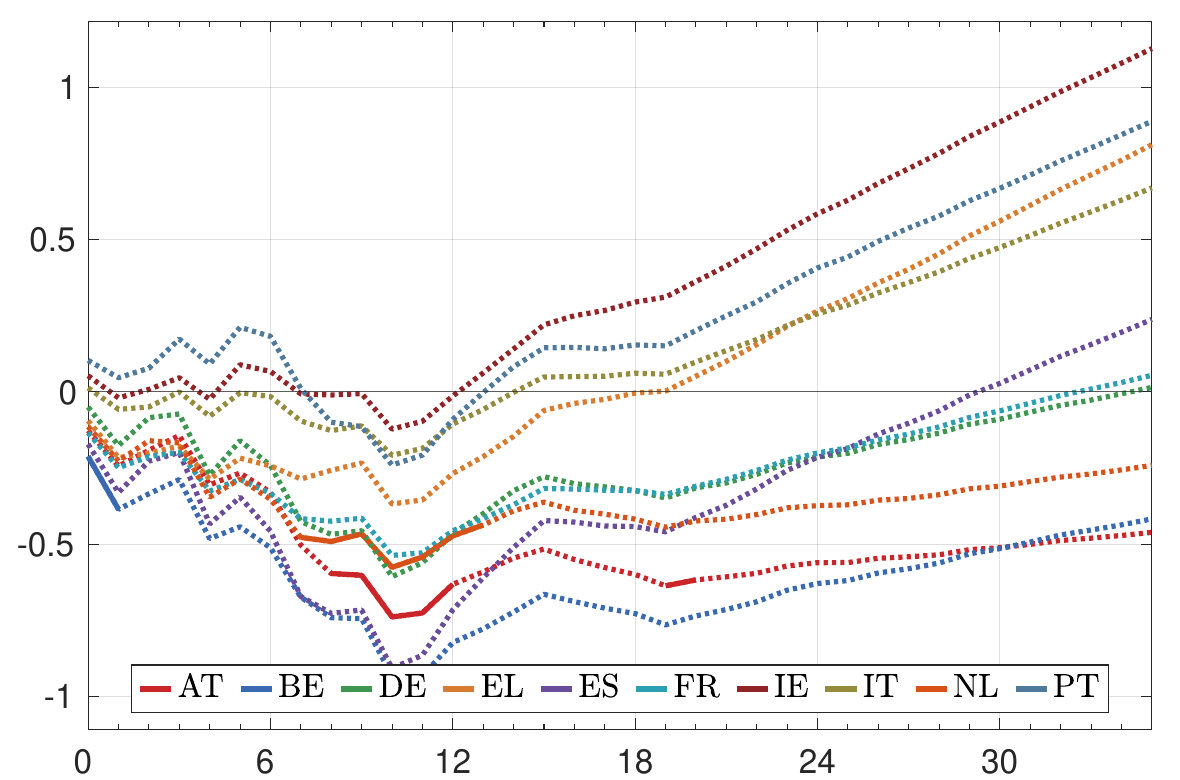} &
\includegraphics[trim= .5cm 8cm .5cm 8.5cm, clip, width=0.32\textwidth]{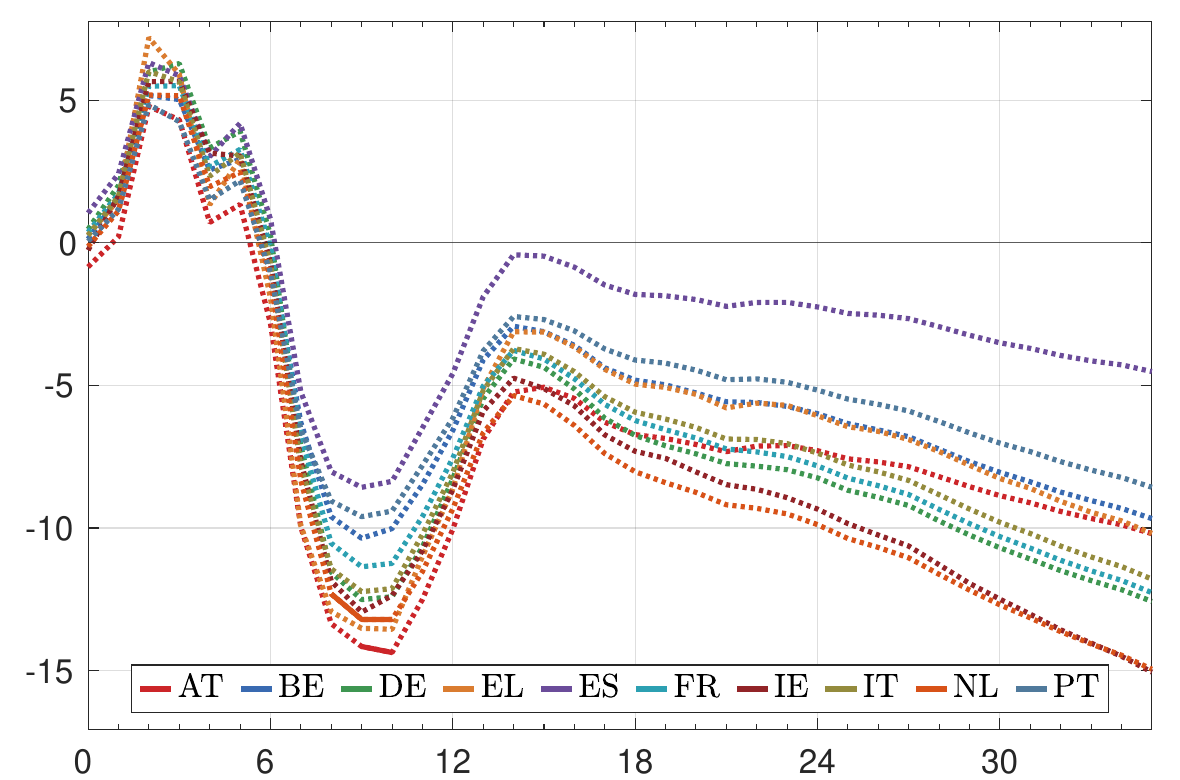} \\[10pt]
\end{tabular}

\begin{tabular}{cc}
\scriptsize10-years Interest Rate (LTIRT) &\scriptsize Unemployment Rate (UNETOT)\\

\includegraphics[trim= .5cm 8cm .5cm 8.5cm, clip, width=0.32\textwidth]{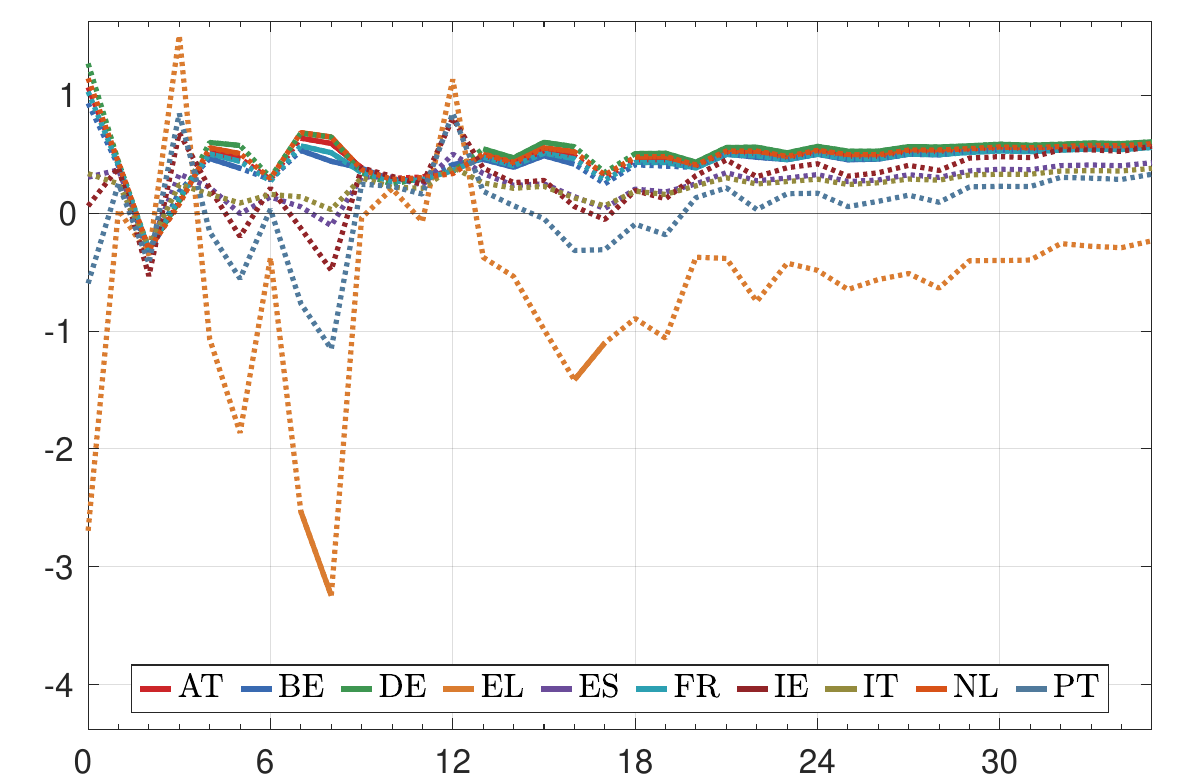} 
&
\includegraphics[trim= .5cm 8cm .5cm 8.5cm, clip, width=0.32\textwidth]{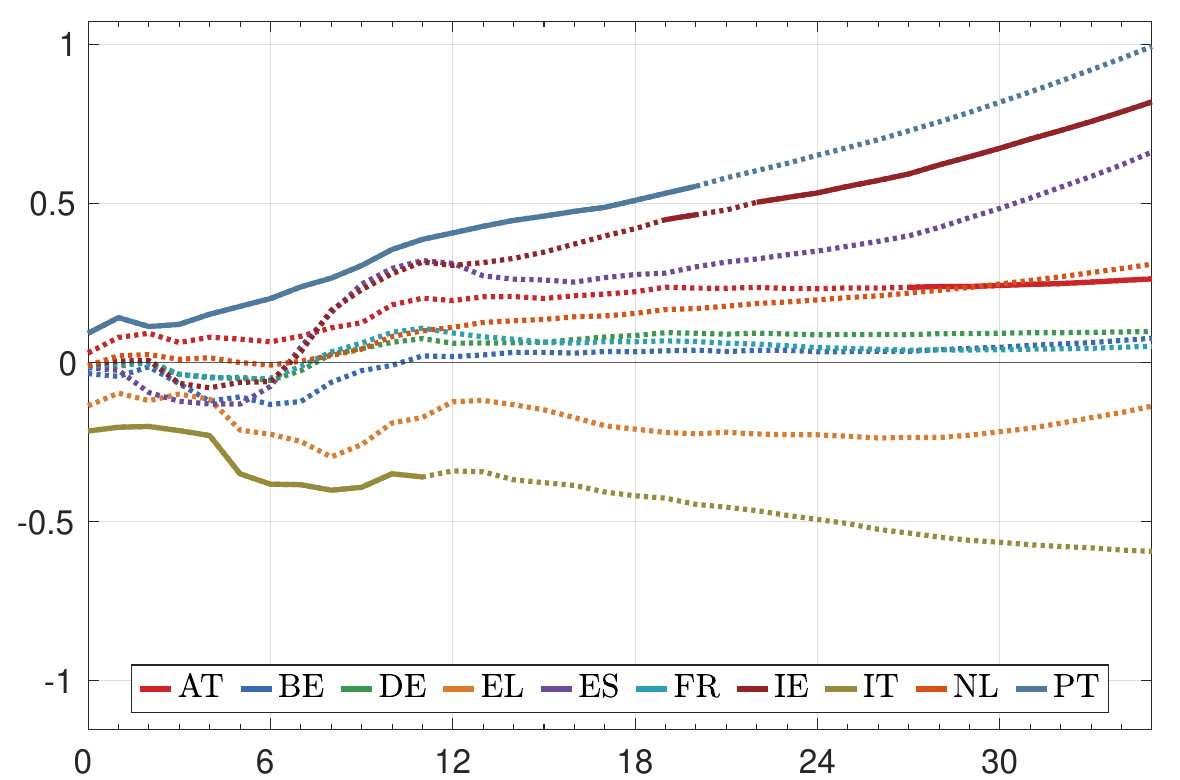}
\end{tabular}
\begin{tabular}{p{.98\textwidth}} \scriptsize Notes: \rm Each sub-figure plots the impulse responses, for all countries, of one variable to a 100bps contractionary monetary policy shock. Within each sub-figure, at each horizon $h=0,\ldots,36$ the country-level impulse responses are denoted with a solid line if the IRF is statistically significant at the 68\% level at that horizon, and with a dotted line otherwise. 
\end{tabular}
\label{fig:national_irf_monthly19}
\end{figure}

\begin{figure}[H]
\setlength{\tabcolsep}{.005\textwidth}
\centering \footnotesize \sc \smallskip
\caption{Difference between EA and country-level IRFs: full sample vs 2019:M12}
\begin{tabular}{lccccc}
& \hspace{5pt}\scriptsize IPMN &  \hspace{5pt}\scriptsize HICPOV & 
 \hspace{5pt}\scriptsize SHIX &  \hspace{5pt}\scriptsize LTIRT &  \hspace{8pt}\scriptsize UNETOT \\[4pt] 
\raisebox{1.5\height}{\rotatebox{90}{\scriptsize AT}} & \includegraphics[trim=5cm 12cm 5cm 12.5cm, clip, width=0.19\textwidth]{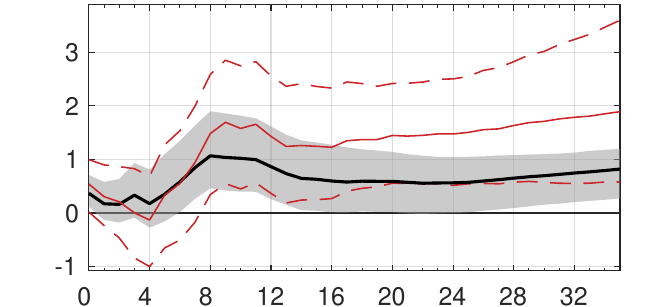} &
\includegraphics[trim=5cm 12cm 5cm 12.5cm, clip, width=0.19\textwidth]{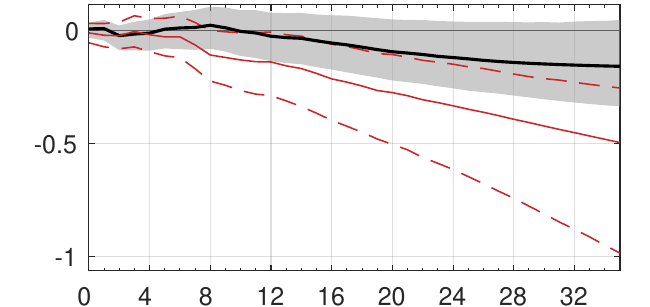} &
\includegraphics[trim=5cm 12cm 5cm 12.5cm, clip, width=0.19\textwidth]{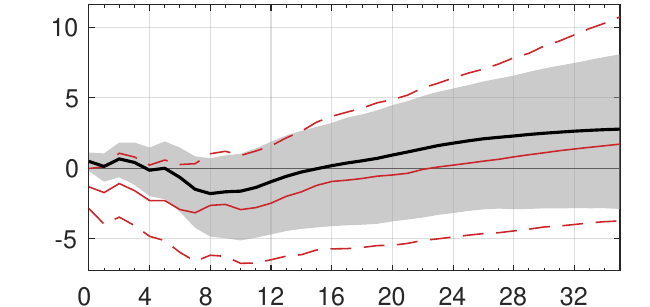} &
\includegraphics[trim=5cm 12cm 5cm 12.5cm, clip, width=0.19\textwidth]{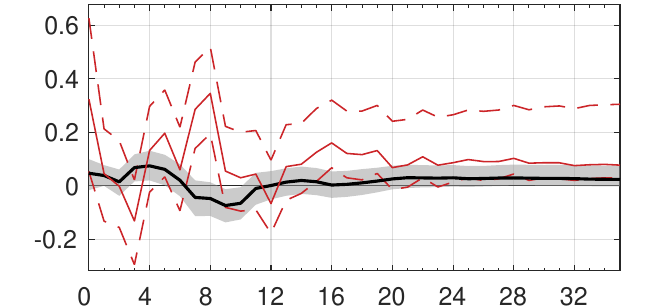} &
\includegraphics[trim=5cm 12cm 5cm 12.5cm, clip, width=0.19\textwidth]{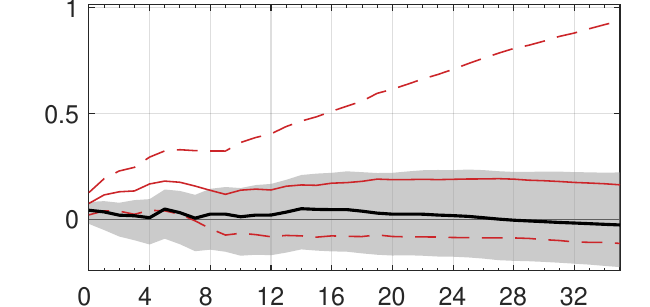} \\

\raisebox{1.5\height}{\rotatebox{90}{\scriptsize BE}} &
\includegraphics[trim=5cm 12cm 5cm 12.5cm, clip, width=0.19\textwidth]{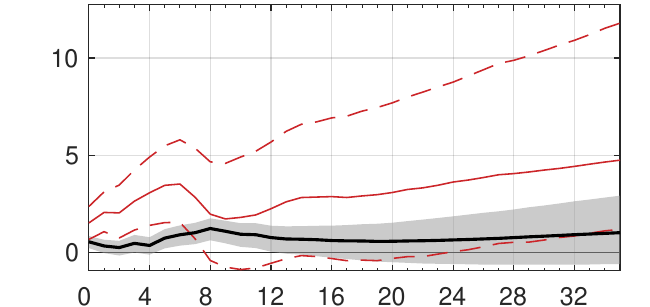} &
\includegraphics[trim=5cm 12cm 5cm 12.5cm, clip, width=0.19\textwidth]{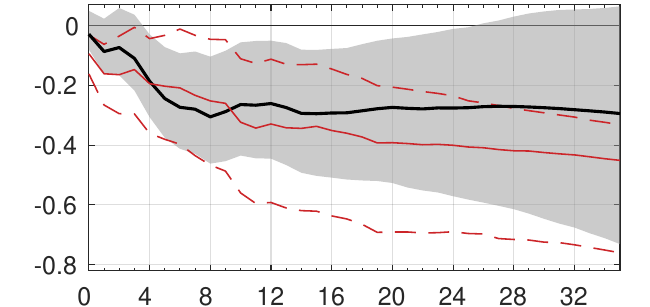} &
\includegraphics[trim=5cm 12cm 5cm 12.5cm, clip, width=0.19\textwidth]{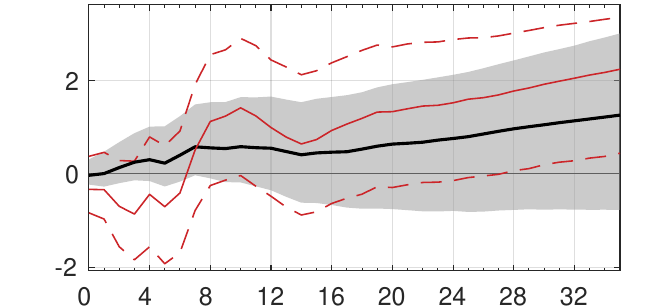} &
\includegraphics[trim=5cm 12cm 5cm 12.5cm, clip, width=0.19\textwidth]{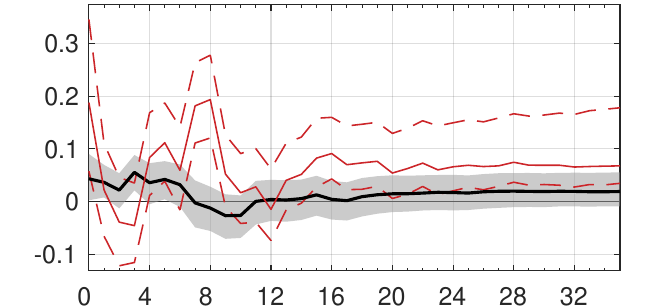} &
\includegraphics[trim=5cm 12cm 5cm 12.5cm, clip, width=0.19\textwidth]{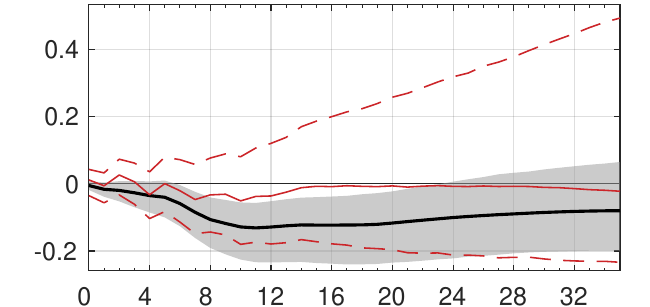} \\

\raisebox{1.5\height}{\rotatebox{90}{\scriptsize DE}} &
\includegraphics[trim=5cm 12cm 5cm 12.5cm, clip, width=0.19\textwidth]{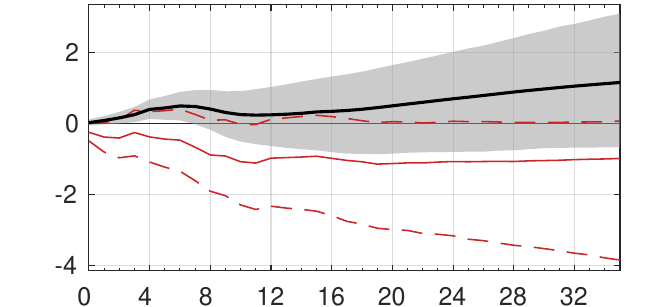} &
\includegraphics[trim=5cm 12cm 5cm 12.5cm, clip, width=0.19\textwidth]{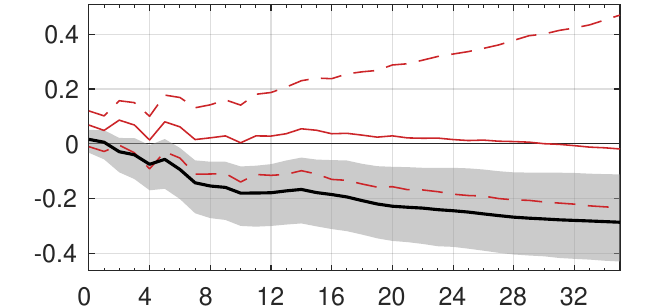} &
\includegraphics[trim=5cm 12cm 5cm 12.5cm, clip, width=0.19\textwidth]{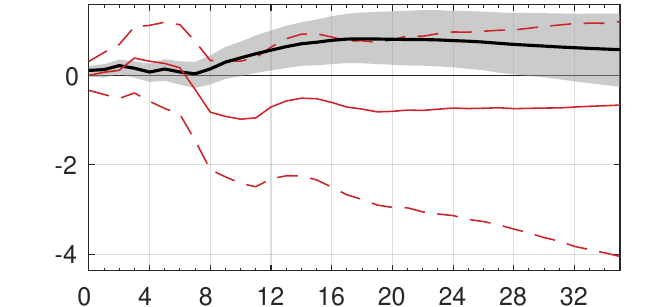} &
\includegraphics[trim=5cm 12cm 5cm 12.5cm, clip, width=0.19\textwidth]{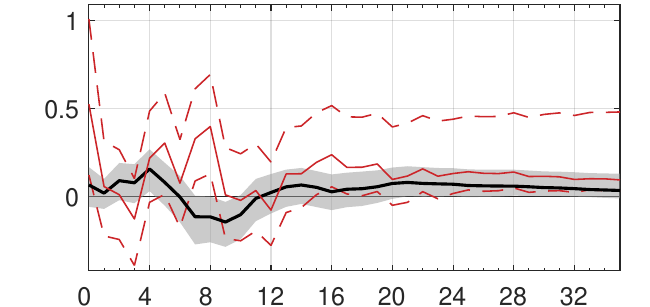} &
\includegraphics[trim=5cm 12cm 5cm 12.5cm, clip, width=0.19\textwidth]{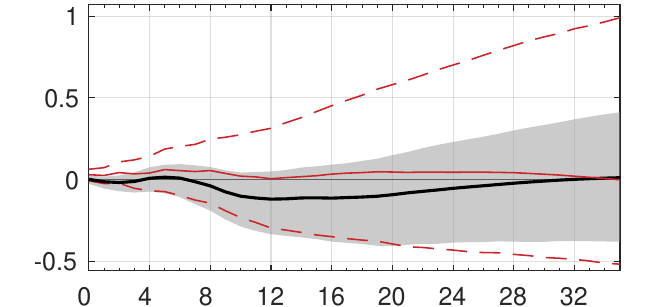} \\

\raisebox{1.5\height}{\rotatebox{90}{\scriptsize FR}} &
\includegraphics[trim=5cm 12cm 5cm 12.5cm, clip, width=0.19\textwidth]{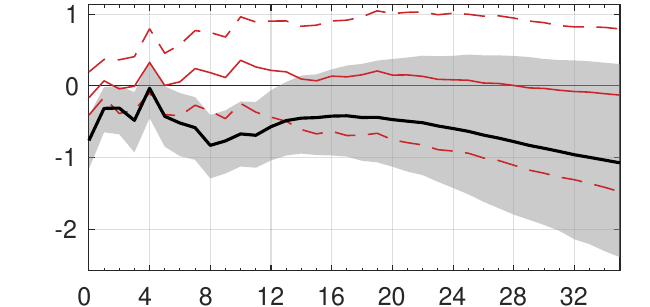} &
\includegraphics[trim=5cm 12cm 5cm 12.5cm, clip, width=0.19\textwidth]{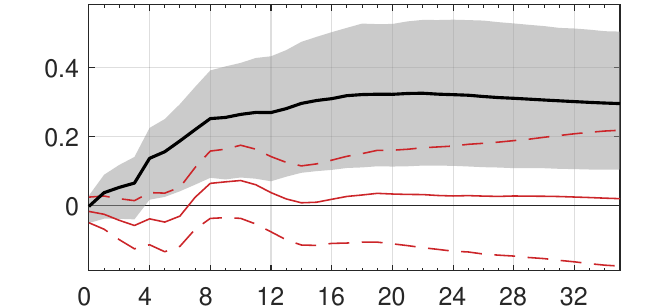} &
\includegraphics[trim=5cm 12cm 5cm 12.5cm, clip, width=0.19\textwidth]{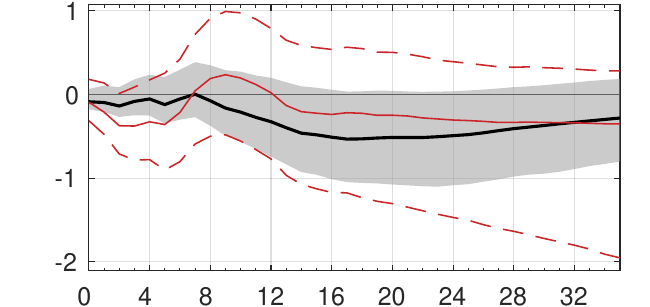} &
\includegraphics[trim=5cm 12cm 5cm 12.5cm, clip, width=0.19\textwidth]{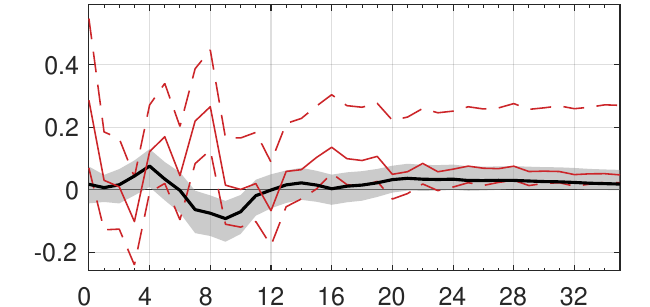} &
\includegraphics[trim=5cm 12cm 5cm 12.5cm, clip, width=0.19\textwidth]{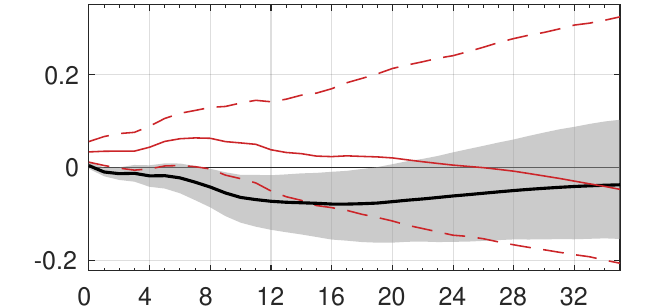} \\

\raisebox{1.5\height}{\rotatebox{90}{\scriptsize NL}} &
\includegraphics[trim=5cm 12cm 5cm 12.5cm, clip, width=0.19\textwidth]{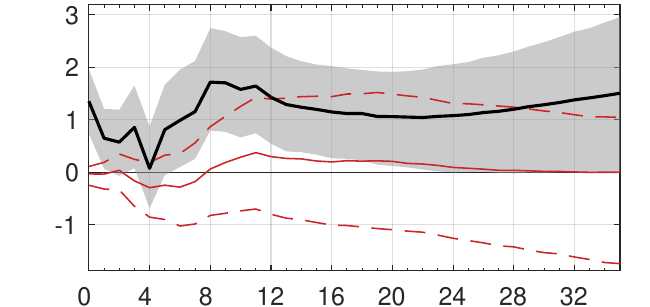} &
\includegraphics[trim=5cm 12cm 5cm 12.5cm, clip, width=0.19\textwidth]{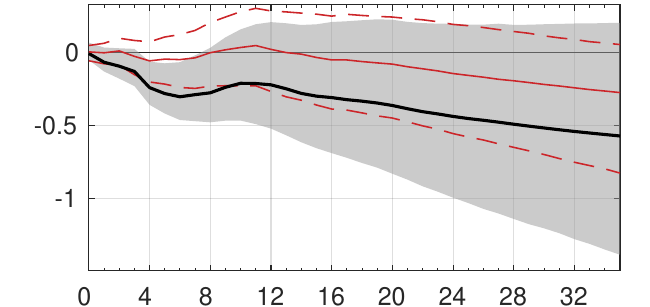} &
\includegraphics[trim=5cm 12cm 5cm 12.5cm, clip, width=0.19\textwidth]{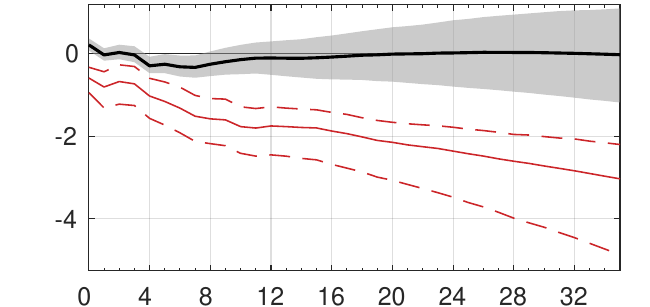} &
\includegraphics[trim=5cm 12cm 5cm 12.5cm, clip, width=0.19\textwidth]{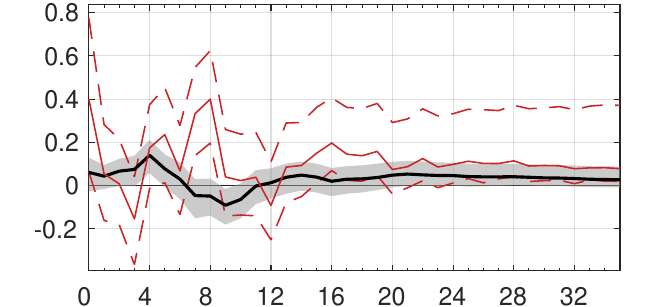} &
\includegraphics[trim=5cm 12cm 5cm 12.5cm, clip, width=0.19\textwidth]{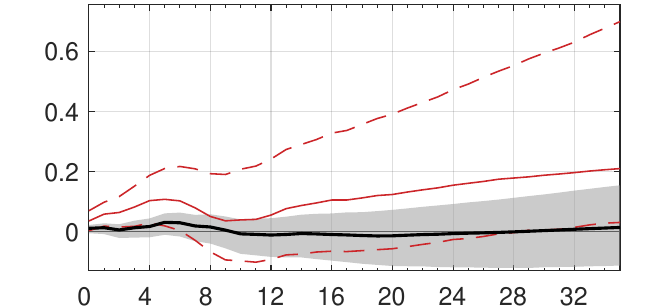} \\

\raisebox{1.7\height}{\rotatebox{90}{\scriptsize EL}} &
\includegraphics[trim=5cm 12cm 5cm 12.5cm, clip, width=0.19\textwidth]{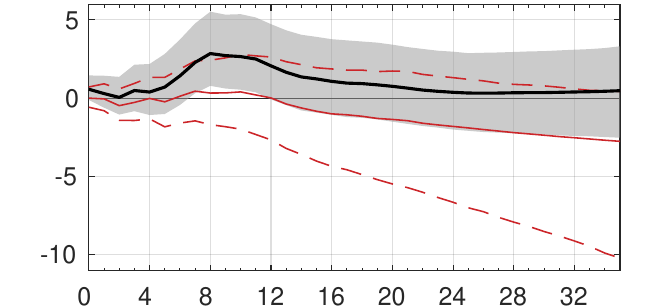} &
\includegraphics[trim=5cm 12cm 5cm 12.5cm, clip, width=0.19\textwidth]{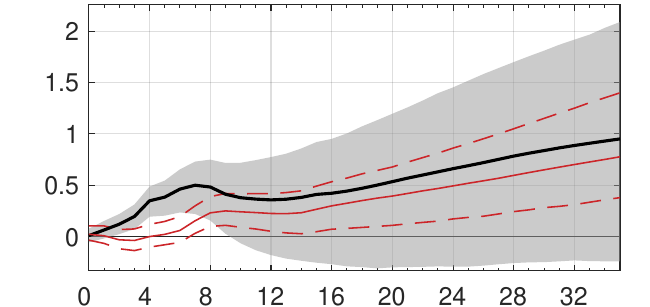} &
\includegraphics[trim=5cm 12cm 5cm 12.5cm, clip, width=0.19\textwidth]{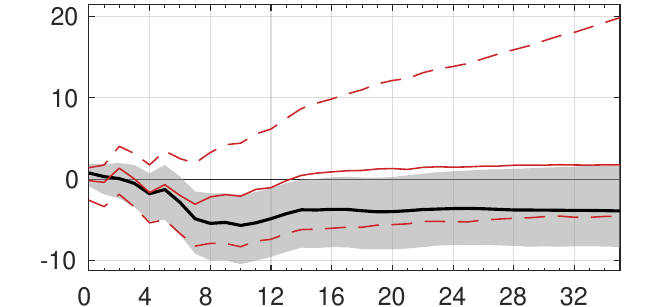} &
\includegraphics[trim=5cm 12cm 5cm 12.5cm, clip, width=0.19\textwidth]{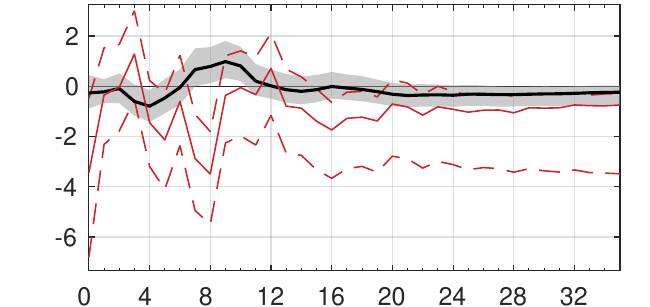} &
\includegraphics[trim=5cm 12cm 5cm 12.5cm, clip, width=0.19\textwidth]{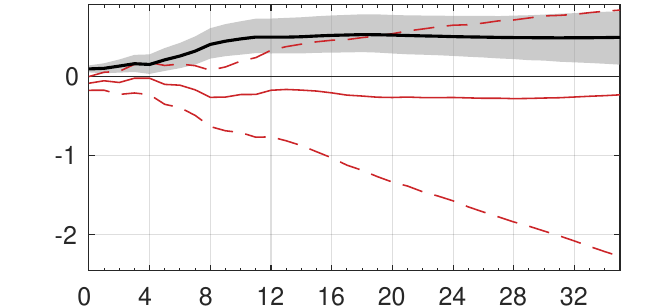} \\

\raisebox{1.7\height}{\rotatebox{90}{\scriptsize ES}} &
\includegraphics[trim=5cm 12cm 5cm 12.5cm, clip, width=0.19\textwidth]{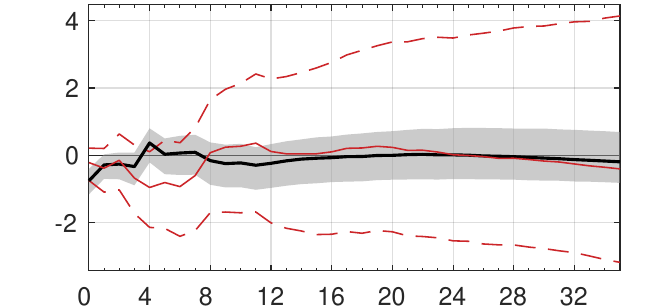} &
\includegraphics[trim=5cm 12cm 5cm 12.5cm, clip, width=0.19\textwidth]{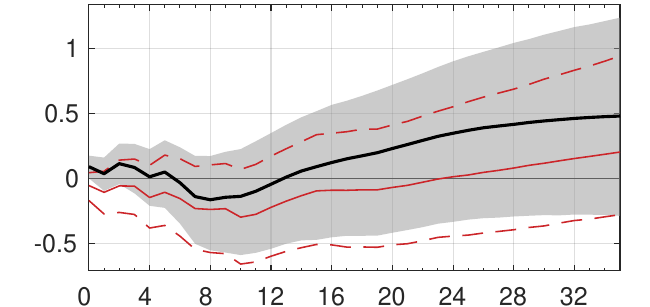} &
\includegraphics[trim=5cm 12cm 5cm 12.5cm, clip, width=0.19\textwidth]{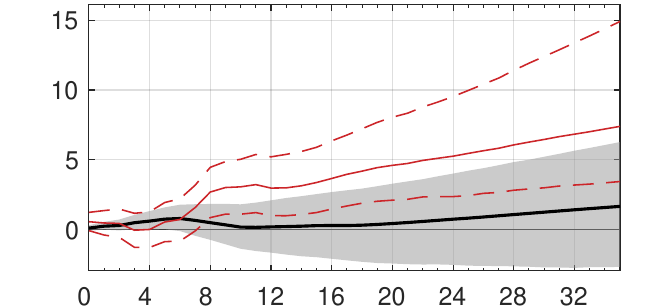} &
\includegraphics[trim=5cm 12cm 5cm 12.5cm, clip, width=0.19\textwidth]{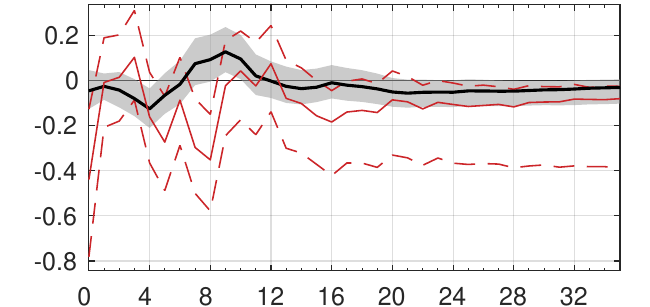} &
\includegraphics[trim=5cm 12cm 5cm 12.5cm, clip, width=0.19\textwidth]{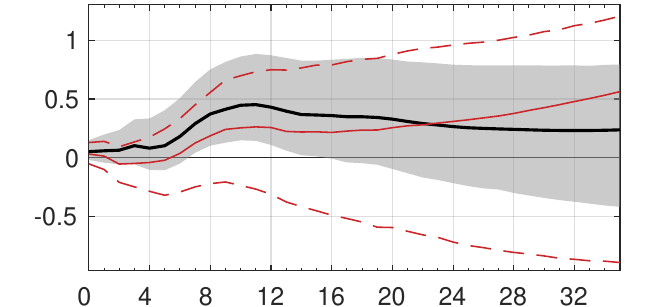} \\

\raisebox{2\height}{\rotatebox{90}{\scriptsize IE}} &
 &
\includegraphics[trim=5cm 12cm 5cm 12.5cm, clip, width=0.19\textwidth]{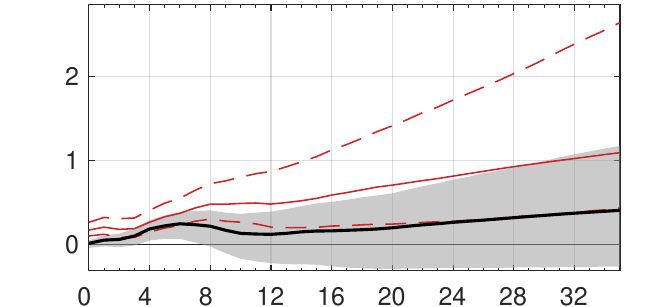} &
\includegraphics[trim=5cm 12cm 5cm 12.5cm, clip, width=0.19\textwidth]{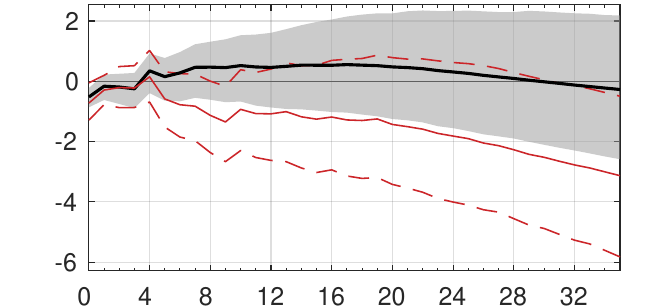} &
\includegraphics[trim=5cm 12cm 5cm 12.5cm, clip, width=0.19\textwidth]{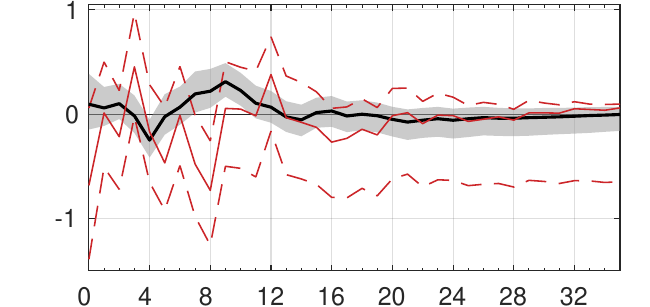} &
\includegraphics[trim=5cm 12cm 5cm 12.5cm, clip, width=0.19\textwidth]{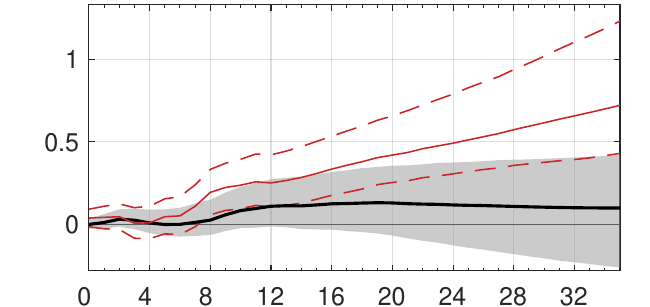} \\

\raisebox{2\height}{\rotatebox{90}{\scriptsize IT}} &
\includegraphics[trim=5cm 12cm 5cm 12.5cm, clip, width=0.19\textwidth]{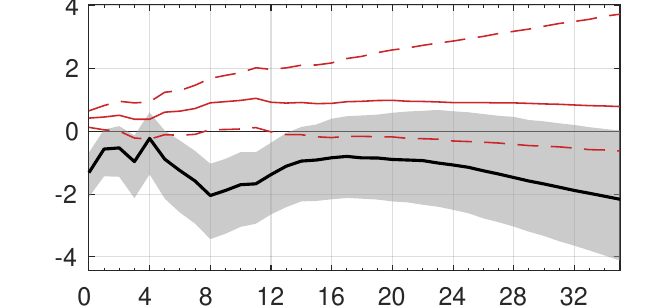} &
\includegraphics[trim=5cm 12cm 5cm 12.5cm, clip, width=0.19\textwidth]{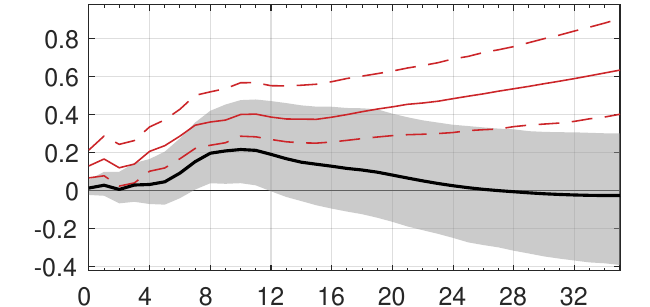} &
\includegraphics[trim=5cm 12cm 5cm 12.5cm, clip, width=0.19\textwidth]{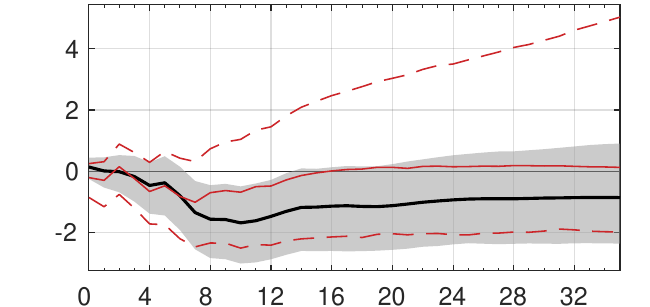} &
\includegraphics[trim=5cm 12cm 5cm 12.5cm, clip, width=0.19\textwidth]{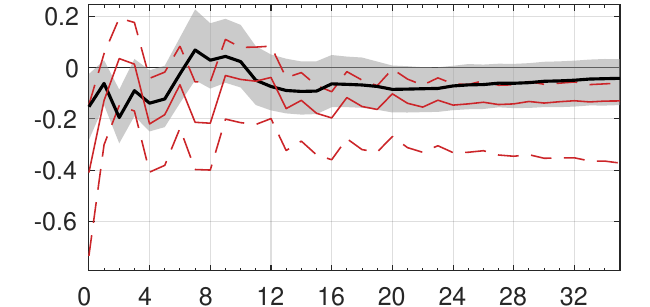} &
\includegraphics[trim=5cm 12cm 5cm 12.5cm, clip, width=0.19\textwidth]{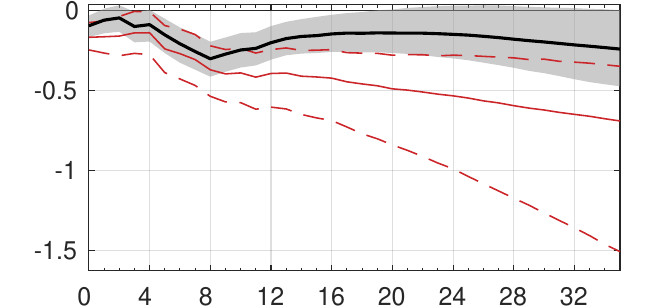} \\

\raisebox{1.5\height}{\rotatebox{90}{\scriptsize PT}} &
\includegraphics[trim=5cm 12cm 5cm 12.5cm, clip, width=0.19\textwidth]{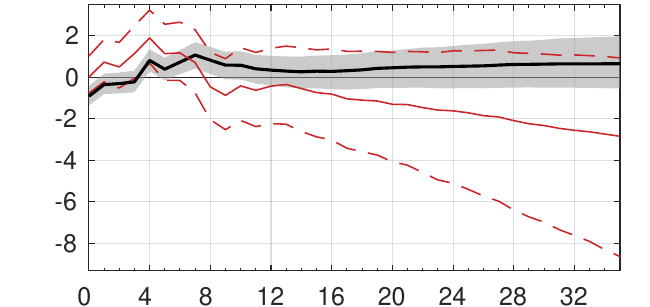} &
\includegraphics[trim=5cm 12cm 5cm 12.5cm, clip, width=0.19\textwidth]{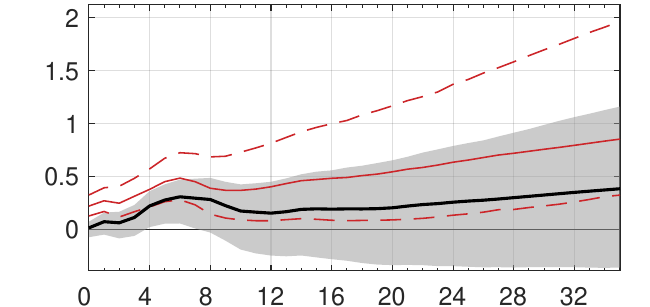} &
\includegraphics[trim=5cm 12cm 5cm 12.5cm, clip, width=0.19\textwidth]{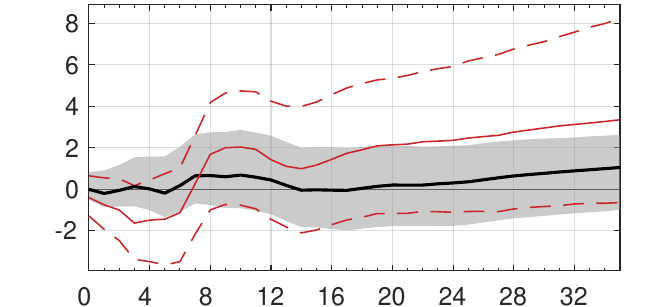} &
\includegraphics[trim=5cm 12cm 5cm 12.5cm, clip, width=0.19\textwidth]{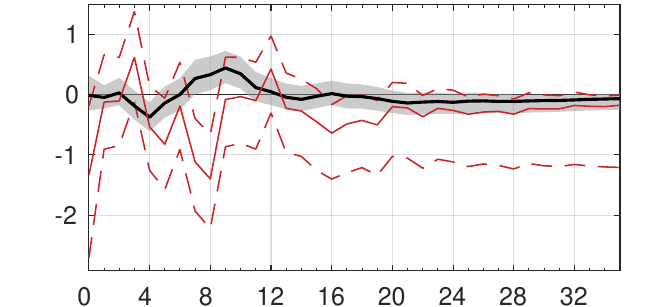} &
\includegraphics[trim=5cm 12cm 5cm 12.5cm, clip, width=0.19\textwidth]{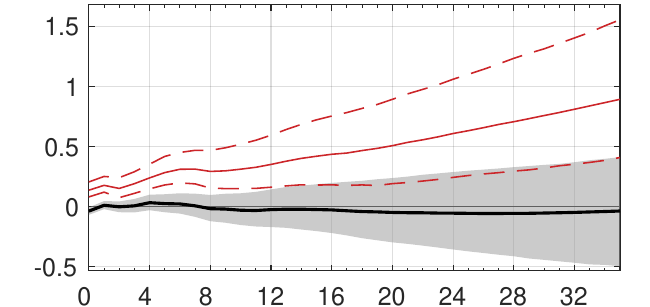} \\
\end{tabular}
\begin{tabular}{p{1\textwidth}} \scriptsize Notes: \rm Each sub-figure plots the difference between the country-level IRF and the corresponding EA counterpart for one variable and one country. Each column of the graph represents a variable, while each row represents a country. The variables considered are: Industrial Production: Manufacturing (IPMN), HICP: Overall (HICPOV), Stock Price Index (SHIX), 10-years Interest Rates (LTIRT) and Unemployment Rate (UNETOT). The black solid line is the point estimate in our baseline setting, while the red solid line is the point estimate obtained with data truncated in 2019:M12. The gray shaded area and red dotted lines are the corresponding 68\% confidence intervals. The scale in the vertical axis differs across variables and countries.
\end{tabular}
\label{fig::diffIRF19}
\end{figure}

\section{Comparison with alternative estimation approaches}
\label{app::altest}
\subsection{VAR}

\begin{table}[h!]
\centering
\caption{Components of $\mathbf Y_t$ in the VAR}\label{tab::VAR}
\footnotesize
\begin{tabular}{l l l}
\hline
\hline
notation & ID & name \\
\hline
${R}_t$ & - & EA 2-years Interest Rate \\
$x_{{\rm IPMN\, EA},t}$ & IPMN & EA IP: Manufacturing \\
$x_{{\rm HICPOV\,EA},t}$ & HICPOV & EA HICP: Overall \\
$x_{{\rm LTIRT\,EA},t}$ & LTIRT & EA 10-years Interest Rate \\
$x_{{\rm SHIX\, EA},t}$ & SHIX & EA Stock Price Index \\
$x_{{\rm UNETOT\, EA},t}$ & UNETOT & EA Unemployment Rate \\
$x_{{\rm nat.},t}$ & - & National variable \\
\hline
\hline
\end{tabular}
\end{table}

\label{sbsec::VAR}
\begin{figure}[H]
\centering \footnotesize \sc \smallskip
\setlength{\tabcolsep}{.005\textwidth}
\caption{EA IRFs: CC-VAR vs VAR}
\begin{tabular}{ccc}
\scriptsize 2-years Interest Rate ($R_t$) &\scriptsize  IP: Manufacturing (IPMN) &\scriptsize  HICP: Overall (HICPOV)\\
\includegraphics[trim= .5cm 8cm .5cm 8.5cm, clip, width=0.32\textwidth]{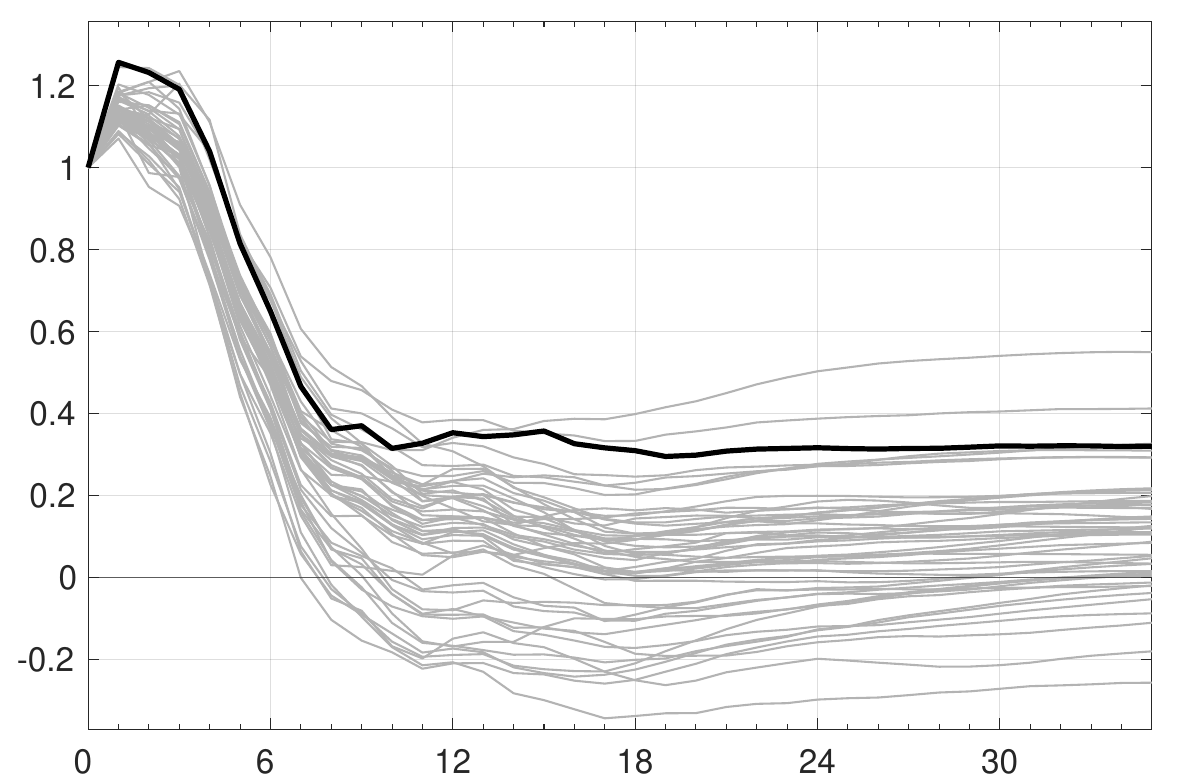} &
\includegraphics[trim= .5cm 8cm .5cm 8.5cm, clip, width=0.32\textwidth]{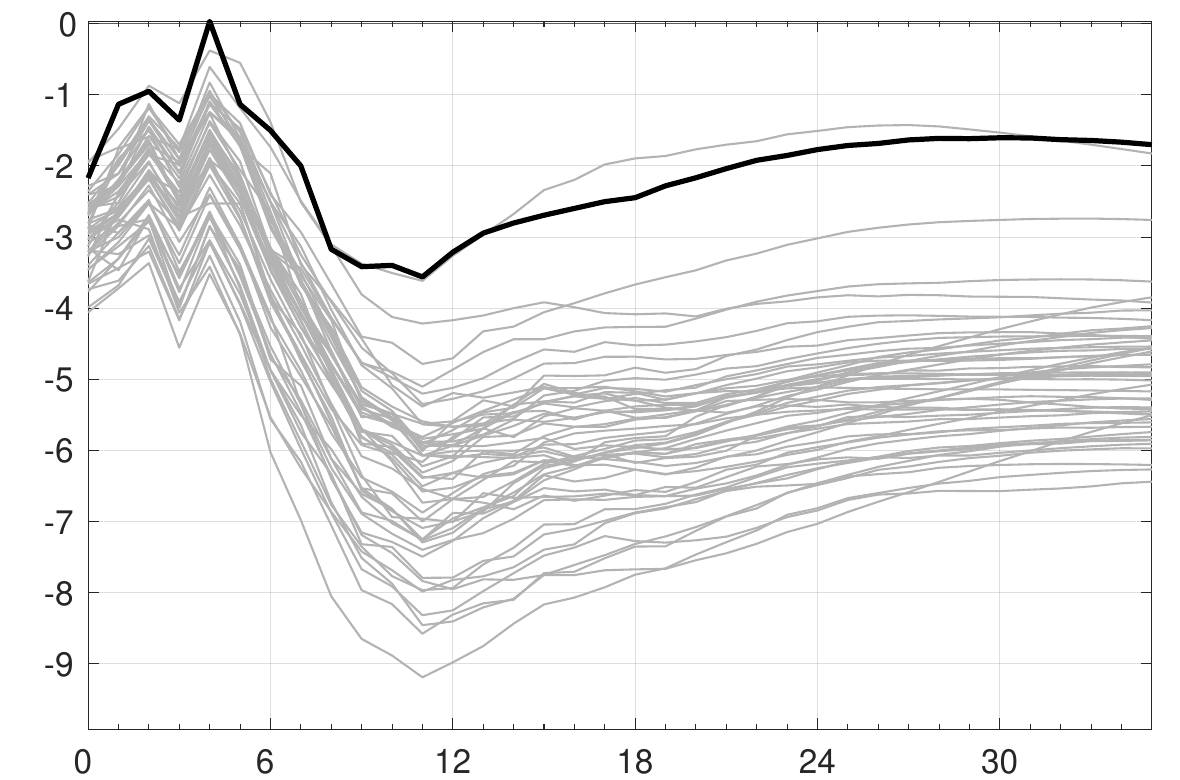} &
\includegraphics[trim= .5cm 8cm .5cm 8.5cm, clip, width=0.32\textwidth]{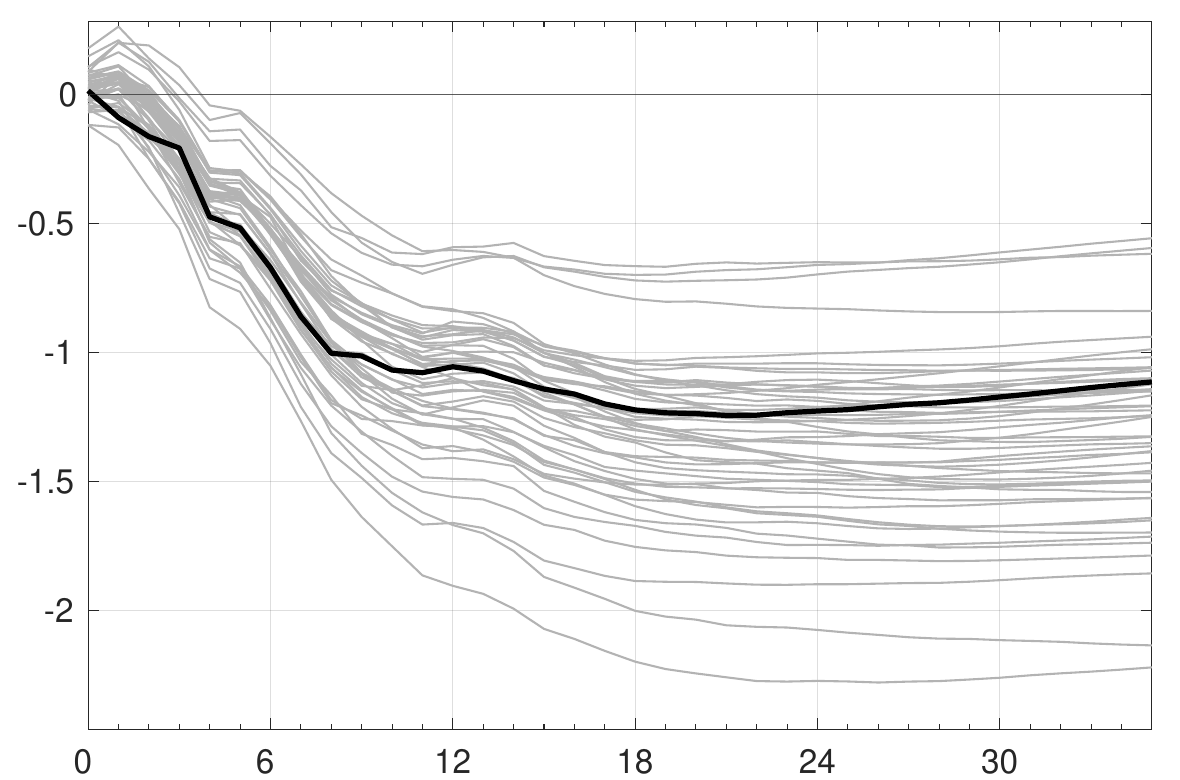} \\
\scriptsize 10-years Interest Rate (LTIRT) &\scriptsize Stock Price Index (SHIX) &\scriptsize Unemployment Rate (UNETOT) \\
\includegraphics[trim= .5cm 8cm .5cm 8.5cm, clip, width=0.32\textwidth]{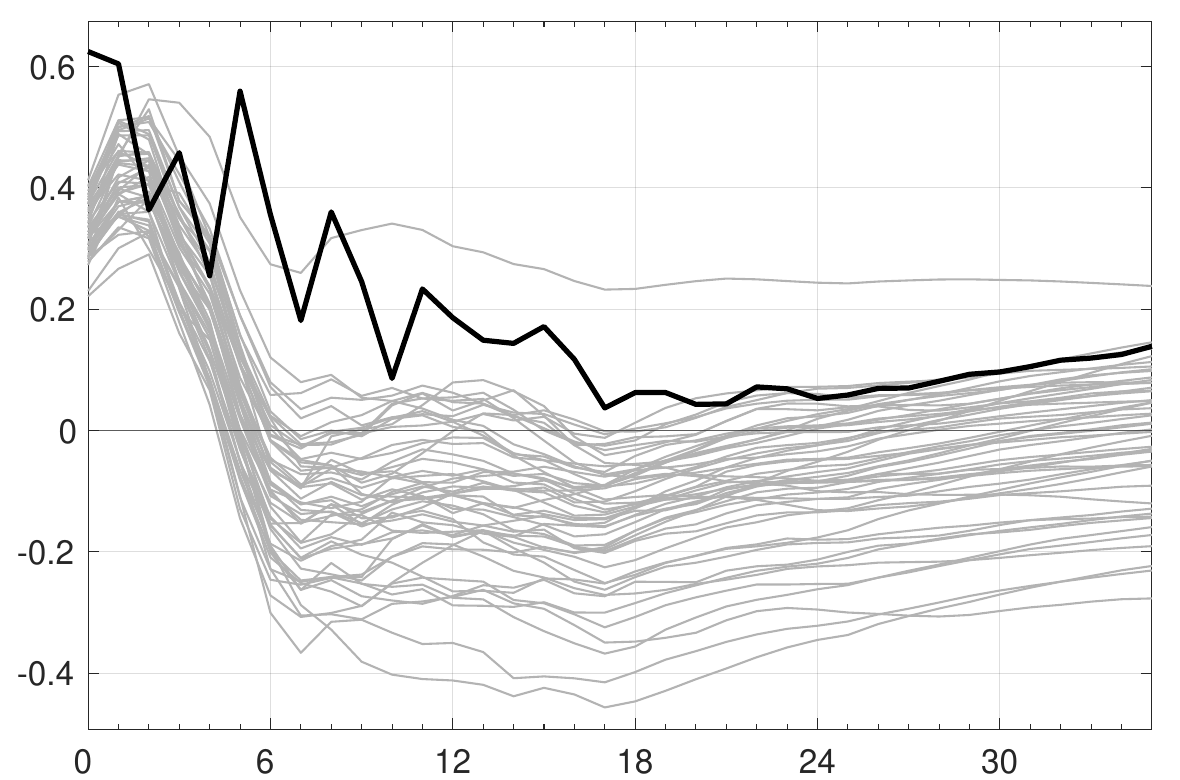} &
\includegraphics[trim= .5cm 8cm .5cm 8.5cm, clip, width=0.32\textwidth]{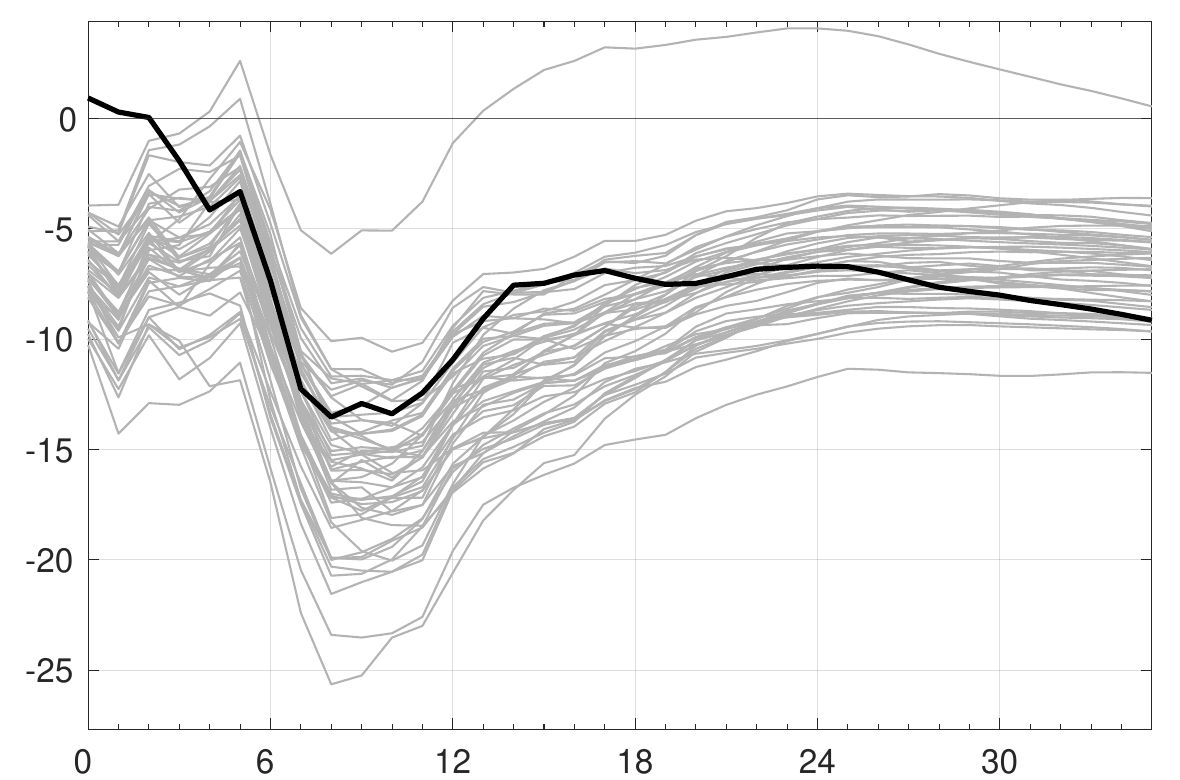} &
\includegraphics[trim= .5cm 8cm .5cm 8.5cm, clip, width=0.32\textwidth]{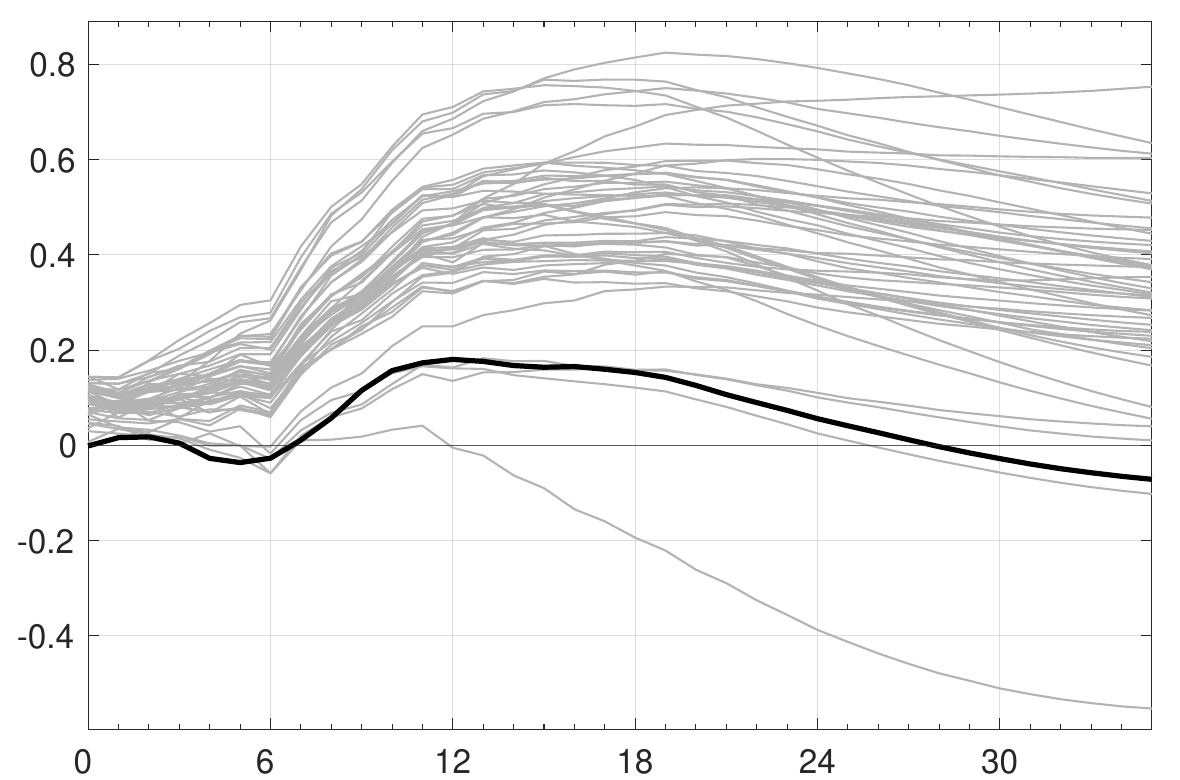} \\
\end{tabular}
\begin{tabular}{p{.98\textwidth}} \scriptsize Notes: \rm Each sub-figure plots the impulse response of one EA variable to a 100bps contractionary monetary policy shock. The black solid line is the point estimate obtained with the CC-VAR, which is the same regardless of the set of common components included in the model. The thin gray lines are the point estimates obtained with the 49 considered VAR models differing only for the national variable included.
\end{tabular}
\label{fig::EAirfs_VAR}
\end{figure}

\newpage

\subsection{FAVAR}
\label{sbsec::FAVAR}

\begin{table}[h!]
\centering
\caption{Components of $\mathbf Y_t$ in the FAVAR}\label{tab::FAVAR}
\footnotesize
\begin{tabular}{l l l}
\hline
\hline
notation & ID & name \\
\hline
${R}_t$ & - & EA 2-years Interest Rate \\
$x_{{\rm IPMN\, EA},t}$ & IPMN & EA IP: Manufacturing \\
$x_{{\rm HICPOV\,EA},t}$ & HICPOV & EA HICP: Overall \\
$x_{{\rm LTIRT\,EA},t}$ & LTIRT & EA 10-years Interest Rate \\
$x_{{\rm SHIX\, EA},t}$ & SHIX & EA Stock Price Index \\
$x_{{\rm UNETOT\, EA},t}$ & UNETOT & EA Unemployment Rate \\
$\widehat{f}_{1t}$&-& First estimated factor\\
$x_{{\rm nat.},t}$ & - & National variable \\
\hline
\hline
\end{tabular}
\end{table}

\begin{figure}[H]
\centering \footnotesize \sc \smallskip
\setlength{\tabcolsep}{.005\textwidth}
\caption{EA IRFs: CC-VAR vs FAVAR}
\begin{tabular}{ccc}
\scriptsize 2-years Interest Rate ($R_t$) &\scriptsize  IP: Manufacturing (IPMN) &\scriptsize  HICP: Overall (HICPOV)\\
\includegraphics[trim= .5cm 8cm .5cm 8.5cm, clip, width=0.32\textwidth]{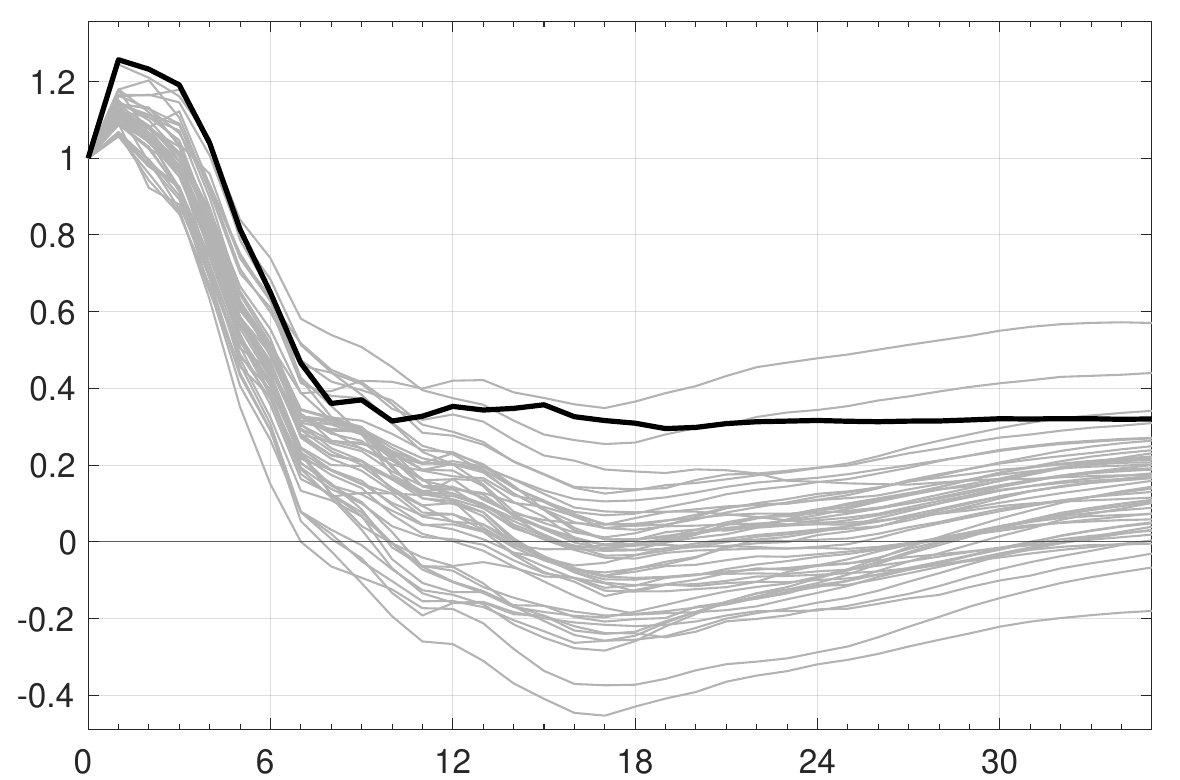} &
\includegraphics[trim= .5cm 8cm .5cm 8.5cm, clip, width=0.32\textwidth]{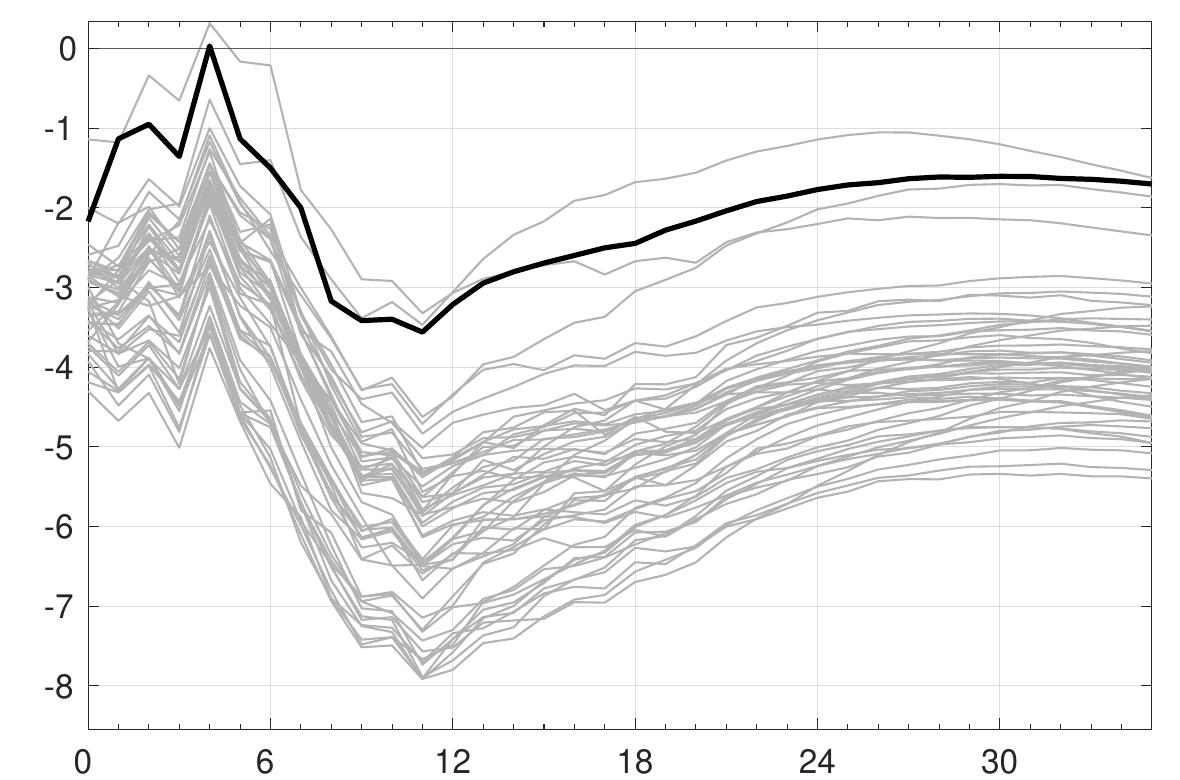} &
\includegraphics[trim= .5cm 8cm .5cm 8.5cm, clip, width=0.32\textwidth]{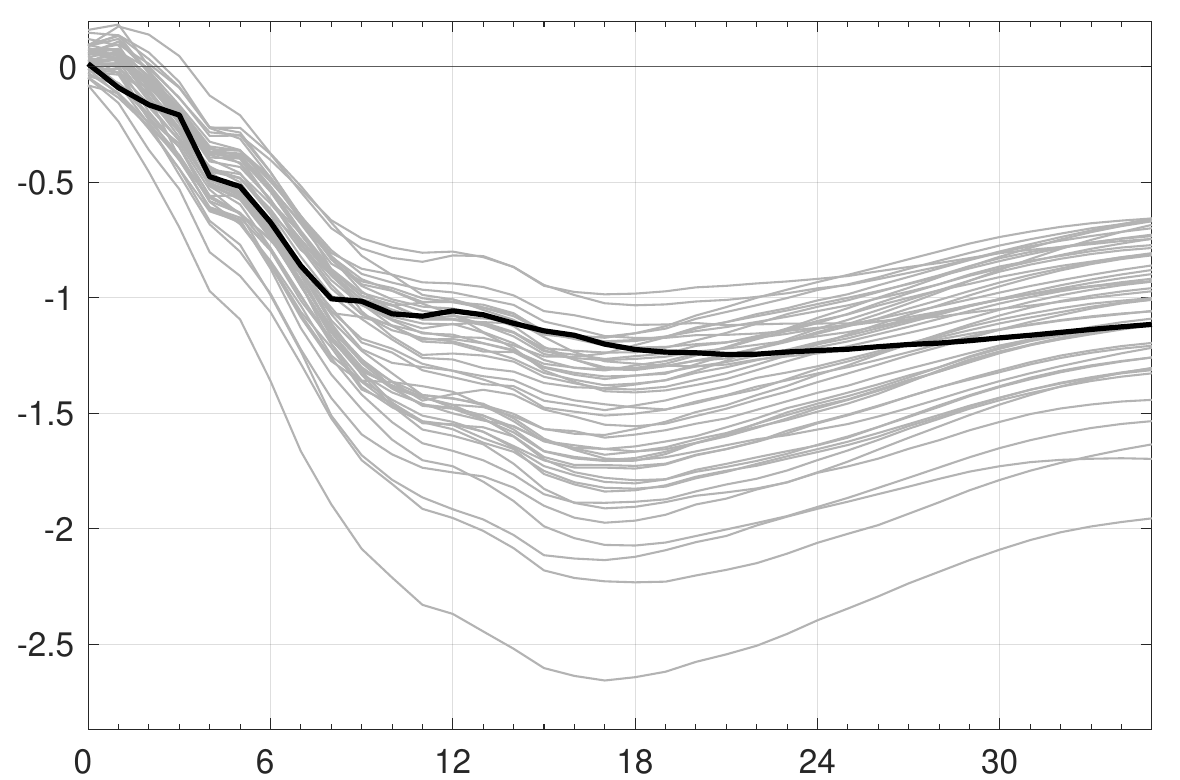} \\
\scriptsize 10-years Interest Rate (LTIRT) &\scriptsize Stock Price Index (SHIX) &\scriptsize Unemployment Rate (UNETOT) \\
\includegraphics[trim= .5cm 8cm .5cm 8.5cm, clip, width=0.32\textwidth]{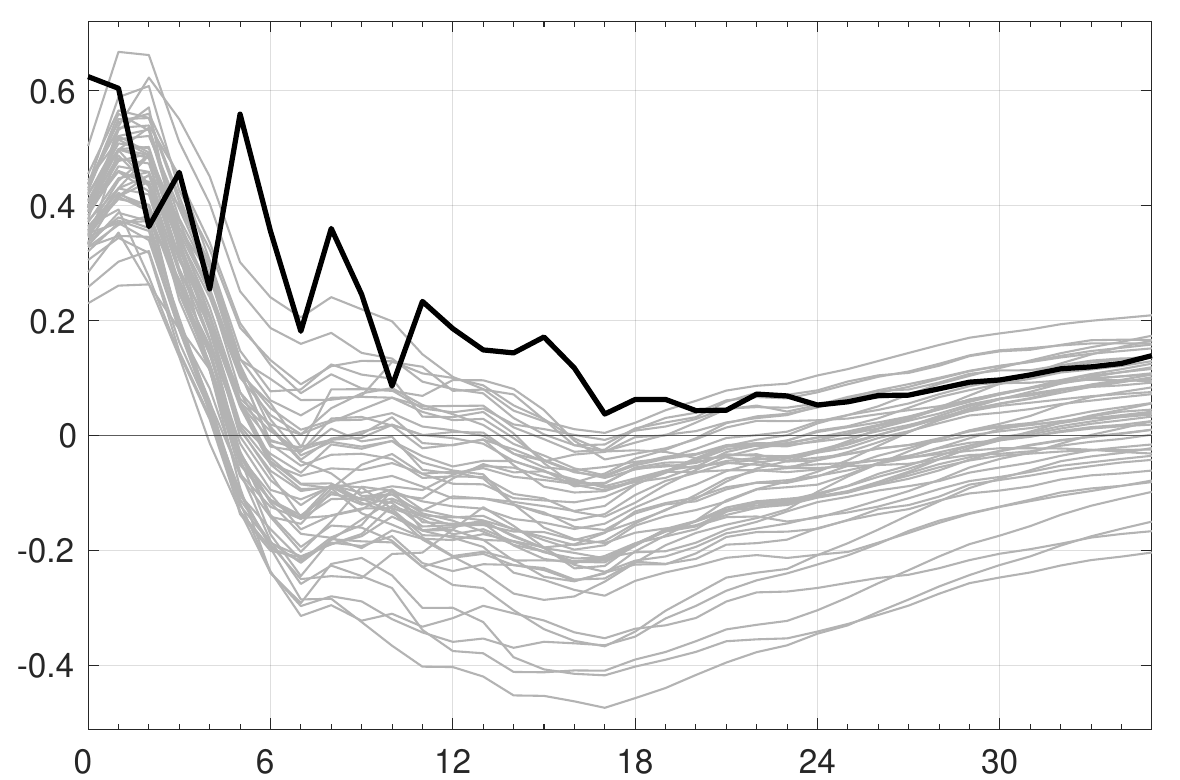} &
\includegraphics[trim= .5cm 8cm .5cm 8.5cm, clip, width=0.32\textwidth]{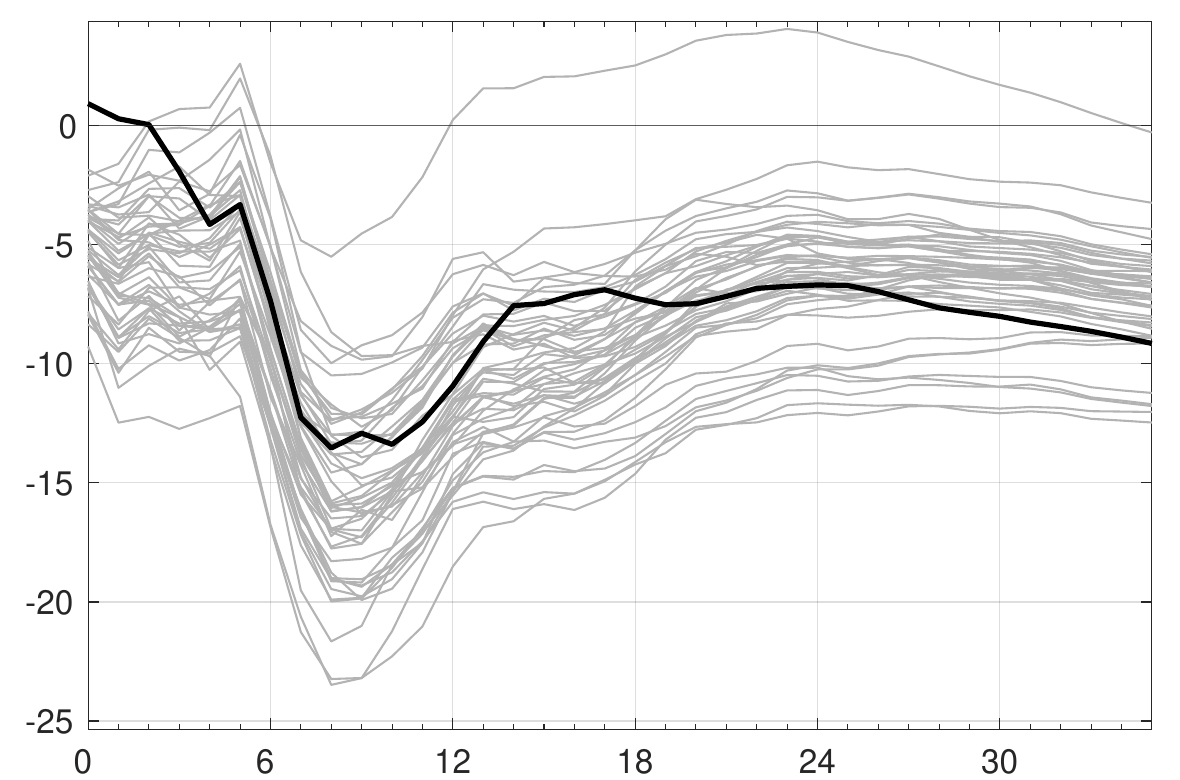} &
\includegraphics[trim= .5cm 8cm .5cm 8.5cm, clip, width=0.32\textwidth]{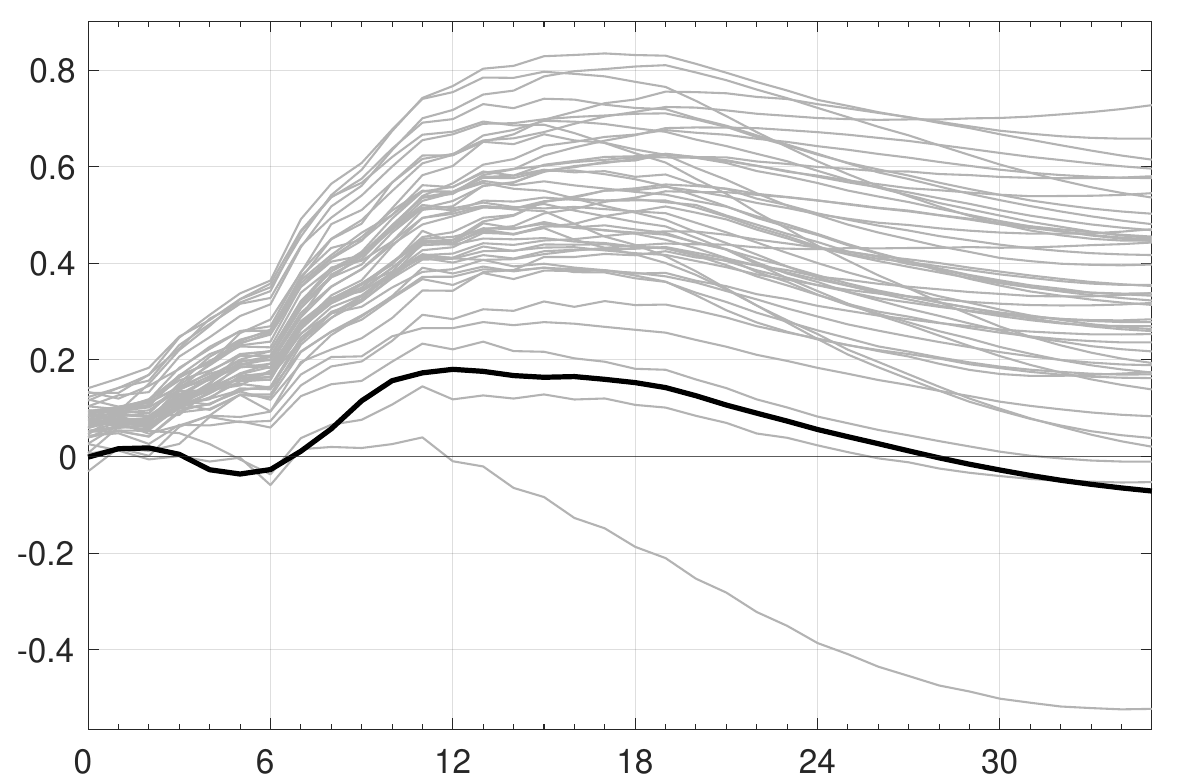} \\
\end{tabular}
\begin{tabular}{p{.98\textwidth}} \scriptsize Notes: \rm Each sub-figure plots the impulse response of one EA variable to a 100bps contractionary monetary policy shock. The black solid line is the point estimate obtained with the CC-VAR, which is the same regardless of the set of common components included in the model. The thin gray lines are the point estimates obtained with the 49 considered FAVAR models differing only for the national variable included.
\end{tabular}
\label{fig::EAirfs_FAVAR}
\end{figure}

\newpage

\section{Alternative transformations: statistical transformations}
\label{app::TR1}
\subsection{EA IRFs}
\begin{figure}[H]
\centering \footnotesize \sc \smallskip
\setlength{\tabcolsep}{.005\textwidth}
\caption{EA IRFs: baseline vs alternative set of transformations}
\begin{tabular}{ccc}
\scriptsize 2-years Interest Rate ($R_t$) &\scriptsize  IP: Manufacturing (IPMN) &\scriptsize  HICP: Overall (HICPOV)\\
\includegraphics[trim= .5cm 8cm .5cm 8.5cm, clip, width=0.32\textwidth]{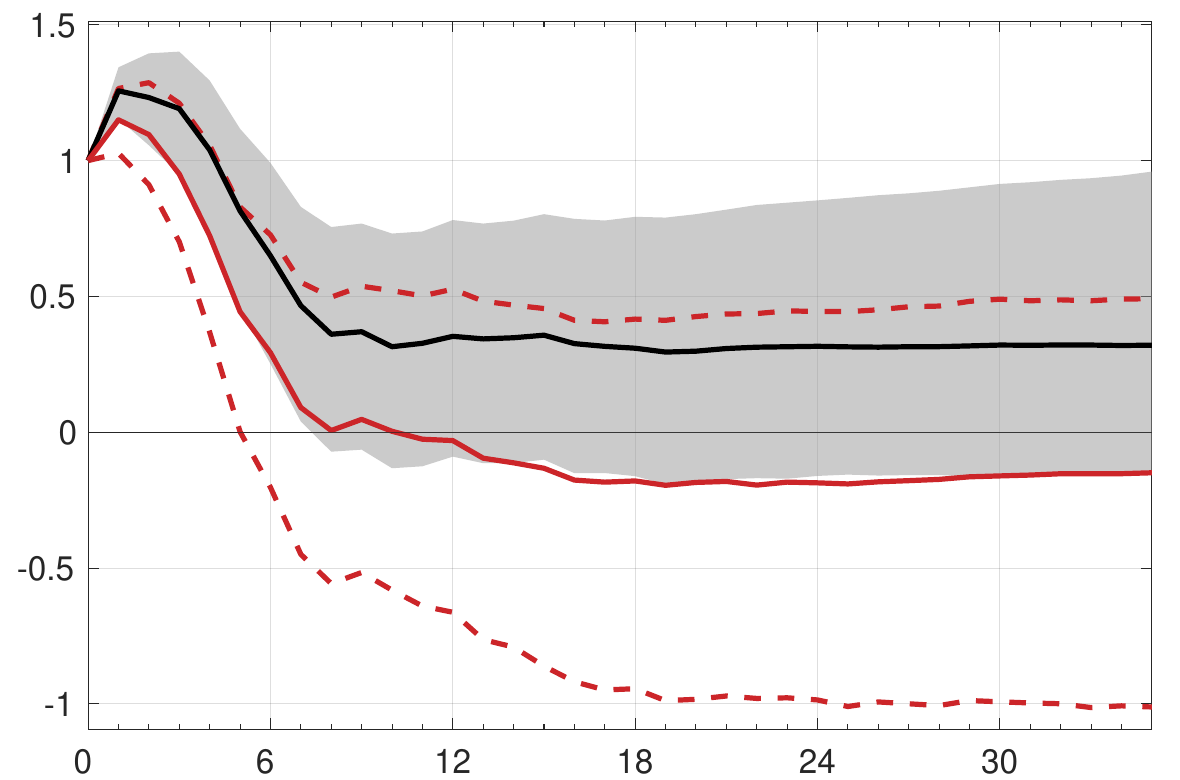} &
\includegraphics[trim= .5cm 8cm .5cm 8.5cm, clip, width=0.32\textwidth]{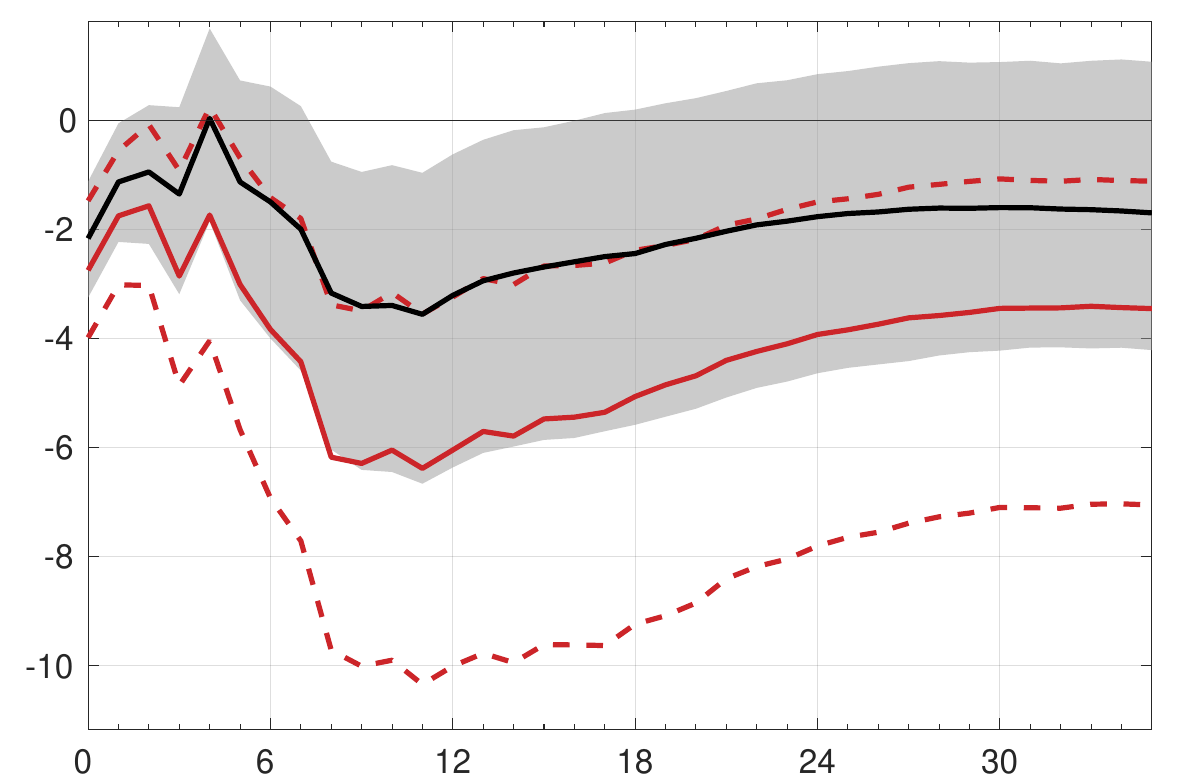} &
\includegraphics[trim= .5cm 8cm .5cm 8.5cm, clip, width=0.32\textwidth]{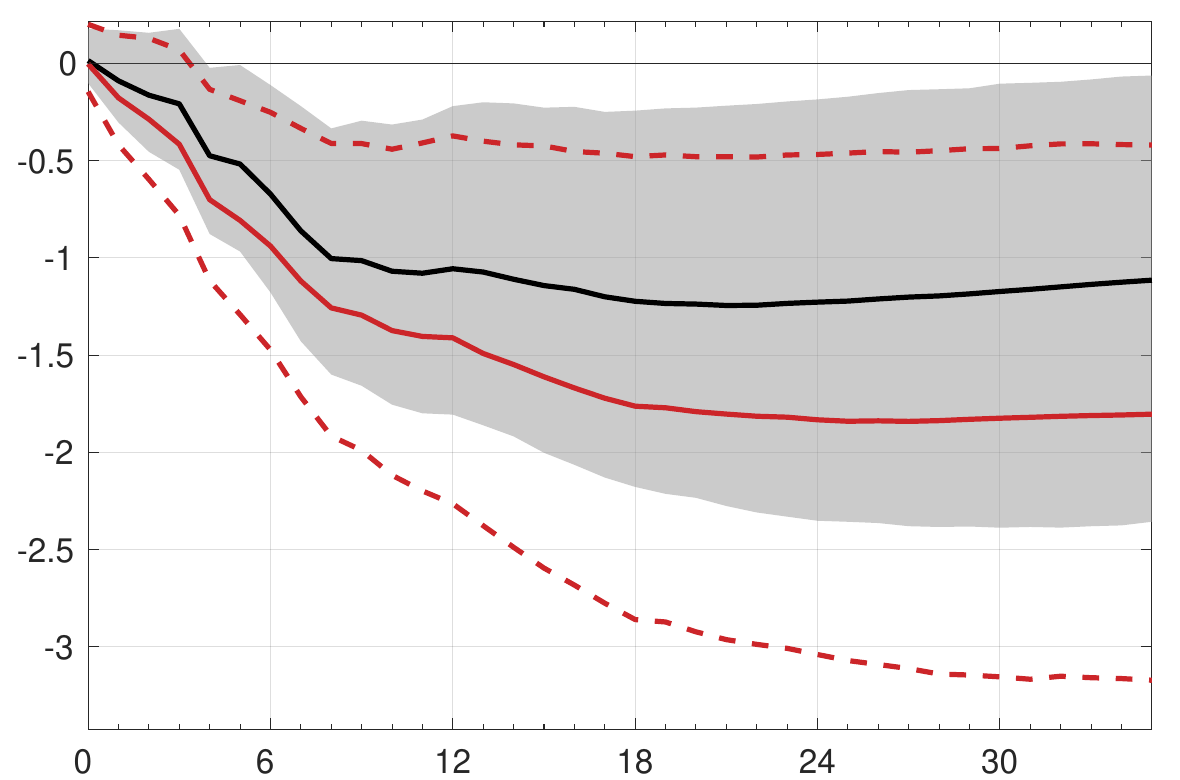} \\
\scriptsize 10-years Interest Rate (LTIRT) &\scriptsize Stock Price Index (SHIX) &\scriptsize Unemployment Rate (UNETOT) \\
\includegraphics[trim= .5cm 8cm .5cm 8.5cm, clip, width=0.32\textwidth]{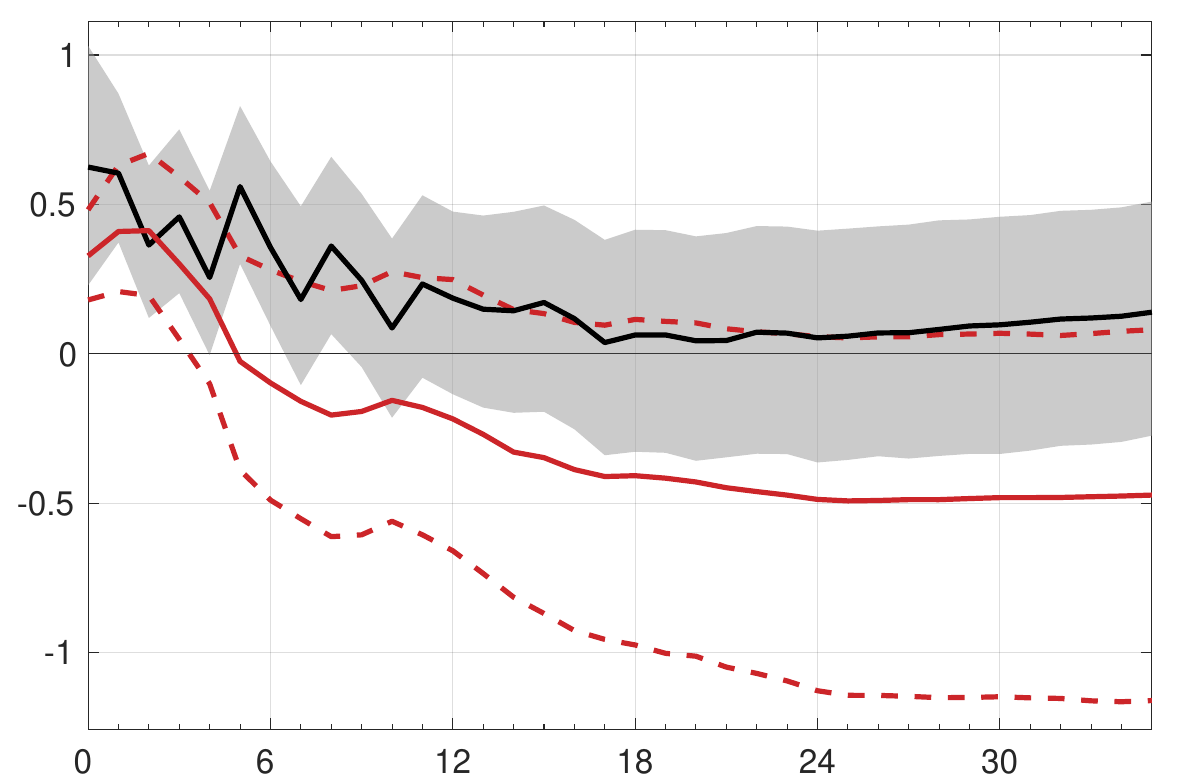} &
\includegraphics[trim= .5cm 8cm .5cm 8.5cm, clip, width=0.32\textwidth]{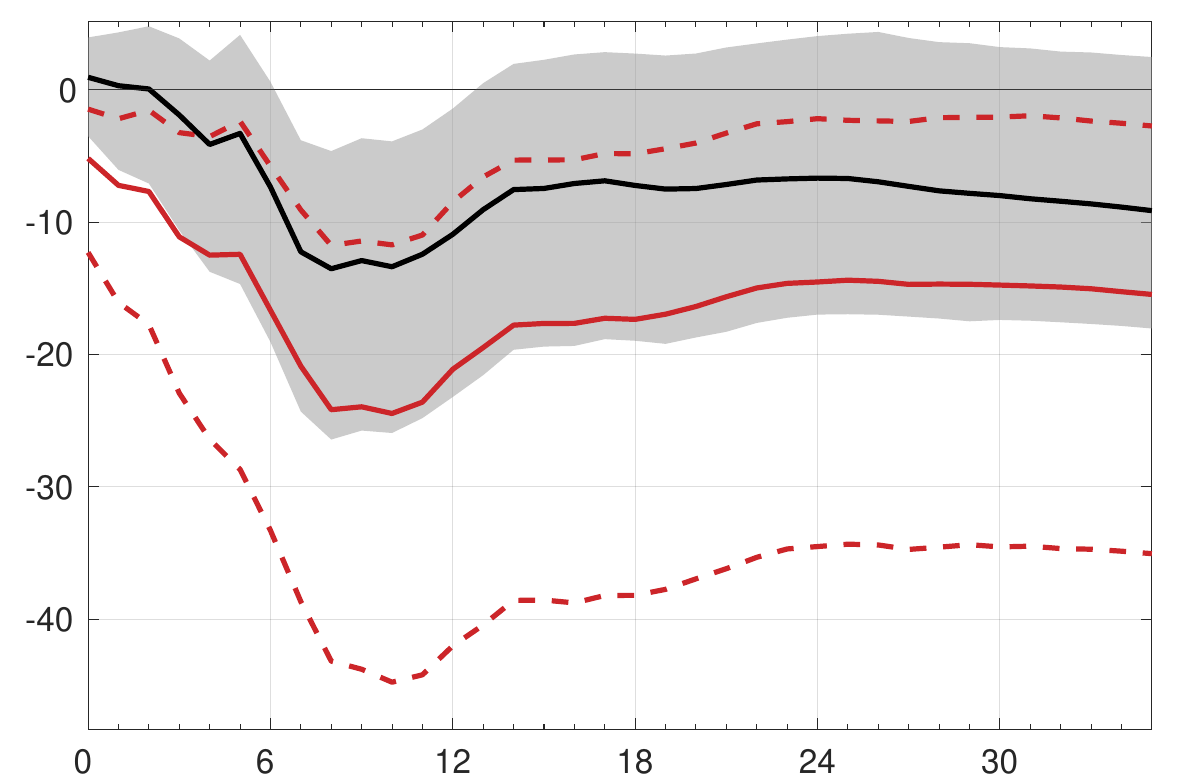} &
\includegraphics[trim= .5cm 8cm .5cm 8.5cm, clip, width=0.32\textwidth]{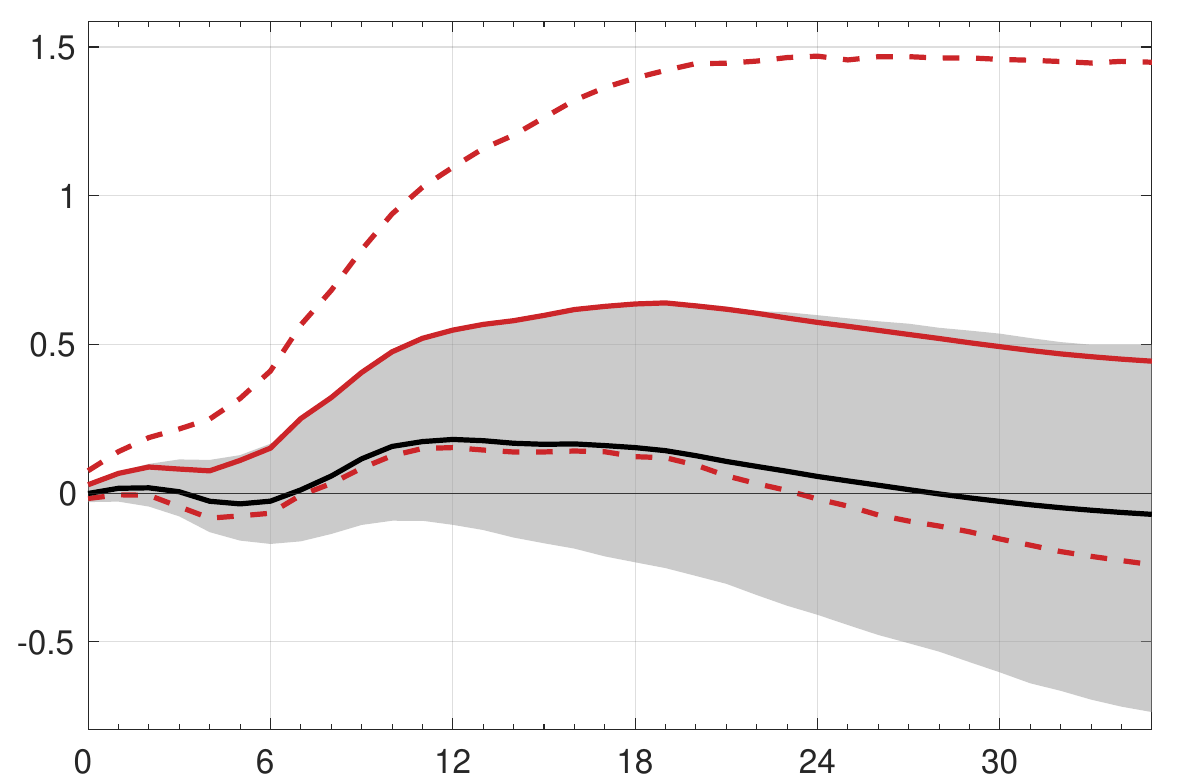} \\
\end{tabular}
\begin{tabular}{p{.98\textwidth}} \scriptsize Notes: \rm Each sub-figure plots the impulse response of one EA variable to a 100bps contractionary monetary policy shock. The black solid line is the point estimate in our baseline setting, while the red solid line is the point estimate obtained with the alternative set of transformations based on statistical criteria. The gray shaded area and red dotted lines are the corresponding 68\% confidence intervals.
\end{tabular}
\label{fig::EAirfs_TR1}
\end{figure}

\subsection{Country-level IRFs}
\begin{figure}[H]
\centering \footnotesize \sc \smallskip
\setlength{\tabcolsep}{.005\textwidth}
\caption{Country-level IRFs:  statistical transformations}
\begin{tabular}{ccc}
\scriptsize IP: Manufacturing (IPMN)&\scriptsize HICP: Overall (HICPOV) &\scriptsize Stock Price Index (SHIX)  \\
\includegraphics[trim= .5cm 8cm .5cm 8.5cm, clip, width=0.32\textwidth]{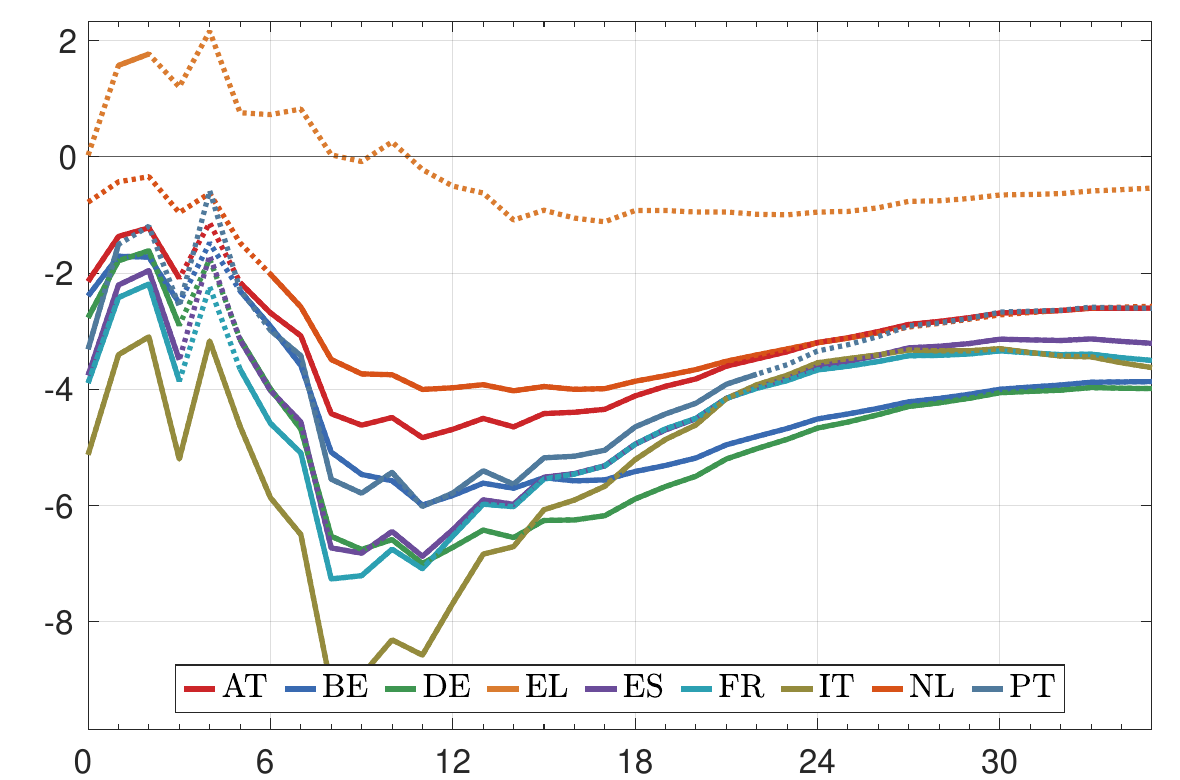} &
\includegraphics[trim= .5cm 8cm .5cm 8.5cm, clip, width=0.32\textwidth]{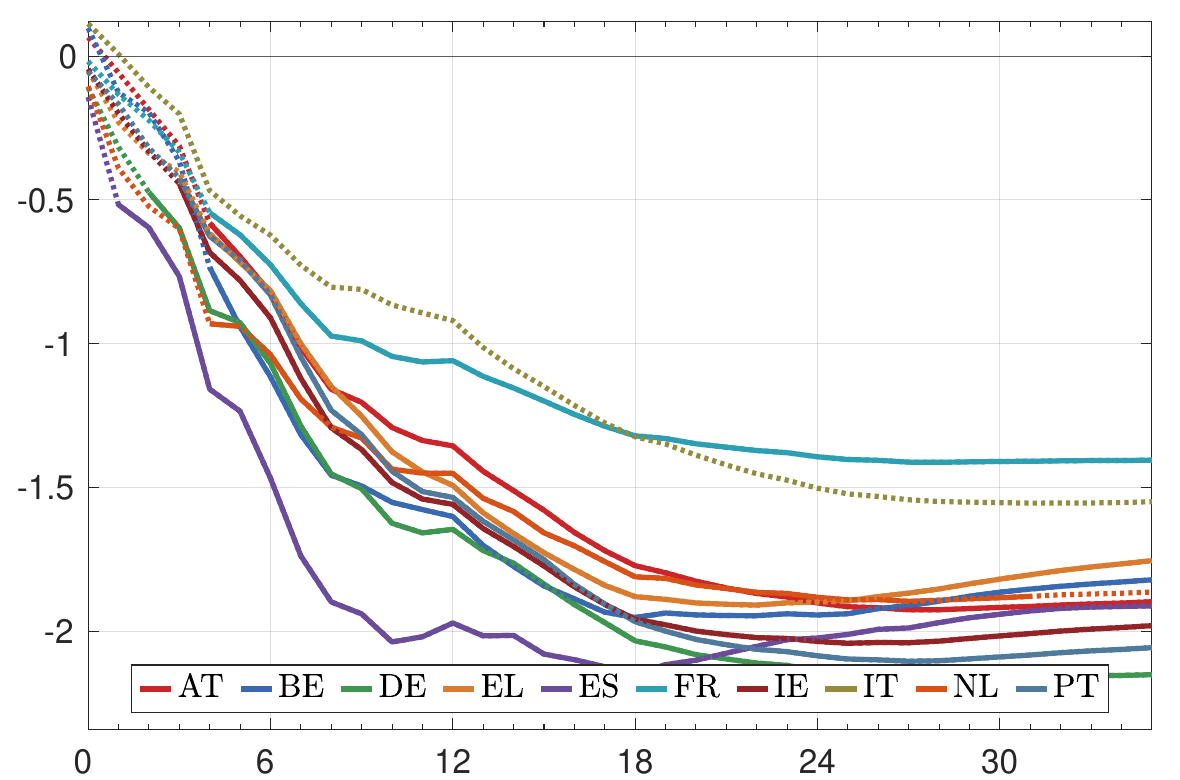} &
\includegraphics[trim= .5cm 8cm .5cm 8.5cm, clip, width=0.32\textwidth]{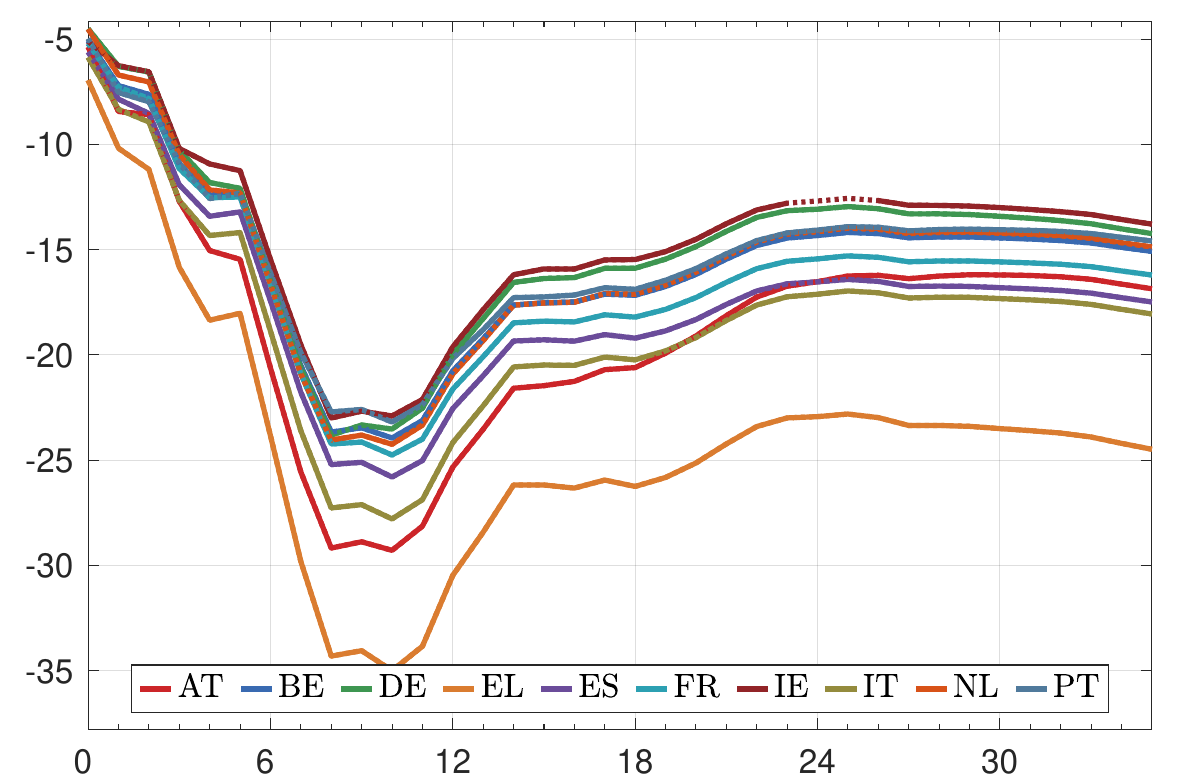} \\[10pt]
\end{tabular}

\begin{tabular}{cc}
\scriptsize10-years Interest Rate (LTIRT) &\scriptsize Unemployment Rate (UNETOT)\\

\includegraphics[trim= .5cm 8cm .5cm 8.5cm, clip, width=0.32\textwidth]{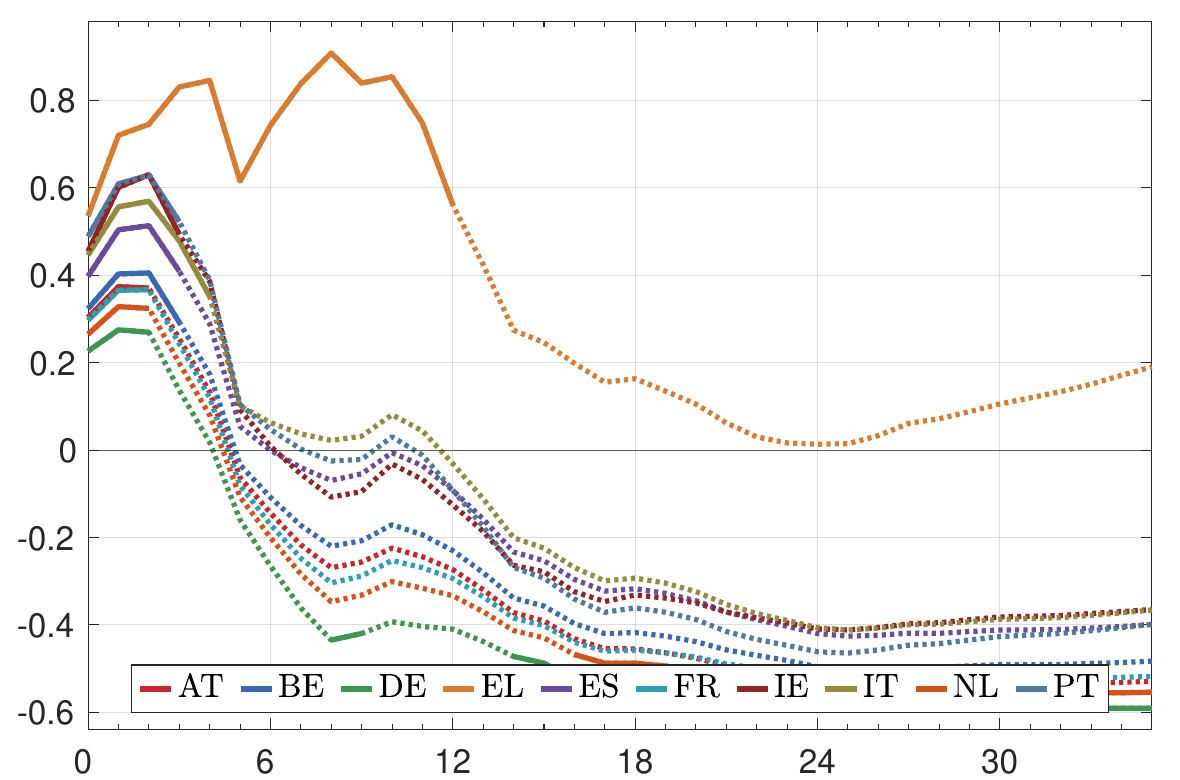} 
&
\includegraphics[trim= .5cm 8cm .5cm 8.5cm, clip, width=0.32\textwidth]{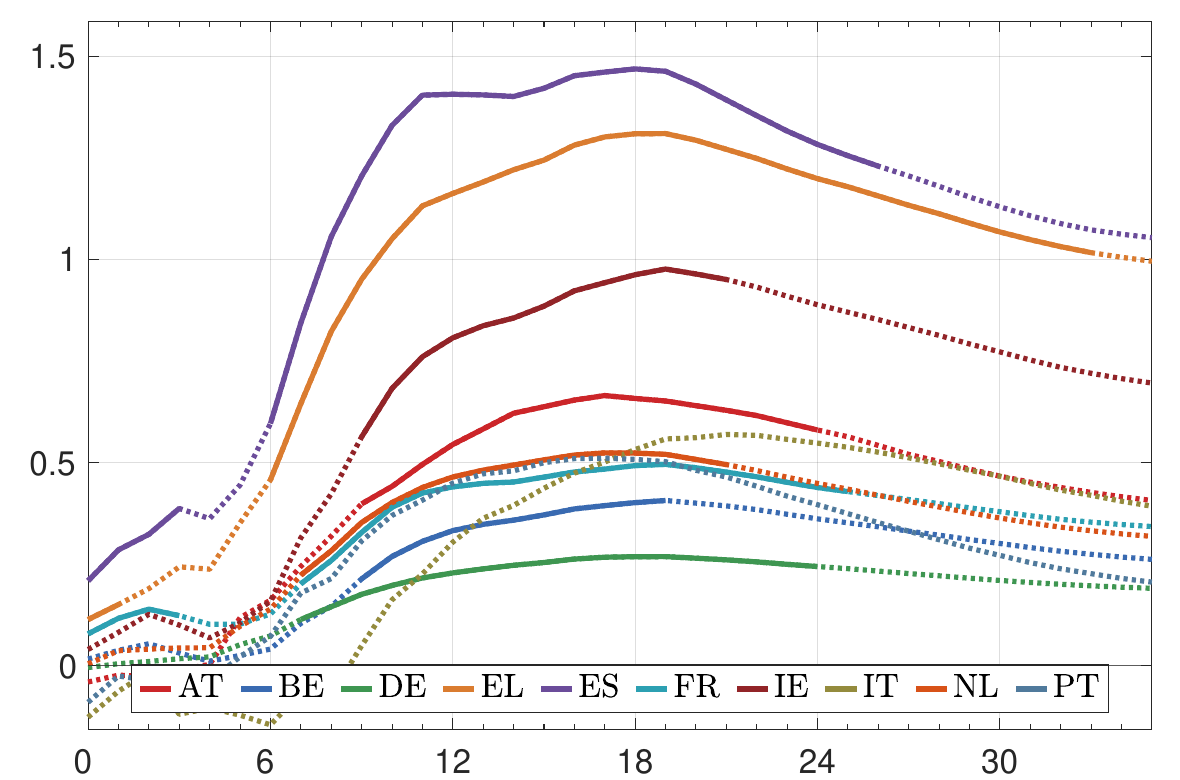}
\end{tabular}
\begin{tabular}{p{.98\textwidth}} \scriptsize Notes: \rm Each sub-figure plots the impulse responses, for all countries, of one variable to a 100bps contractionary monetary policy shock. Within each sub-figure, at each horizon $h=0,\ldots,36$ the country-level impulse responses are denoted with a solid line if the IRF is statistically significant at the 68\% level at that horizon, and with a dotted line otherwise. 
\end{tabular}
\label{fig:national_irf_monthly}
\end{figure}

\newpage
\begin{figure}[H]
\centering
\setlength{\tabcolsep}{.005\textwidth}
\centering \footnotesize \sc \smallskip
\caption{Difference between country-level and EA IRFs: baseline vs. statistical transformations}
\begin{tabular}{lccccc}
& \hspace{5pt}\scriptsize IPMN &  \hspace{5pt}\scriptsize HICPOV & 
 \hspace{5pt}\scriptsize SHIX &  \hspace{5pt}\scriptsize LTIRT &  \hspace{8pt}\scriptsize UNETOT \\[4pt] 
\raisebox{1.5\height}{\rotatebox{90}{\scriptsize AT}} & \includegraphics[trim=5cm 12cm 5cm 12.5cm, clip, width=0.19\textwidth]{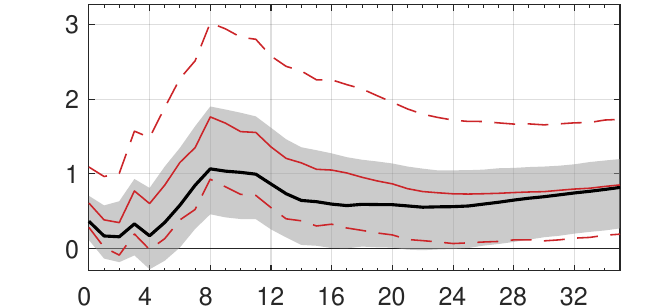} &
\includegraphics[trim=5cm 12cm 5cm 12.5cm, clip, width=0.19\textwidth]{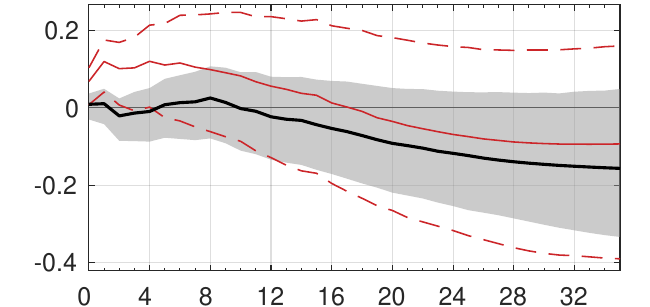} &
\includegraphics[trim=5cm 12cm 5cm 12.5cm, clip, width=0.19\textwidth]{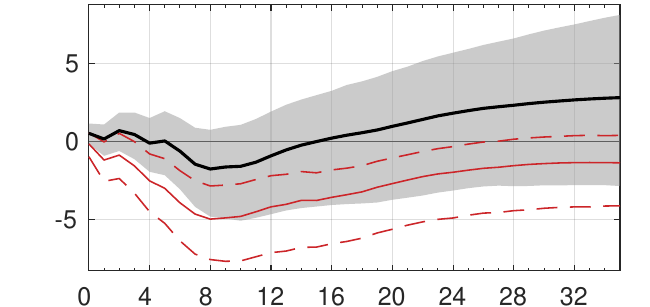} &
\includegraphics[trim=5cm 12cm 5cm 12.5cm, clip, width=0.19\textwidth]{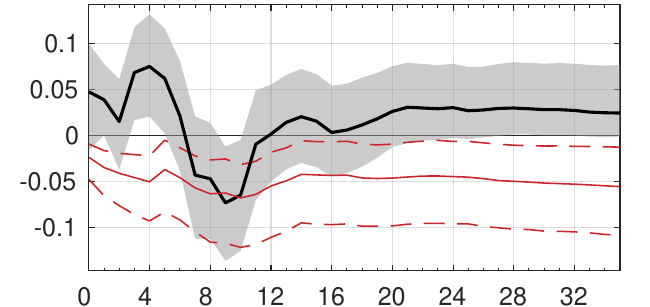} &
\includegraphics[trim=5cm 12cm 5cm 12.5cm, clip, width=0.19\textwidth]{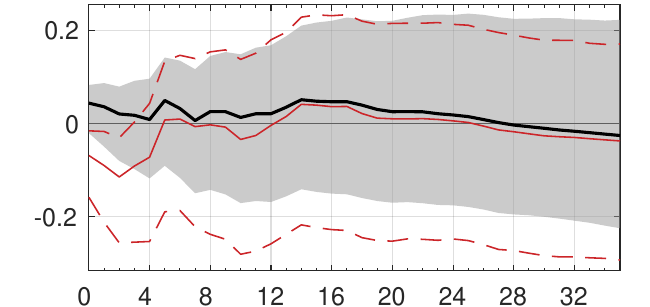} \\

\raisebox{1.5\height}{\rotatebox{90}{\scriptsize BE}} &
\includegraphics[trim=5cm 12cm 5cm 12.5cm, clip, width=0.19\textwidth]{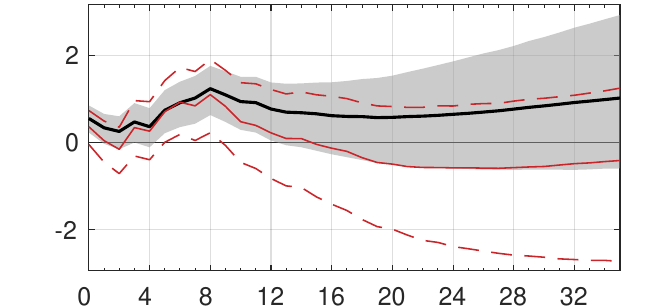} &
\includegraphics[trim=5cm 12cm 5cm 12.5cm, clip, width=0.19\textwidth]{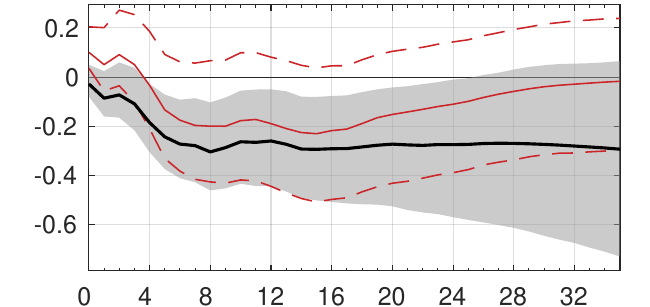} &
\includegraphics[trim=5cm 12cm 5cm 12.5cm, clip, width=0.19\textwidth]{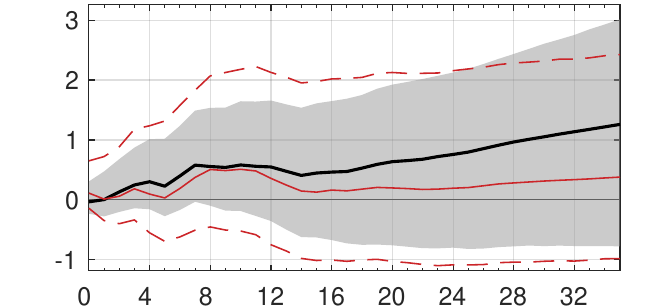} &
\includegraphics[trim=5cm 12cm 5cm 12.5cm, clip, width=0.19\textwidth]{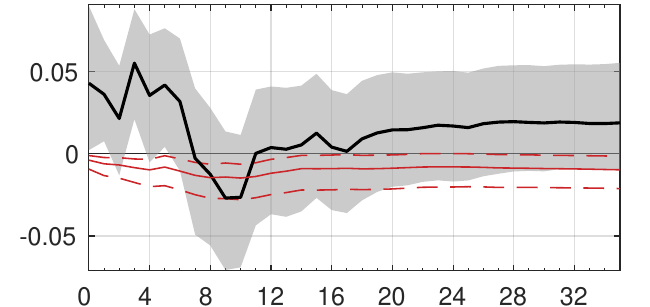} &
\includegraphics[trim=5cm 12cm 5cm 12.5cm, clip, width=0.19\textwidth]{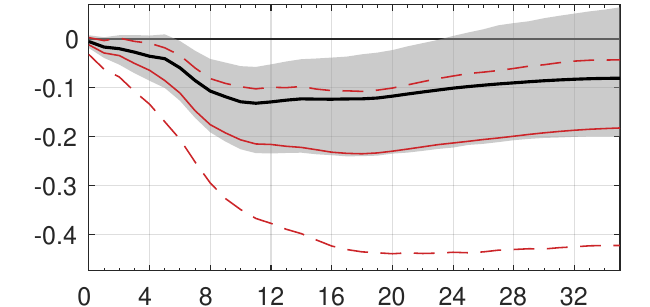} \\

\raisebox{1.5\height}{\rotatebox{90}{\scriptsize DE}} &
\includegraphics[trim=5cm 12cm 5cm 12.5cm, clip, width=0.19\textwidth]{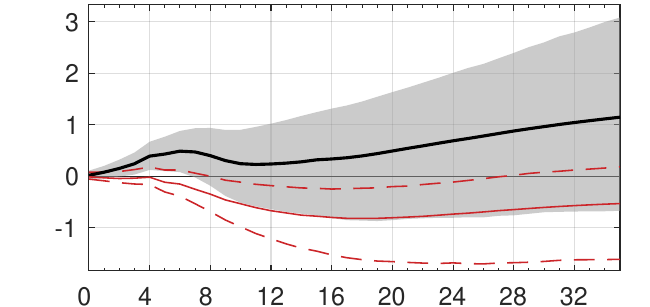} &
\includegraphics[trim=5cm 12cm 5cm 12.5cm, clip, width=0.19\textwidth]{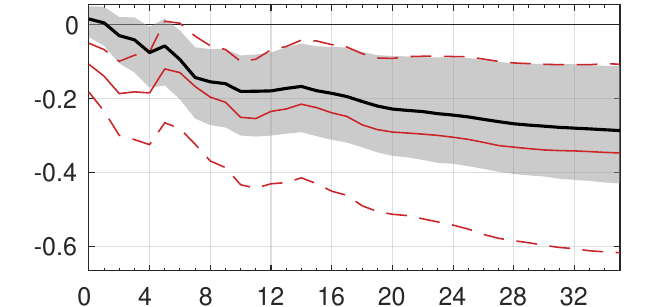} &
\includegraphics[trim=5cm 12cm 5cm 12.5cm, clip, width=0.19\textwidth]{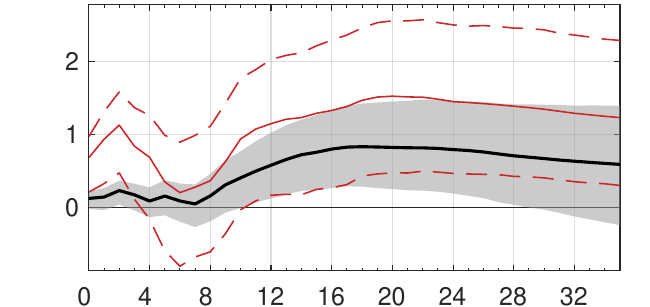} &
\includegraphics[trim=5cm 12cm 5cm 12.5cm, clip, width=0.19\textwidth]{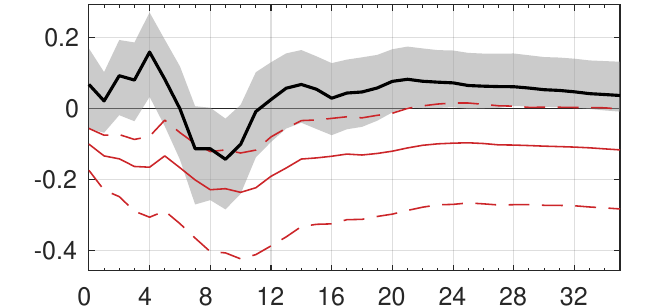} &
\includegraphics[trim=5cm 12cm 5cm 12.5cm, clip, width=0.19\textwidth]{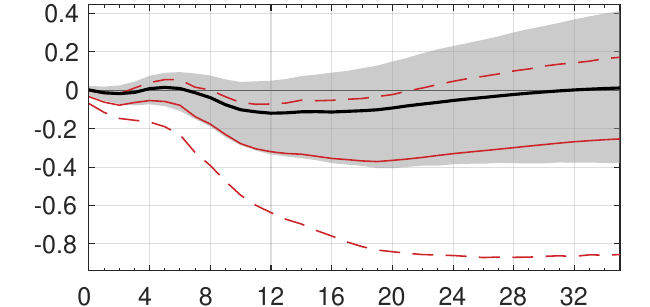} \\

\raisebox{1.5\height}{\rotatebox{90}{\scriptsize FR}} &
\includegraphics[trim=5cm 12cm 5cm 12.5cm, clip, width=0.19\textwidth]{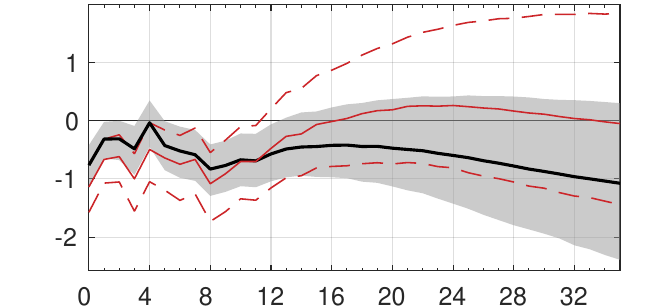} &
\includegraphics[trim=5cm 12cm 5cm 12.5cm, clip, width=0.19\textwidth]{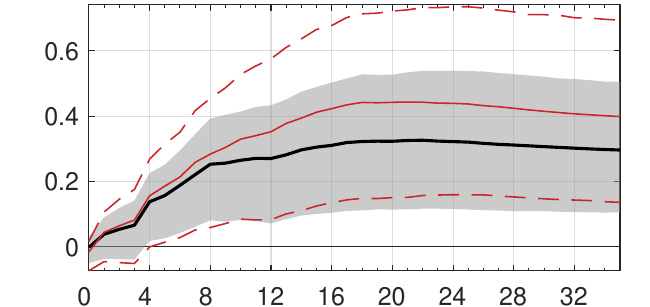} &
\includegraphics[trim=5cm 12cm 5cm 12.5cm, clip, width=0.19\textwidth]{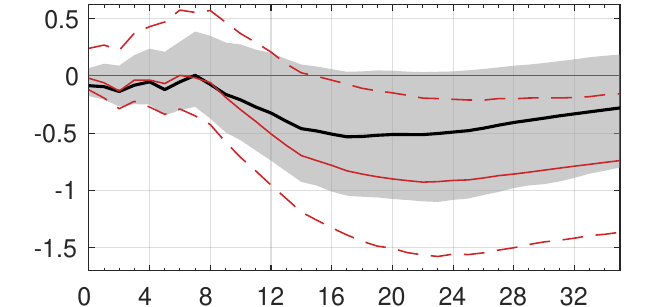} &
\includegraphics[trim=5cm 12cm 5cm 12.5cm, clip, width=0.19\textwidth]{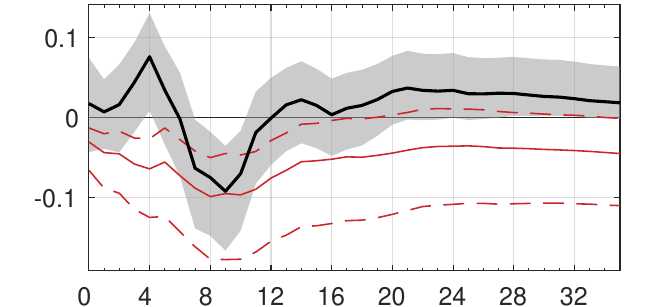} &
\includegraphics[trim=5cm 12cm 5cm 12.5cm, clip, width=0.19\textwidth]{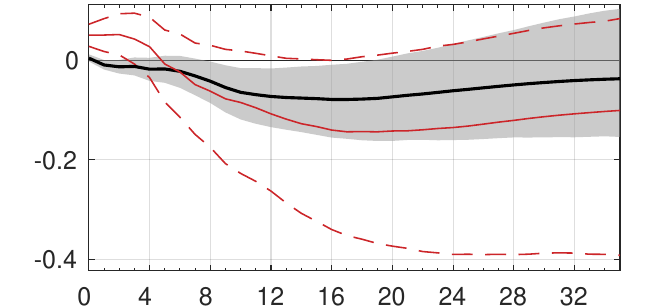} \\

\raisebox{1.5\height}{\rotatebox{90}{\scriptsize NL}} &
\includegraphics[trim=5cm 12cm 5cm 12.5cm, clip, width=0.19\textwidth]{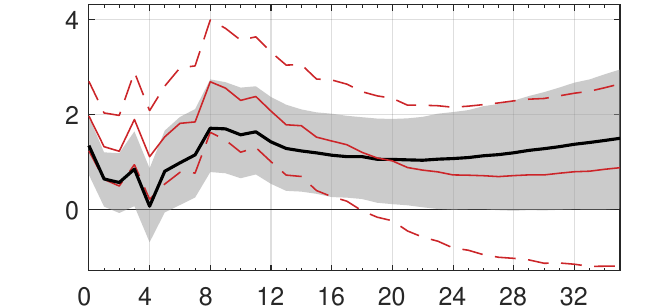} &
\includegraphics[trim=5cm 12cm 5cm 12.5cm, clip, width=0.19\textwidth]{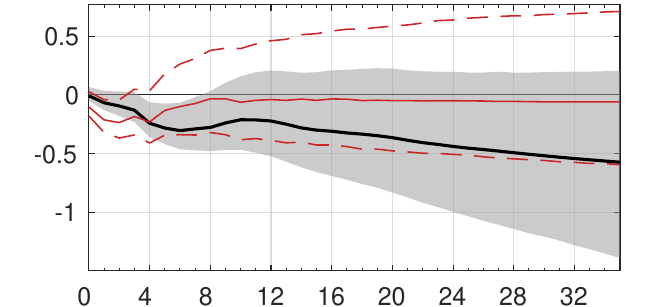} &
\includegraphics[trim=5cm 12cm 5cm 12.5cm, clip, width=0.19\textwidth]{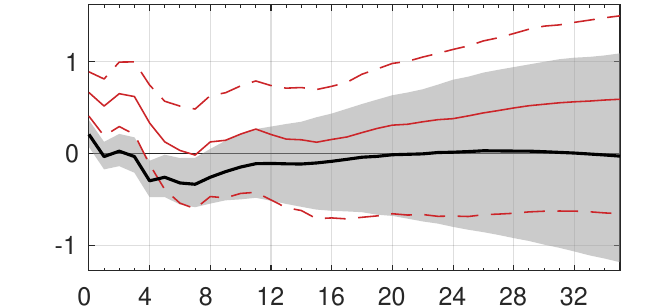} &
\includegraphics[trim=5cm 12cm 5cm 12.5cm, clip, width=0.19\textwidth]{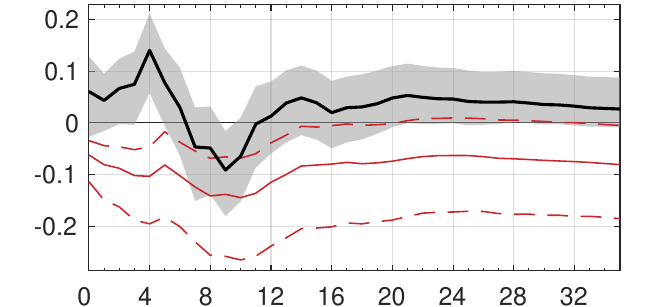} &
\includegraphics[trim=5cm 12cm 5cm 12.5cm, clip, width=0.19\textwidth]{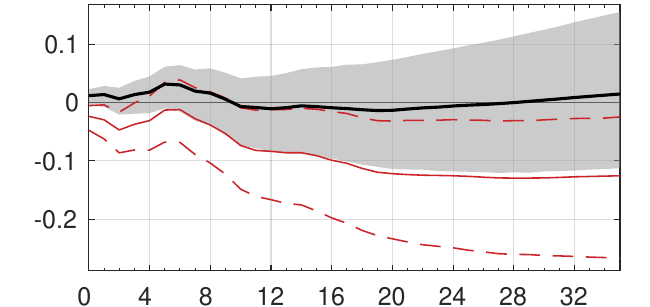} \\

\raisebox{1.7\height}{\rotatebox{90}{\scriptsize EL}} &
\includegraphics[trim=5cm 12cm 5cm 12.5cm, clip, width=0.19\textwidth]{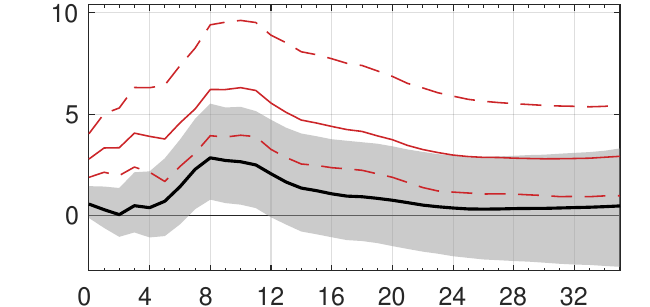} &
\includegraphics[trim=5cm 12cm 5cm 12.5cm, clip, width=0.19\textwidth]{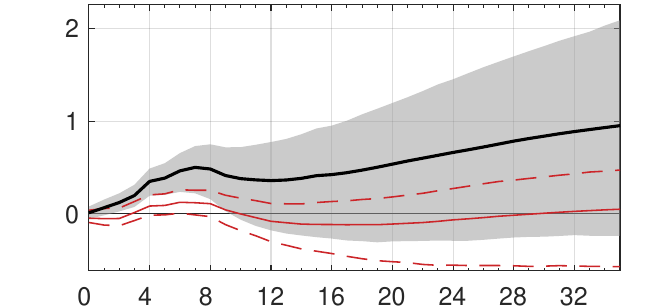} &
\includegraphics[trim=5cm 12cm 5cm 12.5cm, clip, width=0.19\textwidth]{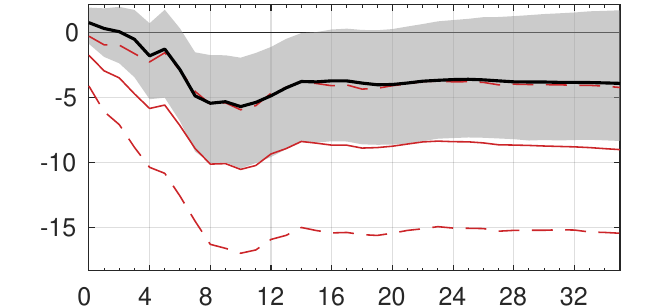} &
\includegraphics[trim=5cm 12cm 5cm 12.5cm, clip, width=0.19\textwidth]{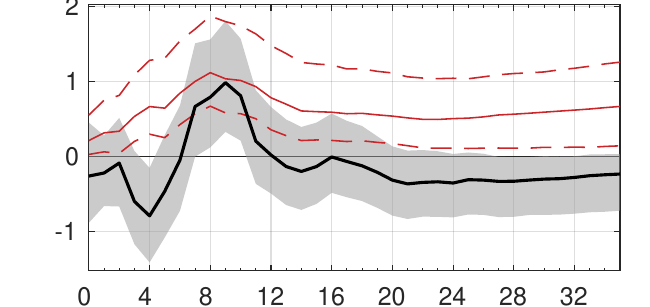} &
\includegraphics[trim=5cm 12cm 5cm 12.5cm, clip, width=0.19\textwidth]{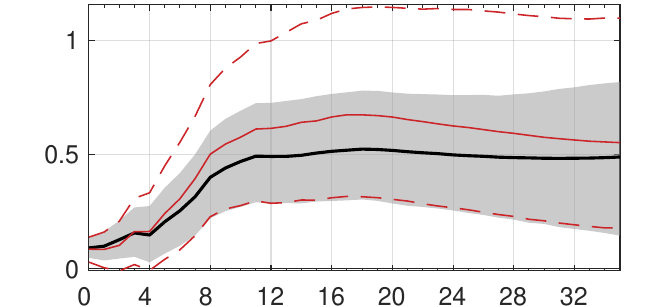} \\

\raisebox{1.7\height}{\rotatebox{90}{\scriptsize ES}} &
\includegraphics[trim=5cm 12cm 5cm 12.5cm, clip, width=0.19\textwidth]{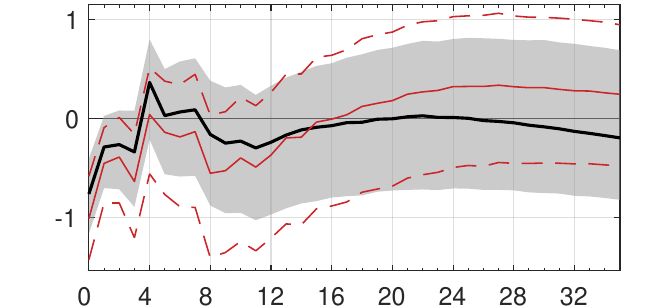} &
\includegraphics[trim=5cm 12cm 5cm 12.5cm, clip, width=0.19\textwidth]{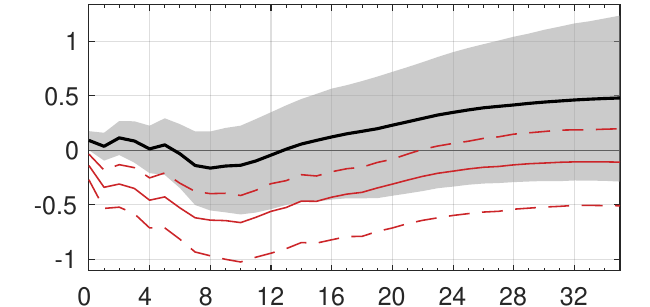} &
\includegraphics[trim=5cm 12cm 5cm 12.5cm, clip, width=0.19\textwidth]{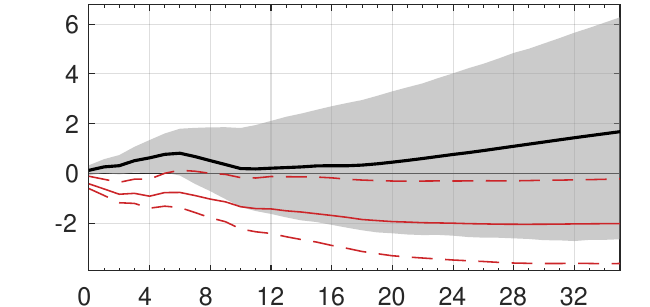} &
\includegraphics[trim=5cm 12cm 5cm 12.5cm, clip, width=0.19\textwidth]{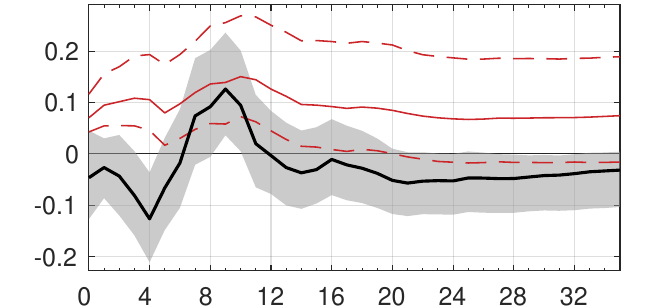} &
\includegraphics[trim=5cm 12cm 5cm 12.5cm, clip, width=0.19\textwidth]{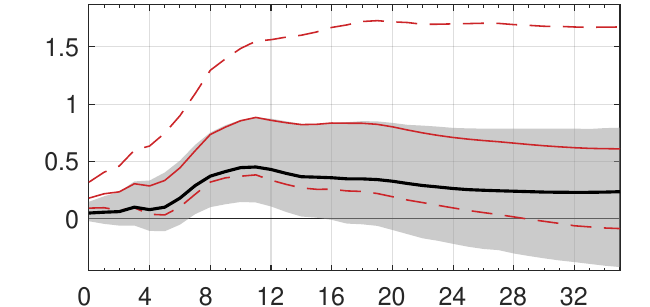} \\

\raisebox{2\height}{\rotatebox{90}{\scriptsize IE}} &
 &
\includegraphics[trim=5cm 12cm 5cm 12.5cm, clip, width=0.19\textwidth]{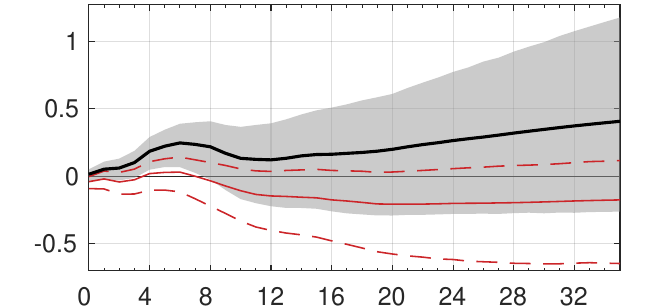} &
\includegraphics[trim=5cm 12cm 5cm 12.5cm, clip, width=0.19\textwidth]{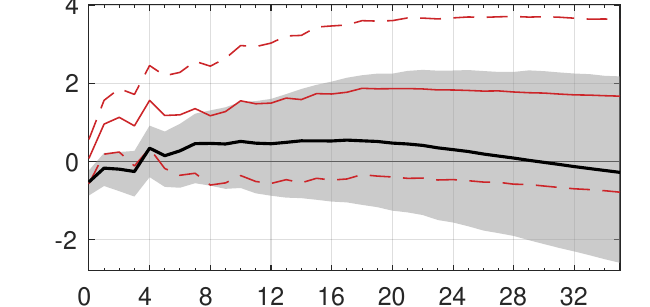} &
\includegraphics[trim=5cm 12cm 5cm 12.5cm, clip, width=0.19\textwidth]{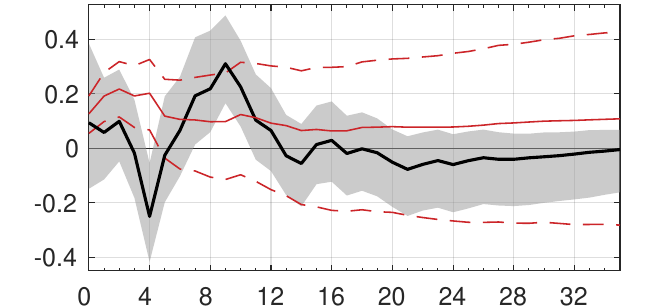} &
\includegraphics[trim=5cm 12cm 5cm 12.5cm, clip, width=0.19\textwidth]{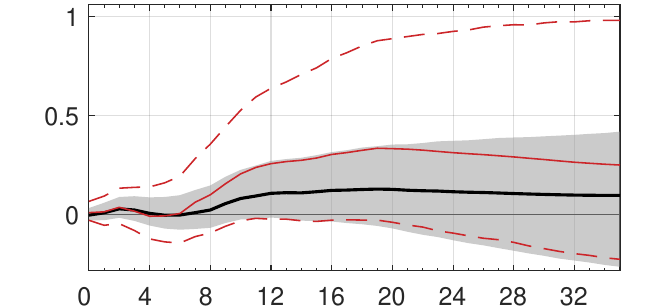} \\

\raisebox{2\height}{\rotatebox{90}{\scriptsize IT}} &
\includegraphics[trim=5cm 12cm 5cm 12.5cm, clip, width=0.19\textwidth]{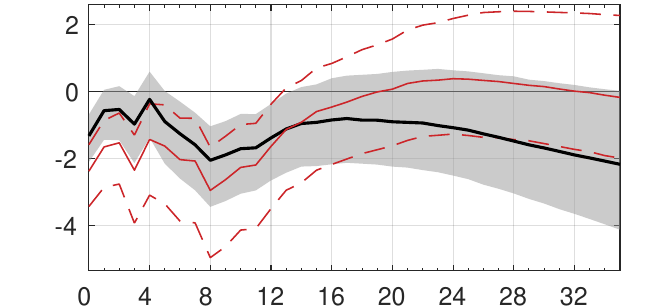} &
\includegraphics[trim=5cm 12cm 5cm 12.5cm, clip, width=0.19\textwidth]{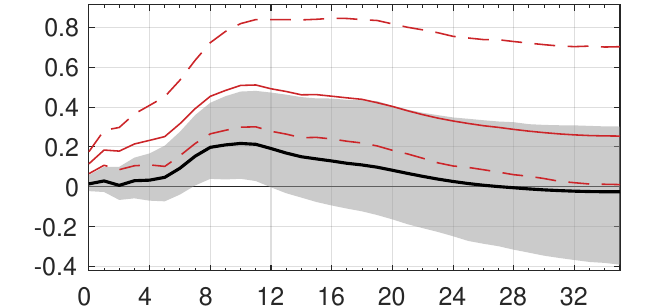} &
\includegraphics[trim=5cm 12cm 5cm 12.5cm, clip, width=0.19\textwidth]{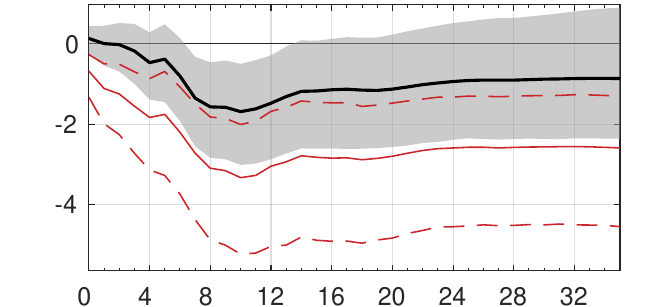} &
\includegraphics[trim=5cm 12cm 5cm 12.5cm, clip, width=0.19\textwidth]{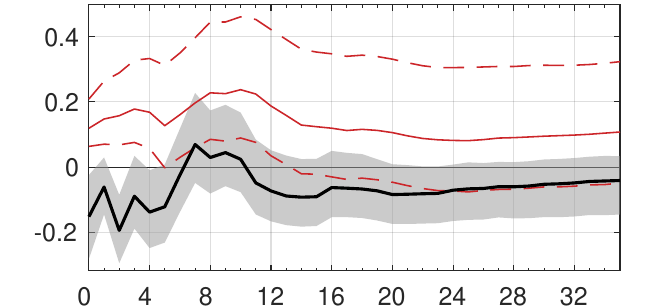} &
\includegraphics[trim=5cm 12cm 5cm 12.5cm, clip, width=0.19\textwidth]{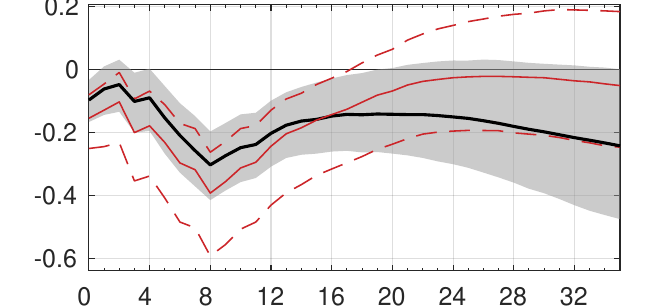} \\

\raisebox{1.5\height}{\rotatebox{90}{\scriptsize PT}} &
\includegraphics[trim=5cm 12cm 5cm 12.5cm, clip, width=0.19\textwidth]{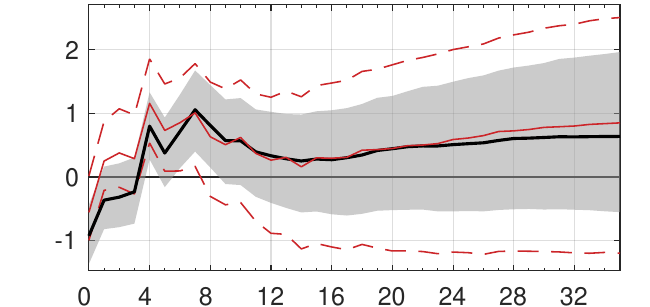} &
\includegraphics[trim=5cm 12cm 5cm 12.5cm, clip, width=0.19\textwidth]{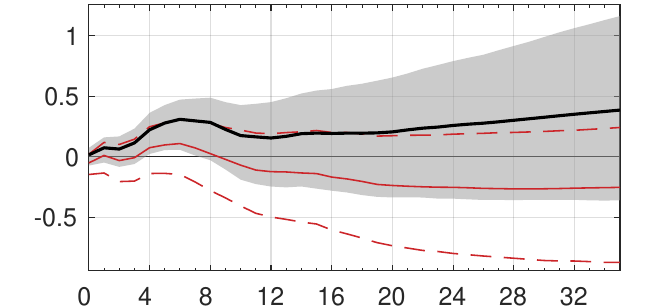} &
\includegraphics[trim=5cm 12cm 5cm 12.5cm, clip, width=0.19\textwidth]{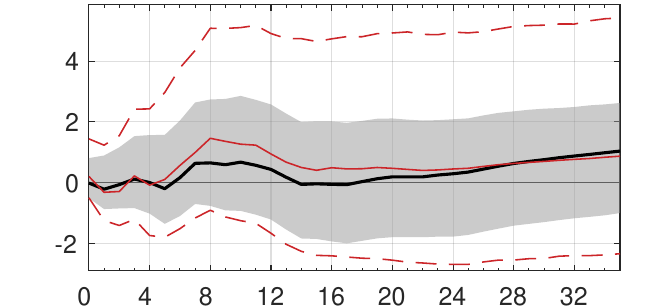} &
\includegraphics[trim=5cm 12cm 5cm 12.5cm, clip, width=0.19\textwidth]{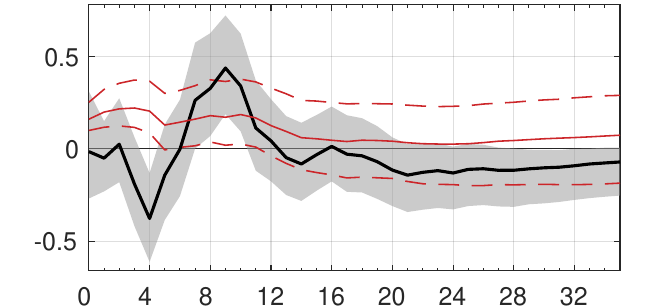} &
\includegraphics[trim=5cm 12cm 5cm 12.5cm, clip, width=0.19\textwidth]{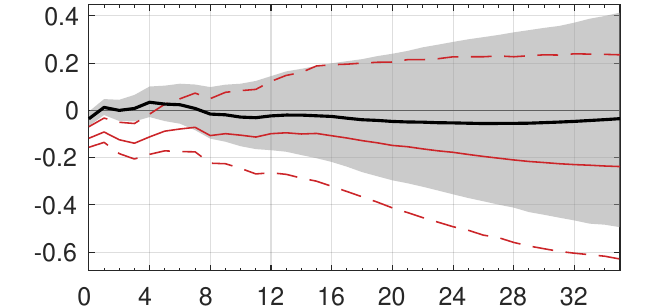} \\
\end{tabular}
\begin{tabular}{p{1\textwidth}} \scriptsize Notes: \rm Each sub-figure plots the difference between the country-level IRF and the corresponding EA counterpart for one variable and one country. Each column of the graph represents a variable, while each row represents a country. The variables considered are: Industrial Production: Manufacturing (IPMN), HICP: Overall (HICPOV), Stock Price Index (SHIX), 10-years Interest Rates (LTIRT) and Unemployment Rate (UNETOT). The black solid line is the point estimate in our baseline setting, while the red solid line is the point estimate obtained with the statistical transformations. The gray shaded area and red dotted lines are the corresponding 68\% confidence intervals. The scale in the vertical axis differs across variables and countries.
\end{tabular}
\label{fig::diffIRF_TR1}
\end{figure}

\newpage

\subsection{Cross-country characteristics and IRF dynamics}
\begin{table}[ht!]
\centering
\caption{Cross-country characteristics and IRFs dynamics: statistical transformations}
\setlength{\tabcolsep}{.03\textwidth}
\resizebox{6.5in}{!}{ 
\begin{tabular}{l|l|ccccc}
\hline
\hline
& & & & & & \\[-12pt]
\textbf{ID} & \textbf{Variable} & \textbf{IPMN}  & \textbf{HICPOV}  & \textbf{LTIRT} & \textbf{SHIX}  & \textbf{UNETOT} \\
 & & & & & & \\[-12pt]
\hline
& & & & & & \\[-10pt]
1 & Employment Protection & -0.05 & 0.17  & -0.09 & -0.05 & -0.44 \\
  & &\footnotesize (0.38) &\footnotesize (0.35) &\footnotesize (0.35) &\footnotesize (0.35) &\footnotesize (0.32) \\

& & & & & & \\[-10pt]
         
2 & Wage Adj. Frequency: more than once a year & 0.16  & 0.28  & -0.52 & 0.38  & -0.51 \\
  & &\footnotesize (0.44) &\footnotesize (0.39) &\footnotesize (0.35) &\footnotesize (0.38) &\footnotesize (0.35) \\

& & & & & & \\[-10pt]
       
3 & Wage Adj. Frequency: once a year & 0.65  & -0.15 & -0.61 & -0.22 & 0.11 \\
  & &\footnotesize (0.34) &\footnotesize (0.40) &\footnotesize (0.32) &\footnotesize (0.40) &\footnotesize (0.41) \\
 
& & & & & & \\[-10pt]
       
4 & Wage Adj. Frequency: less than once a year & -0.65 & 0.38  & 0.42  & -0.16 & -0.22 \\
  & &\footnotesize (0.34) &\footnotesize (0.38) &\footnotesize (0.37) &\footnotesize (0.40) &\footnotesize (0.40) \\

& & & & & & \\[-10pt]
          
5 & Price Flexibility & -0.09 & -0.39 & -0.39 & 0.64  & -0.07 \\
  & &\footnotesize (0.45) &\footnotesize (0.41) &\footnotesize (0.41) &\footnotesize (0.34) &\footnotesize (0.45) \\

& & & & & & \\[-10pt] 
     
6 & NFC Leverage & 0.00  & -0.06 & -0.36 & 0.68** & -0.15 \\
  & &\footnotesize (0.38) &\footnotesize (0.35) &\footnotesize (0.33) &\footnotesize (0.26) &\footnotesize (0.35) \\

& & & & & & \\[-10pt]
                  
7 & Homeownership & 0.09  & -0.02 & 0.67** & -0.09 & 0.57* \\  
  & &\footnotesize (0.38) &\footnotesize (0.35) &\footnotesize (0.26) &\footnotesize (0.35) &\footnotesize (0.29) \\

& & & & & & \\[-10pt]

8 & Homeownership with mortgage & 0.07  & -0.21 & -0.52 & 0.65** & -0.30 \\
  & &\footnotesize (0.38) &\footnotesize (0.35) &\footnotesize (0.30) &\footnotesize (0.27) &\footnotesize (0.34) \\

& & & & & & \\[-10pt]
       
9 & Homeownership without mortgage & -0.01 & 0.17  & 0.81*** & -0.59* & 0.57* \\
  & &\footnotesize (0.38) &\footnotesize (0.35) &\footnotesize (0.21) &\footnotesize (0.28) &\footnotesize (0.29) \\

& & & & & & \\[-10pt]
             
10 & Fixed-Rate Mortgages & -0.28 & 0.33  & -0.77*** & 0.43  & -0.65** \\
   & &\footnotesize (0.36) &\footnotesize (0.33) &\footnotesize (0.23) &\footnotesize (0.32) &\footnotesize (0.27) \\

& & & & & & \\[-10pt]
          
11 & Loan to Value & -0.32 & -0.26 & -0.11 & 0.71** & -0.07 \\
   & &\footnotesize (0.39) &\footnotesize (0.36) &\footnotesize (0.38) &\footnotesize (0.27) &\footnotesize (0.38) \\

& & & & & & \\[-10pt]
          
12 & Share of HtM & 0.39  & -0.53 & 0.49  & 0.17  & 0.35 \\
   & &\footnotesize (0.35) &\footnotesize (0.30) &\footnotesize (0.31) &\footnotesize (0.35) &\footnotesize (0.33) \\

& & & & & & \\[-10pt]
       
13 & Share of WHtM & 0.40  & -0.48 & 0.70** & -0.02 & 0.51 \\
   & &\footnotesize (0.35) &\footnotesize (0.31) &\footnotesize (0.25) &\footnotesize (0.35) &\footnotesize (0.30) \\

& & & & & & \\[-10pt]
          
14 & Saving Rate & -0.37 & -0.12 & -0.85*** & 0.60* & -0.37 \\
   & &\footnotesize (0.35) &\footnotesize (0.35) &\footnotesize (0.19) &\footnotesize (0.28) &\footnotesize (0.33) \\

& & & & & & \\[-10pt]

15 & Explained Variance & -0.75** & 0.13  & -0.95*** & 0.37  & 0.37 \\
   & &\footnotesize (0.25) &\footnotesize (0.35) &\footnotesize (0.11) &\footnotesize (0.33) &\footnotesize (0.33) \\
\hline
\hline
\end{tabular}%
} 
\begin{tabular}{p{.95\textwidth}} \scriptsize {\sc Notes}: \rm Each entry of the table corresponds to the correlation between each indicator (in the row) and the peak of the impulse response for a specific variable (in the columns). The variables considered are: Industrial Production: Manufacturing (IPMN), HICP: Overall (HICPOV), Stock Price Index (SHIX), 10-years Interest Rates (LTIRT) and Unemployment Rate (UNETOT). Standard errors in parentheses, asterisks denote statistical significance at 99\% (***), 95\% (**), and 90\% (*). The first row reports the Employment Protection Index (OECD). Rows 2-4 show the frequency of wage adjustments by firms \citep{branten2018nominal}, while row 1 reports the frequency with which firms adjust output prices \citep{gautier2024new}.
Row 6 shows the leverage of non-financial corporations relative to GDP (Eurostat). Row 7 reports the homeownership rate as a share of the population (Eurostat). Rows 8 and 9 indicate, respectively, the shares of residential properties with and without an outstanding mortgage (Eurostat).
Row 10 reports the share of fixed-rate mortgages \citep{de2025long}. Row 11 shows the average household loan-to-value ratio from the Eurosystem Household Finance and Consumption Survey (HFCS).
Rows 12 and 13 display, respectively, the shares of \enquote{wealthy hand-to-mouth} and \enquote{poor hand-to-mouth} households, constructed as in \citet{slacalek2020household} based on HFCS data.
Row 14 reports the household saving rate as a percentage of GDP (OECD). Finally, row 15 corresponds to the share of variance explained by the common factors for each variable in the columns.
\end{tabular}
\label{tab::CRanalysis_TR1}
\end{table}

\newpage

\section{Alternative transformations: all rates in levels}
\label{app::TR7}
\subsection{EA IRFs}
\begin{figure}[H]
\centering \footnotesize \sc \smallskip
\setlength{\tabcolsep}{.005\textwidth}
\caption{EA IRFs: baseline vs all rates in levels}
\begin{tabular}{ccc}
\scriptsize 2-years Interest Rate ($R_t$) &\scriptsize  IP: Manufacturing (IPMN) &\scriptsize  HICP: Overall (HICPOV) \\
\includegraphics[trim= .5cm 8cm .5cm 8.5cm, clip, width=0.32\textwidth]{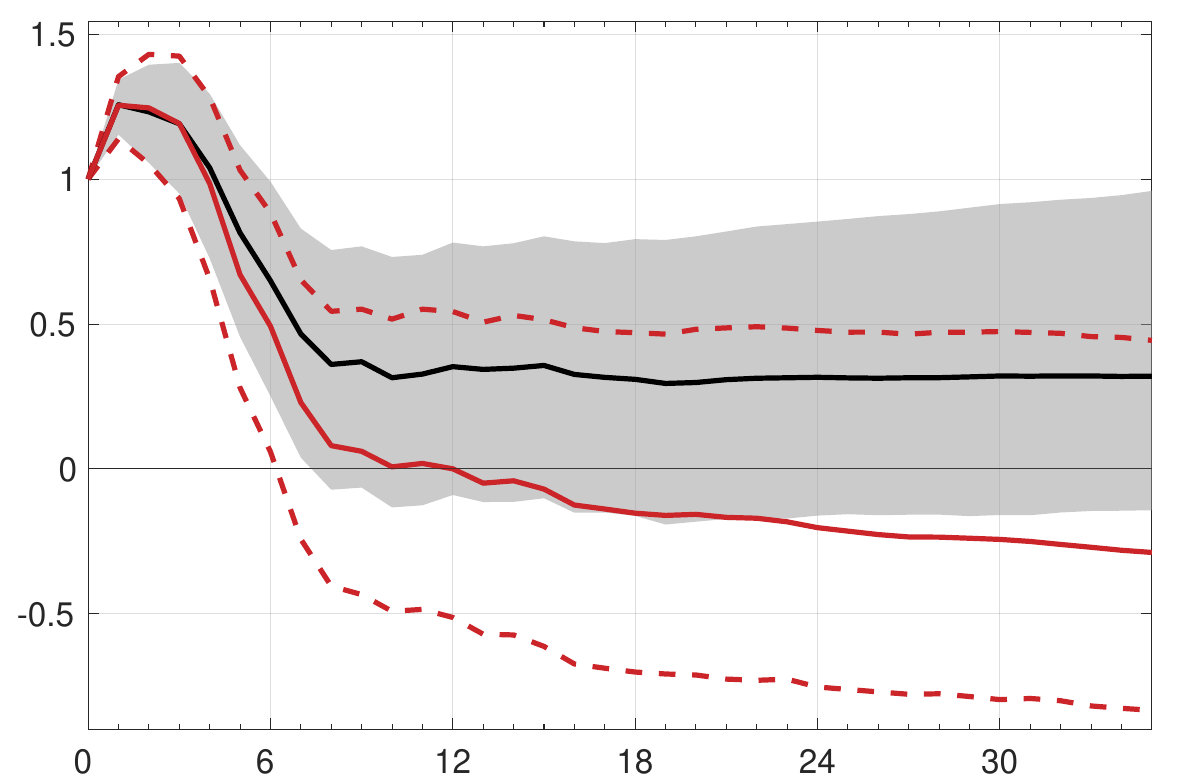} &
\includegraphics[trim= .5cm 8cm .5cm 8.5cm, clip, width=0.32\textwidth]{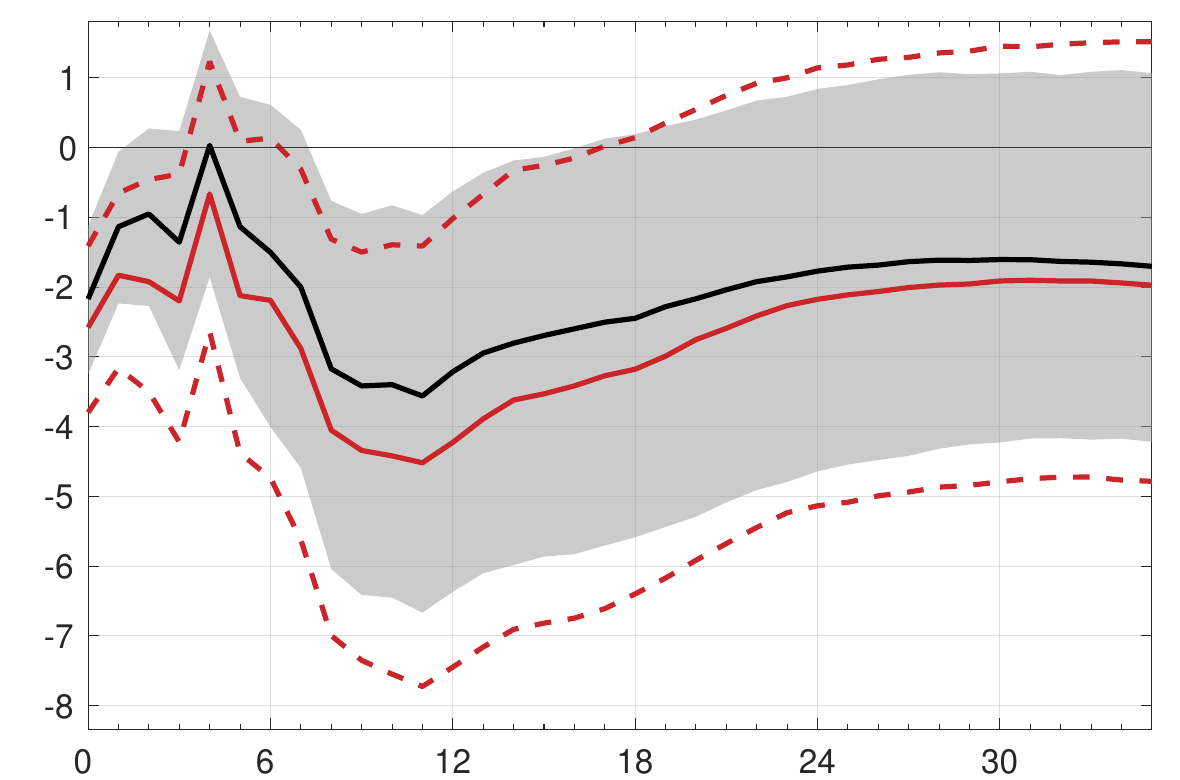} &
\includegraphics[trim= .5cm 8cm .5cm 8.5cm, clip, width=0.32\textwidth]{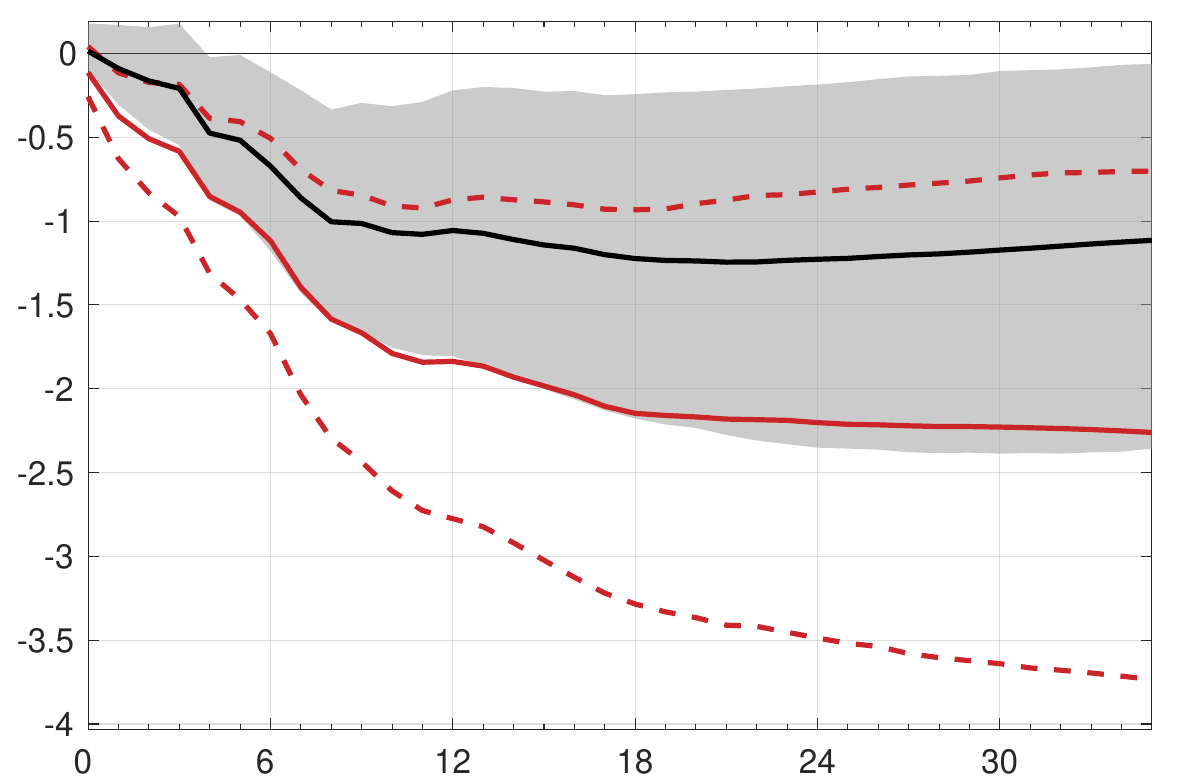} \\
\scriptsize 10-years Interest Rate (LTIRT) &\scriptsize Stock Price Index (SHIX) &\scriptsize Unemployment Rate (UNETOT)\\
\includegraphics[trim= .5cm 8cm .5cm 8.5cm, clip, width=0.32\textwidth]{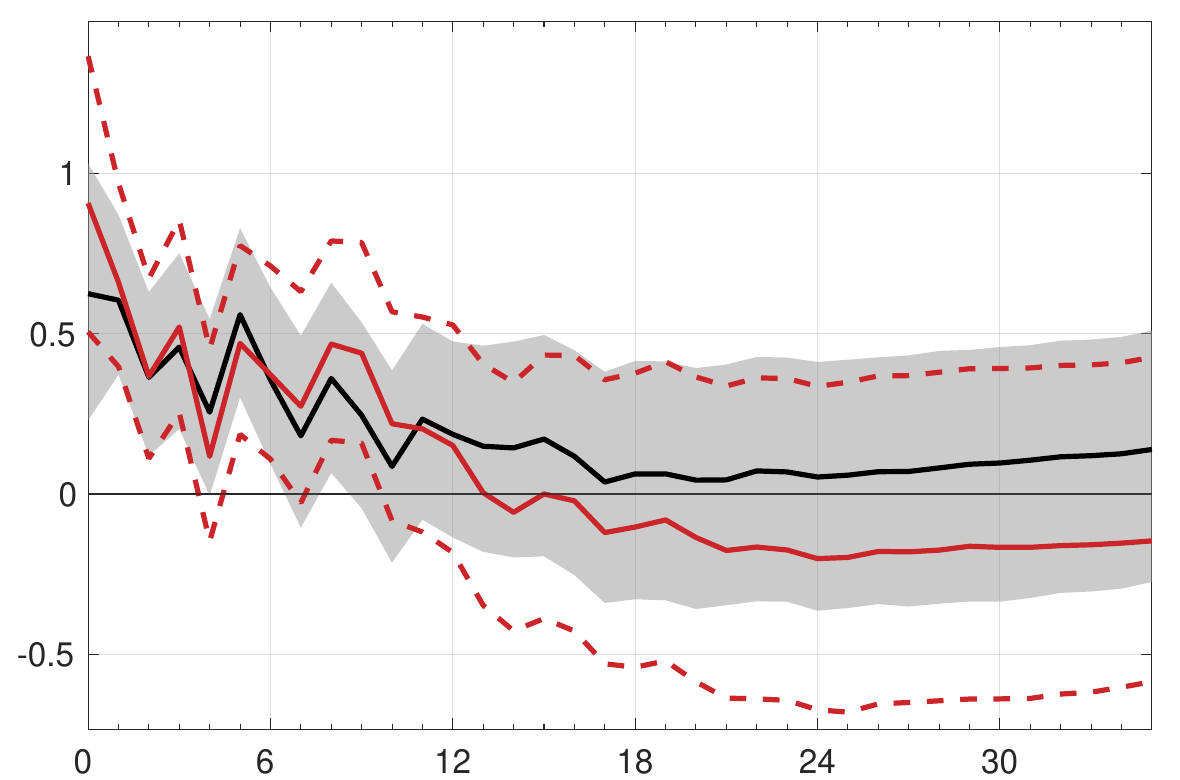} &
\includegraphics[trim= .5cm 8cm .5cm 8.5cm, clip, width=0.32\textwidth]{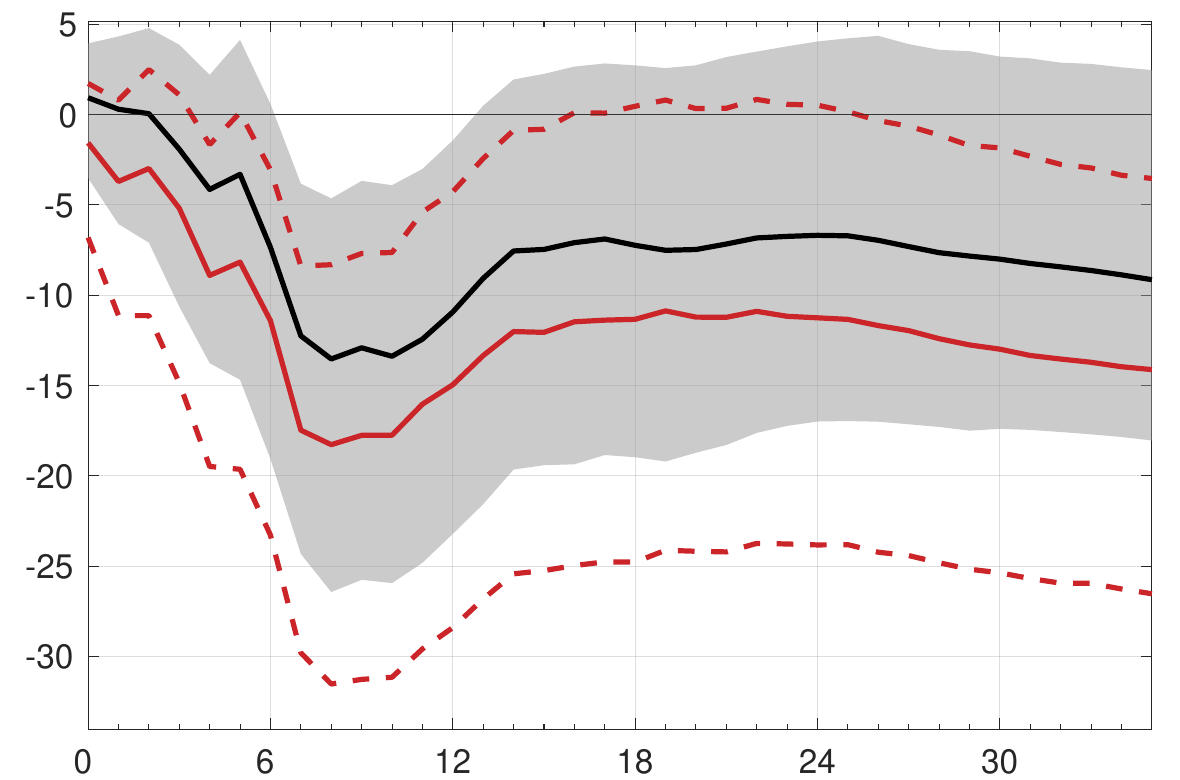} &
\includegraphics[trim= .5cm 8cm .5cm 8.5cm, clip, width=0.32\textwidth]{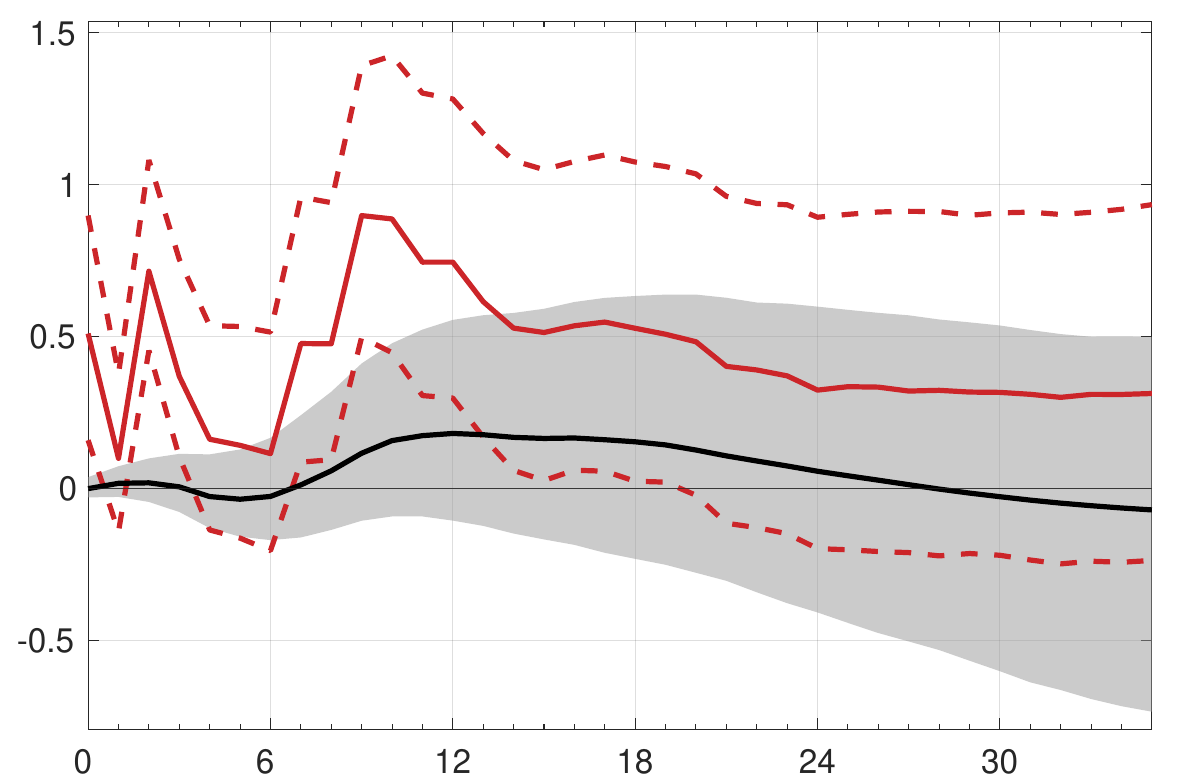} \\
\end{tabular}
\begin{tabular}{p{.98\textwidth}} \scriptsize Notes: \rm Each sub-figure plots the impulse response of one EA variable to a 100bps contractionary monetary policy shock. The black solid line is the point estimate in our baseline setting, while the red solid line is the point estimate obtained with the alternative set of transformations keeping all rates in levels. The gray shaded area and red dotted lines are the corresponding 68\% confidence intervals.
\end{tabular}
\label{fig::EAirfs_TR7}
\end{figure}

\subsection{Country-level IRFs}
\begin{figure}[H]
\centering \footnotesize \sc \smallskip
\setlength{\tabcolsep}{.005\textwidth}
\caption{Country-level IRFs: all rates in levels}
\begin{tabular}{ccc}

\scriptsize{IP: Manufacturing (IPMN)}  
&\scriptsize {HICP: Overall (HICPOV)}      
&\scriptsize {Stock Price Index (SHIX)}  \\

\includegraphics[trim= .5cm 8cm .5cm 8.5cm, clip, width=0.32\textwidth]{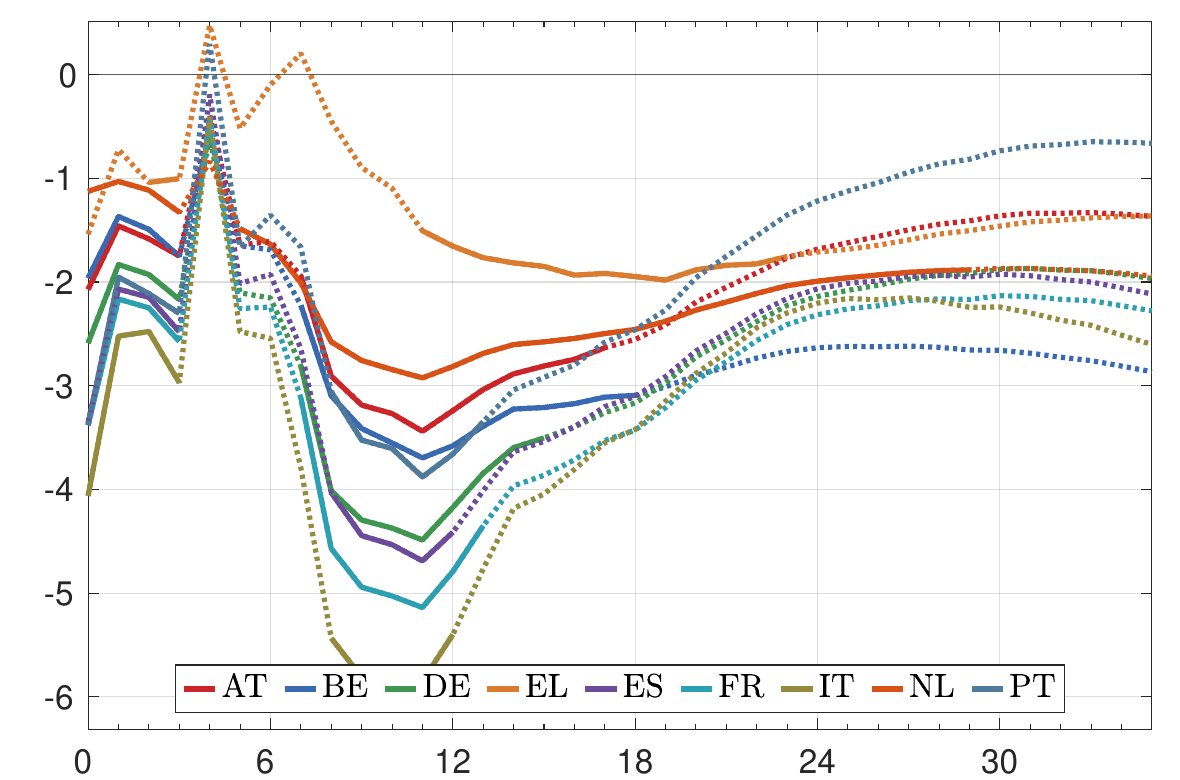} &
\includegraphics[trim= .5cm 8cm .5cm 8.5cm, clip, width=0.32\textwidth]{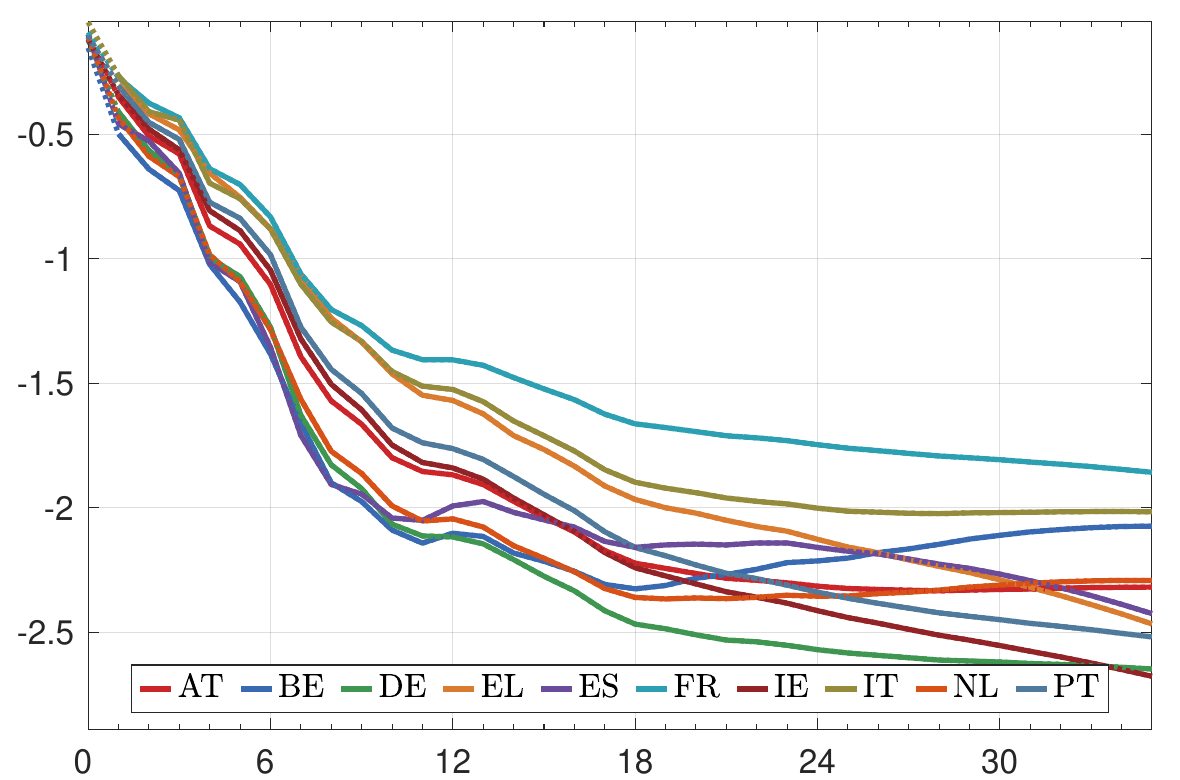} &
\includegraphics[trim= .5cm 8cm .5cm 8.5cm, clip, width=0.32\textwidth]{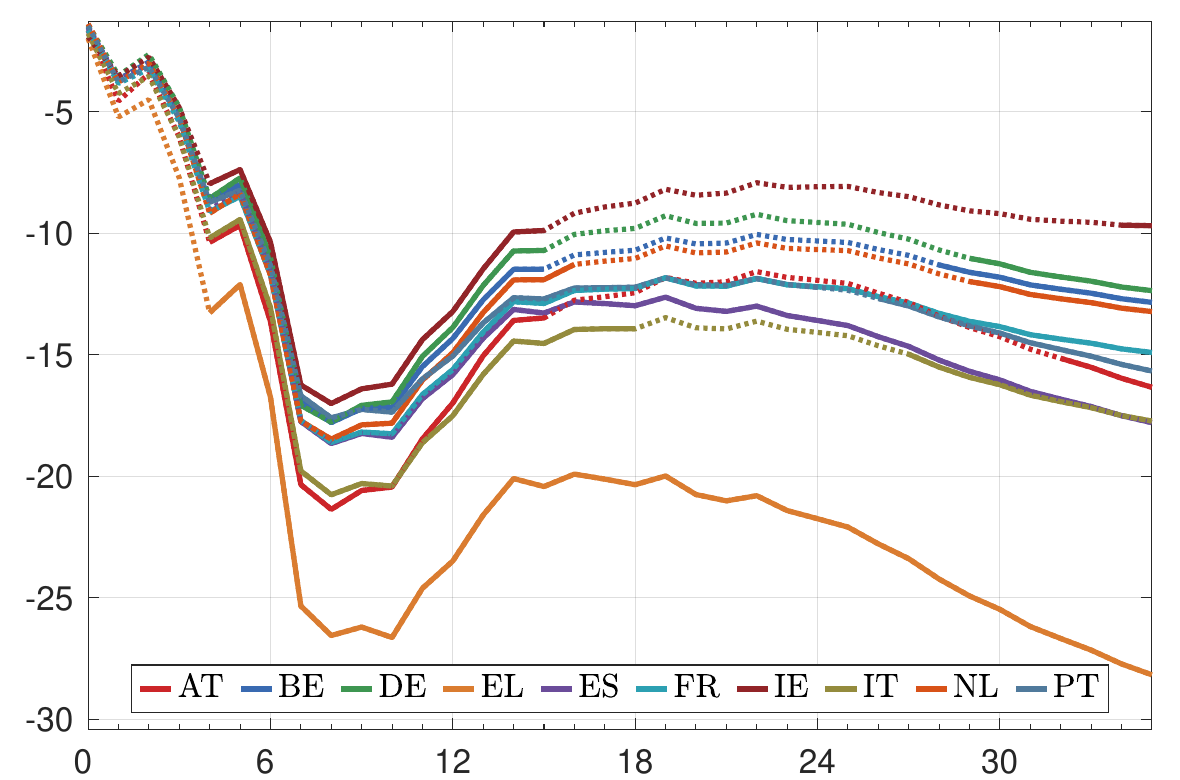} \\[10pt]
\end{tabular}

\begin{tabular}{cc}
\scriptsize {10-years Interest Rate (LTIRT)} 
&\scriptsize {Unemployment Rate (UNETOT)} \\

\includegraphics[trim= .5cm 8cm .5cm 8.5cm, clip, width=0.32\textwidth]{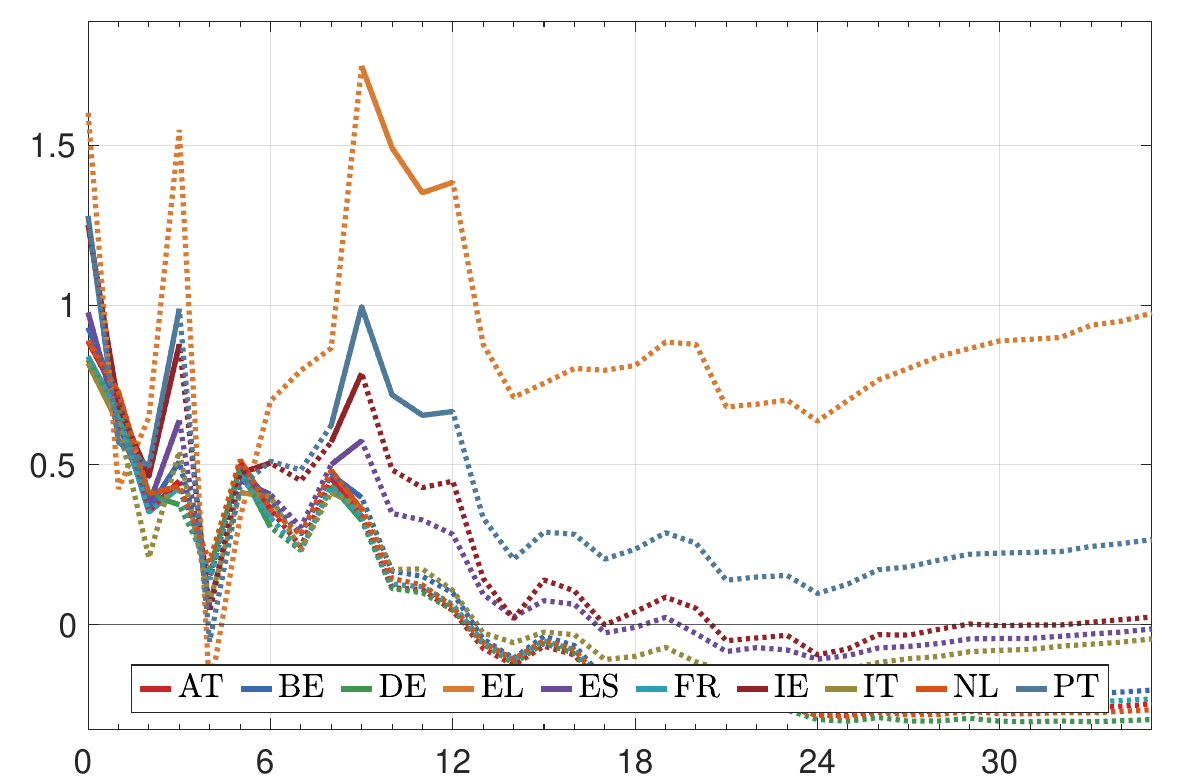} 
&
\includegraphics[trim= .5cm 8cm .5cm 8.5cm, clip, width=0.32\textwidth]{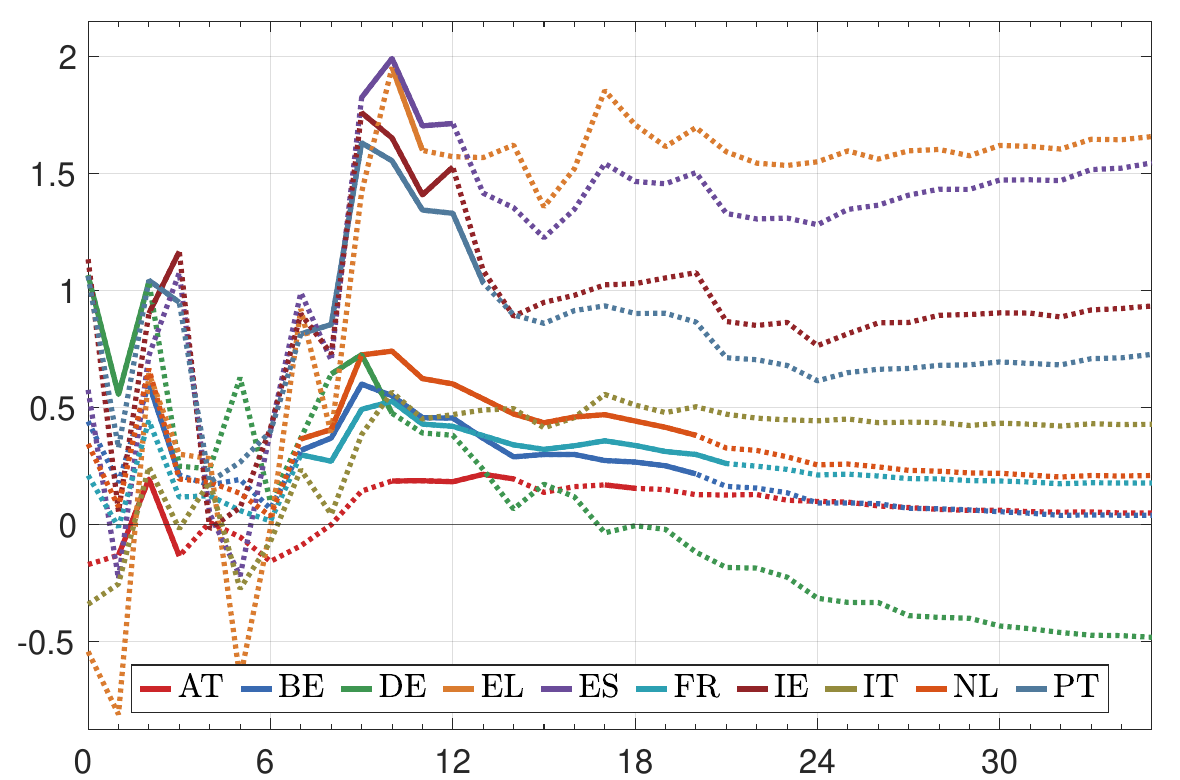}
\end{tabular}
\begin{tabular}{p{.98\textwidth}} \scriptsize Notes: \rm Each sub-figure plots the impulse responses, for all countries, of one variable to a 100bps contractionary monetary policy shock. Within each sub-figure, at each horizon $h=0,\ldots,36$ the country-level impulse responses are denoted with a solid line if the IRF is statistically significant at the 68\% level at that horizon, and with a dotted line otherwise. 
\end{tabular}
\label{fig::nat_IRFs_TR7}
\end{figure}

\newpage

\begin{figure}[H]
\centering \footnotesize \sc \smallskip
\setlength{\tabcolsep}{.005\textwidth}
\caption{Difference between country-level and EA IRFs: baseline vs. all rates in levels}
\begin{tabular}{lccccc}
& \hspace{5pt}\scriptsize IPMN &  \hspace{5pt}\scriptsize HICPOV & 
 \hspace{5pt}\scriptsize SHIX &  \hspace{5pt}\scriptsize LTIRT &  \hspace{8pt}\scriptsize UNETOT \\[4pt] 
\raisebox{1.5\height}{\rotatebox{90}{\scriptsize AT}} & \includegraphics[trim=5cm 12cm 5cm 12.5cm, clip, width=0.19\textwidth]{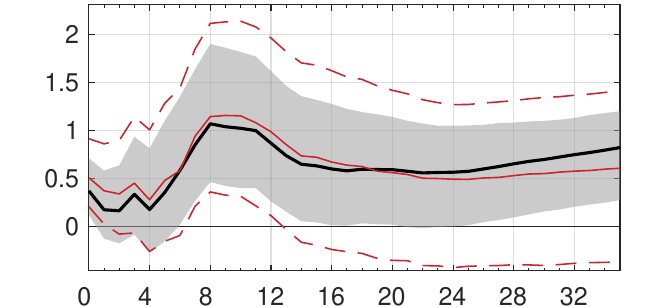} &
\includegraphics[trim=5cm 12cm 5cm 12.5cm, clip, width=0.19\textwidth]{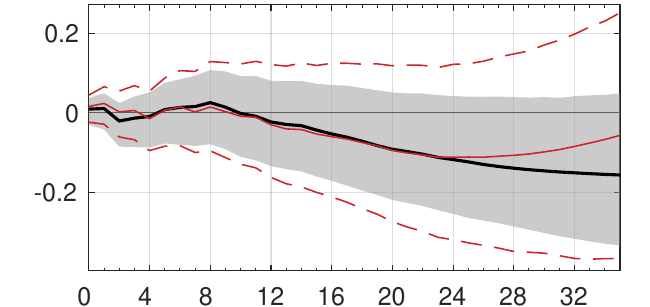} &
\includegraphics[trim=5cm 12cm 5cm 12.5cm, clip, width=0.19\textwidth]{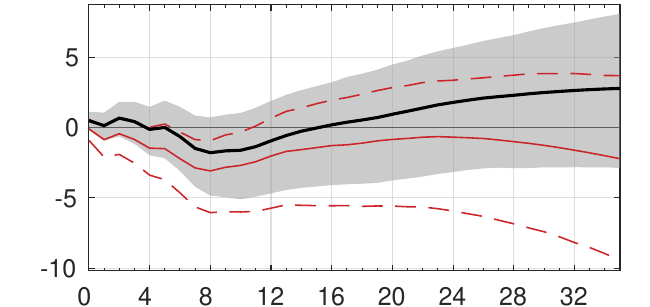} &
\includegraphics[trim=5cm 12cm 5cm 12.5cm, clip, width=0.19\textwidth]{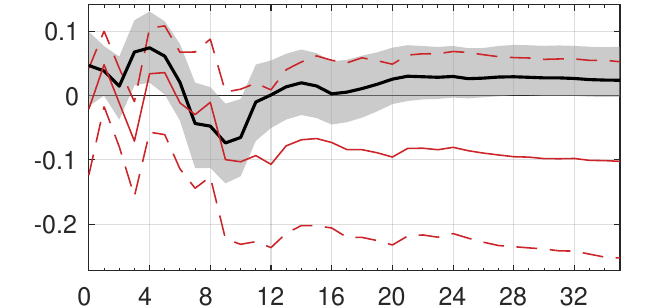} &
\includegraphics[trim=5cm 12cm 5cm 12.5cm, clip, width=0.19\textwidth]{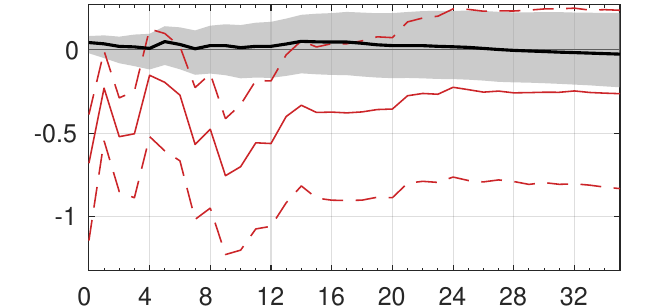} \\

\raisebox{1.5\height}{\rotatebox{90}{\scriptsize BE}} &
\includegraphics[trim=5cm 12cm 5cm 12.5cm, clip, width=0.19\textwidth]{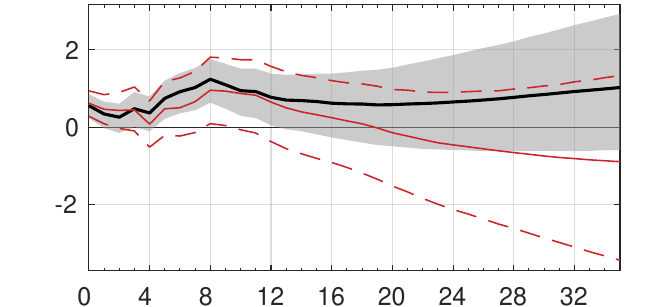} &
\includegraphics[trim=5cm 12cm 5cm 12.5cm, clip, width=0.19\textwidth]{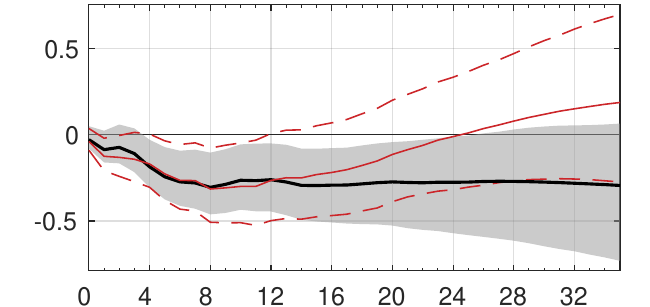} &
\includegraphics[trim=5cm 12cm 5cm 12.5cm, clip, width=0.19\textwidth]{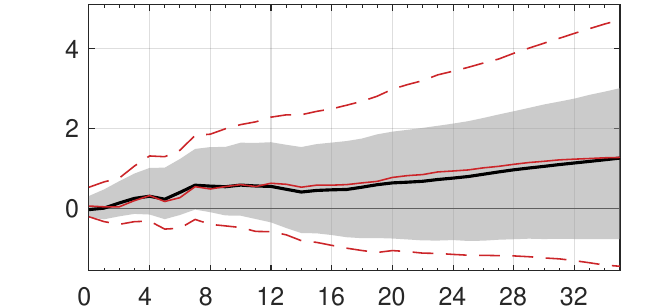} &
\includegraphics[trim=5cm 12cm 5cm 12.5cm, clip, width=0.19\textwidth]{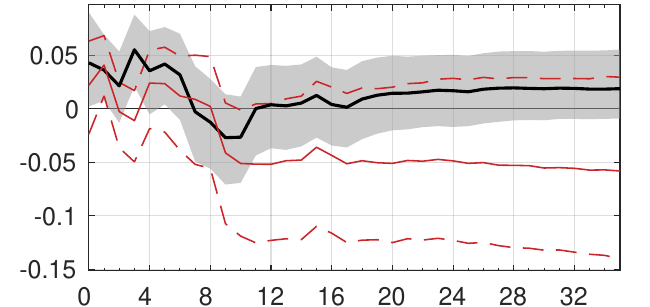} &
\includegraphics[trim=5cm 12cm 5cm 12.5cm, clip, width=0.19\textwidth]{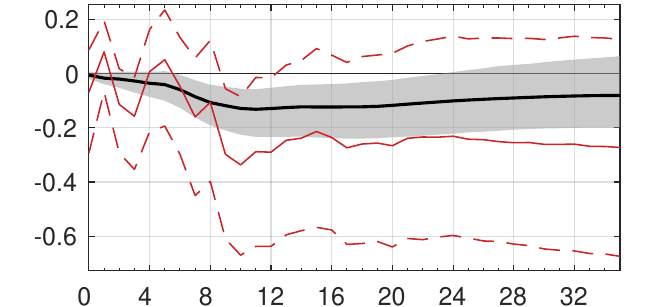} \\

\raisebox{1.5\height}{\rotatebox{90}{\scriptsize DE}} &
\includegraphics[trim=5cm 12cm 5cm 12.5cm, clip, width=0.19\textwidth]{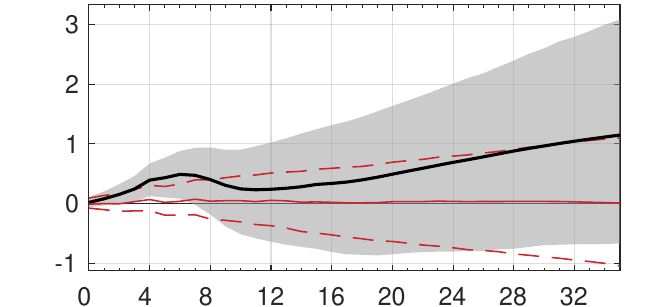} &
\includegraphics[trim=5cm 12cm 5cm 12.5cm, clip, width=0.19\textwidth]{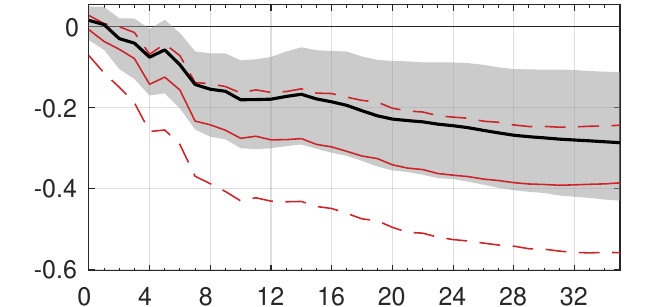} &
\includegraphics[trim=5cm 12cm 5cm 12.5cm, clip, width=0.19\textwidth]{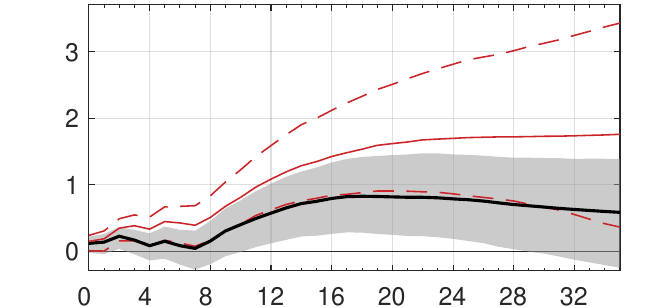} &
\includegraphics[trim=5cm 12cm 5cm 12.5cm, clip, width=0.19\textwidth]{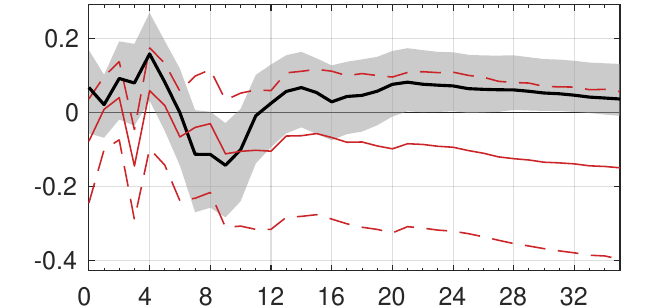} &
\includegraphics[trim=5cm 12cm 5cm 12.5cm, clip, width=0.19\textwidth]{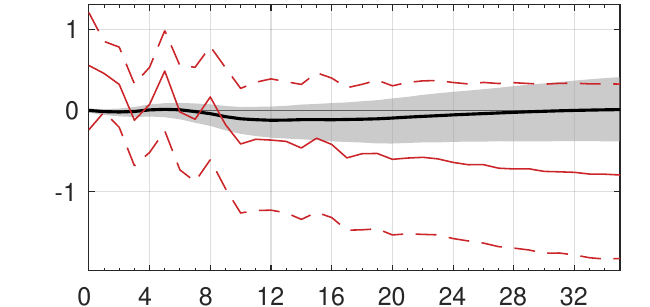} \\

\raisebox{1.5\height}{\rotatebox{90}{\scriptsize FR}} &
\includegraphics[trim=5cm 12cm 5cm 12.5cm, clip, width=0.19\textwidth]{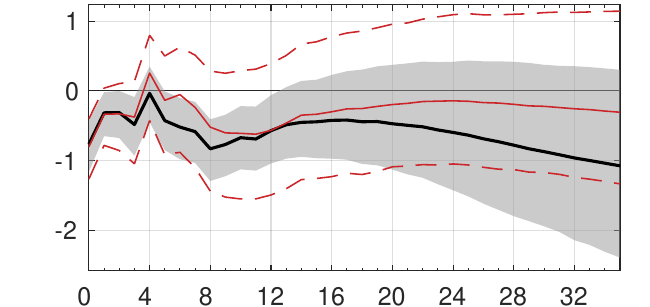} &
\includegraphics[trim=5cm 12cm 5cm 12.5cm, clip, width=0.19\textwidth]{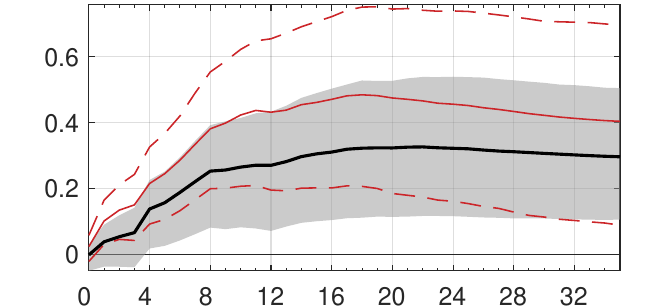} &
\includegraphics[trim=5cm 12cm 5cm 12.5cm, clip, width=0.19\textwidth]{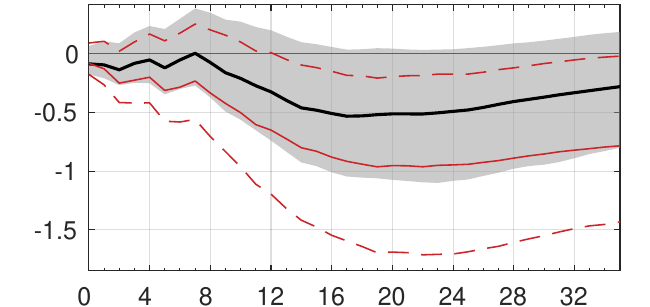} &
\includegraphics[trim=5cm 12cm 5cm 12.5cm, clip, width=0.19\textwidth]{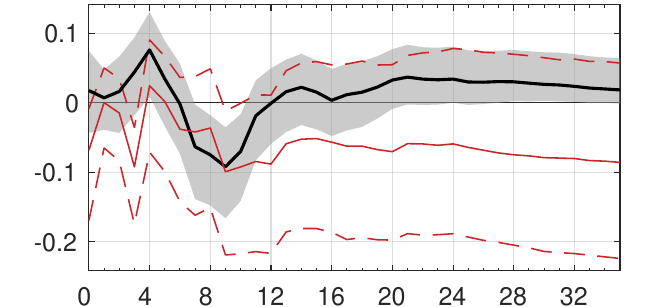} &
\includegraphics[trim=5cm 12cm 5cm 12.5cm, clip, width=0.19\textwidth]{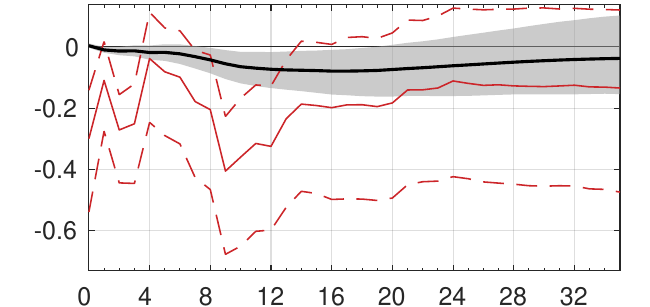} \\

\raisebox{1.5\height}{\rotatebox{90}{\scriptsize NL}} &
\includegraphics[trim=5cm 12cm 5cm 12.5cm, clip, width=0.19\textwidth]{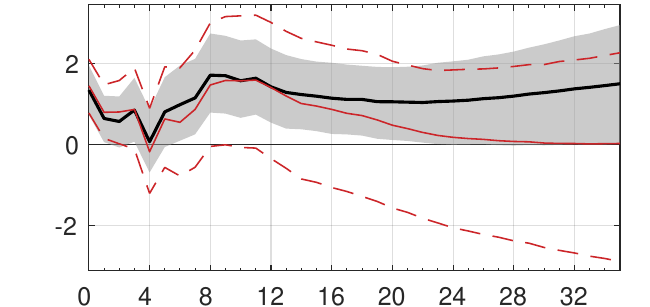} &
\includegraphics[trim=5cm 12cm 5cm 12.5cm, clip, width=0.19\textwidth]{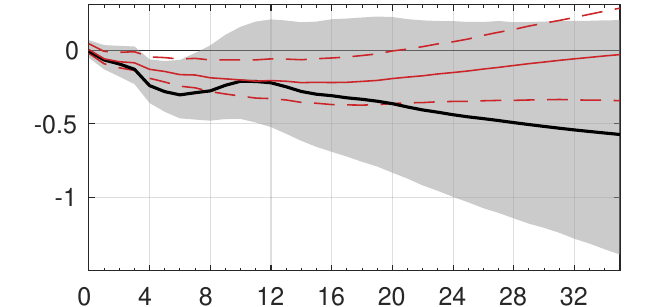} &
\includegraphics[trim=5cm 12cm 5cm 12.5cm, clip, width=0.19\textwidth]{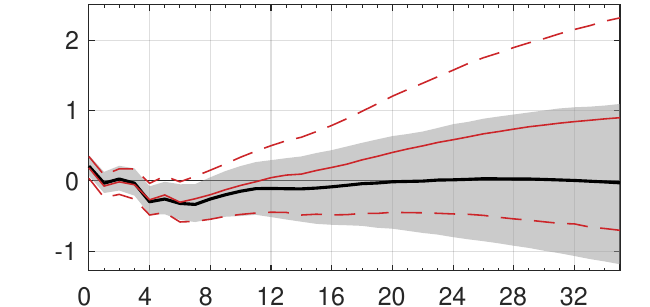} &
\includegraphics[trim=5cm 12cm 5cm 12.5cm, clip, width=0.19\textwidth]{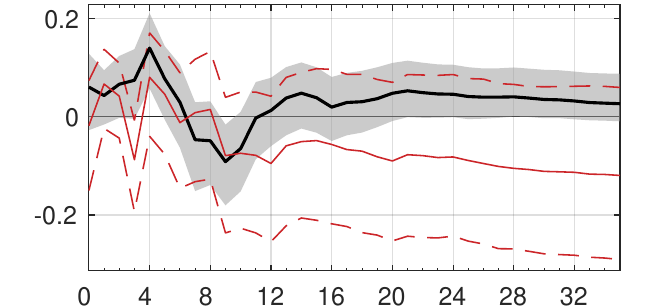} &
\includegraphics[trim=5cm 12cm 5cm 12.5cm, clip, width=0.19\textwidth]{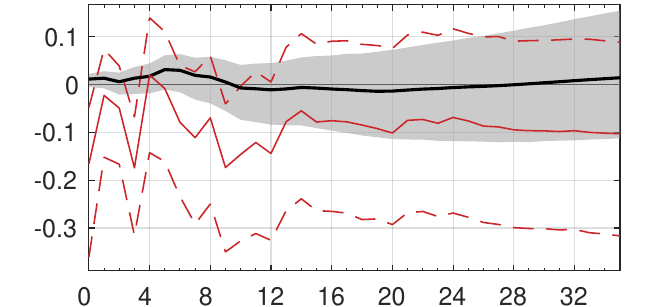} \\

\raisebox{1.7\height}{\rotatebox{90}{\scriptsize EL}} &
\includegraphics[trim=5cm 12cm 5cm 12.5cm, clip, width=0.19\textwidth]{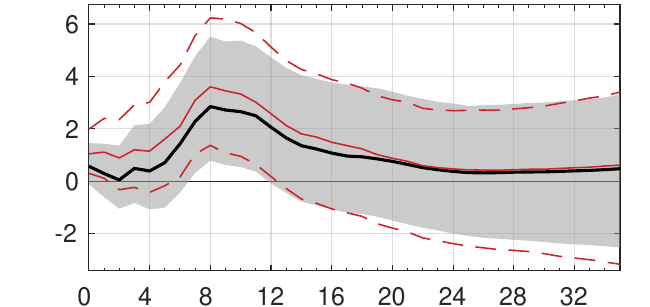} &
\includegraphics[trim=5cm 12cm 5cm 12.5cm, clip, width=0.19\textwidth]{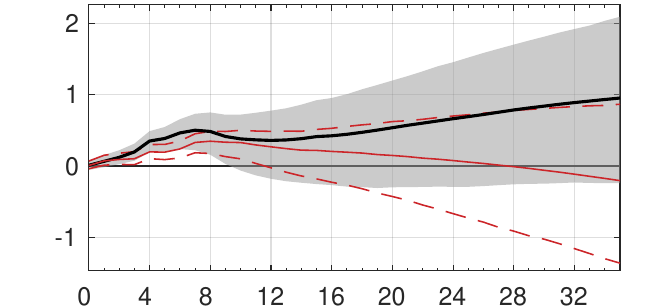} &
\includegraphics[trim=5cm 12cm 5cm 12.5cm, clip, width=0.19\textwidth]{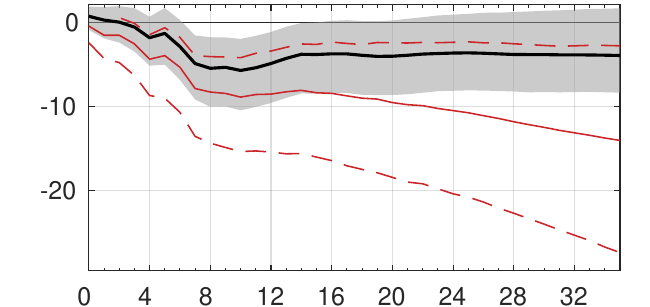} &
\includegraphics[trim=5cm 12cm 5cm 12.5cm, clip, width=0.19\textwidth]{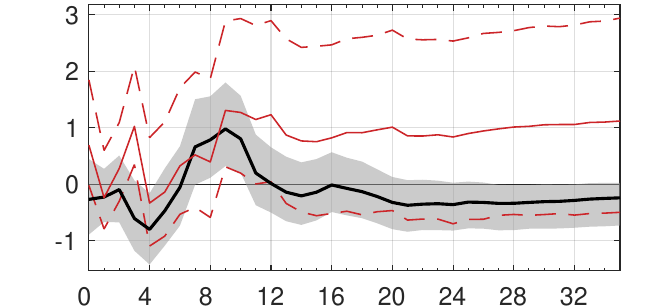} &
\includegraphics[trim=5cm 12cm 5cm 12.5cm, clip, width=0.19\textwidth]{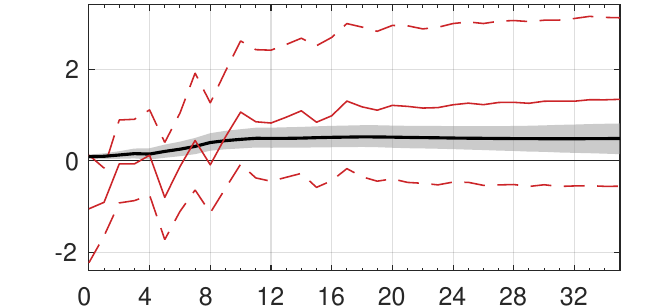} \\

\raisebox{1.7\height}{\rotatebox{90}{\scriptsize ES}} &
\includegraphics[trim=5cm 12cm 5cm 12.5cm, clip, width=0.19\textwidth]{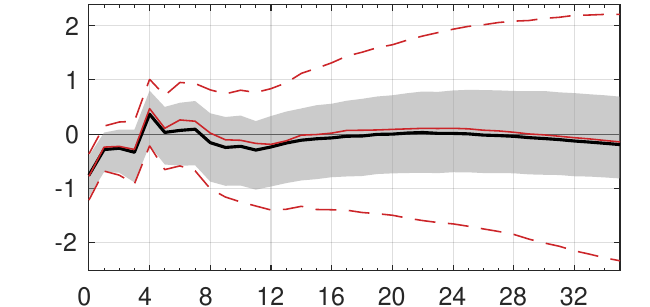} &
\includegraphics[trim=5cm 12cm 5cm 12.5cm, clip, width=0.19\textwidth]{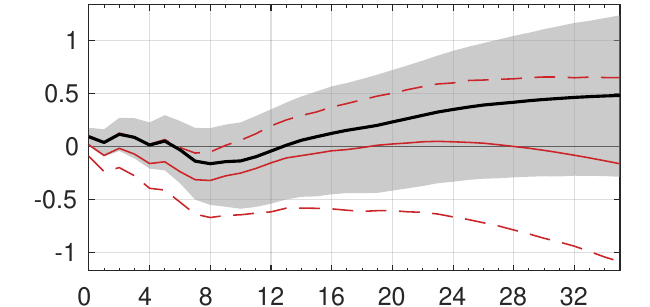} &
\includegraphics[trim=5cm 12cm 5cm 12.5cm, clip, width=0.19\textwidth]{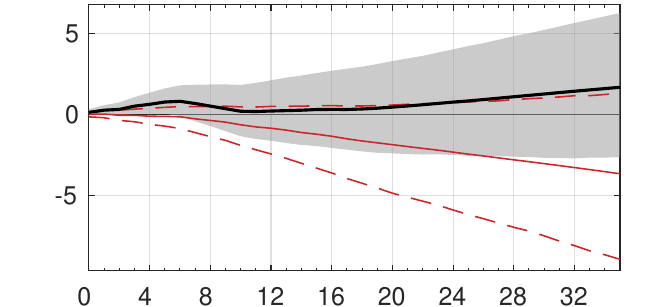} &
\includegraphics[trim=5cm 12cm 5cm 12.5cm, clip, width=0.19\textwidth]{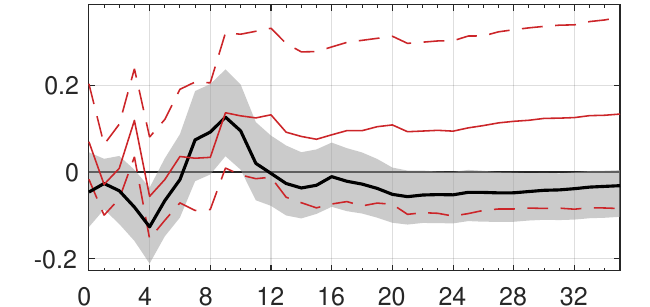} &
\includegraphics[trim=5cm 12cm 5cm 12.5cm, clip, width=0.19\textwidth]{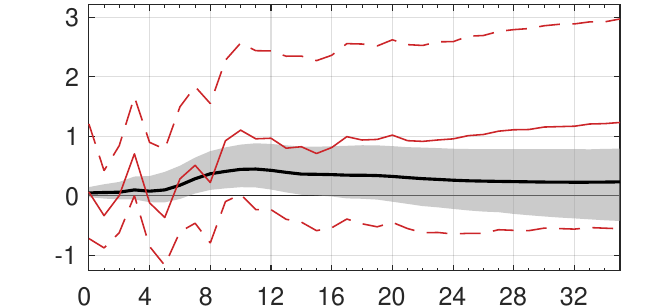} \\

\raisebox{2\height}{\rotatebox{90}{\scriptsize IE}} &
 &
\includegraphics[trim=5cm 12cm 5cm 12.5cm, clip, width=0.19\textwidth]{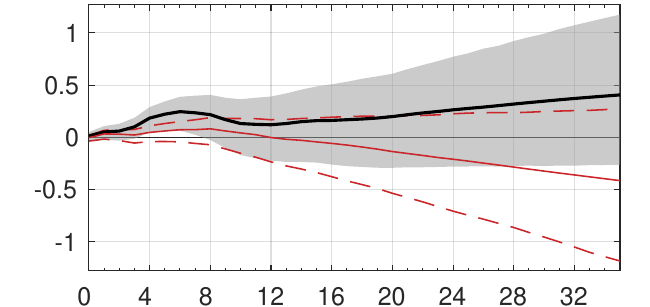} &
\includegraphics[trim=5cm 12cm 5cm 12.5cm, clip, width=0.19\textwidth]{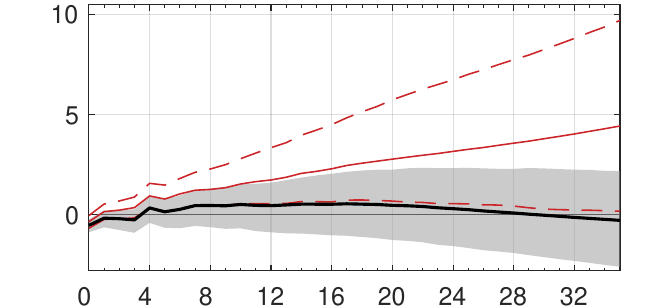} &
\includegraphics[trim=5cm 12cm 5cm 12.5cm, clip, width=0.19\textwidth]{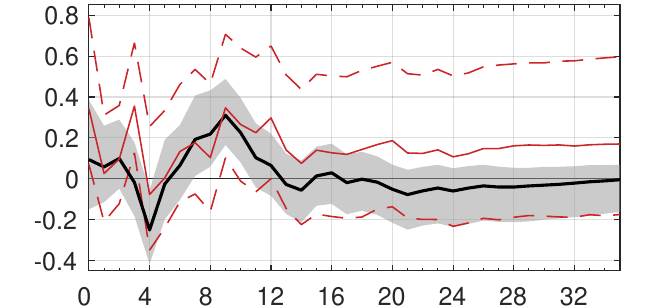} &
\includegraphics[trim=5cm 12cm 5cm 12.5cm, clip, width=0.19\textwidth]{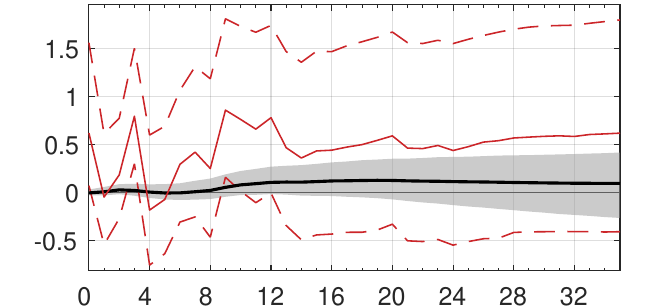} \\

\raisebox{2\height}{\rotatebox{90}{\scriptsize IT}} &
\includegraphics[trim=5cm 12cm 5cm 12.5cm, clip, width=0.19\textwidth]{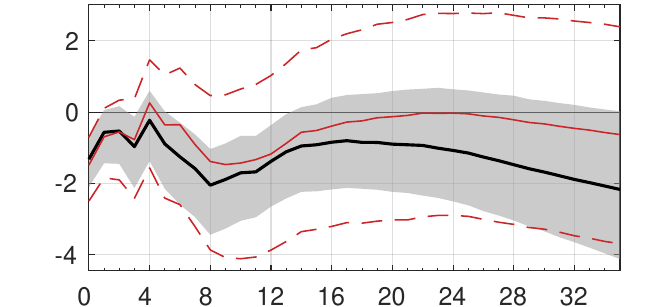} &
\includegraphics[trim=5cm 12cm 5cm 12.5cm, clip, width=0.19\textwidth]{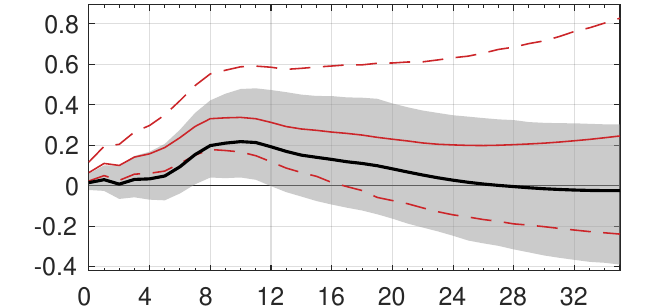} &
\includegraphics[trim=5cm 12cm 5cm 12.5cm, clip, width=0.19\textwidth]{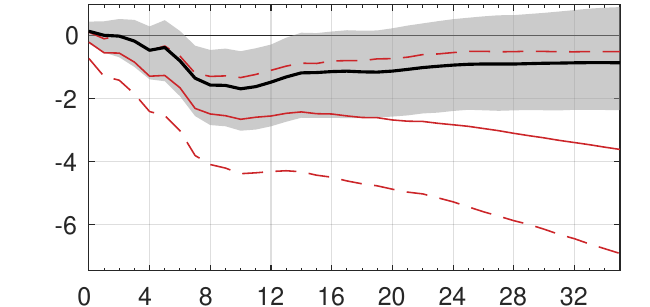} &
\includegraphics[trim=5cm 12cm 5cm 12.5cm, clip, width=0.19\textwidth]{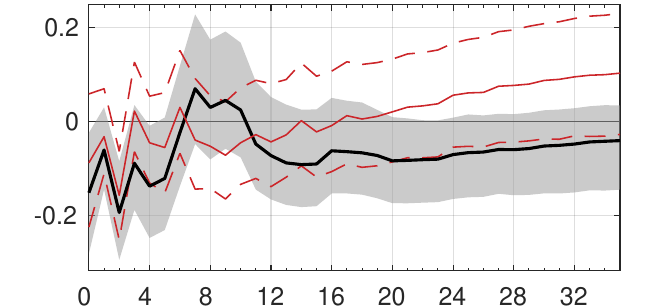} &
\includegraphics[trim=5cm 12cm 5cm 12.5cm, clip, width=0.19\textwidth]{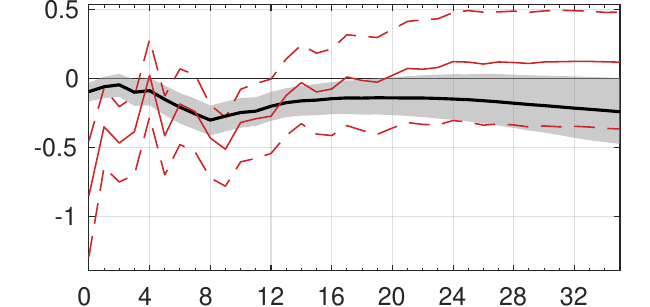} \\

\raisebox{1.5\height}{\rotatebox{90}{\scriptsize PT}} &
\includegraphics[trim=5cm 12cm 5cm 12.5cm, clip, width=0.19\textwidth]{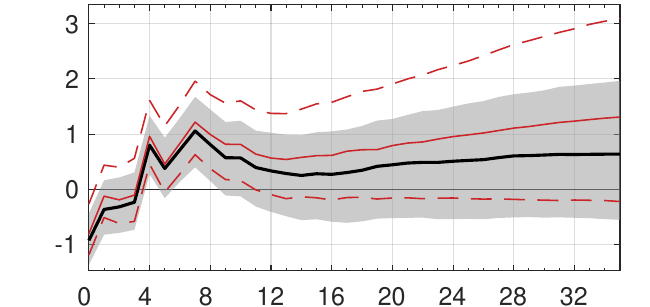} &
\includegraphics[trim=5cm 12cm 5cm 12.5cm, clip, width=0.19\textwidth]{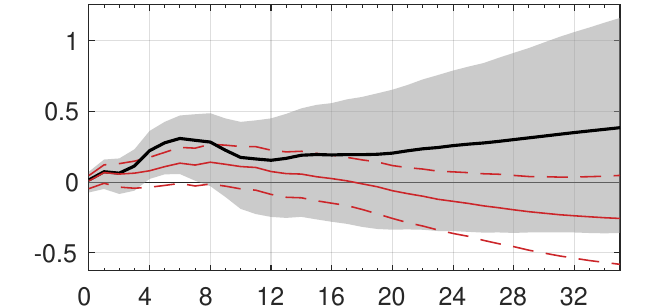} &
\includegraphics[trim=5cm 12cm 5cm 12.5cm, clip, width=0.19\textwidth]{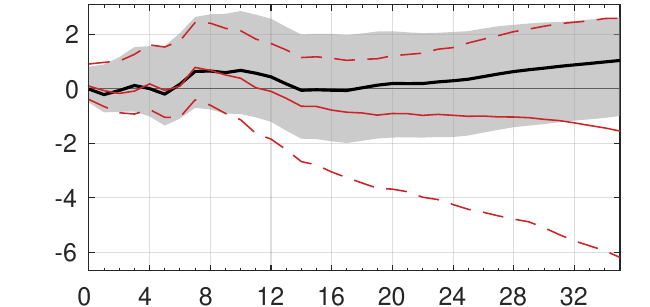} &
\includegraphics[trim=5cm 12cm 5cm 12.5cm, clip, width=0.19\textwidth]{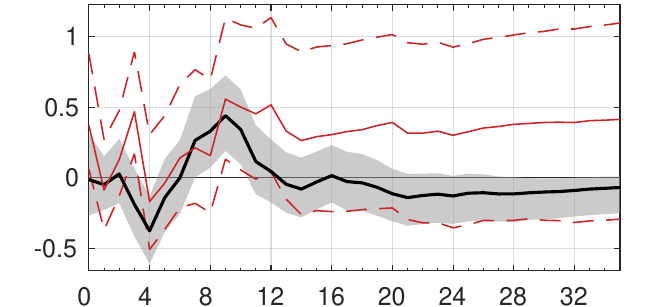} &
\includegraphics[trim=5cm 12cm 5cm 12.5cm, clip, width=0.19\textwidth]{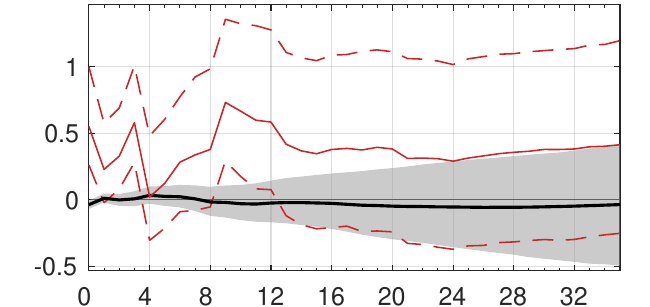} \\
\end{tabular}
\begin{tabular}{p{1\textwidth}} \scriptsize Notes: \rm Each sub-figure plots the difference between the country-level IRF and the corresponding EA counterpart for one variable and one country. Each column of the graph represents a variable, while each row represents a country. The variables considered are: Industrial Production: Manufacturing (IPMN), HICP: Overall (HICPOV), Stock Price Index (SHIX), 10-years Interest Rates (LTIRT) and Unemployment Rate (UNETOT). The black solid line is the point estimate in our baseline setting, while the red solid line is the point estimate obtained with the alternative set of transformation leaving all rates in levels. The gray shaded area and red dotted lines are the corresponding 68\% confidence intervals. The scale in the vertical axis differs across variables and countries.
\end{tabular}
\label{fig::diffIRF_TR7}
\end{figure}
\newpage

\subsection{Cross-country characteristics and IRF dynamics}
\begin{table}[ht!]
\centering
\caption{Cross-country characteristics and IRFs dynamics: all rates in level}
\setlength{\tabcolsep}{.03\textwidth}
\resizebox{6.5in}{!}{ 
\begin{tabular}{l|l|ccccc}
\hline
\hline
& & & & & & \\[-12pt]
\textbf{ID} & \textbf{Variable} & \textbf{IPMN}  & \textbf{HICPOV}  & \textbf{LTIRT} & \textbf{SHIX}  & \textbf{UNETOT} \\
 & & & & & & \\[-12pt]
\hline
& & & & & & \\[-10pt]
1 & Employment Protection & -0.05 & 0.17  & -0.09 & -0.05 & -0.44 \\
  & &\footnotesize (0.38) &\footnotesize (0.35) &\footnotesize (0.35) &\footnotesize (0.35) &\footnotesize (0.32) \\

& & & & & & \\[-10pt]
         
2 & Wage Adj. Frequency: more than once a year & 0.16  & 0.28  & -0.52 & 0.38  & -0.51 \\
  & &\footnotesize (0.44) &\footnotesize (0.39) &\footnotesize (0.35) &\footnotesize (0.38) &\footnotesize (0.35) \\

& & & & & & \\[-10pt]
       
3 & Wage Adj. Frequency: once a year & 0.65  & -0.15 & -0.61 & -0.22 & 0.11 \\
  & &\footnotesize (0.34) &\footnotesize (0.40) &\footnotesize (0.32) &\footnotesize (0.40) &\footnotesize (0.41) \\
 
& & & & & & \\[-10pt]
       
4 & Wage Adj. Frequency: less than once a year & -0.65 & 0.38  & 0.42  & -0.16 & -0.22 \\
  & &\footnotesize (0.34) &\footnotesize (0.38) &\footnotesize (0.37) &\footnotesize (0.40) &\footnotesize (0.40) \\

& & & & & & \\[-10pt]
          
5 & Price Flexibility & -0.09 & -0.39 & -0.39 & 0.64  & -0.07 \\
  & &\footnotesize (0.45) &\footnotesize (0.41) &\footnotesize (0.41) &\footnotesize (0.34) &\footnotesize (0.45) \\

& & & & & & \\[-10pt] 
     
6 & NFC Leverage & 0.00  & -0.06 & -0.36 & 0.68** & -0.15 \\
  & &\footnotesize (0.38) &\footnotesize (0.35) &\footnotesize (0.33) &\footnotesize (0.26) &\footnotesize (0.35) \\

& & & & & & \\[-10pt]
                  
7 & Homeownership & 0.09  & -0.02 & 0.67** & -0.09 & 0.57* \\  
  & &\footnotesize (0.38) &\footnotesize (0.35) &\footnotesize (0.26) &\footnotesize (0.35) &\footnotesize (0.29) \\

& & & & & & \\[-10pt]

8 & Homeownership with mortgage & 0.07  & -0.21 & -0.52 & 0.65** & -0.30 \\
  & &\footnotesize (0.38) &\footnotesize (0.35) &\footnotesize (0.30) &\footnotesize (0.27) &\footnotesize (0.34) \\

& & & & & & \\[-10pt]
       
9 & Homeownership without mortgage & -0.01 & 0.17  & 0.81*** & -0.59* & 0.57* \\
  & &\footnotesize (0.38) &\footnotesize (0.35) &\footnotesize (0.21) &\footnotesize (0.28) &\footnotesize (0.29) \\

& & & & & & \\[-10pt]
             
10 & Fixed-Rate Mortgages & -0.28 & 0.33  & -0.77*** & 0.43  & -0.65** \\
   & &\footnotesize (0.36) &\footnotesize (0.33) &\footnotesize (0.23) &\footnotesize (0.32) &\footnotesize (0.27) \\

& & & & & & \\[-10pt]
          
11 & Loan to Value & -0.32 & -0.26 & -0.11 & 0.71** & -0.07 \\
   & &\footnotesize (0.39) &\footnotesize (0.36) &\footnotesize (0.38) &\footnotesize (0.27) &\footnotesize (0.38) \\

& & & & & & \\[-10pt]
          
12 & Share of HtM & 0.39  & -0.53 & 0.49  & 0.17  & 0.35 \\
   & &\footnotesize (0.35) &\footnotesize (0.30) &\footnotesize (0.31) &\footnotesize (0.35) &\footnotesize (0.33) \\

& & & & & & \\[-10pt]
       
13 & Share of WHtM & 0.40  & -0.48 & 0.70** & -0.02 & 0.51 \\
   & &\footnotesize (0.35) &\footnotesize (0.31) &\footnotesize (0.25) &\footnotesize (0.35) &\footnotesize (0.30) \\

& & & & & & \\[-10pt]
          
14 & Saving Rate & -0.37 & -0.12 & -0.85*** & 0.60* & -0.37 \\
   & &\footnotesize (0.35) &\footnotesize (0.35) &\footnotesize (0.19) &\footnotesize (0.28) &\footnotesize (0.33) \\

& & & & & & \\[-10pt]

15 & Explained Variance & -0.75** & 0.13  & -0.95*** & 0.37  & 0.37 \\
   & &\footnotesize (0.25) &\footnotesize (0.35) &\footnotesize (0.11) &\footnotesize (0.33) &\footnotesize (0.33) \\
\hline
\hline
\end{tabular}%
} 
\begin{tabular}{p{.95\textwidth}} \scriptsize {\sc Notes}: \rm Each entry of the table corresponds to the correlation between each indicator (in the row) and the peak of the impulse response for a specific variable (in the columns). The variables considered are: Industrial Production: Manufacturing (IPMN), HICP: Overall (HICPOV), Stock Price Index (SHIX), 10-years Interest Rates (LTIRT) and Unemployment Rate (UNETOT). Standard errors in parentheses, asterisks denote statistical significance at 99\% (***), 95\% (**), and 90\% (*). The first row reports the Employment Protection Index (OECD). Rows 2-4 show the frequency of wage adjustments by firms \citep{branten2018nominal}, while row 1 reports the frequency with which firms adjust output prices \citep{gautier2024new}.
Row 6 shows the leverage of non-financial corporations relative to GDP (Eurostat). Row 7 reports the homeownership rate as a share of the population (Eurostat). Rows 8 and 9 indicate, respectively, the shares of residential properties with and without an outstanding mortgage (Eurostat).
Row 10 reports the share of fixed-rate mortgages \citep{de2025long}. Row 11 shows the average household loan-to-value ratio from the Eurosystem Household Finance and Consumption Survey (HFCS).
Rows 12 and 13 display, respectively, the shares of \enquote{wealthy hand-to-mouth} and \enquote{poor hand-to-mouth} households, constructed as in \citet{slacalek2020household} based on HFCS data.
Row 14 reports the household saving rate as a percentage of GDP (OECD). Finally, row 15 corresponds to the share of variance explained by the common factors for each variable in the columns.
\end{tabular}
\label{tab::CRanalysis_TR7}
\end{table}

\newpage

\section{Monthly data: identification via sign restrictions}
\label{app::signres}
Sign restrictions are imposed as summarized in Table \ref{tab::signM}. For the policy rate, $R_t$, we use the shadow rate of \citet{Wu}, as it is common in the literature on identification with sign restrictions to use a very short-term interest rate. Specifically, we impose a positive response of short- and long-term EA Interest Rates in the first two periods following the shock. For Industrial Production, Overall HICP, and the Stock Price Index, we impose a negative response in the second and third periods. Finally, for the Unemployment Rate, we impose a positive response in the second and third periods.  This specification accommodates nominal rigidities and delayed real adjustments, allowing for a gradual transmission of monetary policy effects. Restrictions on Industrial Production, the Unemployment Rate, Overall HICP, and Interest Rates are consistent with the predictions of a standard New Keynesian DSGE model and with the conventional sign restrictions typically imposed to identify a monetary policy shock \citep[e.g.,][]{peersman2005caused, barigozzi2014euro}. The restriction on the Stock Price Index is consistent with the \citet{jarocinski2020deconstructing} approach used to disentangle a standard monetary policy shock from an information shock.

\begin{table}[H]
  \centering
  \caption{Sign restrictions: monthly data}
  \footnotesize
  \begin{tabular}{c | c c c c c c}
\hline
\hline
Horizon & $R_t$ & $\widehat\chi_{{\rm IPMN\, EA},t}$ & $\widehat\chi_{{\rm HICPOV\, EA},t}$& $\widehat\chi_{{\rm LTIRT\, EA},t}$ & $\widehat\chi_{{\rm SHIX\, EA},t}$ & $\widehat\chi_{{\rm UNETOT\, EA},t}$\\
\hline
0 & + &	    &	  & + &  	&     \\
1 & + & $-$	& $-$ & + & $-$	& $+$ \\
2 &	  & $-$	& $-$ &	  & $-$ & $+$ \\
\hline
\hline
  \end{tabular}
\begin{tabular}{p{.85\textwidth}} \scriptsize Notes: \rm Each column of the table represents the common component of one variable, except for the EA Shadow Rate ($R_t$). The EA variables considered are: Industrial Production: Manufacturing (IPMN), HICP: Overall (HICPOV), Stock Price Index (SHIX), 10-years Interest Rates (LTIRT) and Unemployment Rate (UNETOT).
\end{tabular}
\label{tab::signM}
\end{table}
 Confidence intervals and point estimates are obtained using a bootstrap procedure with 10,000 replications. For each iteration, we proceed as follows. We draw an $n \times n$ matrix $\mathbf{W}$ whose rows are independently sampled from a $\mathcal{N}(\mathbf{0}, \mathbf{I}_n)$ distribution. We then compute the QR decomposition $\mathbf{W} = \mathbf{Q}\mathbf{R}$, where $\mathbf{Q}$ is orthogonal and $\mathbf{R}$ is upper triangular \citep{rubio2010structural}. The candidate structural impact matrix is given by $\mathbf{S} = \mathbf{L}\mathbf{Q}'$, where $\mathbf{L}$ denotes the lower triangular Cholesky factor of the reduced-form innovation covariance matrix $\mathbf{\Sigma}$. 

If the resulting impulse responses satisfy the sign restrictions in Table \ref{tab::signM}, the draw is accepted; otherwise, it is discarded. Within each bootstrap replication, we continue this process until $K$ admissible draws are obtained. Following \citet{fry2011sign}, we retain only the IRF that is closest to the median across the $K$ accepted draws. Here we set $K=30$. The same procedure is applied to the observed data. This yields 10,000+1 admissible IRFs, from which we construct the empirical distribution of responses. We take the median of this distribution as the point estimate, while the 16th and 84th percentiles serve as the lower and upper bounds of the confidence interval, respectively.
\subsection{EA IRFs}
\begin{figure}[H]
\centering \footnotesize \sc \smallskip
\setlength{\tabcolsep}{.005\textwidth}
\caption{EA IRF: monthly data and sign restrictions}
\begin{tabular}{ccc}
\scriptsize Shadow Rate ($R_t$) &\scriptsize  IP: Manufacturing (IPMN) &\scriptsize  HICP: Overall (HICPOV) \\
\includegraphics[trim= .5cm 8cm .5cm 8.5cm, clip, width=0.32\textwidth]{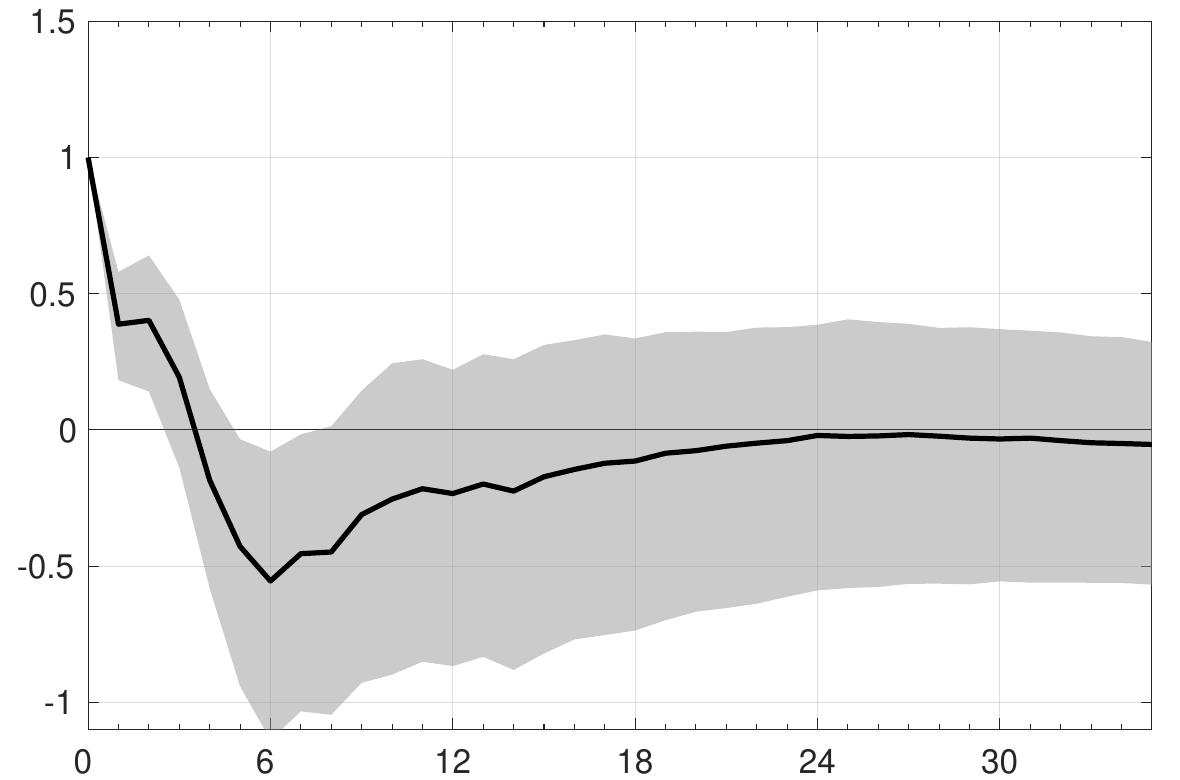} &
\includegraphics[trim= .5cm 8cm .5cm 8.5cm, clip, width=0.32\textwidth]{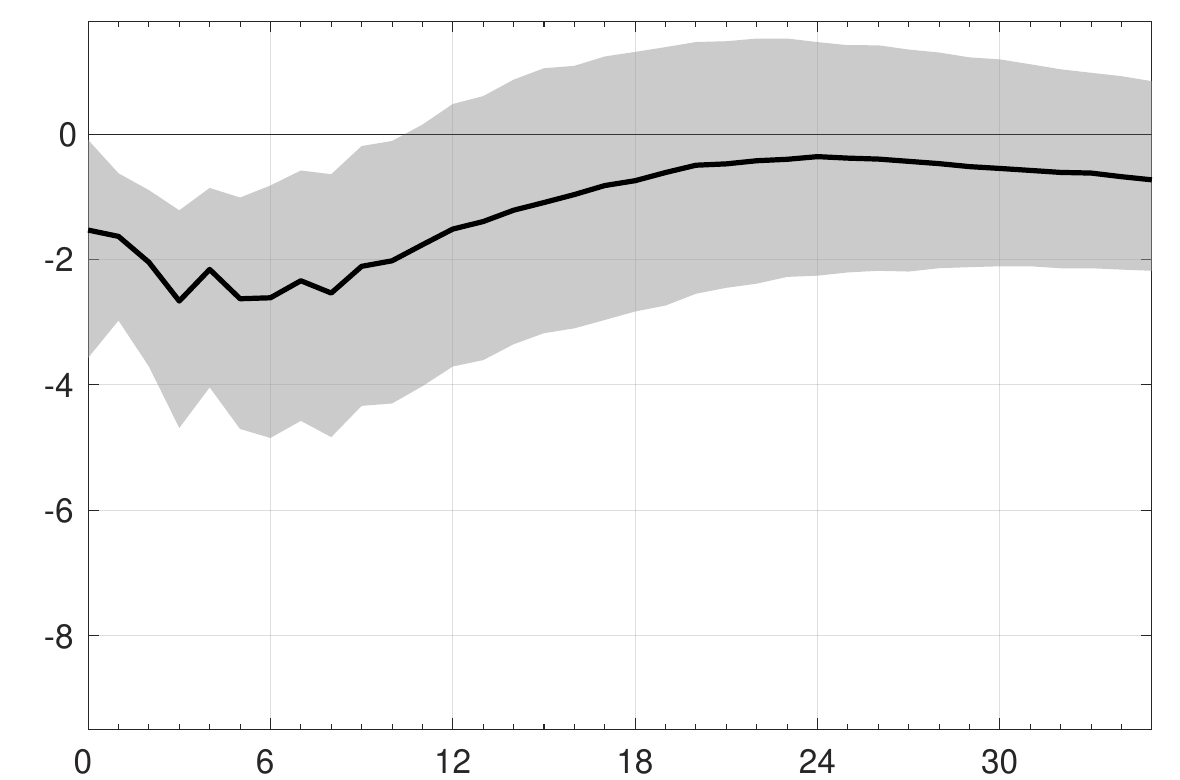} &
\includegraphics[trim= .5cm 8cm .5cm 8.5cm, clip, width=0.32\textwidth]{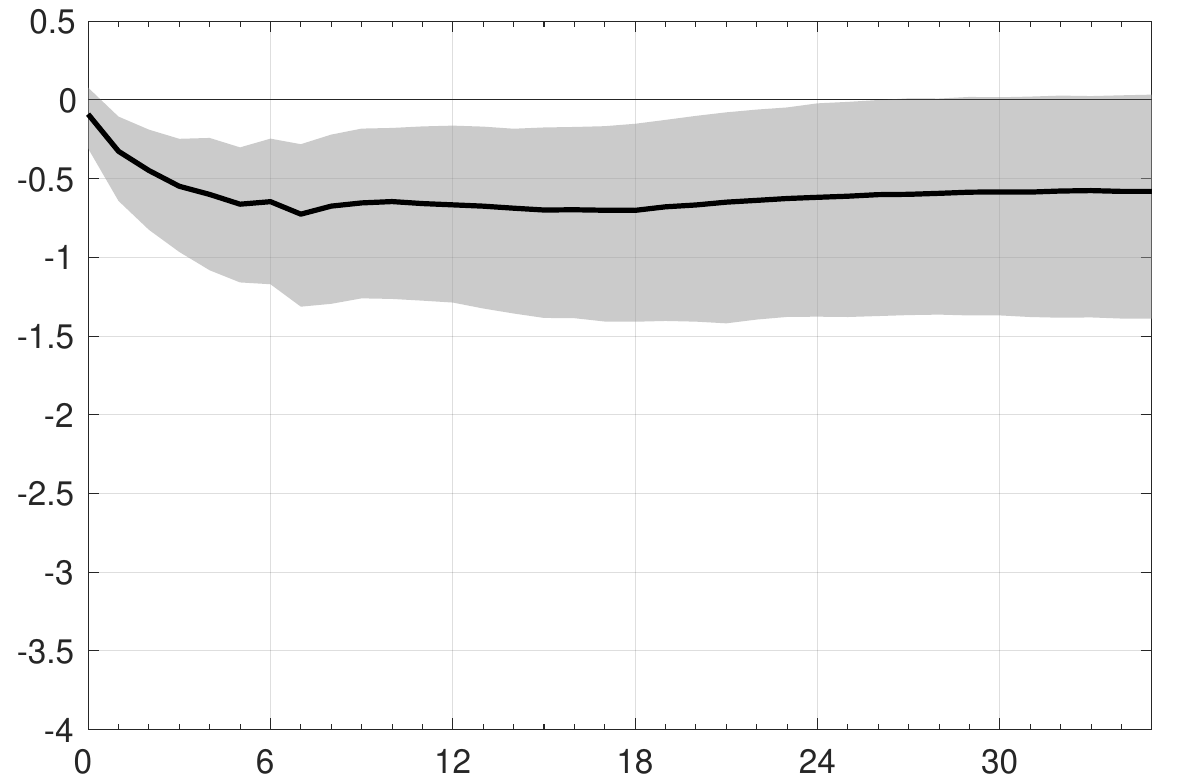} \\
\scriptsize 10-years Interest Rate (LTIRT) &\scriptsize Stock Price Index (SHIX) &\scriptsize Unemployment Rate (UNETOT)\\
\includegraphics[trim= .5cm 8cm .5cm 8.5cm, clip, width=0.32\textwidth]{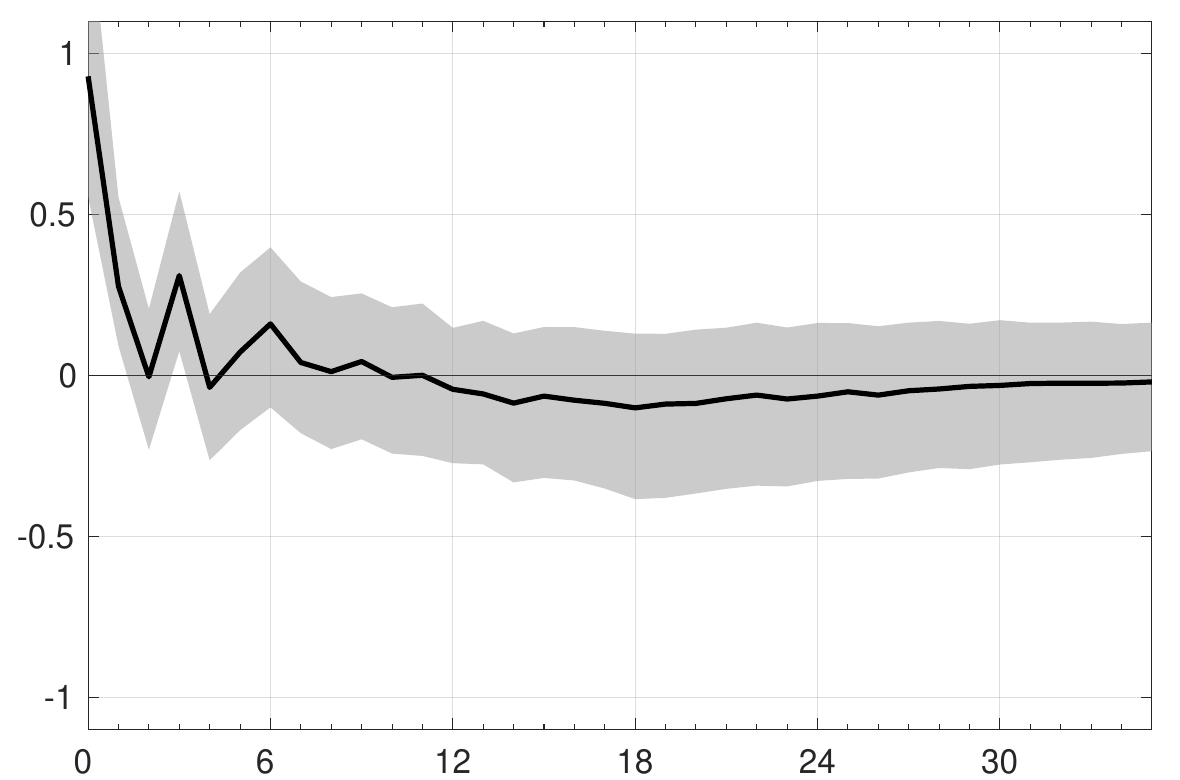} &
\includegraphics[trim= .5cm 8cm .5cm 8.5cm, clip, width=0.32\textwidth]{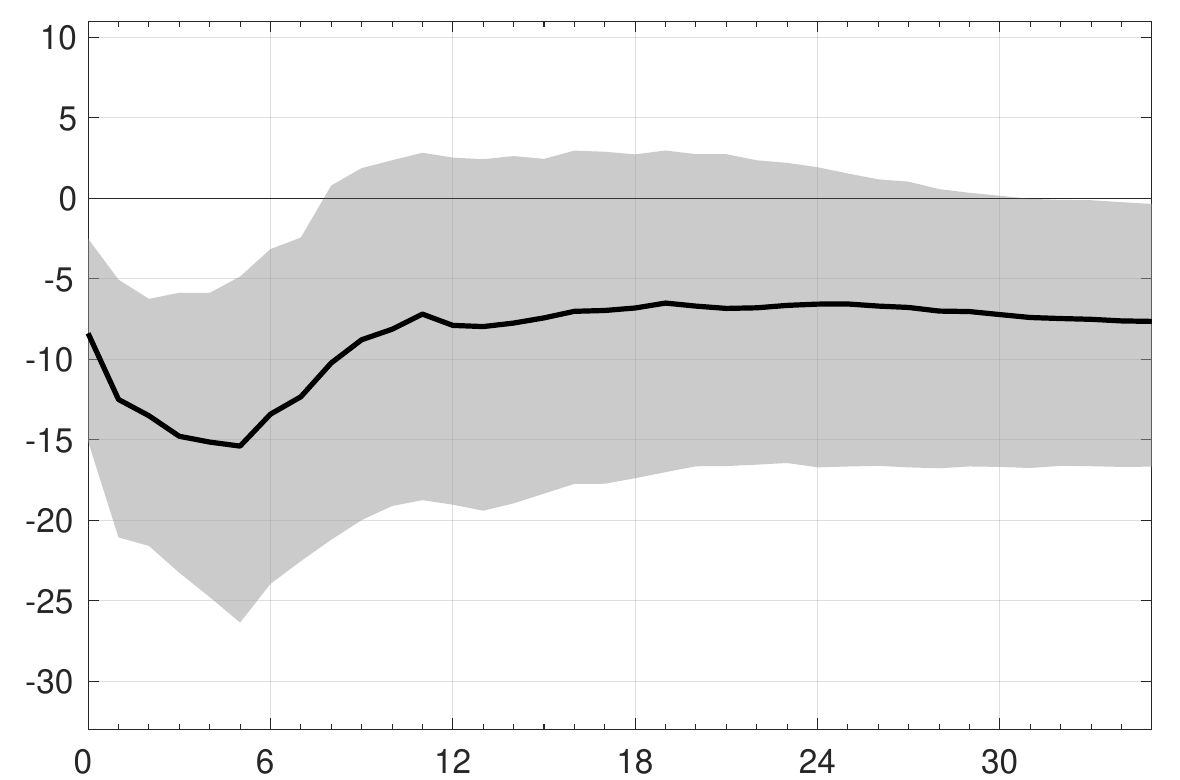} &
\includegraphics[trim= .5cm 8cm .5cm 8.5cm, clip, width=0.32\textwidth]{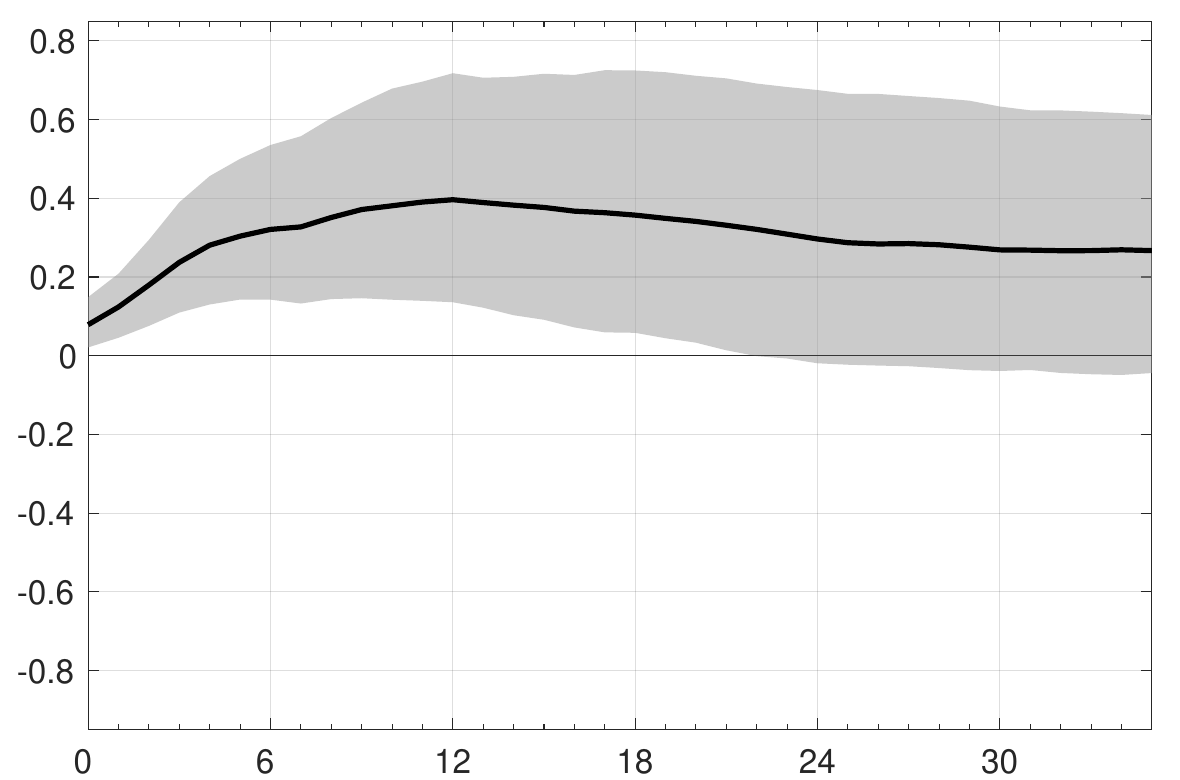} \\
\end{tabular}
\begin{tabular}{p{.98\textwidth}} \scriptsize Notes: \rm Each sub-figure plots the impulse response of one EA variable to a 100bps contractionary monetary policy shock. The black solid line is the point estimate in our baseline setting, while the gray shaded area is the corresponding 68\% confidence interval.
\end{tabular}
\label{fig::EA_IRFs_sign}
\end{figure}

\subsection{Country-level IRFs}

\begin{figure}[H]
\centering \footnotesize \sc \smallskip
\setlength{\tabcolsep}{.005\textwidth}
\caption{Country-level IRFs: monthly data and sign restrictions}
\begin{tabular}{ccc}
\scriptsize{IP: Manufacturing (IPMN)}  
&\scriptsize {HICP: Overall (HICPOV)}      
&\scriptsize {Stock Price Index (SHIX)}  \\

\includegraphics[trim= .5cm 8cm .5cm 8.5cm, clip, width=0.32\textwidth]{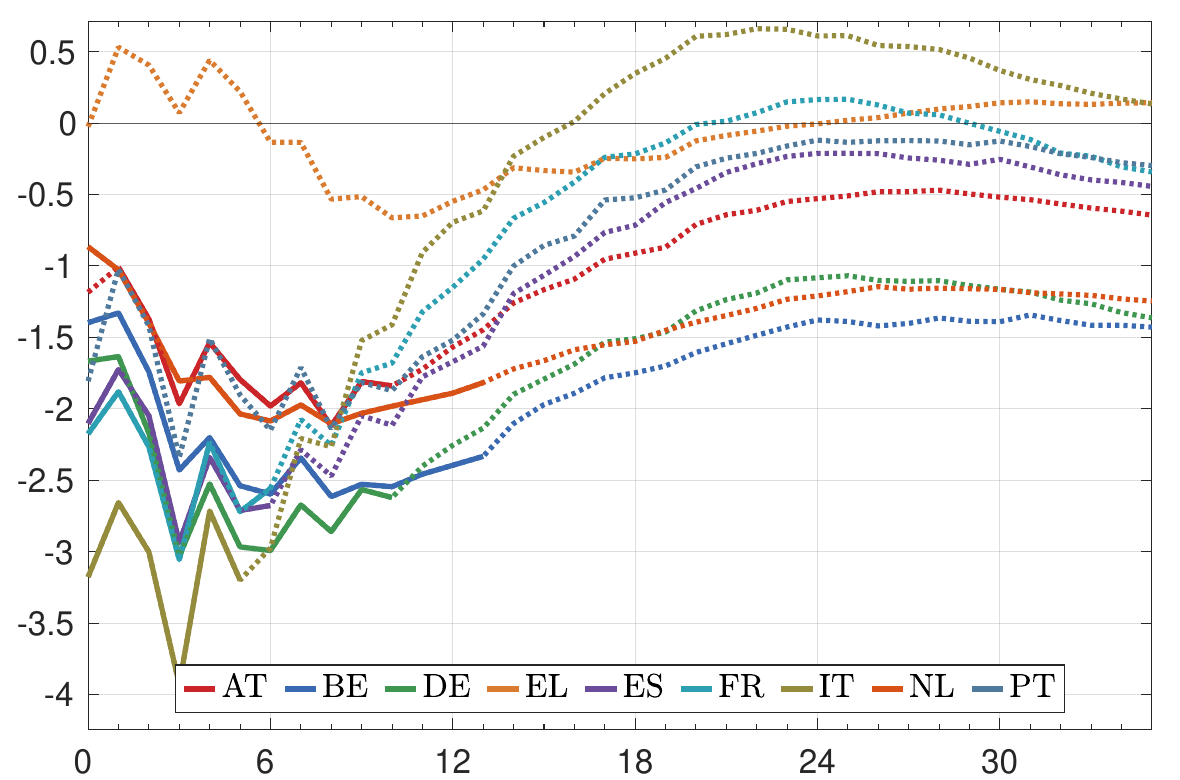} &
\includegraphics[trim= .5cm 8cm .5cm 8.5cm, clip, width=0.32\textwidth]{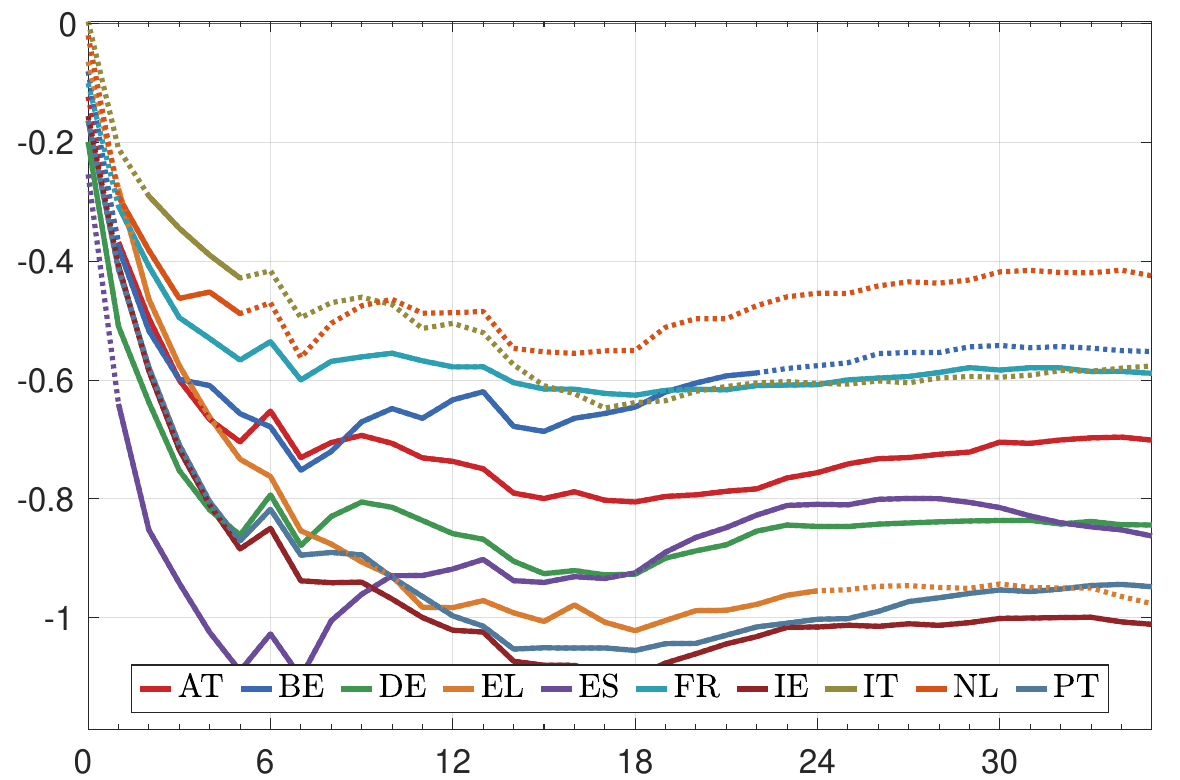} &
\includegraphics[trim= .5cm 8cm .5cm 8.5cm, clip, width=0.32\textwidth]{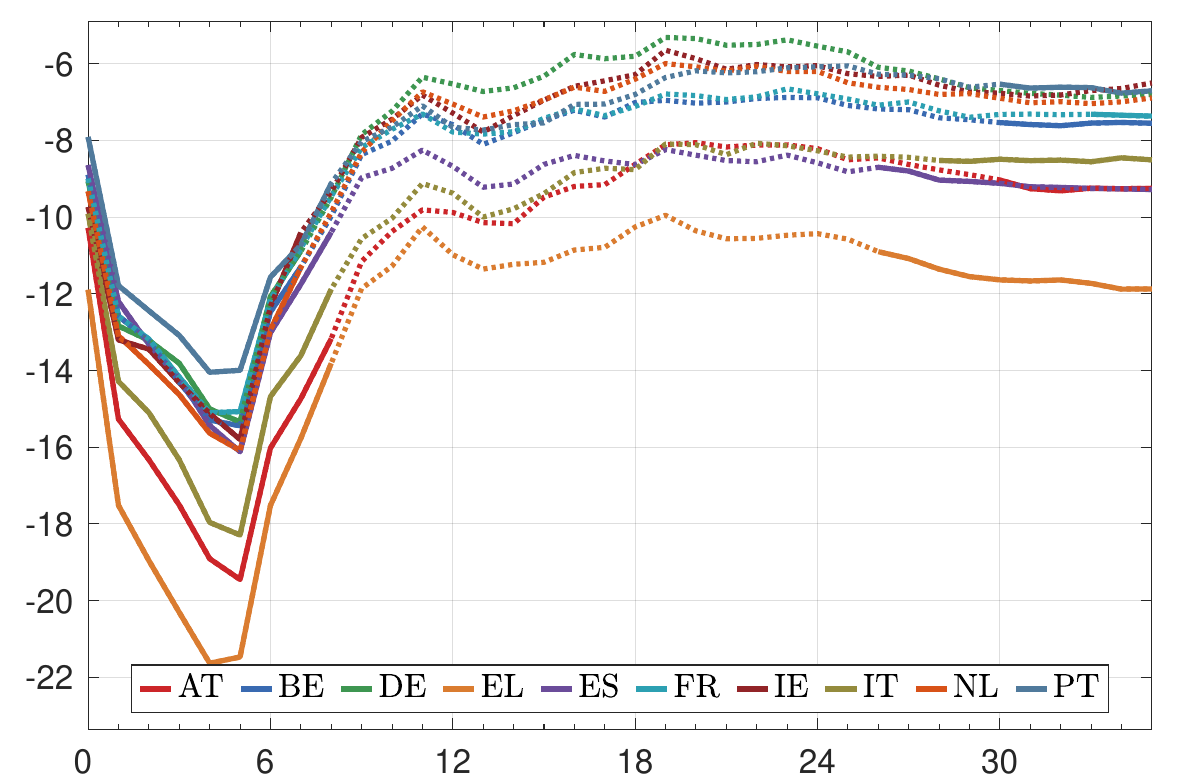} \\[10pt]
\end{tabular}

\begin{tabular}{cc}
\scriptsize {10-years Interest Rate (LTIRT)} 
&\scriptsize {Unemployment Rate (UNETOT)} \\

\includegraphics[trim= .5cm 8cm .5cm 8.5cm, clip, width=0.32\textwidth]{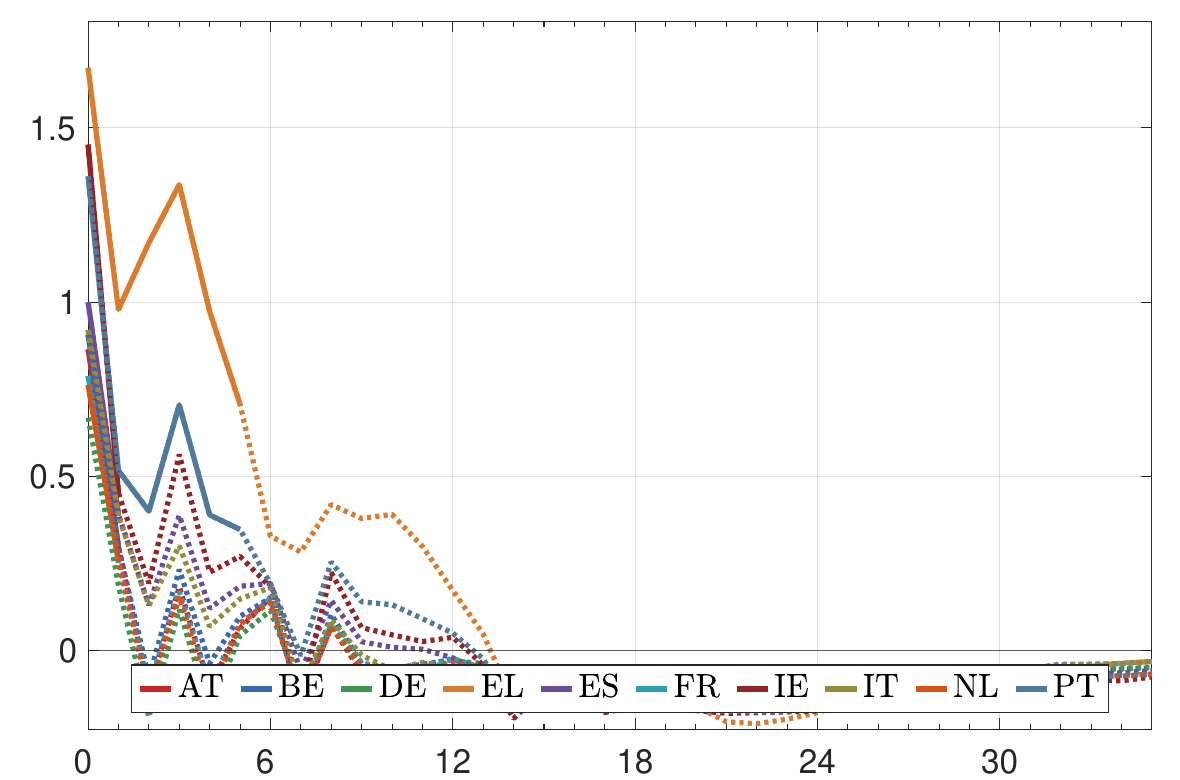} 
&
\includegraphics[trim= .5cm 8cm .5cm 8.5cm, clip, width=0.32\textwidth]{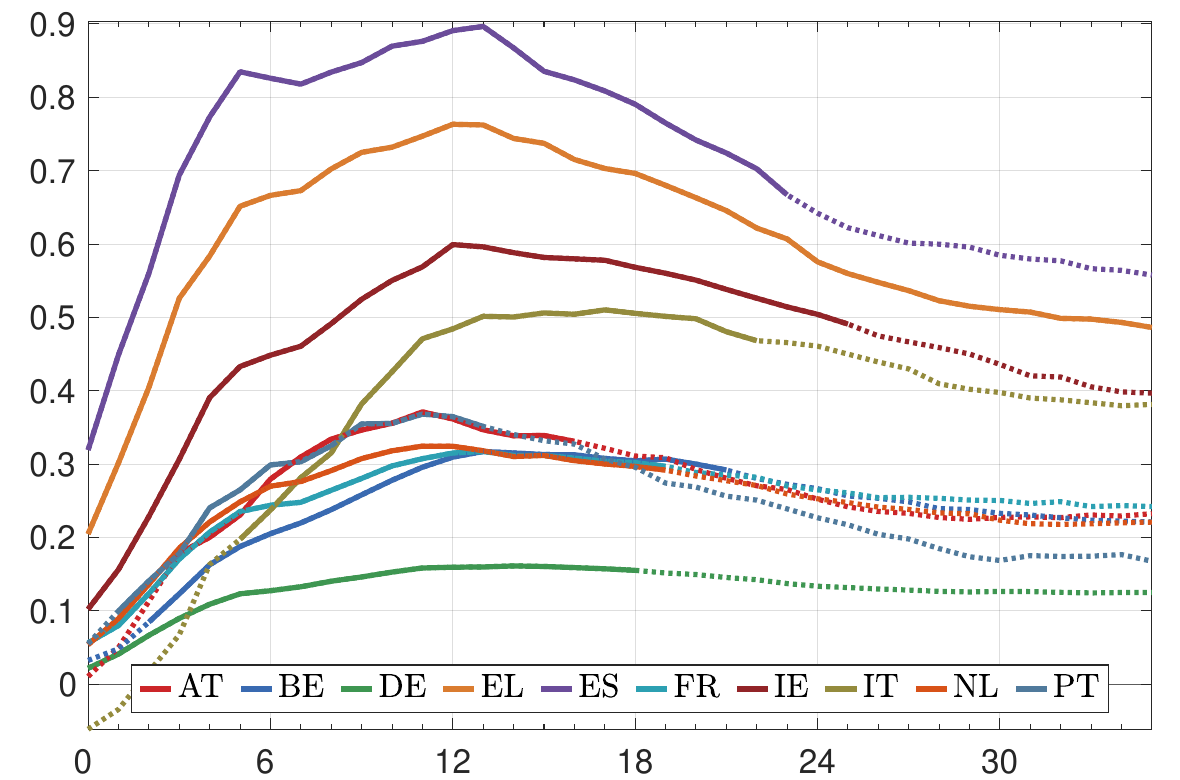}
\end{tabular}
\begin{tabular}{p{.98\textwidth}} \scriptsize Notes: \rm Each sub-figure plots the impulse responses, for all countries, of one variable to a 100bps contractionary monetary policy shock. Within each sub-figure, at each horizon $h=0,\ldots,36$ the country-level impulse responses are denoted with a solid line if the IRF is statistically significant at the 68\% level at that horizon, and with a dotted line otherwise. 
\end{tabular}
\label{fig::nat_IRFs_sign}
\end{figure}

\begin{figure}[H]
\centering \footnotesize \sc \smallskip
\setlength{\tabcolsep}{.005\textwidth}
\caption{Difference between country-specific and EA IRFs: monthly data and restrictions}
\begin{tabular}{lccccc}
& \hspace{5pt}\scriptsize IPMN &  \hspace{5pt}\scriptsize HICPOV & 
 \hspace{5pt}\scriptsize SHIX &  \hspace{5pt}\scriptsize LTIRT &  \hspace{8pt}\scriptsize UNETOT \\[4pt] 
\raisebox{1.3\height}{\rotatebox{90}{\scriptsize AT}} & \includegraphics[trim=5cm 12cm 5cm 12.5cm, clip, width=0.19\textwidth]{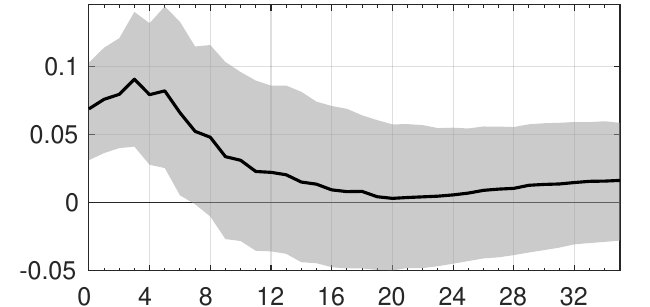} &
\includegraphics[trim=5cm 12cm 5cm 12.5cm, clip, width=0.19\textwidth]{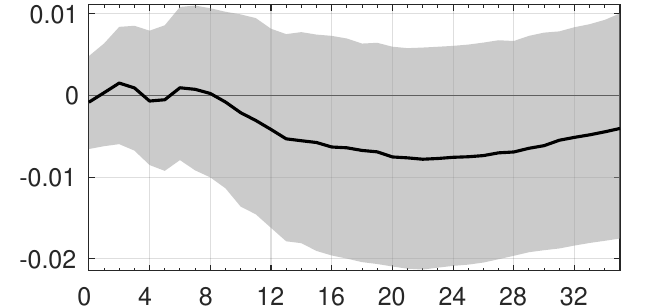} &
\includegraphics[trim=5cm 12cm 5cm 12.5cm, clip, width=0.19\textwidth]{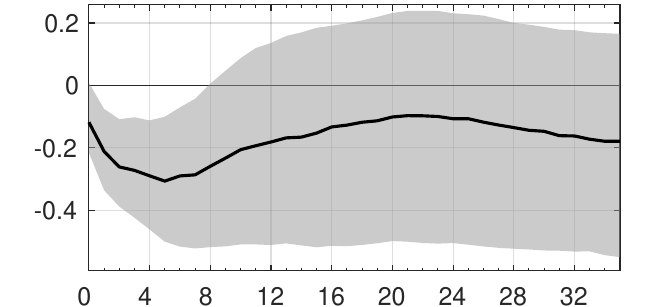} &
\includegraphics[trim=5cm 12cm 5cm 12.5cm, clip, width=0.19\textwidth]{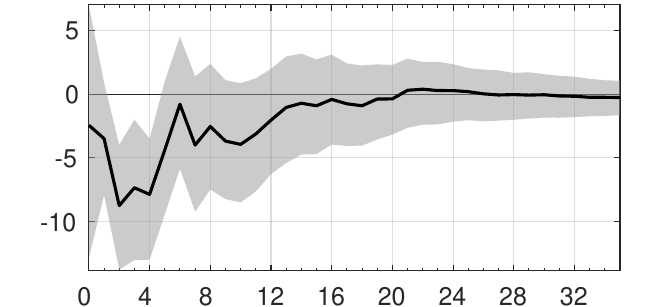} &
\includegraphics[trim=5cm 12cm 5cm 12.5cm, clip, width=0.19\textwidth]{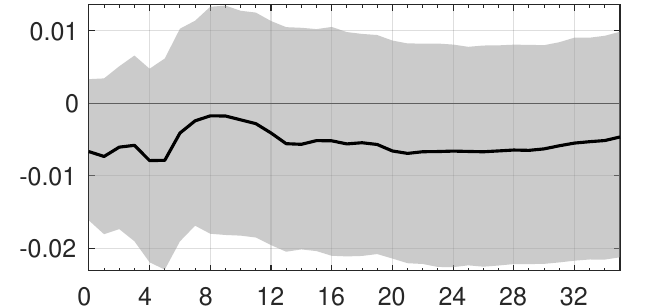} \\

\raisebox{1.3\height}{\rotatebox{90}{\scriptsize BE}} &
\includegraphics[trim=5cm 12cm 5cm 12.5cm, clip, width=0.19\textwidth]{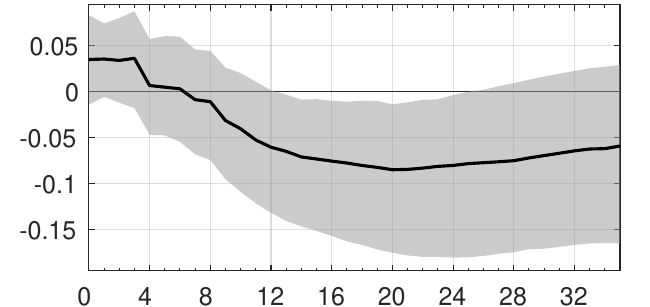} &
\includegraphics[trim=5cm 12cm 5cm 12.5cm, clip, width=0.19\textwidth]{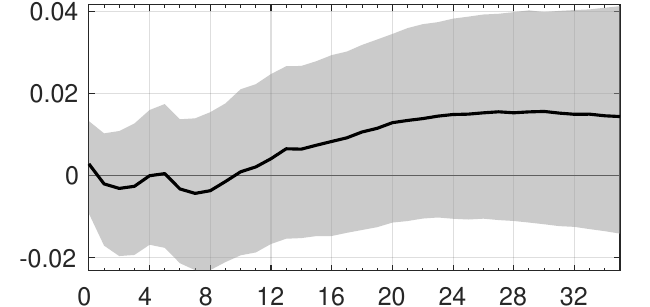} &
\includegraphics[trim=5cm 12cm 5cm 12.5cm, clip, width=0.19\textwidth]{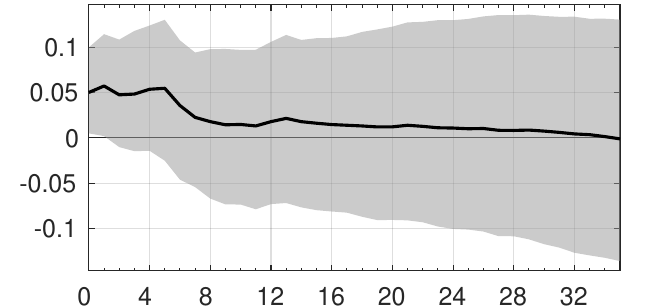} &
\includegraphics[trim=5cm 12cm 5cm 12.5cm, clip, width=0.19\textwidth]{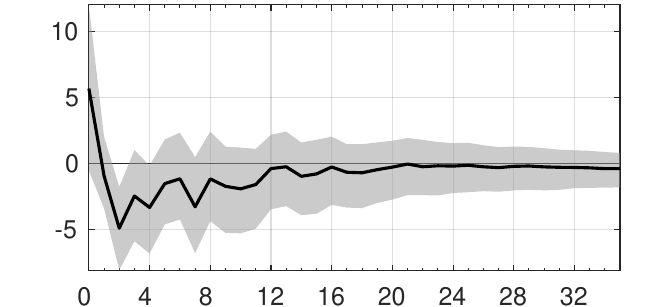} &
\includegraphics[trim=5cm 12cm 5cm 12.5cm, clip, width=0.19\textwidth]{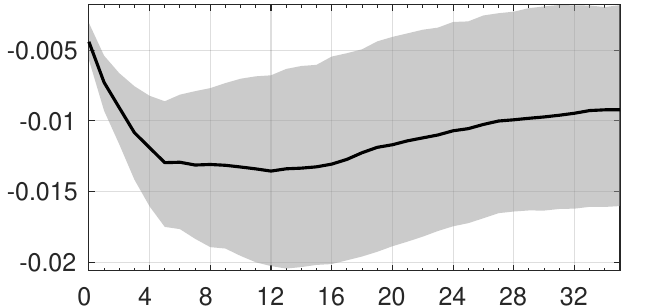} \\

\raisebox{1.3\height}{\rotatebox{90}{\scriptsize DE}} &
\includegraphics[trim=5cm 12cm 5cm 12.5cm, clip, width=0.19\textwidth]{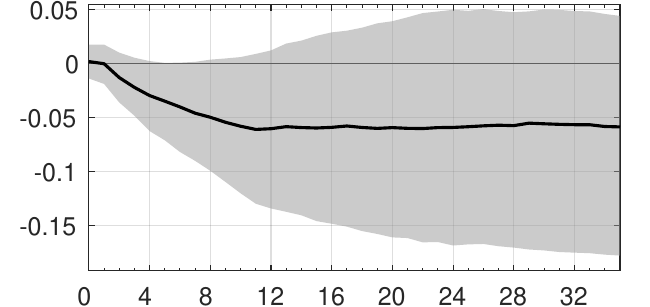} &
\includegraphics[trim=5cm 12cm 5cm 12.5cm, clip, width=0.19\textwidth]{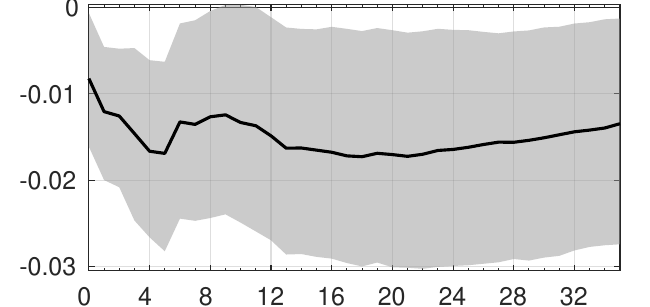} &
\includegraphics[trim=5cm 12cm 5cm 12.5cm, clip, width=0.19\textwidth]{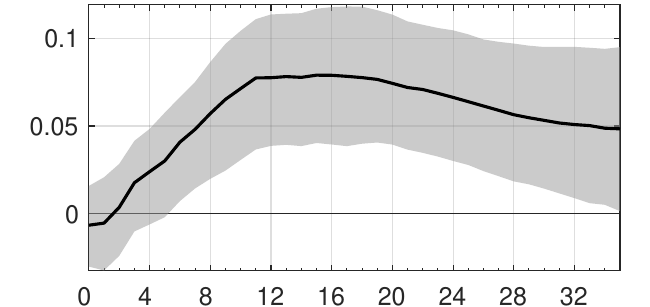} &
\includegraphics[trim=5cm 12cm 5cm 12.5cm, clip, width=0.19\textwidth]{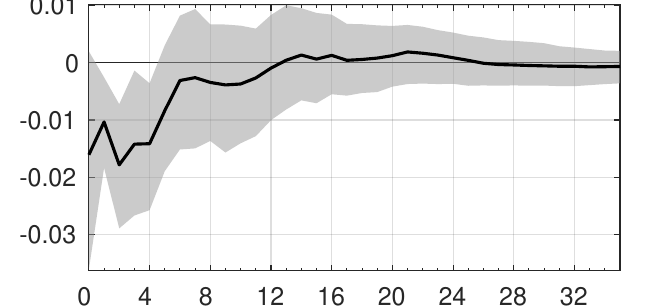} &
\includegraphics[trim=5cm 12cm 5cm 12.5cm, clip, width=0.19\textwidth]{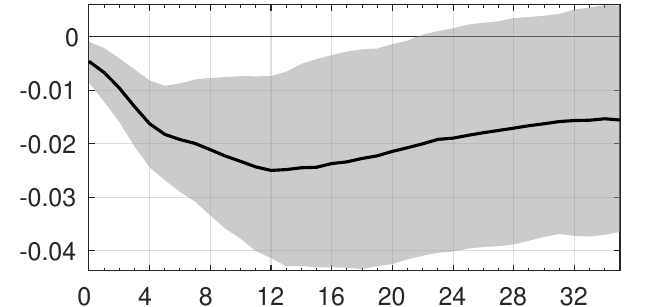} \\

\raisebox{1.3\height}{\rotatebox{90}{\scriptsize FR}} &
\includegraphics[trim=5cm 12cm 5cm 12.5cm, clip, width=0.19\textwidth]{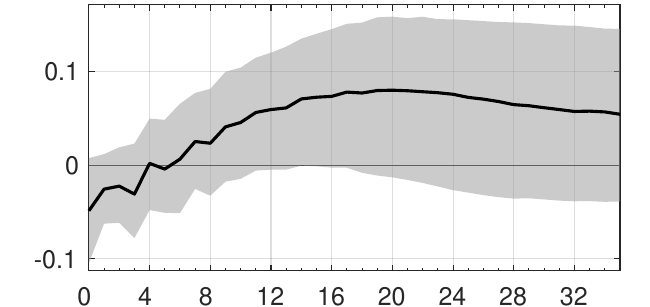} &
\includegraphics[trim=5cm 12cm 5cm 12.5cm, clip, width=0.19\textwidth]{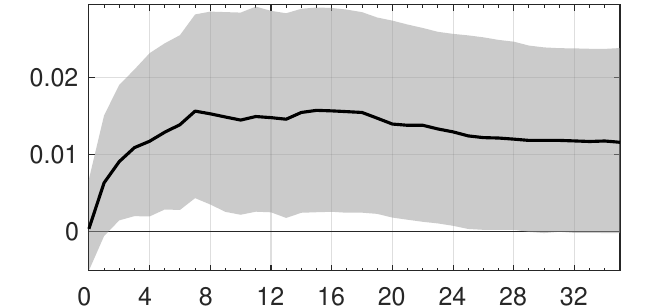} &
\includegraphics[trim=5cm 12cm 5cm 12.5cm, clip, width=0.19\textwidth]{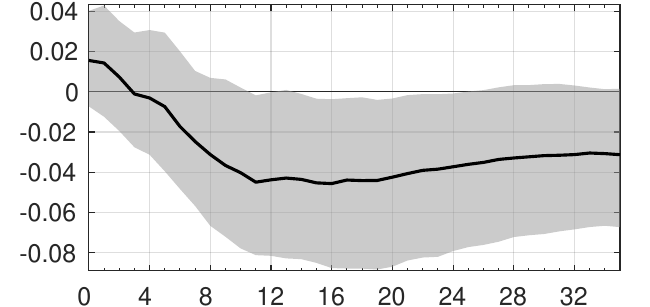} &
\includegraphics[trim=5cm 12cm 5cm 12.5cm, clip, width=0.19\textwidth]{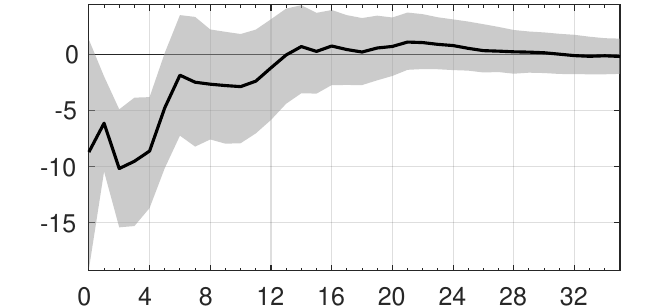} &
\includegraphics[trim=5cm 12cm 5cm 12.5cm, clip, width=0.19\textwidth]{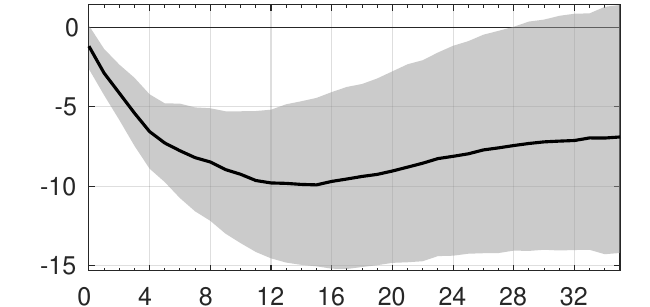} \\

\raisebox{1.3\height}{\rotatebox{90}{\scriptsize NL}} &
\includegraphics[trim=5cm 12cm 5cm 12.5cm, clip, width=0.19\textwidth]{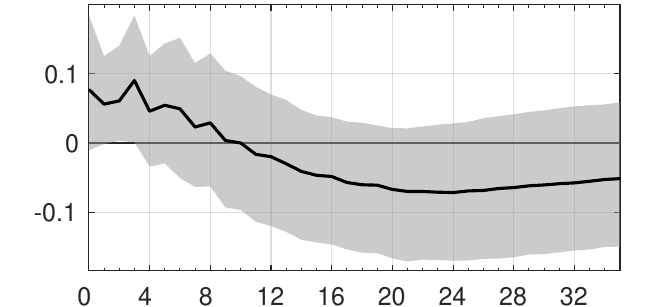} &
\includegraphics[trim=5cm 12cm 5cm 12.5cm, clip, width=0.19\textwidth]{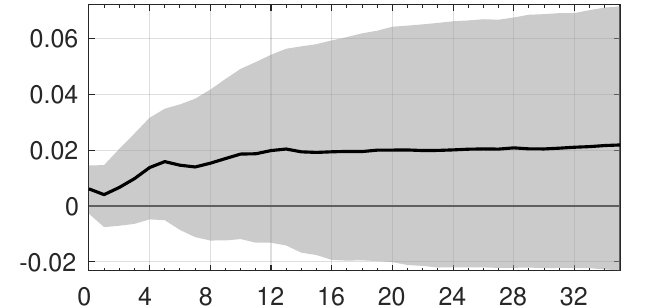} &
\includegraphics[trim=5cm 12cm 5cm 12.5cm, clip, width=0.19\textwidth]{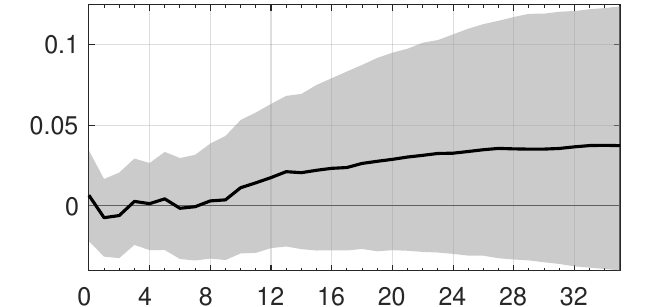} &
\includegraphics[trim=5cm 12cm 5cm 12.5cm, clip, width=0.19\textwidth]{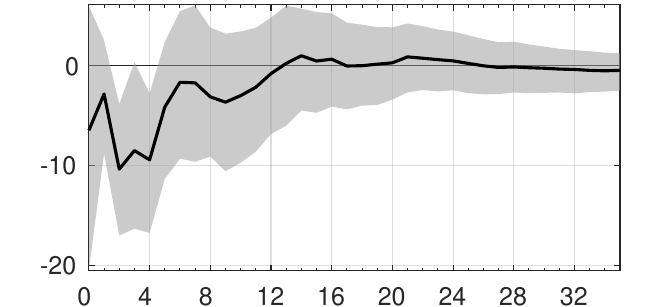} &
\includegraphics[trim=5cm 12cm 5cm 12.5cm, clip, width=0.19\textwidth]{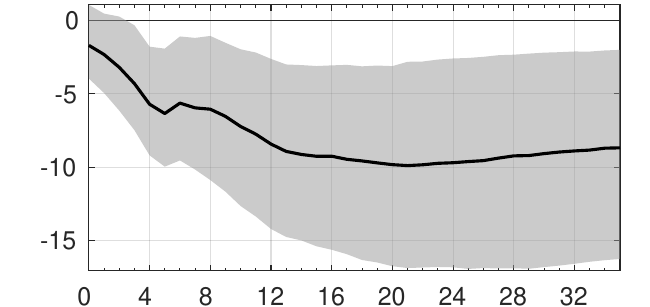} \\

\raisebox{1.5\height}{\rotatebox{90}{\scriptsize EL}} &
\includegraphics[trim=5cm 12cm 5cm 12.5cm, clip, width=0.19\textwidth]{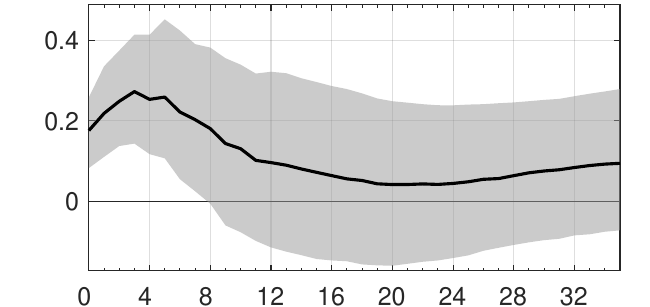} &
\includegraphics[trim=5cm 12cm 5cm 12.5cm, clip, width=0.19\textwidth]{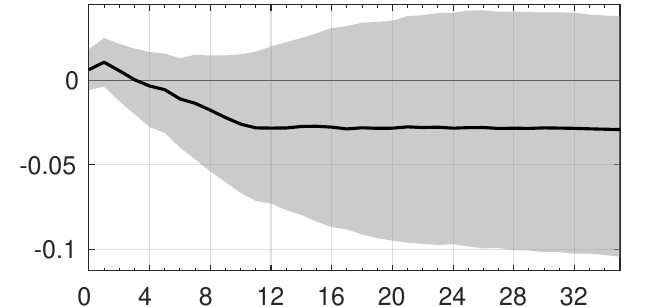} &
\includegraphics[trim=5cm 12cm 5cm 12.5cm, clip, width=0.19\textwidth]{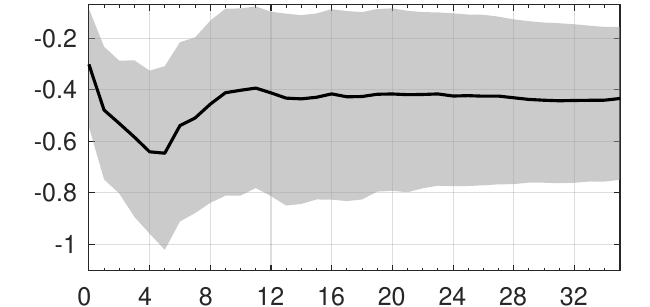} &
\includegraphics[trim=5cm 12cm 5cm 12.5cm, clip, width=0.19\textwidth]{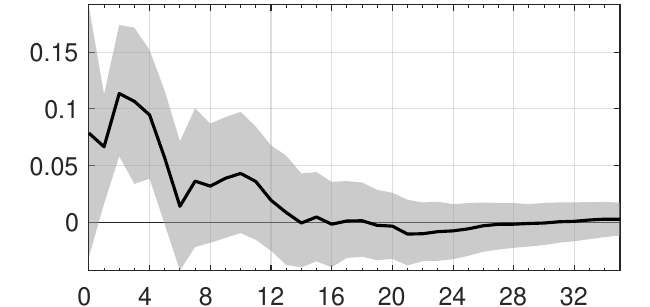} &
\includegraphics[trim=5cm 12cm 5cm 12.5cm, clip, width=0.19\textwidth]{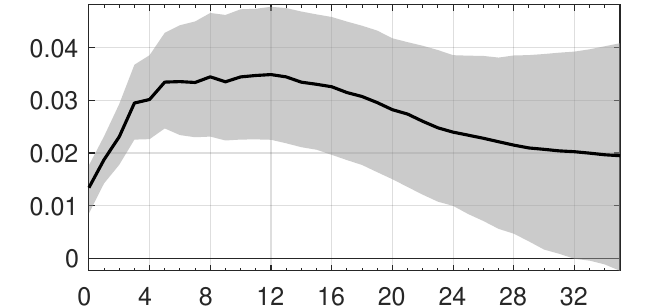} \\

\raisebox{1.5\height}{\rotatebox{90}{\scriptsize ES}} &
\includegraphics[trim=5cm 12cm 5cm 12.5cm, clip, width=0.19\textwidth]{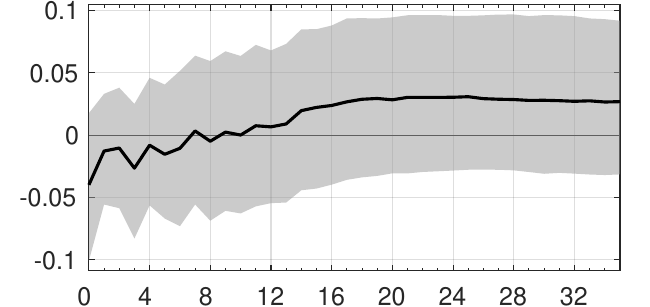} &
\includegraphics[trim=5cm 12cm 5cm 12.5cm, clip, width=0.19\textwidth]{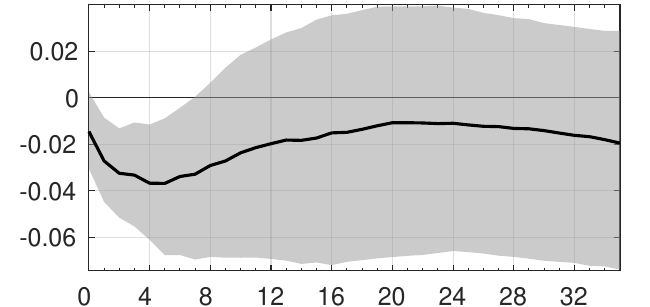} &
\includegraphics[trim=5cm 12cm 5cm 12.5cm, clip, width=0.19\textwidth]{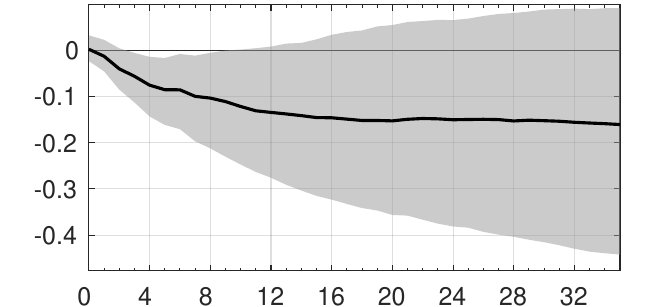} &
\includegraphics[trim=5cm 12cm 5cm 12.5cm, clip, width=0.19\textwidth]{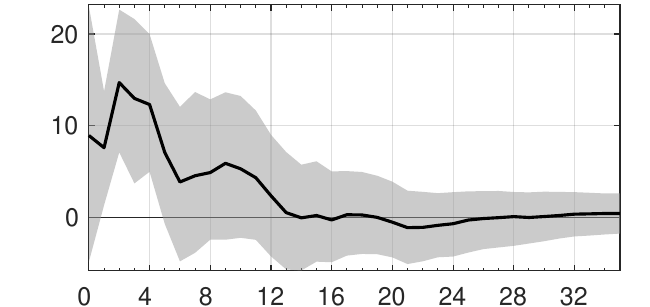} &
\includegraphics[trim=5cm 12cm 5cm 12.5cm, clip, width=0.19\textwidth]{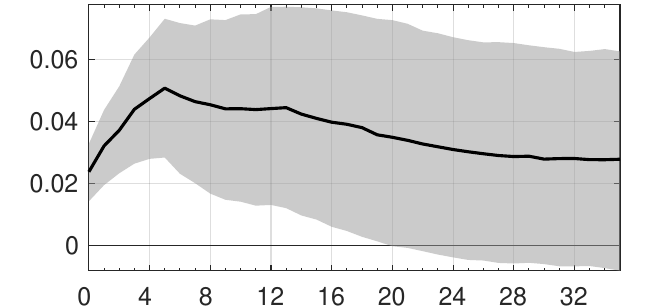} \\

\raisebox{1.8\height}{\rotatebox{90}{\scriptsize IE}} &
 &
\includegraphics[trim=5cm 12cm 5cm 12.5cm, clip, width=0.19\textwidth]{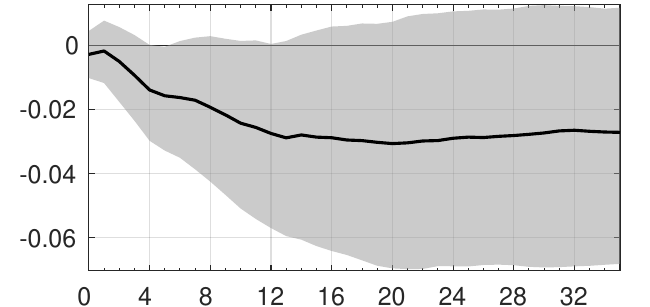} &
\includegraphics[trim=5cm 12cm 5cm 12.5cm, clip, width=0.19\textwidth]{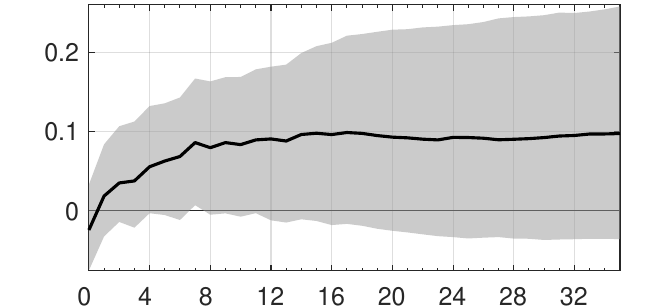} &
\includegraphics[trim=5cm 12cm 5cm 12.5cm, clip, width=0.19\textwidth]{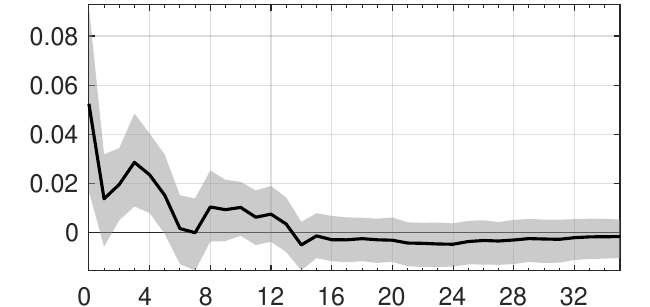} &
\includegraphics[trim=5cm 12cm 5cm 12.5cm, clip, width=0.19\textwidth]{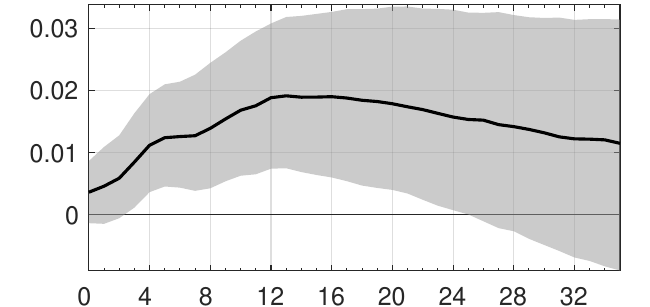} \\

\raisebox{1.8\height}{\rotatebox{90}{\scriptsize IT}} &
\includegraphics[trim=5cm 12cm 5cm 12.5cm, clip, width=0.19\textwidth]{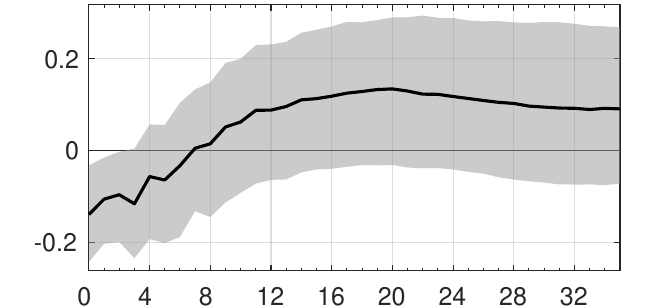} &
\includegraphics[trim=5cm 12cm 5cm 12.5cm, clip, width=0.19\textwidth]{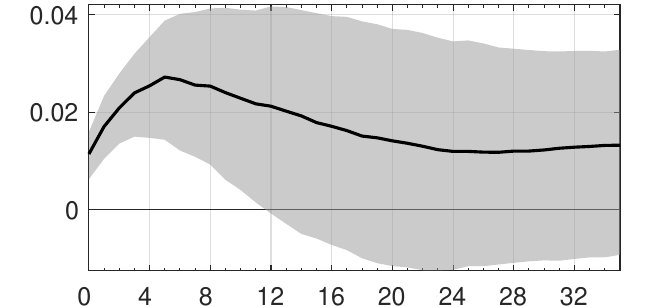} &
\includegraphics[trim=5cm 12cm 5cm 12.5cm, clip, width=0.19\textwidth]{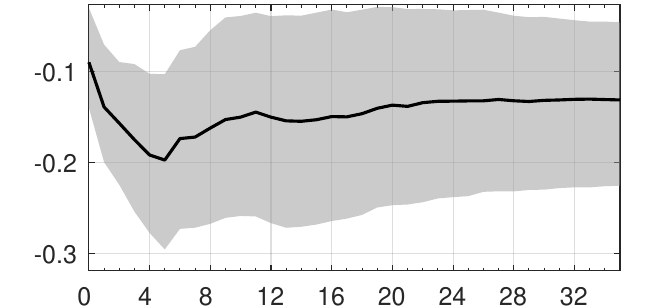} &
\includegraphics[trim=5cm 12cm 5cm 12.5cm, clip, width=0.19\textwidth]{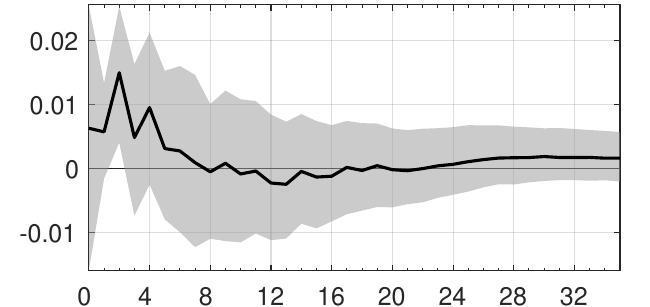} &
\includegraphics[trim=5cm 12cm 5cm 12.5cm, clip, width=0.19\textwidth]{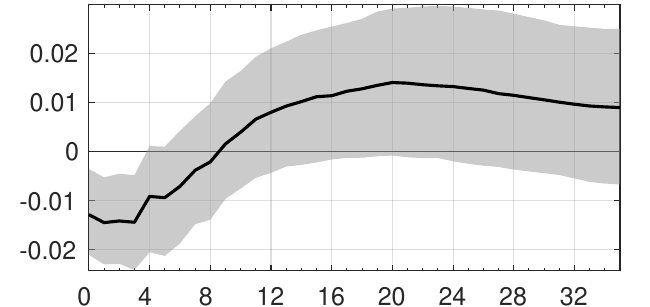} \\

\raisebox{1.3\height}{\rotatebox{90}{\scriptsize PT}} &
\includegraphics[trim=5cm 12cm 5cm 12.5cm, clip, width=0.19\textwidth]{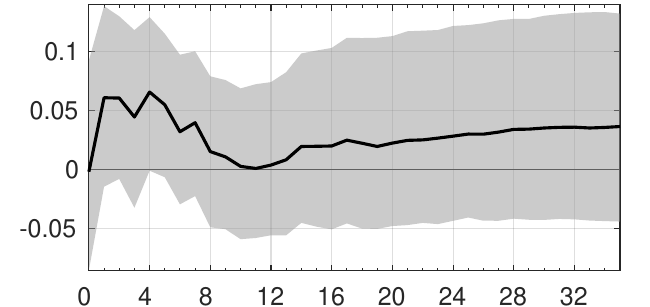} &
\includegraphics[trim=5cm 12cm 5cm 12.5cm, clip, width=0.19\textwidth]{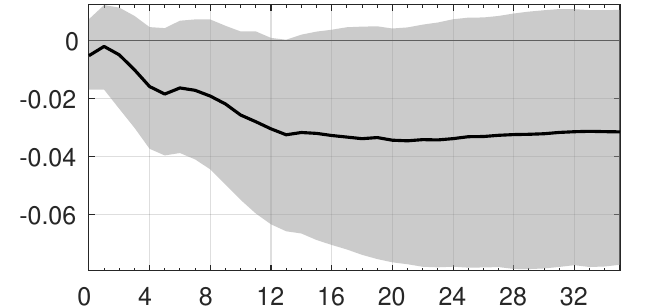} &
\includegraphics[trim=5cm 12cm 5cm 12.5cm, clip, width=0.19\textwidth]{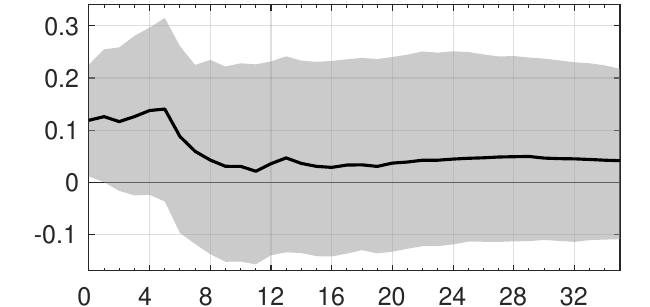} &
\includegraphics[trim=5cm 12cm 5cm 12.5cm, clip, width=0.19\textwidth]{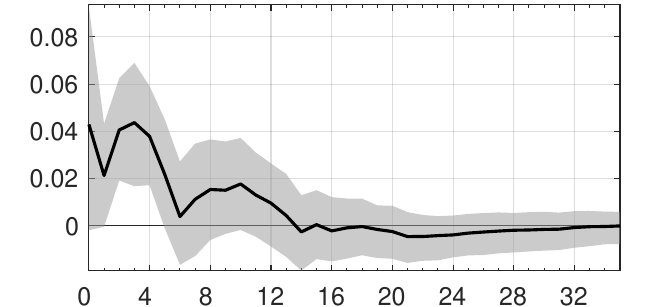} &
\includegraphics[trim=5cm 12cm 5cm 12.5cm, clip, width=0.19\textwidth]{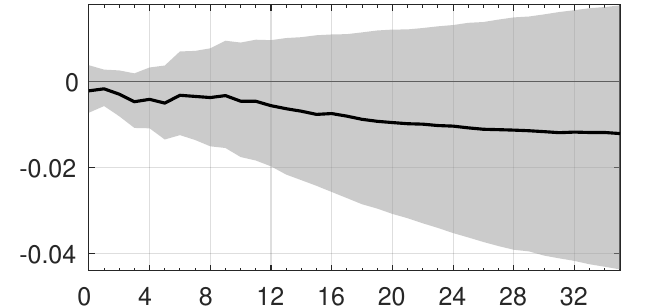} \\
\end{tabular}
\begin{tabular}{p{1\textwidth}} \scriptsize Notes: \rm Each sub-figure plots the impulse response of one variable for each country to a 100bps contractionary monetary policy shock. Each column of the graph represents a variable, while each row represents a country. The variables considered are: Industrial Production: Manufacturing (IPMN), HICP: Overall (HICPOV), Stock Price Index (SHIX), 10-years Interest Rates (LTIRT) and Unemployment Rate (UNETOT). The black solid line is the point estimate obtained identifying the monetary policy shock via sign restrictions, while gray shaded area are the corresponding 68\% confidence intervals. The scale in the vertical axis differs across variables and countries.
\end{tabular}
\label{fig::diffIRF_sign}
\end{figure}

\newpage

\section{Quarterly data: identification via sign restrictions}
\label{app::quarterly}

\subsection{Factor analysis and comovements across EA countries}
In this section, we present the results obtained using quarterly data. Throughout, use $r=6$ factors, according to the results of the results presented in Table \ref{tab:numfQ}.

\begin{table}[ht!]
\begin{center}
\caption{Estimated number of factors} \label{tab:numfQ}
\footnotesize
\begin{tabular}{l | c }
\hline
\hline
Method & {Number of factors $r$}\\
\hline
\cite{bai2002determining}			&	8	\\
\cite{alessi2010improved}			&	6	\\
\cite{onatski2010determining}		&	9	\\
\cite{ahn2013eigenvalue}			&	1	\\
\hline
\hline
\end{tabular}
\end{center}
\end{table}

Table \ref{tab::comovementsQ} reports the share of variance for each variable explained by the common factors extracted using quarterly data.

\begin{table}[ht!]
  \centering \scriptsize
  \caption{Share of Explained Variance by the common factors: quarterly data}
  \resizebox{\textwidth}{!}{  
    \begin{tabular}{c|C{.2\textwidth}|C{.2\textwidth}|C{.2\textwidth}|C{.2\textwidth}|C{.2\textwidth}}
\hline
\hline
    & & & & & \\[-10pt]
    \multirow{2}{*}{Country} & GDP & HICP: Overall & 10-years Interest Rate & Stock Price Index & Unemployment Rate \\
    &  (GDP) &  (HICPOV) & (LTIRT)  & (SHIX) & (UNETOT) \\
    & & & & & \\[-10pt]
\hline
\hline
    & & & & & \\[-10pt]
    EA    & 0.98  & 0.93  & 0.87  & 0.87  & 0.88 \\
          & (0.97-0.99) & (0.90-0.95) & (0.84-0.93) & (0.81-0.89) & (0.84-0.92) \\
    AT    & 0.90  & 0.89  & 0.92  & 0.84  & 0.50 \\
          & (0.85-0.92) & (0.85-0.92) & (0.89-0.94) & (0.77-0.86) & (0.42-0.62) \\
    BE    & 0.95  & 0.82  & 0.91  & 0.84  & 0.19 \\
          & (0.93-0.96) & (0.77-0.86) & (0.88-0.94) & (0.77-0.86) & (0.16-0.31) \\
    DE    & 0.88  & 0.85  & 0.92  & 0.77  & 0.13 \\
          & (0.84-0.91) & (0.81-0.90) & (0.88-0.93) & (0.71-0.82) & (0.19-0.43) \\
    EL    & 0.63  & 0.73  & 0.27  & 0.60  & 0.62 \\
          & (0.56-0.74) & (0.67-0.80) & (0.26-0.57) & (0.52-0.69) & (0.57-0.75) \\
    ES    & 0.95  & 0.83  & 0.73  & 0.75  & 0.75 \\
          & (0.93-0.97) & (0.78-0.87) & (0.70-0.85) & (0.67-0.79) & (0.69-0.81) \\
    FR    & 0.95  & 0.84  & 0.92  & 0.87  & 0.60 \\
          & (0.93-0.96) & (0.80-0.89) & (0.89-0.94) & (0.80-0.89) & (0.54-0.69) \\
    IE    & 0.27  & 0.79  & 0.64  & 0.79  & 0.41 \\
          & (0.22-0.39) & (0.74-0.85) & (0.61-0.79) & (0.72-0.82) & (0.37-0.58) \\
    IT    & 0.94  & 0.77  & 0.76  & 0.85  & 0.65 \\
          & (0.92-0.96) & (0.71-0.83) & (0.74-0.86) & (0.78-0.88) & (0.58-0.71) \\
    NL    & 0.91  & 0.66  & 0.92  & 0.83  & 0.47 \\
          & (0.87-0.94) & (0.58-0.75) & (0.88-0.93) & (0.76-0.85) & (0.41-0.59) \\
    PT    & 0.91  & 0.81  & 0.44  & 0.63  & 0.52 \\
          & (0.87-0.93) & (0.76-0.86) & (0.43-0.68) & (0.55-0.70) & (0.46-0.64) \\
    \hline
    \hline
    \end{tabular}%
    }
\begin{tabular}{p{.98\textwidth}}
\scriptsize {\sc Notes:} \rm Each entry in the table corresponds to the share of variability within each variable (in the columns) for each country (in the rows) explained by the common component, $\widehat{\chi}_{i,t}$. Numbers in parentheses indicate the lower and upper bounds of the 68\% confidence interval, computed using the bootstrap procedure described in Appendix \ref{app::pseudocode}.
\end{tabular}
  \label{tab::comovementsQ}
\end{table}

\subsection{Ireland GDP}

\begin{figure}[H]
\centering \footnotesize \sc 
\setlength{\tabcolsep}{.005\textwidth}
\caption{Ireland's idiosyncratic dynamics vs. real GDP and GNP}
\begin{tabular}{ccc}
Real GDP and GNP &  IE idiosyncratic component \\
\includegraphics[trim= .5cm 8cm .5cm 8.3cm, clip, width=0.475\textwidth]{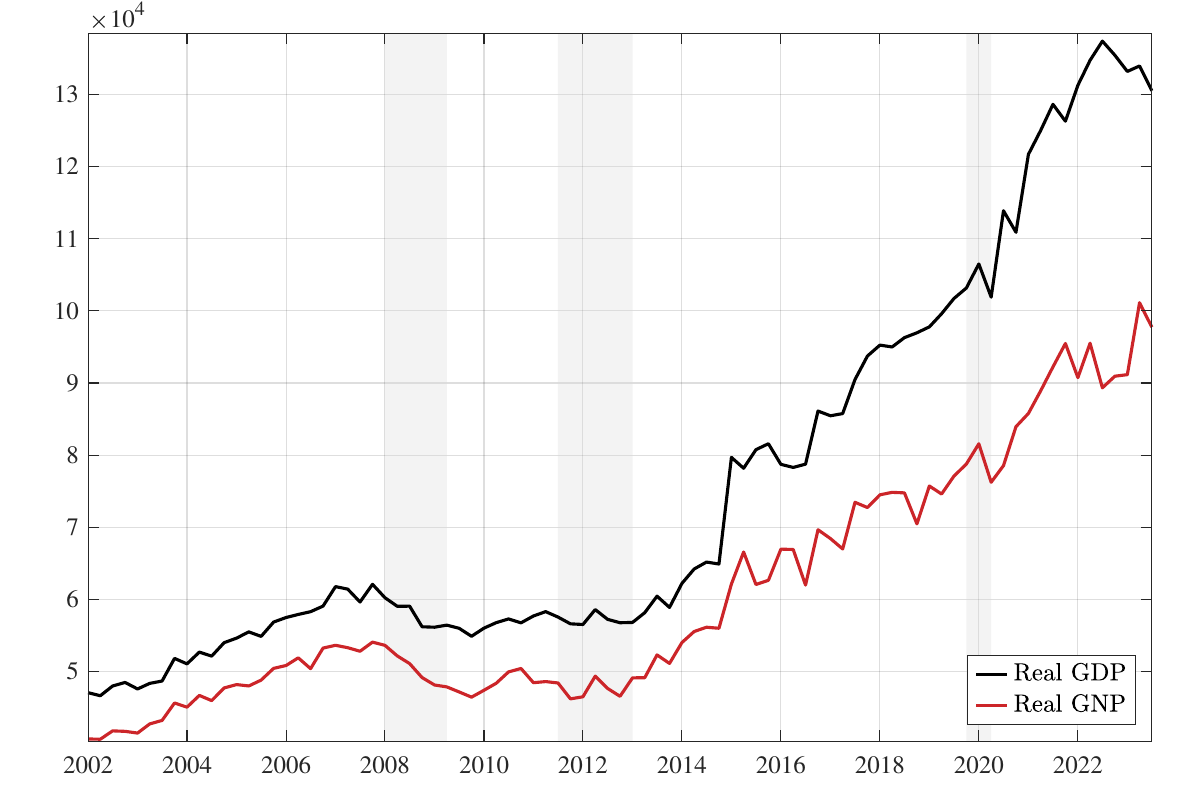} &
\includegraphics[trim= .5cm 8cm .5cm 8.3cm, clip, width=0.475\textwidth]{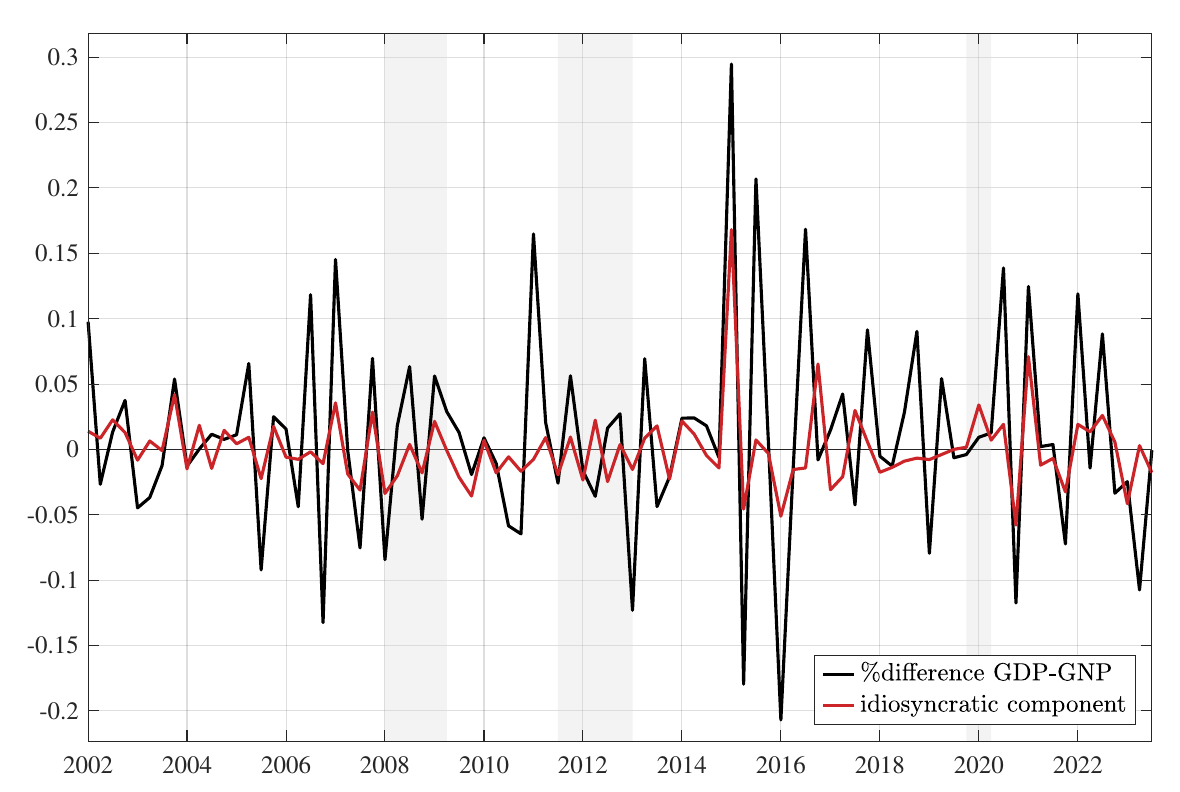} \\[5pt]
\end{tabular}
\begin{tabular}{p{.98\textwidth}} \scriptsize Notes: \rm The left panel plots the real GDP and GNP for Ireland in millions of euros (black and red lines, respectively). The right panel compares the log-difference of real GDP and GNP (black line) with the idiosyncratic component of Ireland's GDP obtained in our setting using quarterly data. 
\end{tabular}
\label{fig::IE_idio}
\end{figure}

From Table \ref{tab::comovementsQ} we notice that the explained variance for Ireland's GDP is particularly low. This, is due to Ireland being the legal domicile of many foreign large firms, creating remarkable distortions in Irish national accounts. To corroborate this finding, we computed the correlation between the idiosyncratic component of Ireland’s real GDP growth and the difference between the growth rates of real GDP and real GNP. Due to the strong presence of foreign-owned multinational corporations, a sizable portion of domestic income generated in Ireland is repatriated abroad, creating a persistent
wedge between GDP and GNP. As a result, other countries’ variables may comove more closely with Ireland’s GNP than with its GDP. If this is the case, the idiosyncratic component of Ireland’s GDP should correlate with the GDP-GNP growth differential.  Figure \ref{fig::IE_idio} reports the results of this exercise. The left panel confirms a sizable and persistent wedge between Ireland’s real GDP and GNP, as expected. The right panel compares
the log-difference of this wedge with the idiosyncratic component of real GDP growth from our model. The two series are positively and strongly correlated, with a correlation of approximately 0.6
over the full sample. This result suggests that the distortion in Ireland’s national accounts–arising from the repatriation of income by multinational firms–is indeed one of the main determinants of Ireland’s idiosyncratic behaviour in our model. 

\subsection{Identification via sign restrictions}

\begin{table}[h!]
\centering
\caption{Components of $\mathbf Y_t$ in the CC-VAR with quarterly data}\label{tab::CCVARQ}
\footnotesize
\begin{tabular}{l l l}
\hline
\hline
notation & ID & name \\
\hline
${R}_t$ & - & EA 2-years Interest Rate \\
$\widehat{\chi}_{{\rm GDP\, EA},t}$ & GDP & EA GDP \\
$\widehat{\chi}_{{\rm HICPOV\,EA},t}$ & HICPOV & EA HICP: Overall \\
$\widehat{\chi}_{{\rm LTIRT\,EA},t}$ & LTIRT & EA 10-years Interest Rate \\
$\widehat{\chi}_{{\rm SHIX\, EA},t}$ & SHIX & EA Stock Price Index \\
$\widehat{\chi}_{{\rm UNETOT\, EA},t}$ & UNETOT & EA Unemployment Rate \\
$\widehat{\chi}_{{\rm nat.},t}$ & - & National variable \\
\hline
\hline
\end{tabular}
\end{table}

In the IRF analysis, we use the same sets of EA common component as in the monthly case, with the exception of Industrial Production which is substituted with GDP (see Table \ref{tab::CCVARQ}).
Following the same procedure described in Appendix \ref{app::signres}, we impose the sign restrictions summarized in Table \ref{tab::signQ}. Specifically, we require a positive response of short- and long-term EA Interest Rates during the first two periods after the shock. For GDP, the Overall HICP, and the Stock Price Index, we impose a negative response in the second and third periods. Finally, for the Unemployment Rate, we impose a positive response in the second and third periods. Confidence intervals and point estimates are obtained using the same procedure described in Appendix \ref{app::signres}.

We do not report results based on identification via instrumental variables, which are unreliable due to the extremely low explanatory power of the instrument at the quarterly frequency. As noted by \cite{kilian2024construct}, time aggregation of high-frequency instruments can become problematic as the sampling frequency decreases.

\begin{table}[H]
  \centering
  \caption{Sign restrictions: quarterly data}
  \footnotesize
  \begin{tabular}{c | c c c c c c}
\hline
\hline
Horizon & $R_t$ & $\widehat\chi_{{\rm GDP\, EA},t}$ & $\widehat\chi_{{\rm HICPOV\, EA},t}$& $\widehat\chi_{{\rm LTIRT\, EA},t}$ & $\widehat\chi_{{\rm SHIX\, EA},t}$ & $\widehat\chi_{{\rm UNETOT\, EA},t}$\\
\hline
0 & + &	    &	  & + &     & 	  \\
1 & + & $-$	& $-$ & + & $-$ & $+$ \\
2 &	  & $-$	& $-$ &	  & $-$ & $+$ \\
\hline
\hline
  \end{tabular}
\label{tab::signQ}
\begin{tabular}{p{.85\textwidth}} \scriptsize Notes: \rm Each column of the table represents the common component of one variable, except for the EA Shadow Rate ($R_t$). The EA variables considered are: GDP, HICP: Overall (HICPOV), Stock Price Index (SHIX), 10-years Interest Rates (LTIRT) and Unemployment Rate (UNETOT).
\end{tabular}
\end{table}

\subsection{EA IRFs}
\begin{figure}[H]
\centering \footnotesize \sc \smallskip
\setlength{\tabcolsep}{.005\textwidth}
\caption{EA IRFs: quarterly data  and sign restrictions}
\begin{tabular}{ccc}
\scriptsize Shadow Rate ($R_t$) &\scriptsize  IP: Manufacturing (IPMN) &\scriptsize  HICP: Overall (HICPOV) \\
\includegraphics[trim= .5cm 8cm .5cm 8.5cm, clip, width=0.32\textwidth]{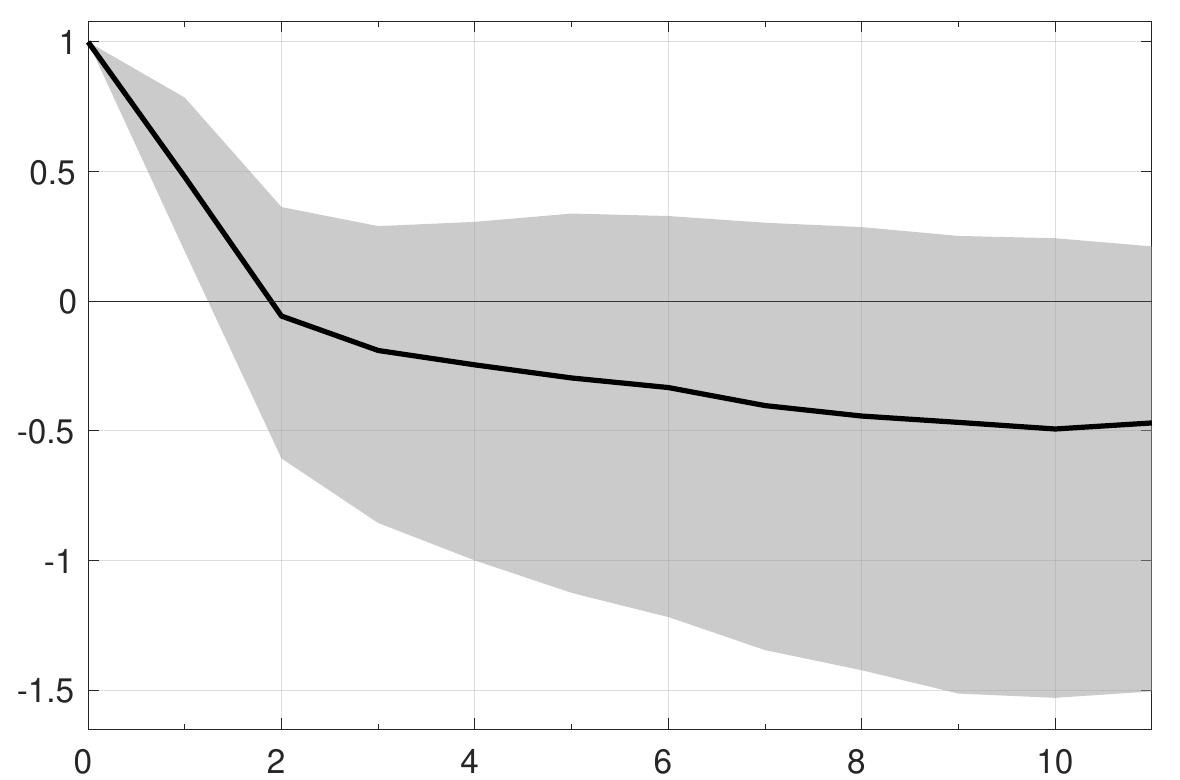} &
\includegraphics[trim= .5cm 8cm .5cm 8.5cm, clip, width=0.32\textwidth]{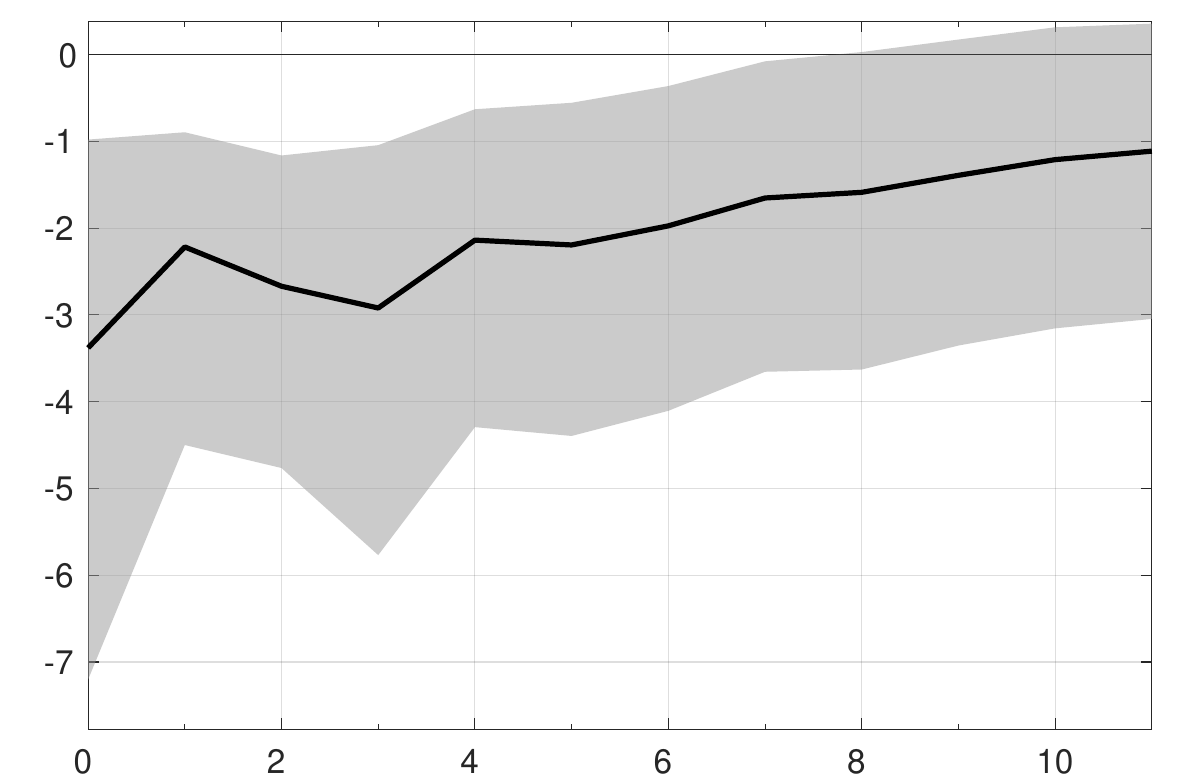} &
\includegraphics[trim= .5cm 8cm .5cm 8.5cm, clip, width=0.32\textwidth]{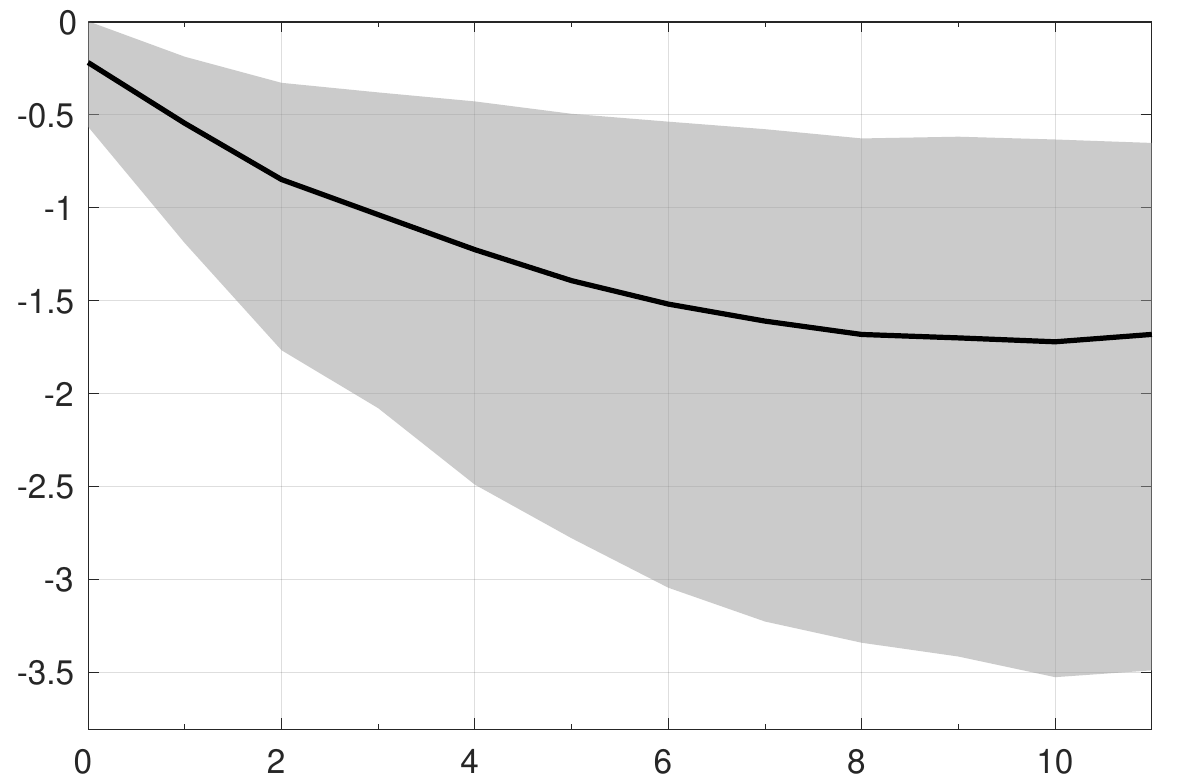} \\
\scriptsize 10-years Interest Rate (LTIRT) &\scriptsize Stock Price Index (SHIX) &\scriptsize Unemployment Rate (UNETOT)\\
\includegraphics[trim= .5cm 8cm .5cm 8.5cm, clip, width=0.32\textwidth]{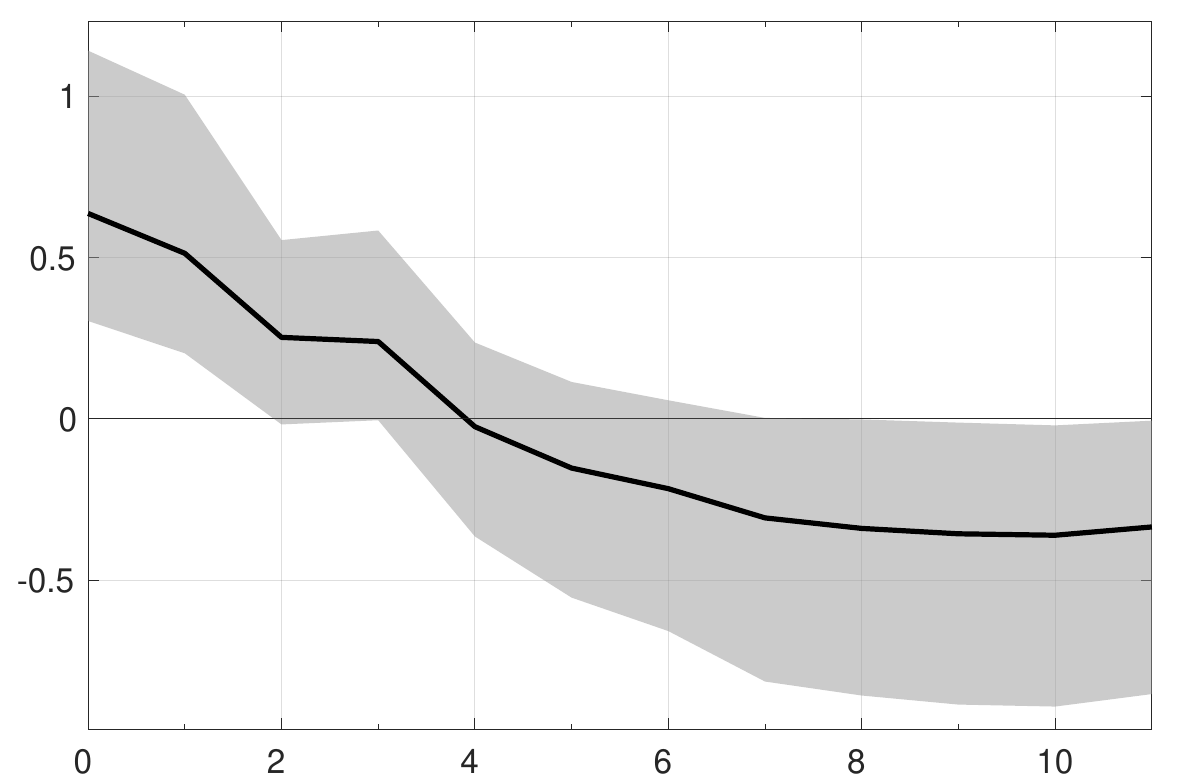} &
\includegraphics[trim= .5cm 8cm .5cm 8.5cm, clip, width=0.32\textwidth]{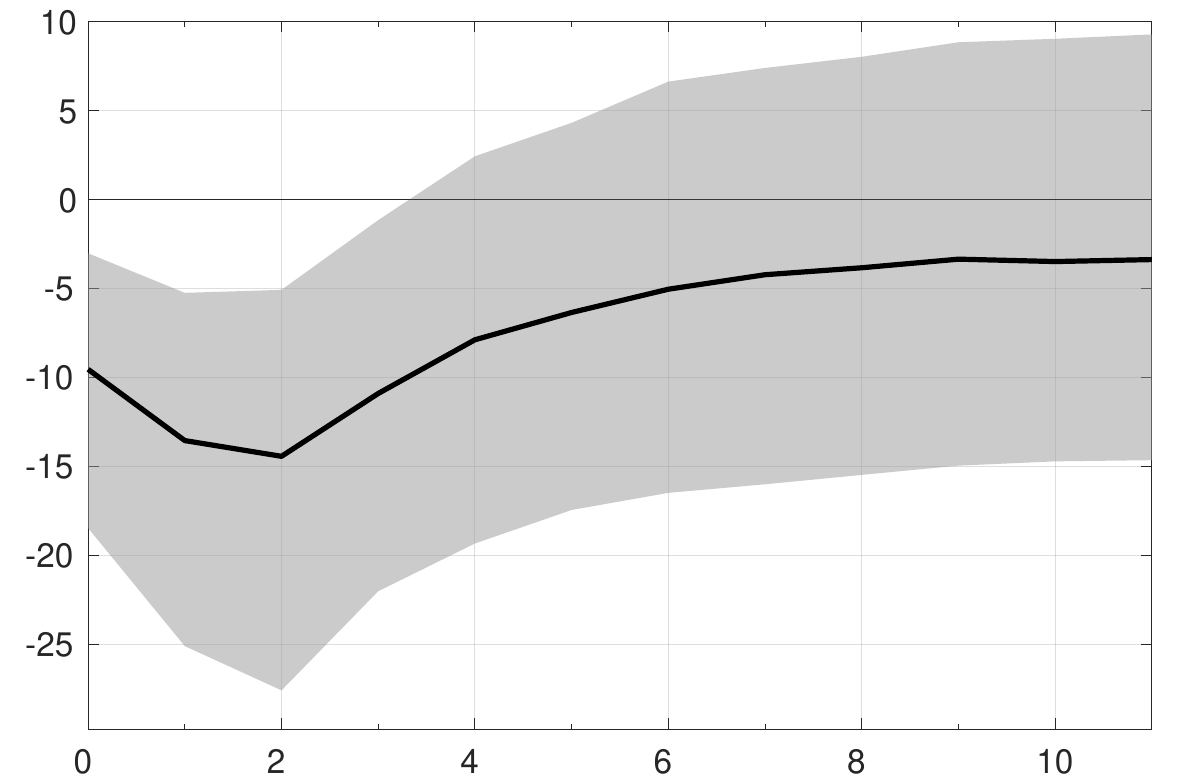} &
\includegraphics[trim= .5cm 8cm .5cm 8.5cm, clip, width=0.32\textwidth]{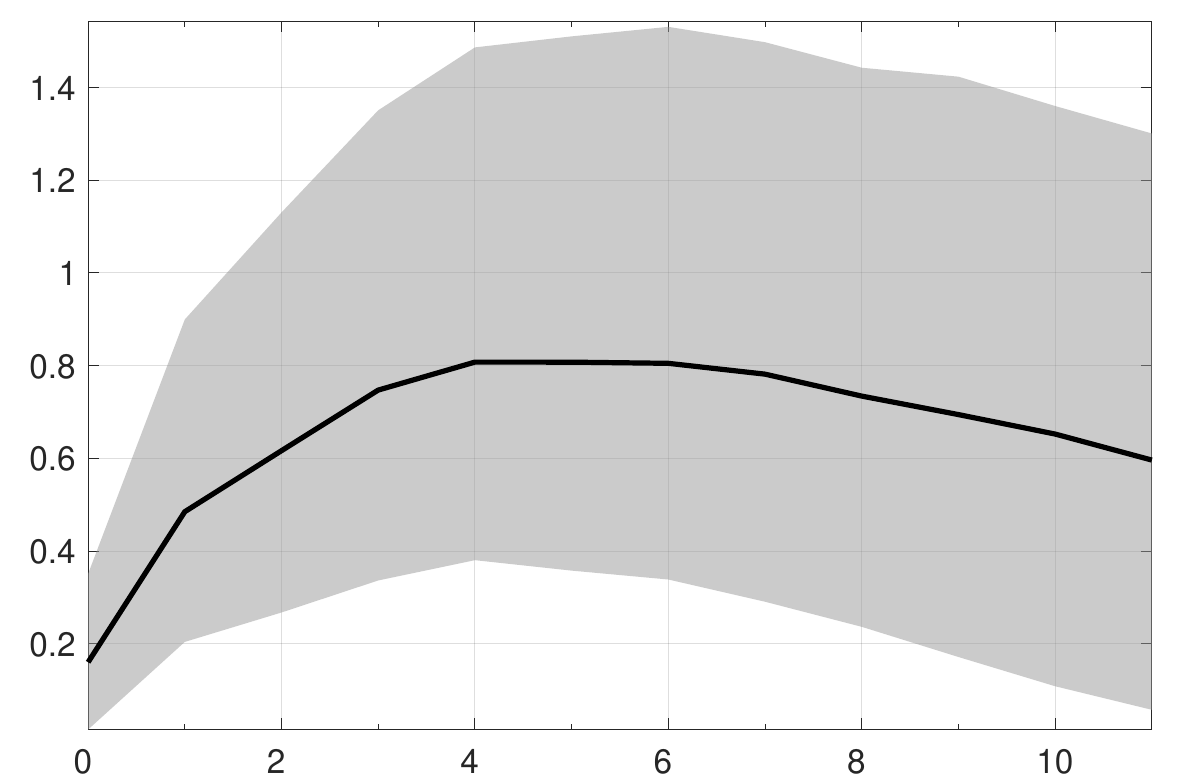} \\
\end{tabular}
\begin{tabular}{p{.98\textwidth}} \scriptsize Notes: \rm Each sub-figure plots the impulse response of one EA variable to a 100bps contractionary monetary policy shock. The black solid line is the point estimate in our baseline setting, while the gray shaded area is the corresponding 68\% confidence interval.
\end{tabular}
\label{fig:EAirfs}
\end{figure}

\subsection{Country-level IRFs}

\begin{figure}[H]
\centering \footnotesize \sc \smallskip
\setlength{\tabcolsep}{.005\textwidth}
\caption{Country-level IRFs: quarterly data  and sign restrictions}
\begin{tabular}{ccc}
\scriptsize{GDP}  
&\scriptsize {HICP: Overall (HICPOV)}      
&\scriptsize {Stock Price Index (SHIX)}  \\

\includegraphics[trim= .5cm 8cm .5cm 8.5cm, clip, width=0.32\textwidth]{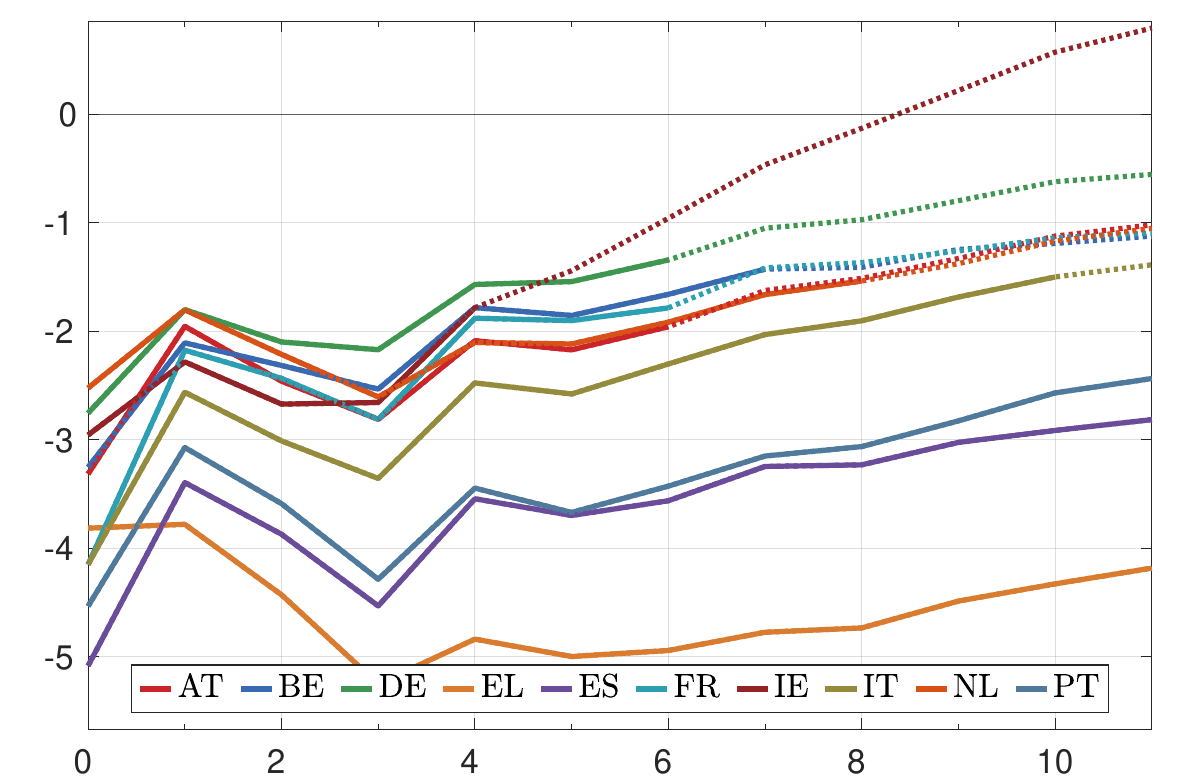} &
\includegraphics[trim= .5cm 8cm .5cm 8.5cm, clip, width=0.32\textwidth]{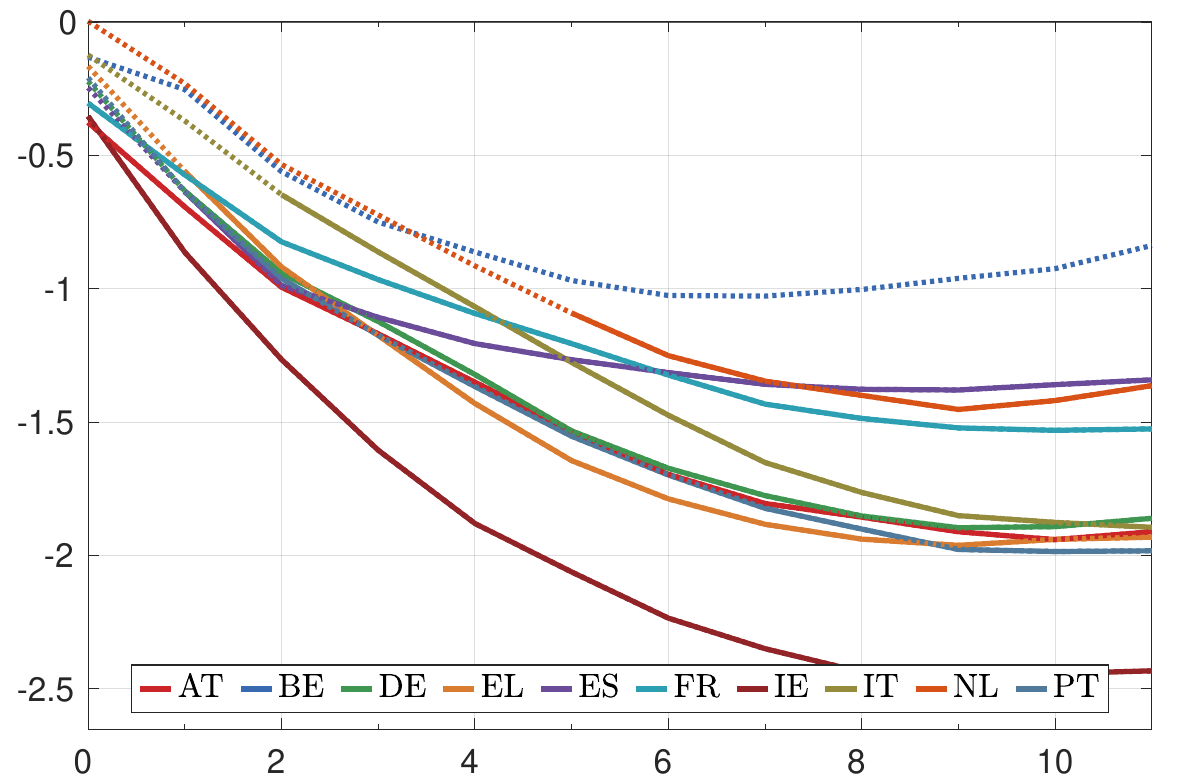} &
\includegraphics[trim= .5cm 8cm .5cm 8.5cm, clip, width=0.32\textwidth]{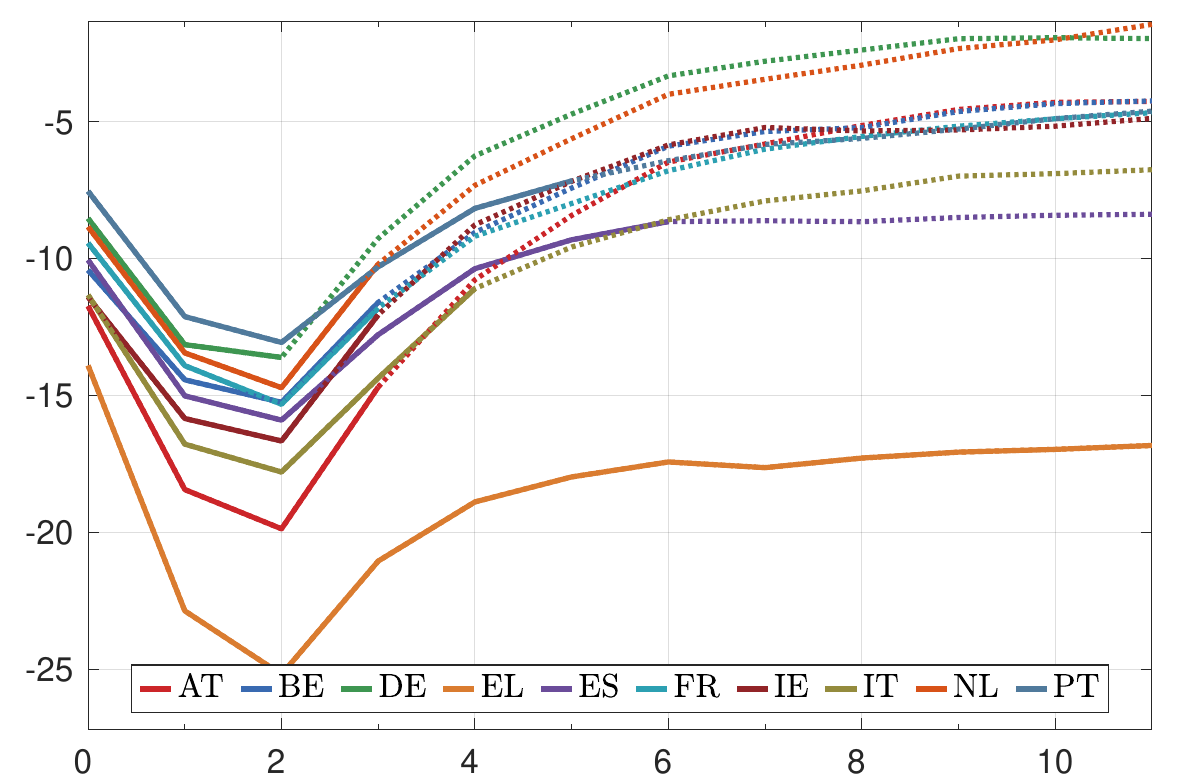} \\[10pt]
\end{tabular}

\begin{tabular}{cc}
\scriptsize {10-years Interest Rate (LTIRT)} 
&\scriptsize {Unemployment Rate (UNETOT)} \\

\includegraphics[trim= .5cm 8cm .5cm 8.5cm, clip, width=0.32\textwidth]{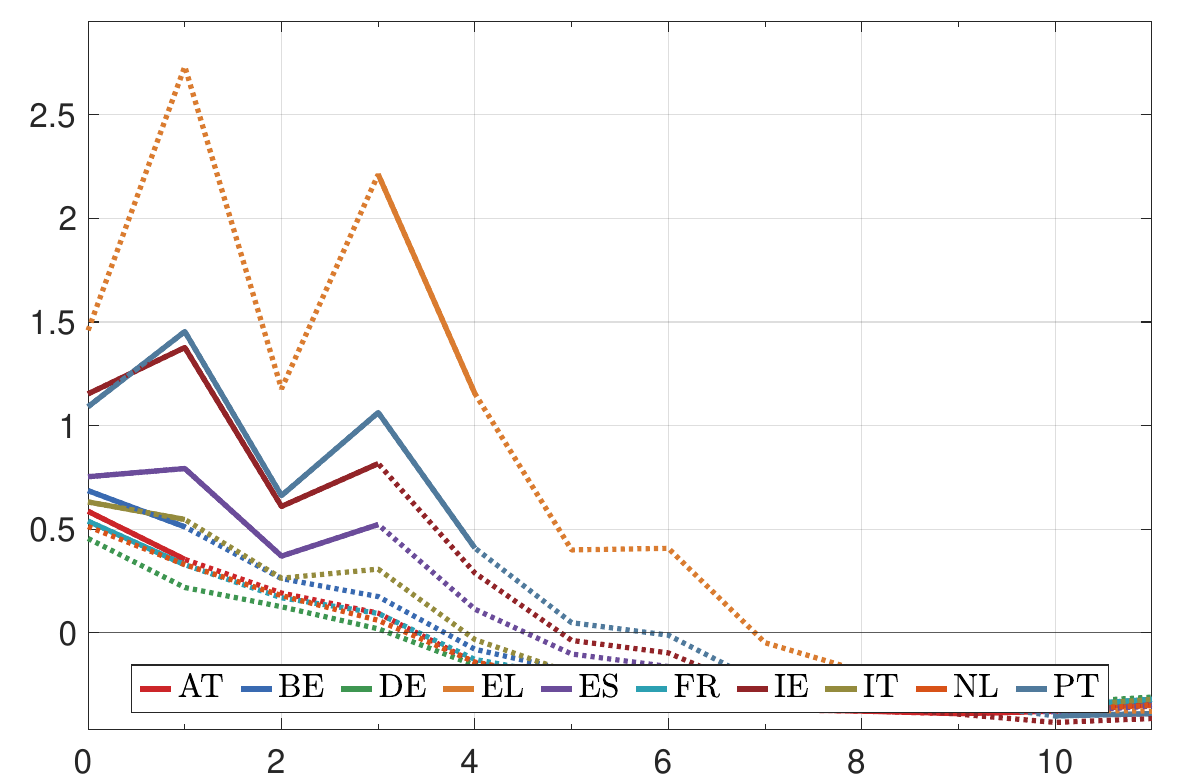} 
&
\includegraphics[trim= .5cm 8cm .5cm 8.5cm, clip, width=0.32\textwidth]{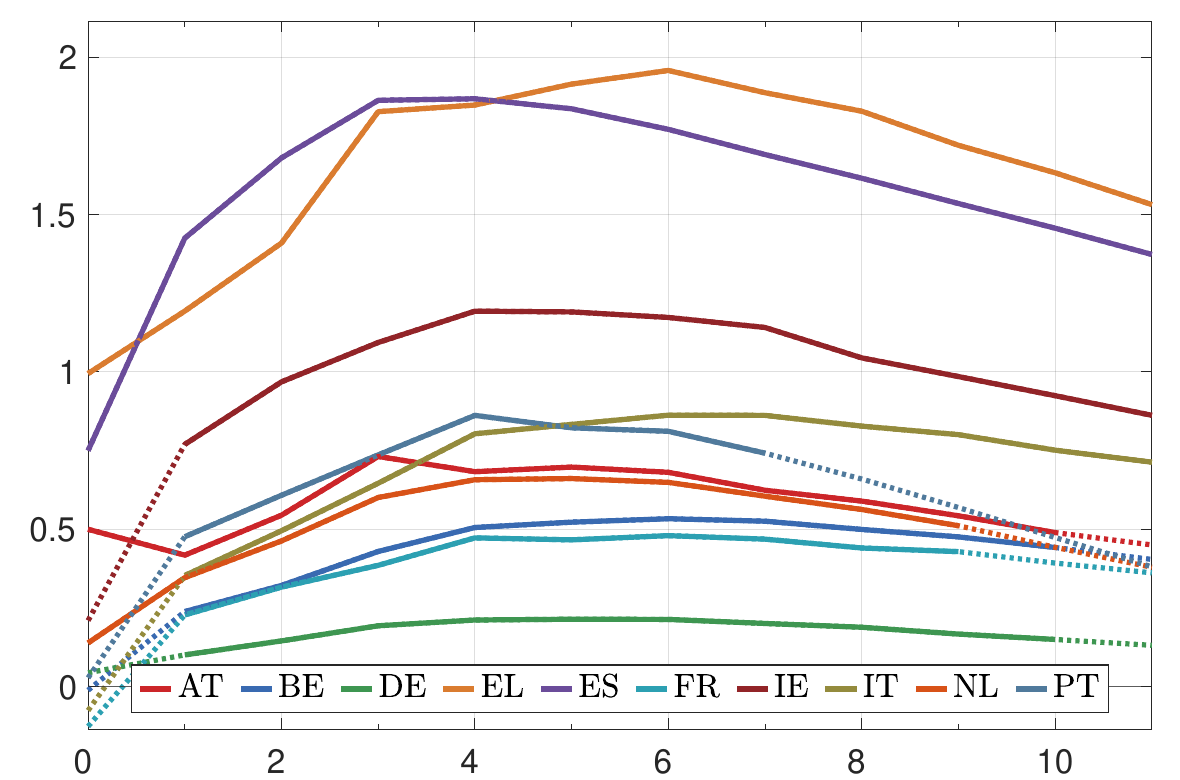}
\end{tabular}
\begin{tabular}{p{.98\textwidth}} \scriptsize Notes: \rm Each sub-figure plots the impulse responses, for all countries, of one variable to a 100bps contractionary monetary policy shock. Within each sub-figure, at each horizon $h=0,\ldots,36$ the country-level impulse responses are denoted with a solid line if the IRF is statistically significant at the 68\% level at that horizon, and with a dotted line otherwise. 
\end{tabular}
\label{fig::nat_IRFs_Q}
\end{figure}

\begin{figure}[H]
\caption{Difference between country-specific and EA IRFs: quarterly data and sign restrictions}
\centering \footnotesize \sc \smallskip
\setlength{\tabcolsep}{.005\textwidth}
\begin{tabular}{lccccc}
& \hspace{5pt}\scriptsize IPMN &  \hspace{5pt}\scriptsize HICPOV & 
 \hspace{5pt}\scriptsize SHIX &  \hspace{5pt}\scriptsize LTIRT &  \hspace{8pt}\scriptsize UNETOT \\[4pt] 
\raisebox{1.5\height}{\rotatebox{90}{\scriptsize AT}} & \includegraphics[trim=5cm 12cm 5cm 12.5cm, clip, width=0.19\textwidth]{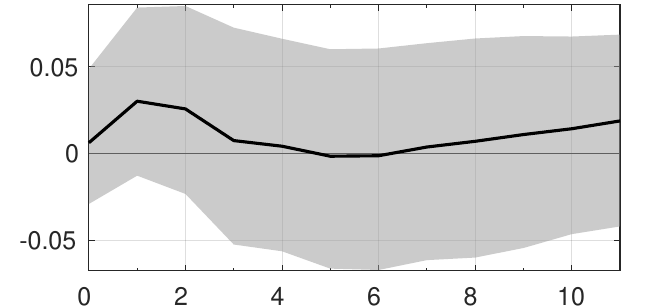} &
\includegraphics[trim=5cm 12cm 5cm 12.5cm, clip, width=0.19\textwidth]{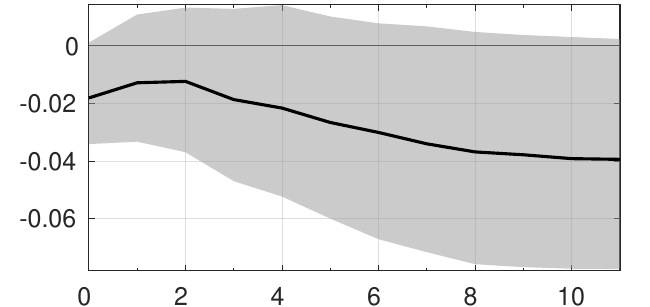} &
\includegraphics[trim=5cm 12cm 5cm 12.5cm, clip, width=0.19\textwidth]{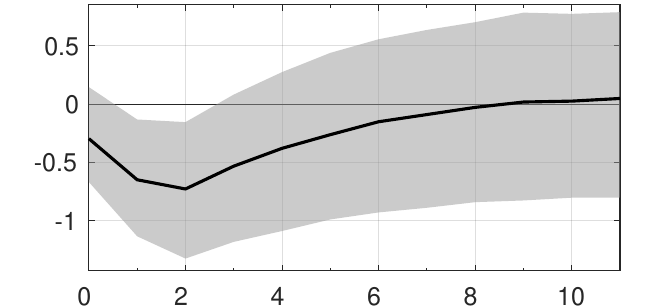} &
\includegraphics[trim=5cm 12cm 5cm 12.5cm, clip, width=0.19\textwidth]{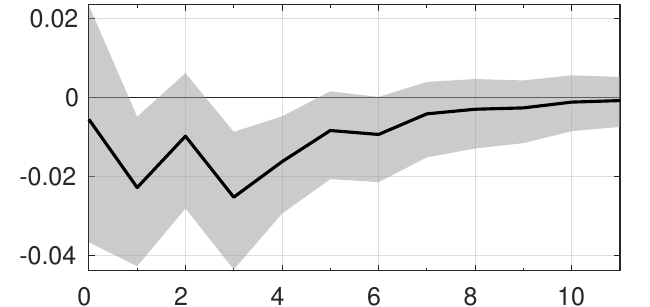} &
\includegraphics[trim=5cm 12cm 5cm 12.5cm, clip, width=0.19\textwidth]{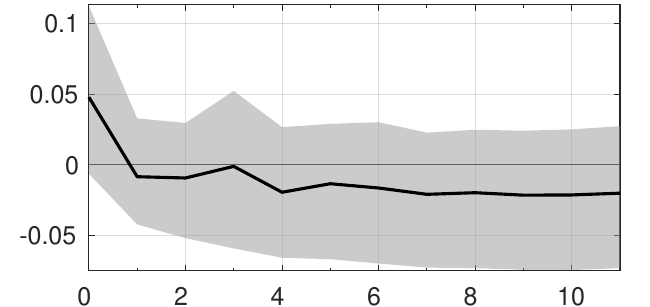} \\

\raisebox{1.5\height}{\rotatebox{90}{\scriptsize BE}} &
\includegraphics[trim=5cm 12cm 5cm 12.5cm, clip, width=0.19\textwidth]{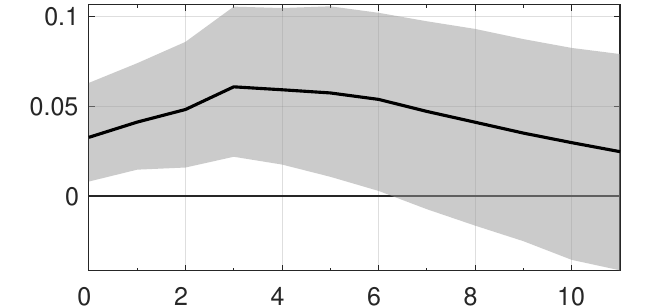} &
\includegraphics[trim=5cm 12cm 5cm 12.5cm, clip, width=0.19\textwidth]{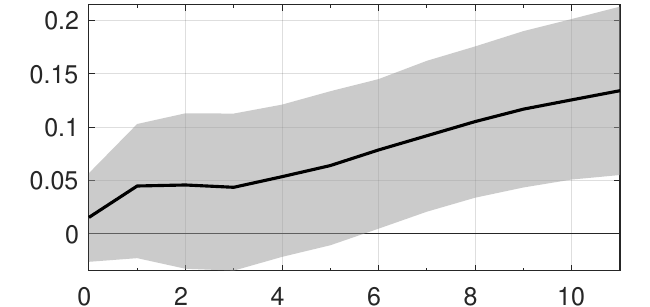} &
\includegraphics[trim=5cm 12cm 5cm 12.5cm, clip, width=0.19\textwidth]{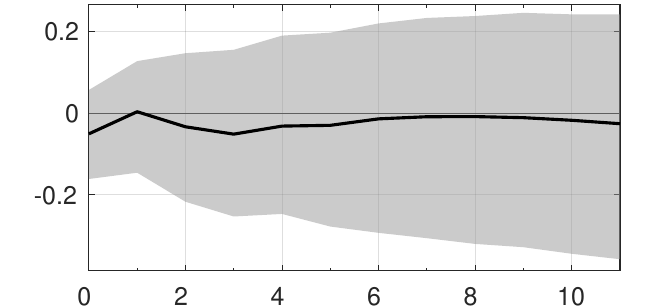} &
\includegraphics[trim=5cm 12cm 5cm 12.5cm, clip, width=0.19\textwidth]{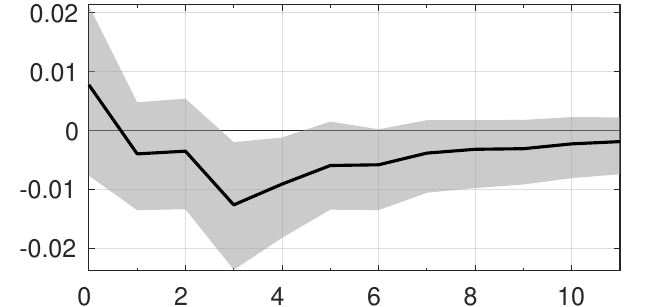} &
\includegraphics[trim=5cm 12cm 5cm 12.5cm, clip, width=0.19\textwidth]{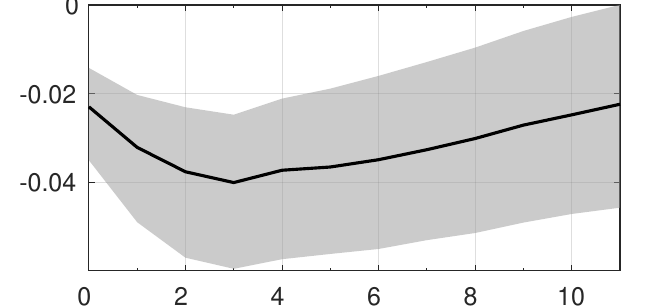} \\

\raisebox{1.5\height}{\rotatebox{90}{\scriptsize DE}} &
\includegraphics[trim=5cm 12cm 5cm 12.5cm, clip, width=0.19\textwidth]{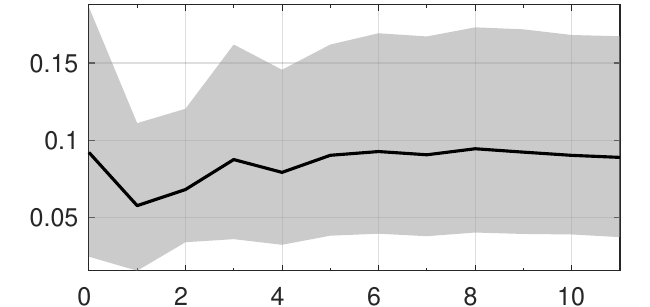} &
\includegraphics[trim=5cm 12cm 5cm 12.5cm, clip, width=0.19\textwidth]{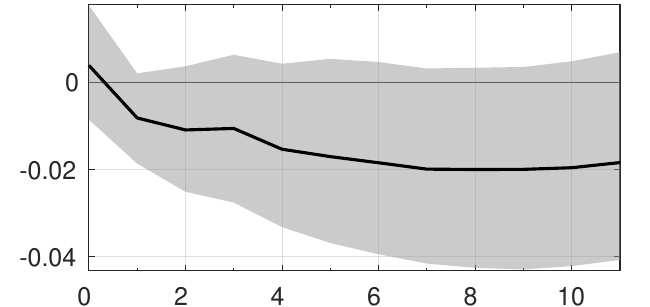} &
\includegraphics[trim=5cm 12cm 5cm 12.5cm, clip, width=0.19\textwidth]{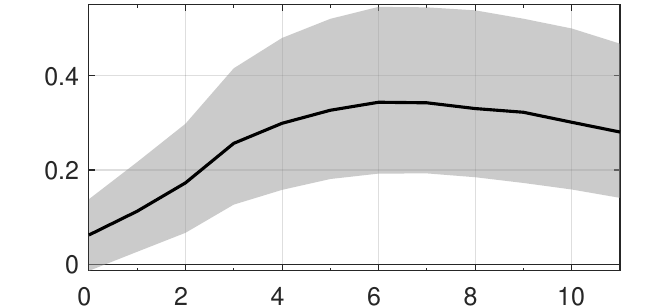} &
\includegraphics[trim=5cm 12cm 5cm 12.5cm, clip, width=0.19\textwidth]{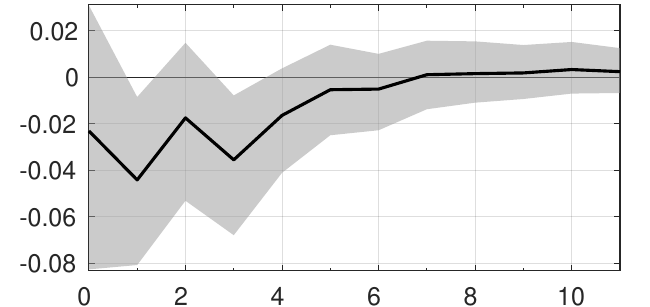} &
\includegraphics[trim=5cm 12cm 5cm 12.5cm, clip, width=0.19\textwidth]{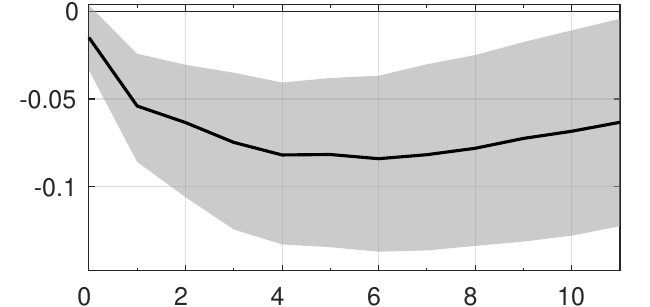} \\

\raisebox{1.5\height}{\rotatebox{90}{\scriptsize FR}} &
\includegraphics[trim=5cm 12cm 5cm 12.5cm, clip, width=0.19\textwidth]{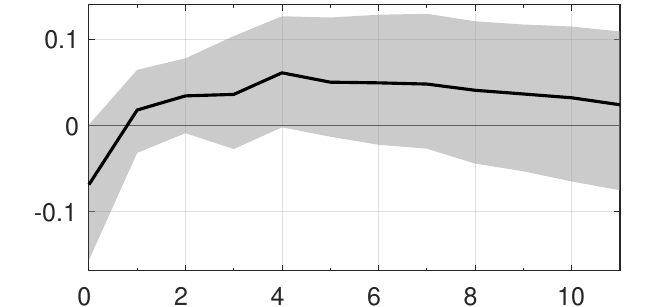} &
\includegraphics[trim=5cm 12cm 5cm 12.5cm, clip, width=0.19\textwidth]{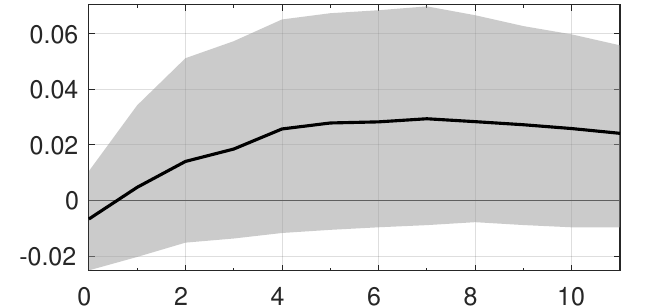} &
\includegraphics[trim=5cm 12cm 5cm 12.5cm, clip, width=0.19\textwidth]{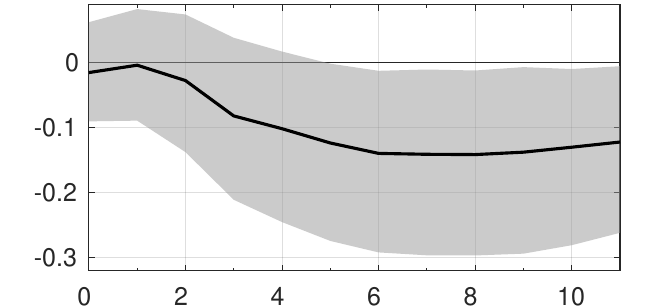} &
\includegraphics[trim=5cm 12cm 5cm 12.5cm, clip, width=0.19\textwidth]{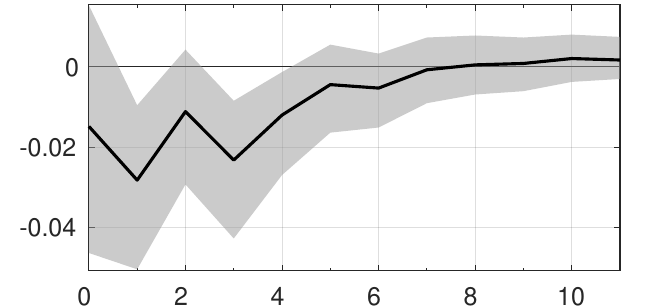} &
\includegraphics[trim=5cm 12cm 5cm 12.5cm, clip, width=0.19\textwidth]{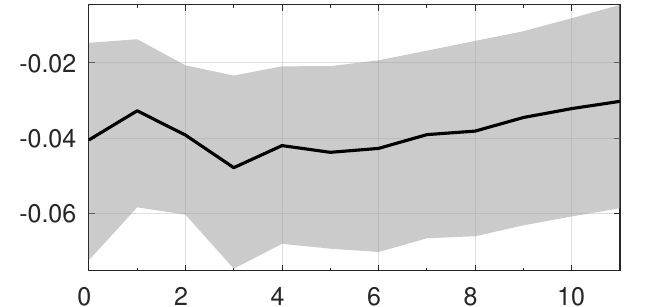} \\

\raisebox{1.5\height}{\rotatebox{90}{\scriptsize NL}} &
\includegraphics[trim=5cm 12cm 5cm 12.5cm, clip, width=0.19\textwidth]{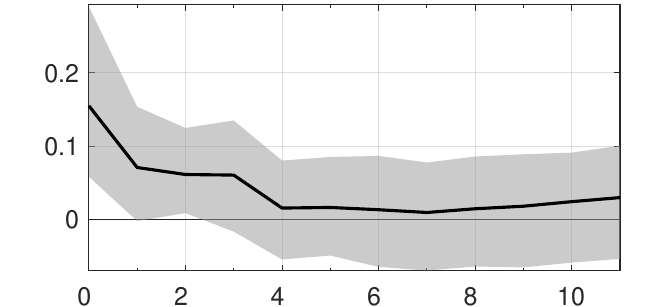} &
\includegraphics[trim=5cm 12cm 5cm 12.5cm, clip, width=0.19\textwidth]{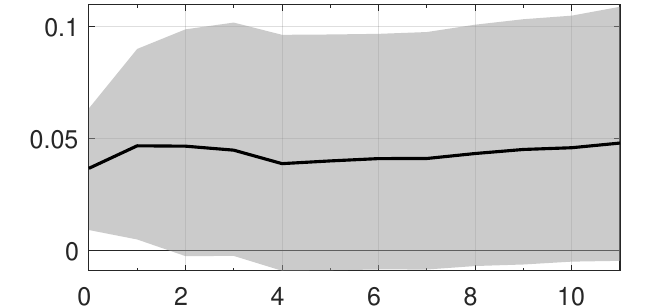} &
\includegraphics[trim=5cm 12cm 5cm 12.5cm, clip, width=0.19\textwidth]{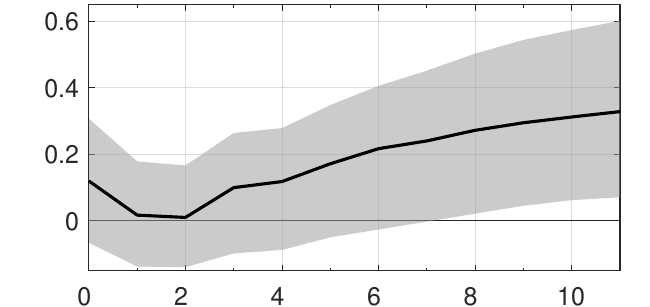} &
\includegraphics[trim=5cm 12cm 5cm 12.5cm, clip, width=0.19\textwidth]{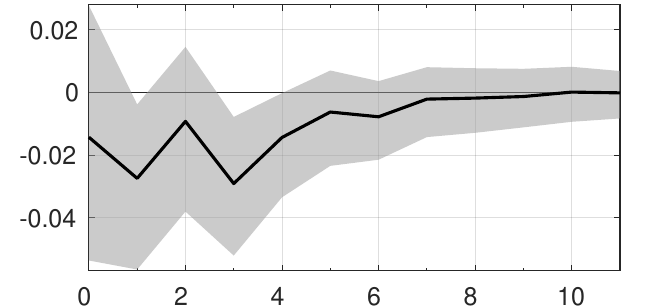} &
\includegraphics[trim=5cm 12cm 5cm 12.5cm, clip, width=0.19\textwidth]{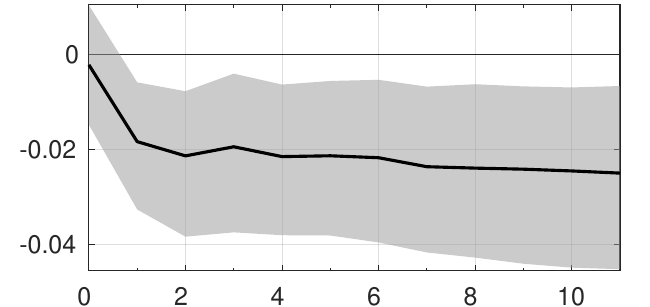} \\

\raisebox{1.7\height}{\rotatebox{90}{\scriptsize EL}} &
\includegraphics[trim=5cm 12cm 5cm 12.5cm, clip, width=0.19\textwidth]{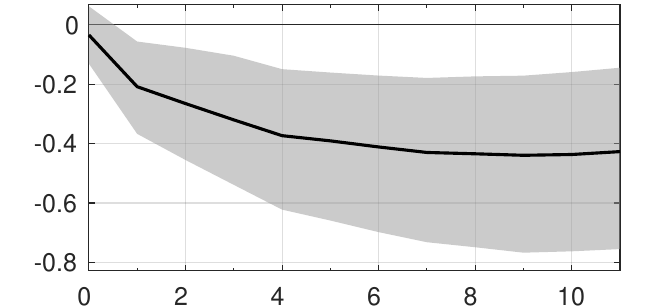} &
\includegraphics[trim=5cm 12cm 5cm 12.5cm, clip, width=0.19\textwidth]{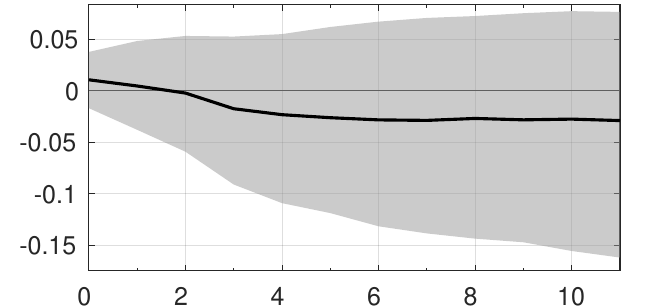} &
\includegraphics[trim=5cm 12cm 5cm 12.5cm, clip, width=0.19\textwidth]{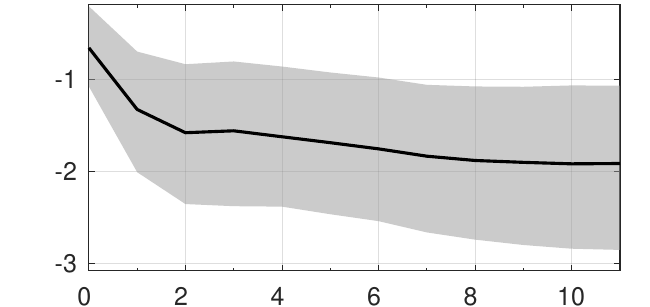} &
\includegraphics[trim=5cm 12cm 5cm 12.5cm, clip, width=0.19\textwidth]{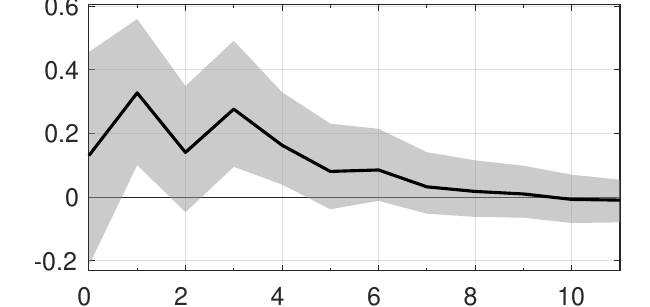} &
\includegraphics[trim=5cm 12cm 5cm 12.5cm, clip, width=0.19\textwidth]{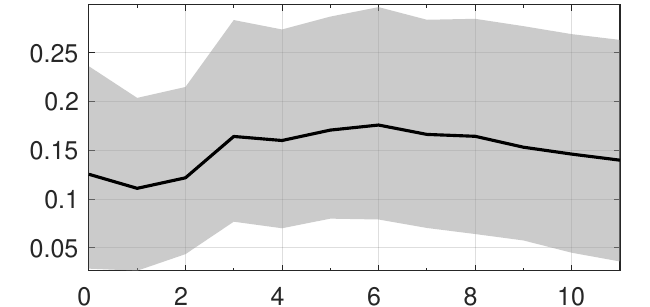} \\

\raisebox{1.7\height}{\rotatebox{90}{\scriptsize ES}} &
\includegraphics[trim=5cm 12cm 5cm 12.5cm, clip, width=0.19\textwidth]{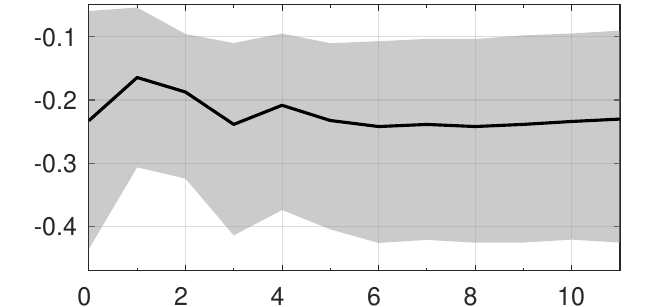} &
\includegraphics[trim=5cm 12cm 5cm 12.5cm, clip, width=0.19\textwidth]{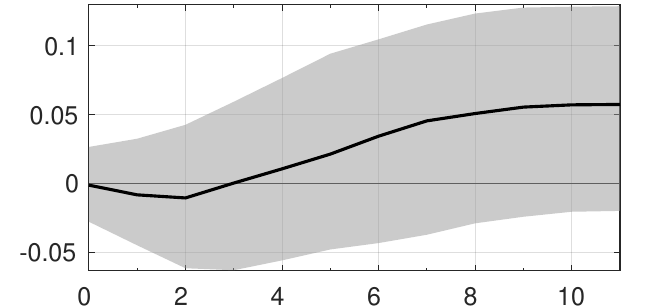} &
\includegraphics[trim=5cm 12cm 5cm 12.5cm, clip, width=0.19\textwidth]{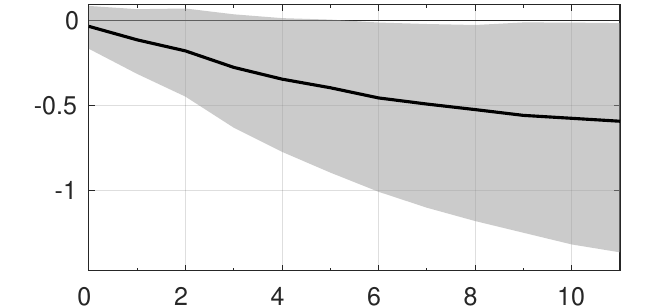} &
\includegraphics[trim=5cm 12cm 5cm 12.5cm, clip, width=0.19\textwidth]{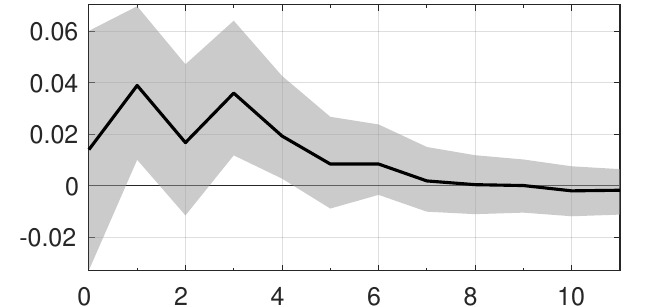} &
\includegraphics[trim=5cm 12cm 5cm 12.5cm, clip, width=0.19\textwidth]{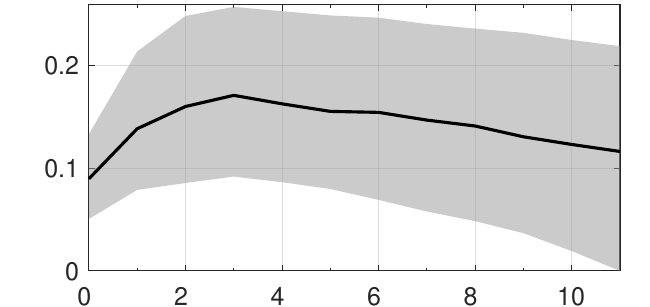} \\

\raisebox{2\height}{\rotatebox{90}{\scriptsize IE}} &
\includegraphics[trim=5cm 12cm 5cm 12.5cm, clip, width=0.19\textwidth]{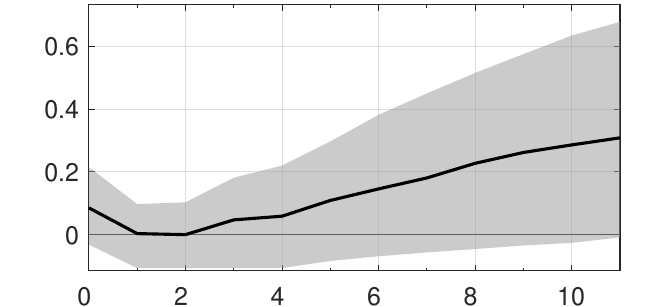} &
\includegraphics[trim=5cm 12cm 5cm 12.5cm, clip, width=0.19\textwidth]{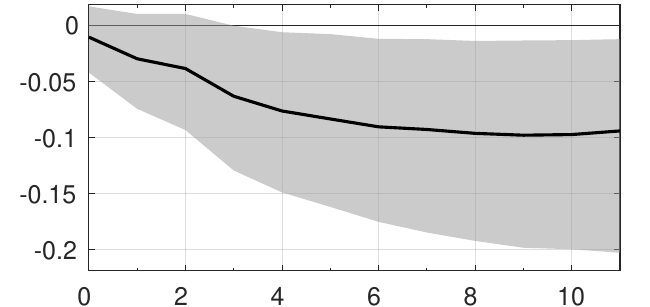} &
\includegraphics[trim=5cm 12cm 5cm 12.5cm, clip, width=0.19\textwidth]{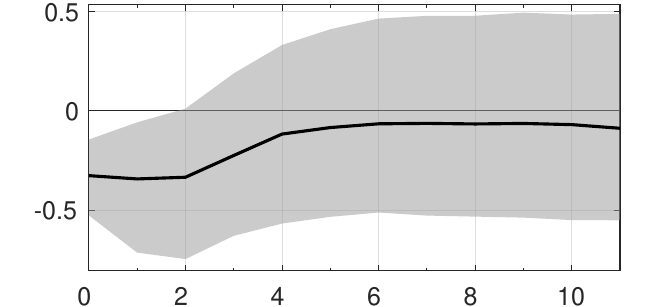} &
\includegraphics[trim=5cm 12cm 5cm 12.5cm, clip, width=0.19\textwidth]{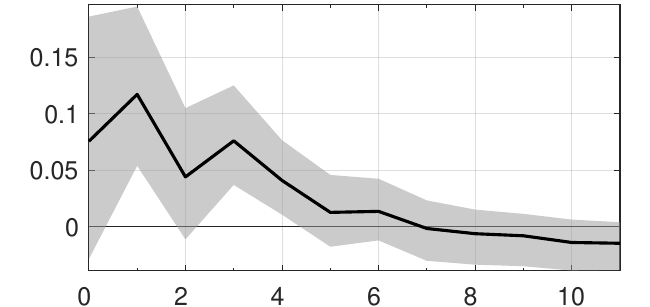} &
\includegraphics[trim=5cm 12cm 5cm 12.5cm, clip, width=0.19\textwidth]{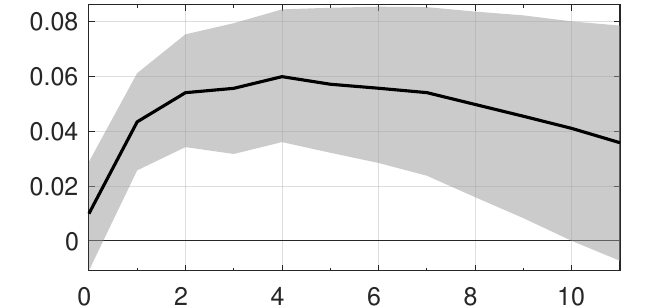} \\

\raisebox{2\height}{\rotatebox{90}{\scriptsize IT}} &
\includegraphics[trim=5cm 12cm 5cm 12.5cm, clip, width=0.19\textwidth]{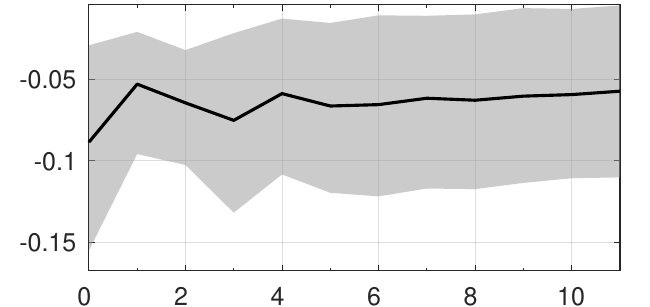} &
\includegraphics[trim=5cm 12cm 5cm 12.5cm, clip, width=0.19\textwidth]{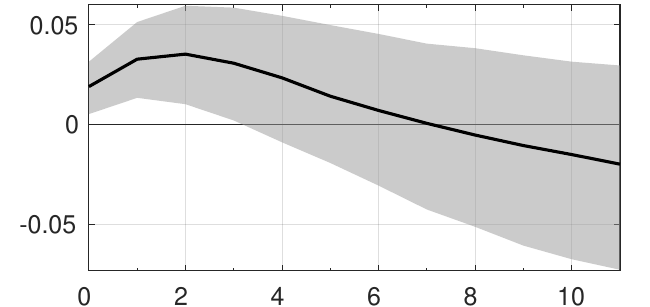} &
\includegraphics[trim=5cm 12cm 5cm 12.5cm, clip, width=0.19\textwidth]{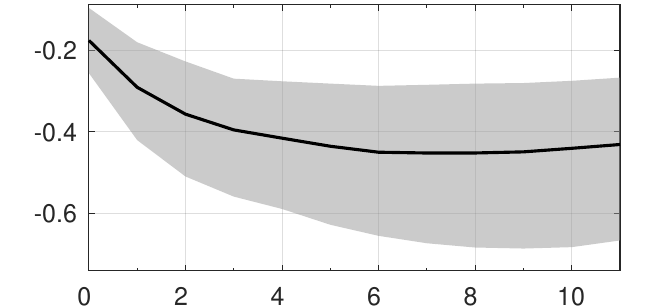} &
\includegraphics[trim=5cm 12cm 5cm 12.5cm, clip, width=0.19\textwidth]{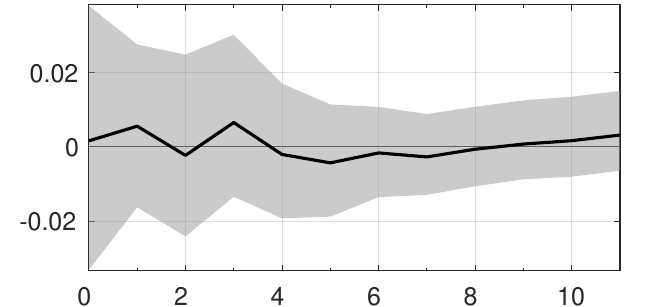} &
\includegraphics[trim=5cm 12cm 5cm 12.5cm, clip, width=0.19\textwidth]{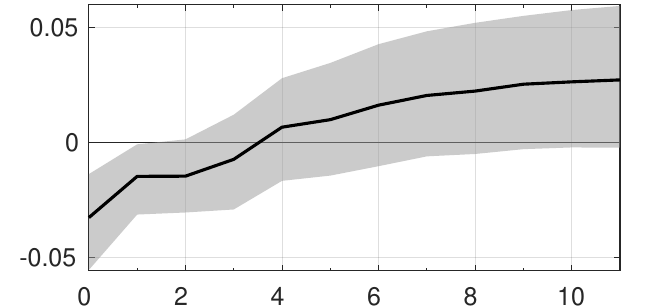} \\

\raisebox{1.5\height}{\rotatebox{90}{\scriptsize PT}} &
\includegraphics[trim=5cm 12cm 5cm 12.5cm, clip, width=0.19\textwidth]{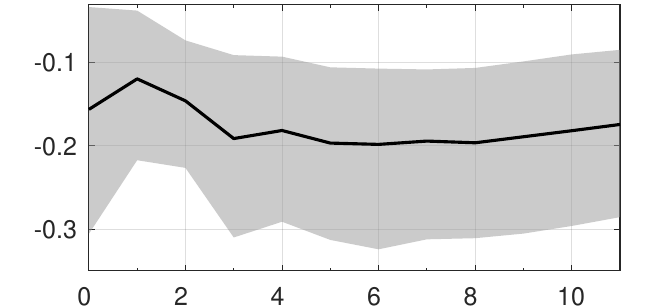} &
\includegraphics[trim=5cm 12cm 5cm 12.5cm, clip, width=0.19\textwidth]{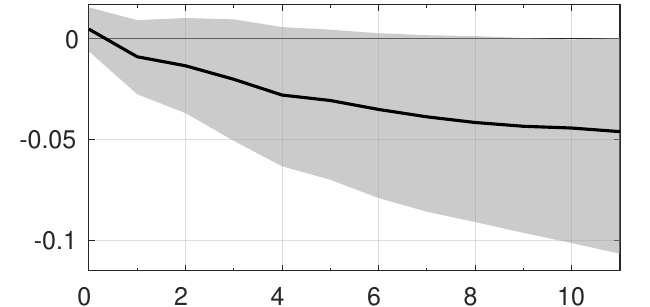} &
\includegraphics[trim=5cm 12cm 5cm 12.5cm, clip, width=0.19\textwidth]{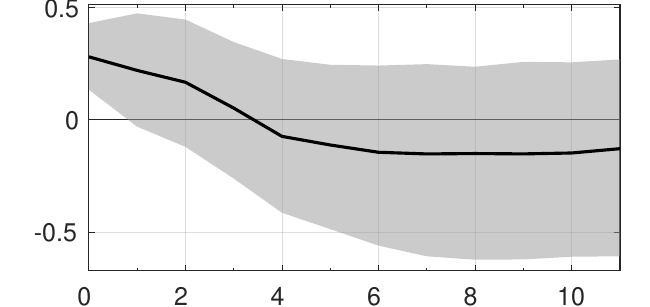} &
\includegraphics[trim=5cm 12cm 5cm 12.5cm, clip, width=0.19\textwidth]{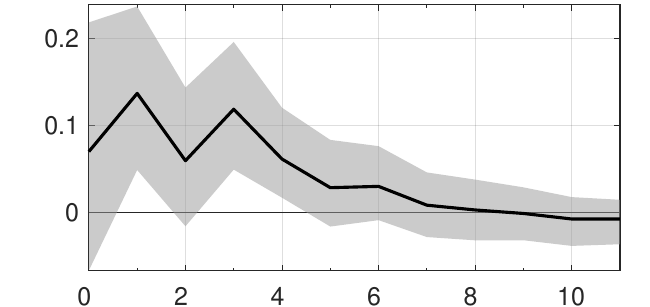} &
\includegraphics[trim=5cm 12cm 5cm 12.5cm, clip, width=0.19\textwidth]{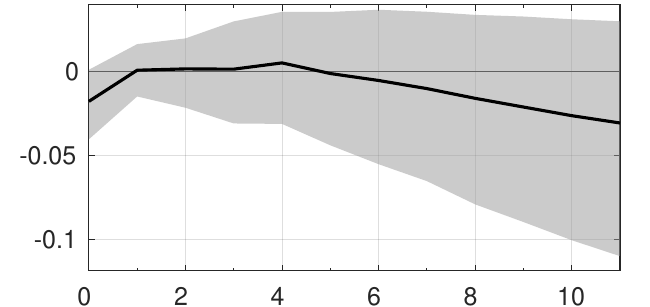} \\
\end{tabular}
\begin{tabular}{p{1\textwidth}} \scriptsize Notes: \rm Each sub-figure plots the difference between the country-level IRF and the corresponding EA counterpart for one variable and one country. Each column of the graph represents a variable, while each row represents a country. The variables considered are: GDP, HICP: Overall (HICPOV), Stock Price Index (SHIX), 10-years Interest Rates (LTIRT) and Unemployment Rate (UNETOT). The black solid line is the point estimate in our baseline setting, while the gray shaded area is the corresponding 68\% confidence interval. The scale in the vertical axis differs across variables and countries.
\end{tabular}
\label{fig::diffIRF_Q}
\end{figure}

\newpage

\section{Mixed frequency data: identification via Instrumental Variables}
\subsection{Estimation}
In order to handle mixed frequency data (monthly data as in the baseline specification plus EA and national quarterly GDPs) estimation of the factor model is via the EM algorithm by \cite{banbura2011nowcasting}.  The estimated common components are all at the monthly frequency. The variable used in the CC-VAR are listed in Table \ref{tab::CCVARMIX}. 
Identification is via Instrumental Variables.

\begin{table}[h!]
\centering
\caption{Components of $\mathbf Y_t$ in the CC-VAR with quarterly data}\label{tab::CCVARMIX}
\footnotesize
\begin{tabular}{l l l}
\hline
\hline
notation & ID & name \\
\hline
${R}_t$ & - & EA 2-years Interest Rate - monthly \\
$\widehat{\chi}_{{\rm GDP\, EA},t}$ & GDP & EA GDP \\
$\widehat{\chi}_{{\rm HICPOV\,EA},t}$ & HICPOV & EA HICP: Overall  \\
$\widehat{\chi}_{{\rm LTIRT\,EA},t}$ & LTIRT & EA 10-years Interest Rate  \\
$\widehat{\chi}_{{\rm SHIX\, EA},t}$ & SHIX & EA Stock Price Index  \\
$\widehat{\chi}_{{\rm UNETOT\, EA},t}$ & UNETOT & EA Unemployment Rate  \\
$\widehat{\chi}_{{\rm nat.},t}$ & - & National variable  \\
\hline
\hline
\end{tabular}
\end{table}

\subsection{EA IRFs}

\begin{figure}[H]
\centering \footnotesize \sc \smallskip
\setlength{\tabcolsep}{.005\textwidth}
\caption{EA IRFs: mixed-frequency data}
\begin{tabular}{ccc}
\scriptsize 2-years Interest Rate ($R_t$) &\scriptsize  GDP &\scriptsize  HICP: Overall (HICPOV) \\
\includegraphics[trim= .5cm 8cm .5cm 8.5cm, clip, width=0.32\textwidth]{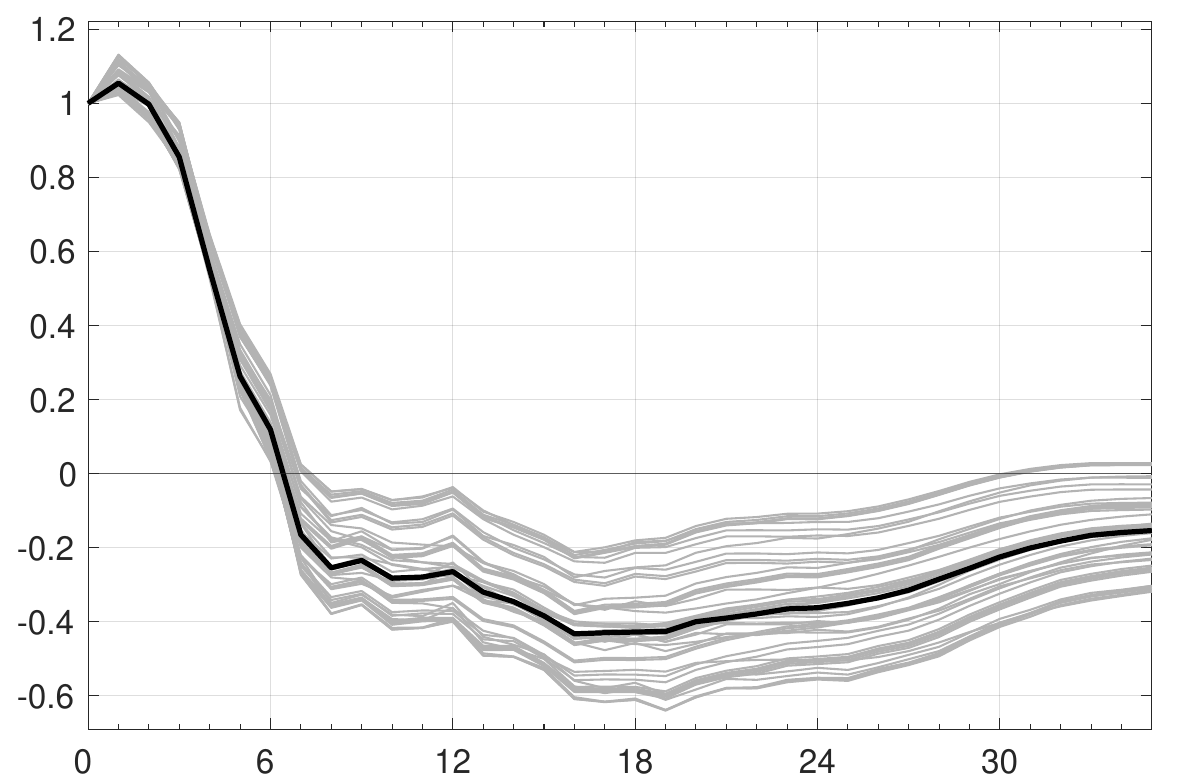} &
\includegraphics[trim= .5cm 8cm .5cm 8.5cm, clip, width=0.32\textwidth]{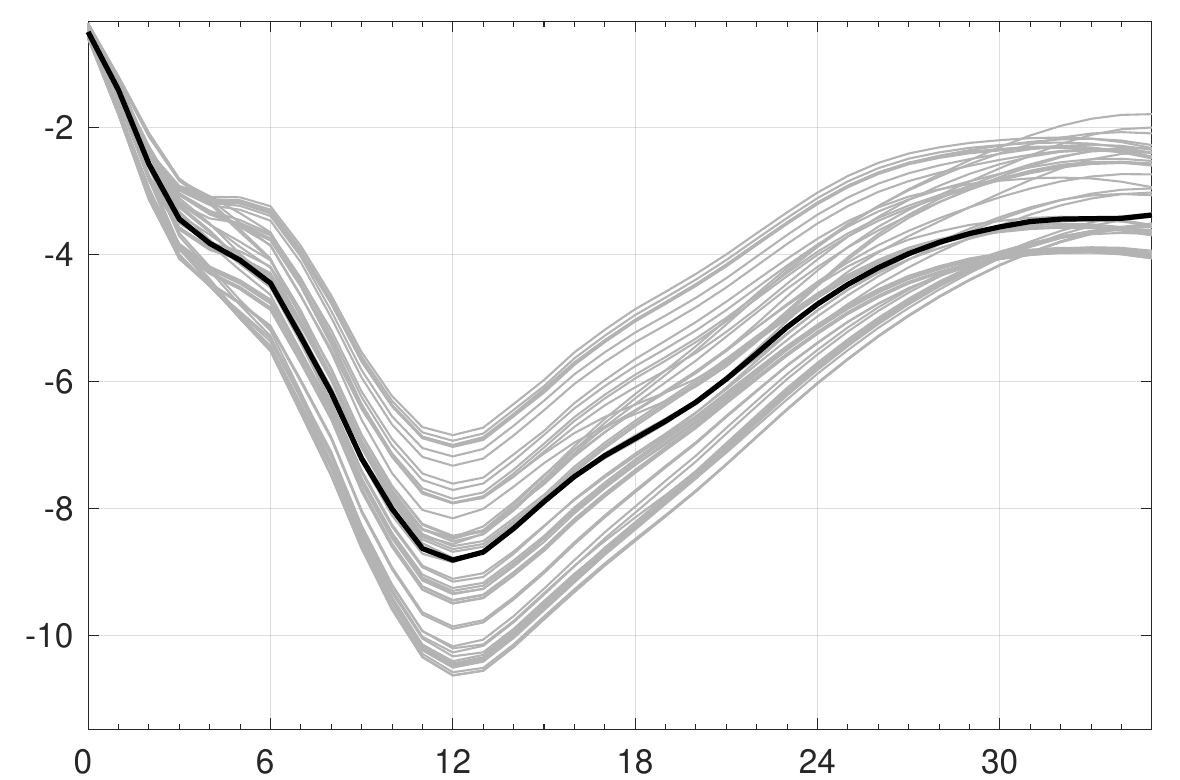} &
\includegraphics[trim= .5cm 8cm .5cm 8.5cm, clip, width=0.32\textwidth]{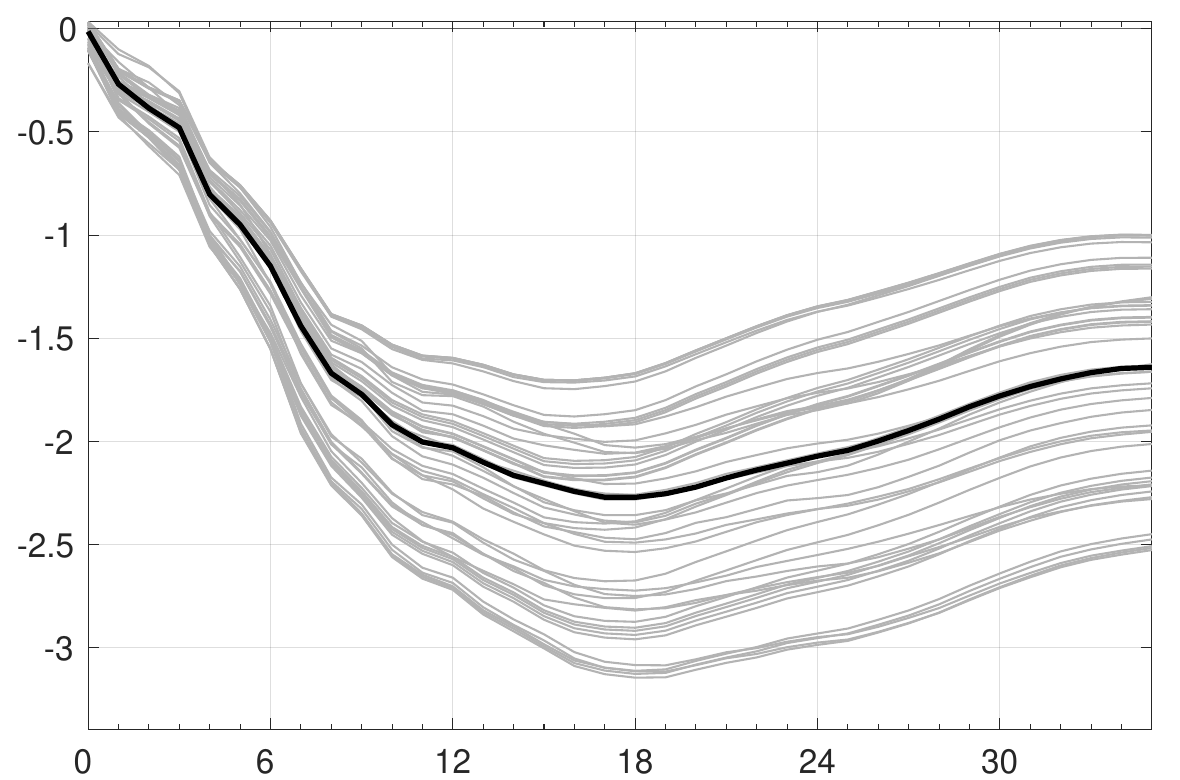} \\
\scriptsize 10-years Interest Rate (LTIRT) &\scriptsize Stock Price Index (SHIX) &\scriptsize Unemployment Rate (UNETOT)\\
\includegraphics[trim= .5cm 8cm .5cm 8.5cm, clip, width=0.32\textwidth]{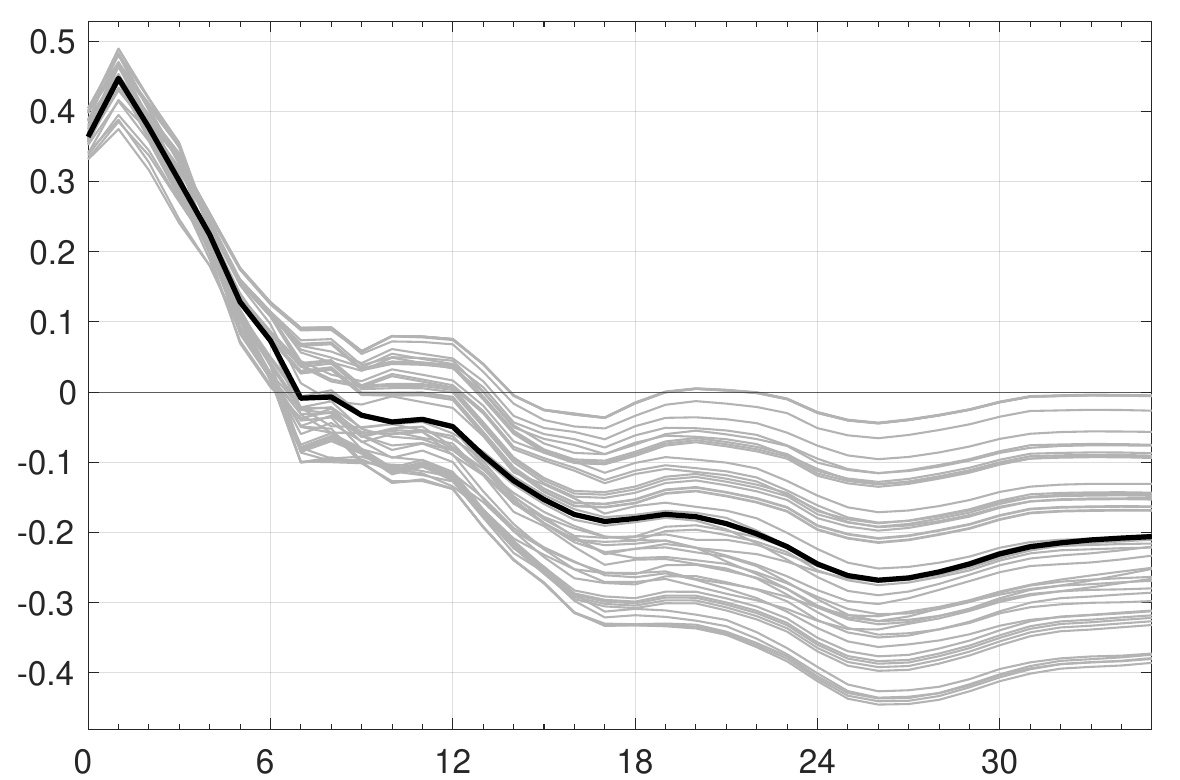} &
\includegraphics[trim= .5cm 8cm .5cm 8.5cm, clip, width=0.32\textwidth]{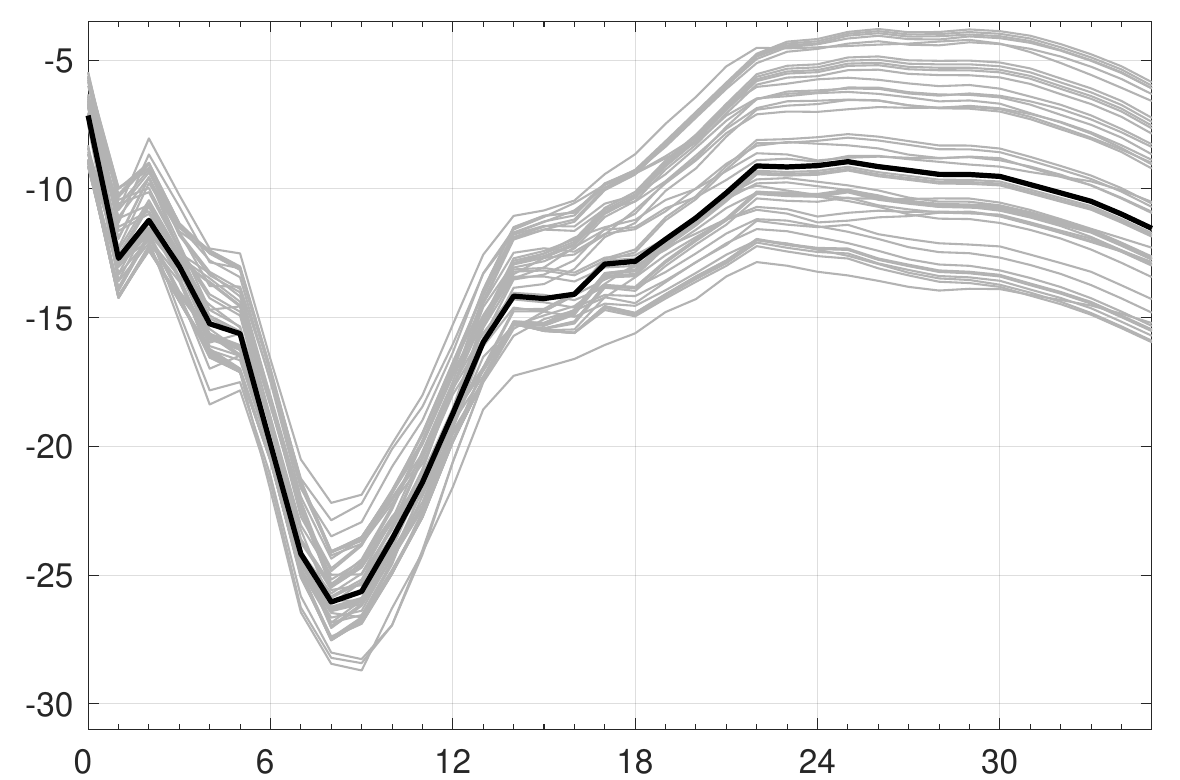} &
\includegraphics[trim= .5cm 8cm .5cm 8.5cm, clip, width=0.32\textwidth]{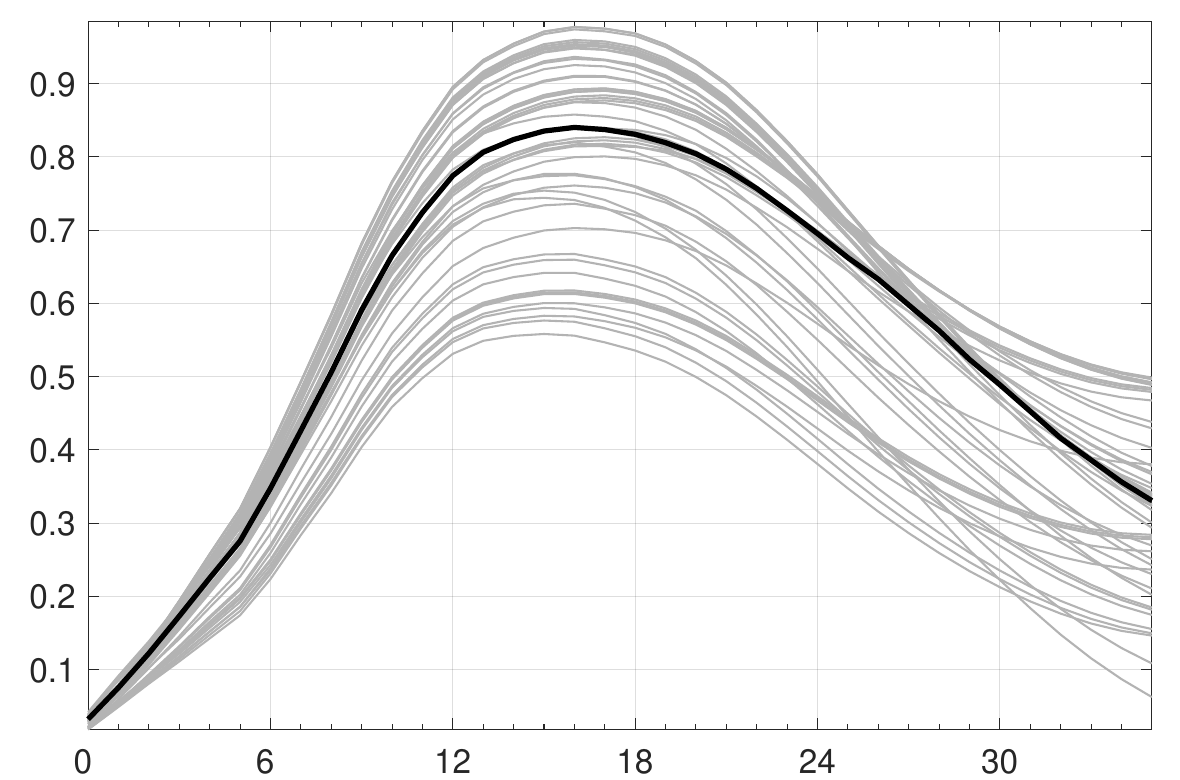} \\
\end{tabular}
\begin{tabular}{p{.98\textwidth}} \scriptsize Notes: \rm Each sub-figure plots the impulse response of one EA variable to a 100bps contractionary monetary policy shock. The thin gray lines are the point estimates obtained with the 50 considered CC-VAR models differing only for the national variable included. The black solid line is the median estimate. 
\end{tabular}
\label{fig::EAirfs_MF}
\end{figure}

\subsection{Country-level IRFs}

\begin{figure}[H]
\centering \footnotesize \sc \smallskip
\setlength{\tabcolsep}{.005\textwidth}
\caption{Country-level IRFs: mixed-frequency data}
\begin{tabular}{ccc}
\scriptsize{GDP}  
&\scriptsize {HICP: Overall (HICPOV)}      
&\scriptsize {Stock Price Index (SHIX)}  \\

\includegraphics[trim= .5cm 8cm .5cm 8.5cm, clip, width=0.32\textwidth]{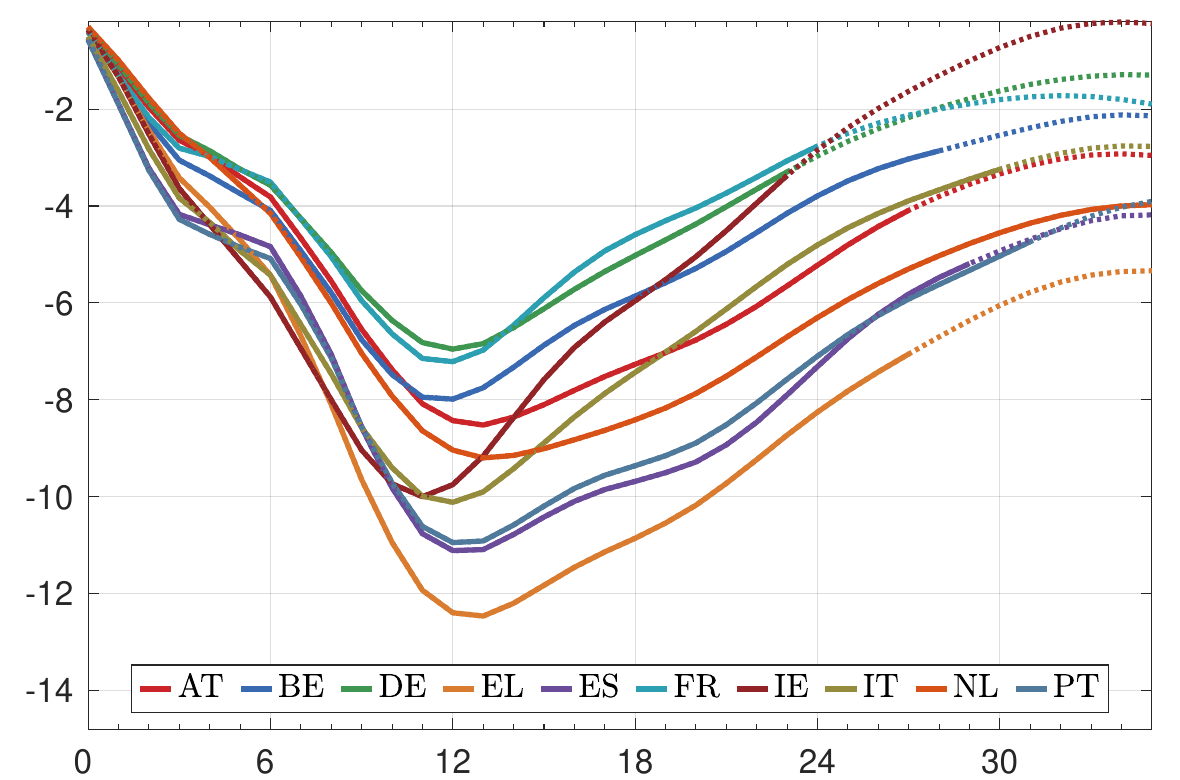} &
\includegraphics[trim= .5cm 8cm .5cm 8.5cm, clip, width=0.32\textwidth]{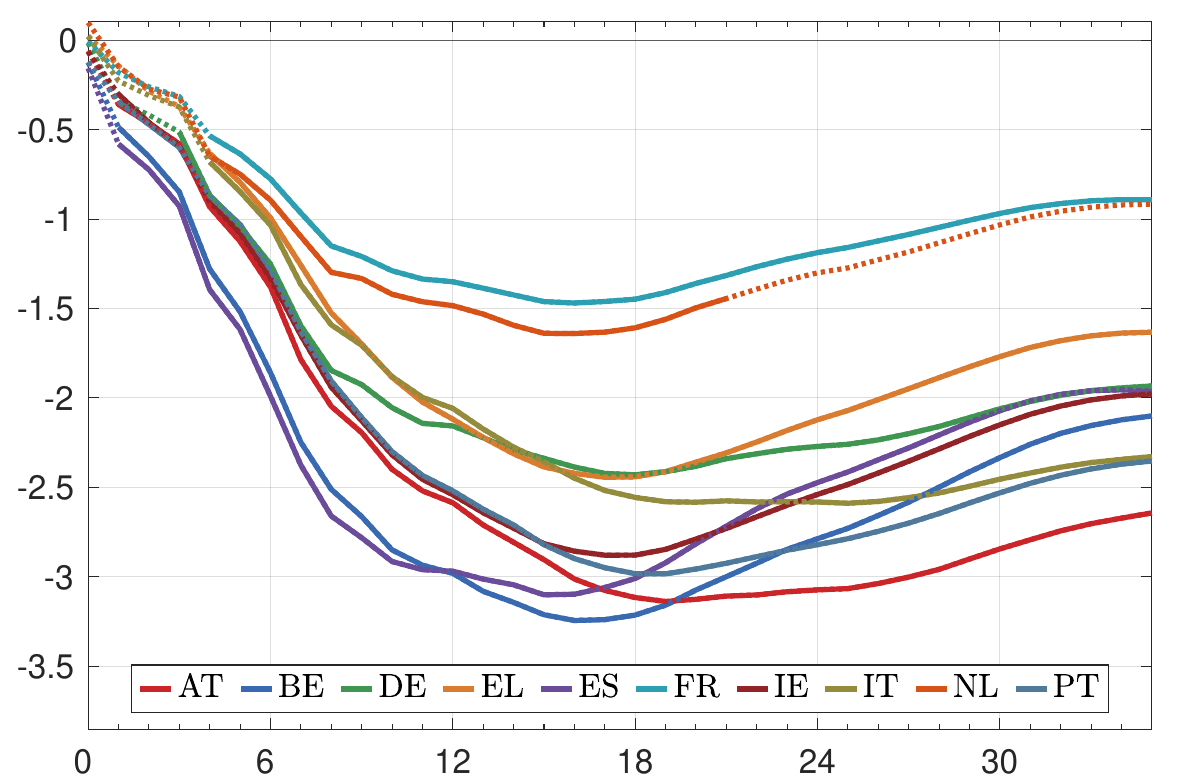} &
\includegraphics[trim= .5cm 8cm .5cm 8.5cm, clip, width=0.32\textwidth]{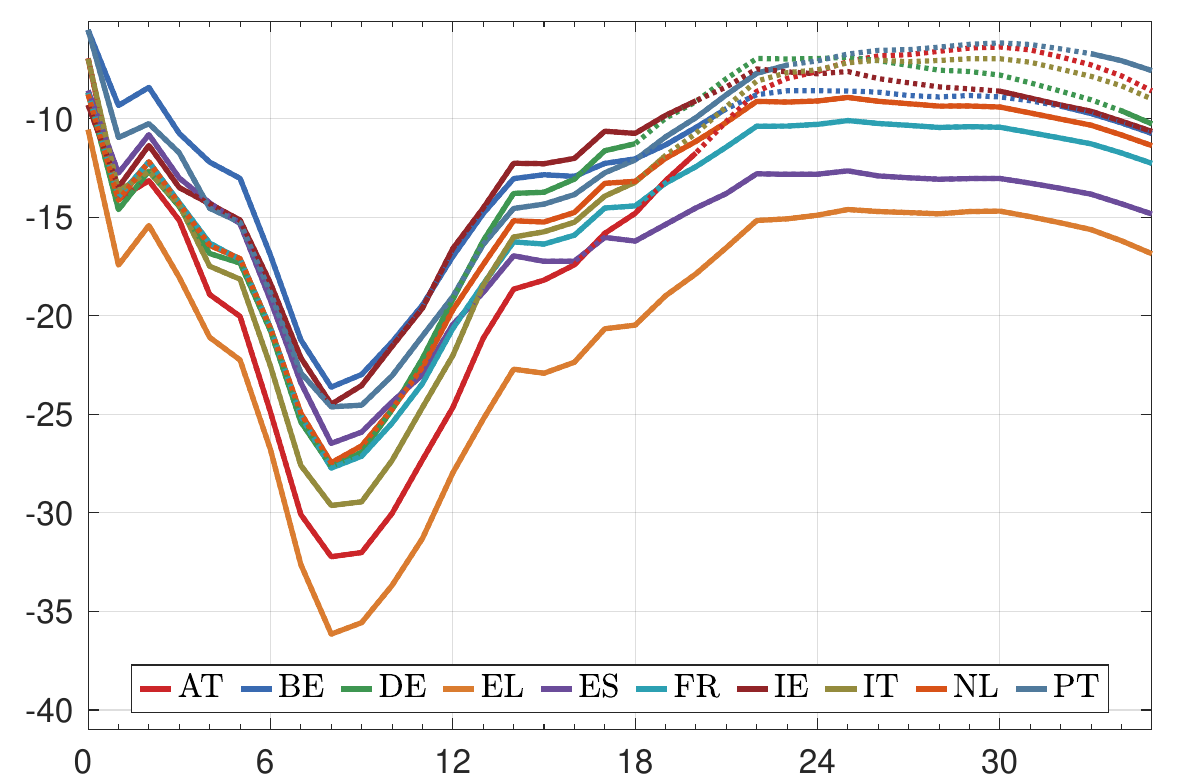} \\[10pt]
\end{tabular}

\begin{tabular}{cc}
\scriptsize {10-years Interest Rate (LTIRT)} 
&\scriptsize {Unemployment Rate (UNETOT)} \\

\includegraphics[trim= .5cm 8cm .5cm 8.5cm, clip, width=0.32\textwidth]{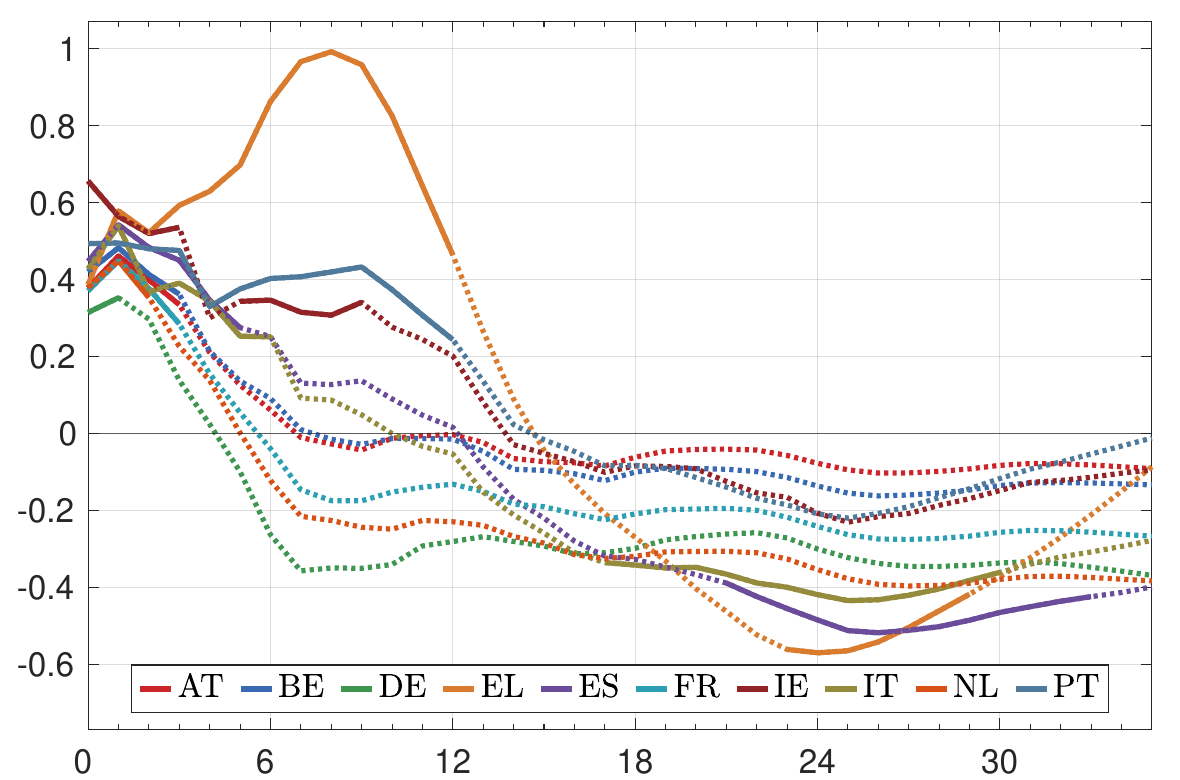} 
&
\includegraphics[trim= .5cm 8cm .5cm 8.5cm, clip, width=0.32\textwidth]{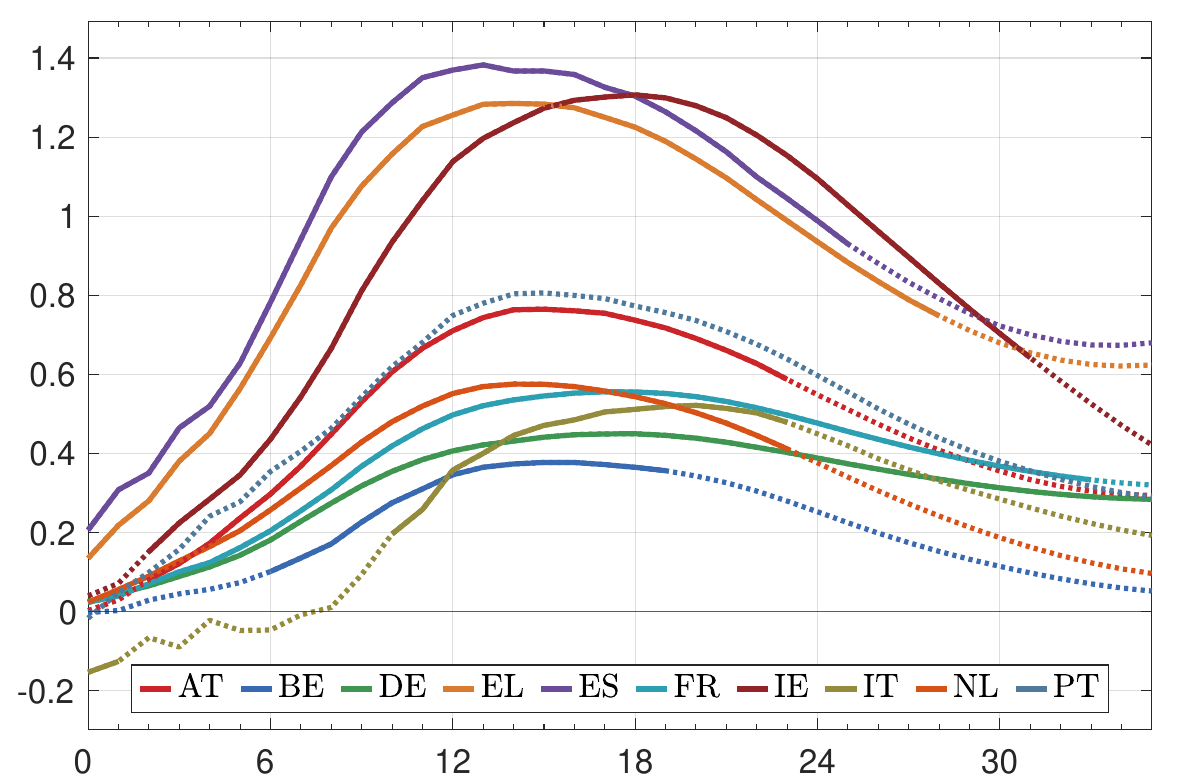}
\end{tabular}
\begin{tabular}{p{.98\textwidth}} \scriptsize Notes: \rm Each sub-figure plots the impulse responses, for all countries, of one variable to a 100bps contractionary monetary policy shock. Within each sub-figure, at each horizon $h=0,\ldots,36$ the country-level impulse responses are denoted with a solid line if the IRF is statistically significant at the 68\% level at that horizon, and with a dotted line otherwise. 
\end{tabular}
\label{fig::nat_IRFs_MF}
\end{figure}

\begin{figure}[H]
\centering \footnotesize \sc \smallskip
\setlength{\tabcolsep}{.005\textwidth}
\centering \footnotesize \sc \smallskip
\caption{Difference between country-level and EA IRFs: mixed-frequency data}
\begin{tabular}{lccccc}
& \hspace{5pt}\scriptsize GDP &  \hspace{5pt}\scriptsize HICPOV & 
 \hspace{5pt}\scriptsize SHIX &  \hspace{5pt}\scriptsize LTIRT &  \hspace{8pt}\scriptsize UNETOT \\[4pt] 
\raisebox{1.5\height}{\rotatebox{90}{\scriptsize AT}} & \includegraphics[trim=5cm 12cm 5cm 12.5cm, clip, width=0.19\textwidth]{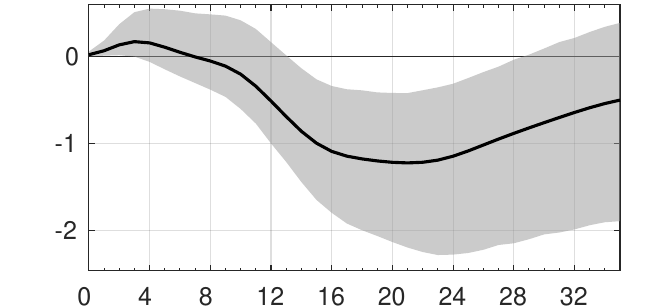} &
\includegraphics[trim=5cm 12cm 5cm 12.5cm, clip, width=0.19\textwidth]{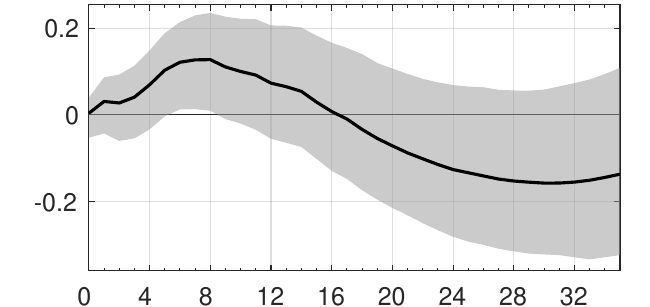} &
\includegraphics[trim=5cm 12cm 5cm 12.5cm, clip, width=0.19\textwidth]{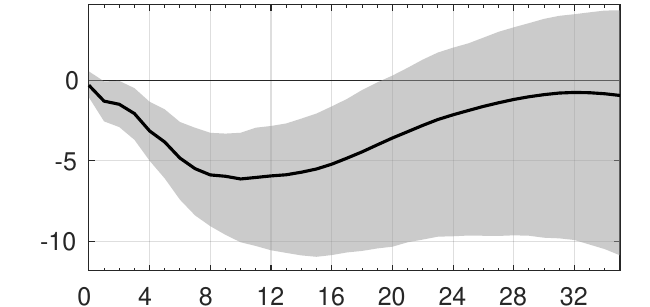} &
\includegraphics[trim=5cm 12cm 5cm 12.5cm, clip, width=0.19\textwidth]{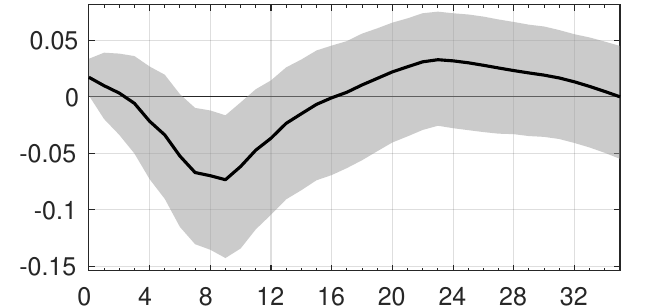} &
\includegraphics[trim=5cm 12cm 5cm 12.5cm, clip, width=0.19\textwidth]{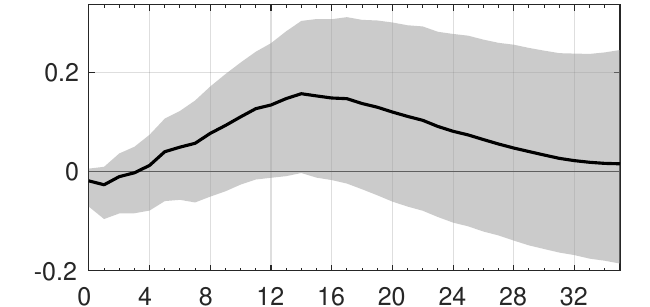} \\

\raisebox{1.5\height}{\rotatebox{90}{\scriptsize BE}} &
\includegraphics[trim=5cm 12cm 5cm 12.5cm, clip, width=0.19\textwidth]{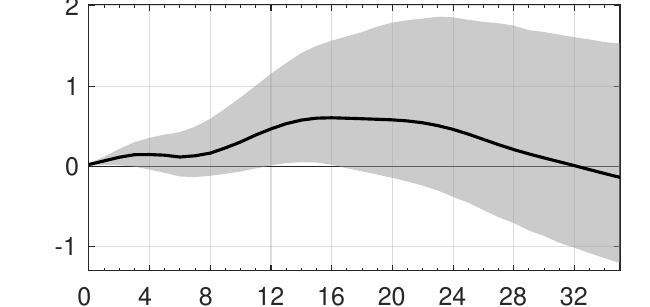} &
\includegraphics[trim=5cm 12cm 5cm 12.5cm, clip, width=0.19\textwidth]{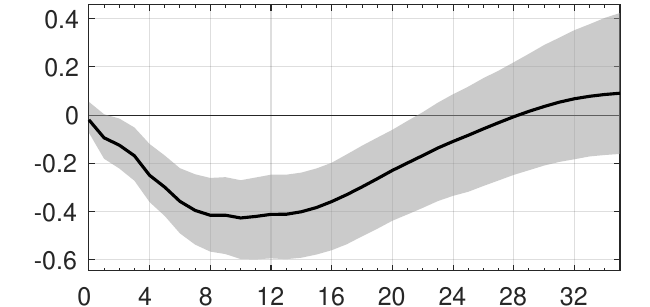} &
\includegraphics[trim=5cm 12cm 5cm 12.5cm, clip, width=0.19\textwidth]{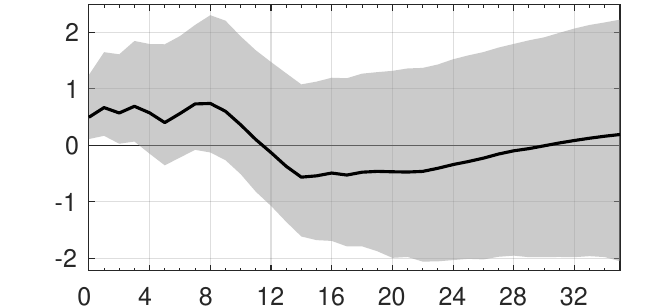} &
\includegraphics[trim=5cm 12cm 5cm 12.5cm, clip, width=0.19\textwidth]{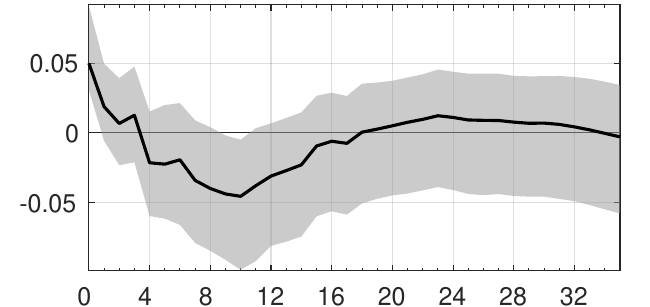} &
\includegraphics[trim=5cm 12cm 5cm 12.5cm, clip, width=0.19\textwidth]{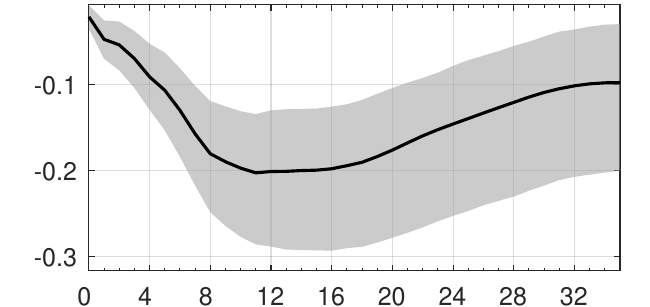} \\

\raisebox{1.5\height}{\rotatebox{90}{\scriptsize DE}} &
\includegraphics[trim=5cm 12cm 5cm 12.5cm, clip, width=0.19\textwidth]{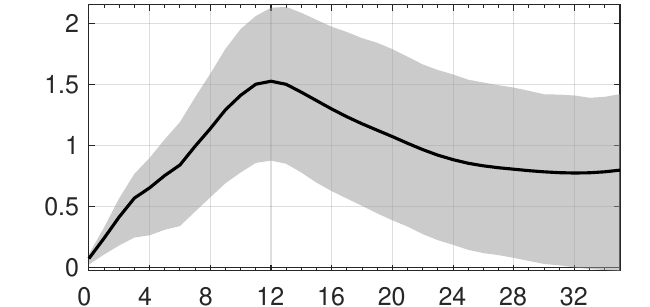} &
\includegraphics[trim=5cm 12cm 5cm 12.5cm, clip, width=0.19\textwidth]{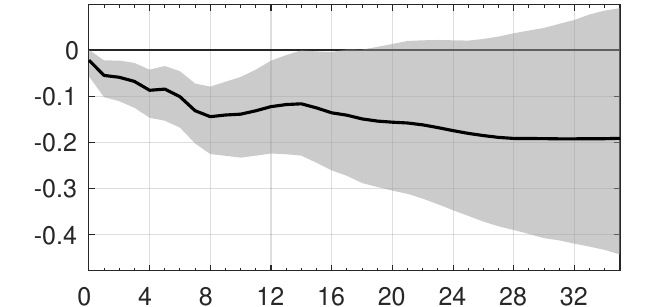} &
\includegraphics[trim=5cm 12cm 5cm 12.5cm, clip, width=0.19\textwidth]{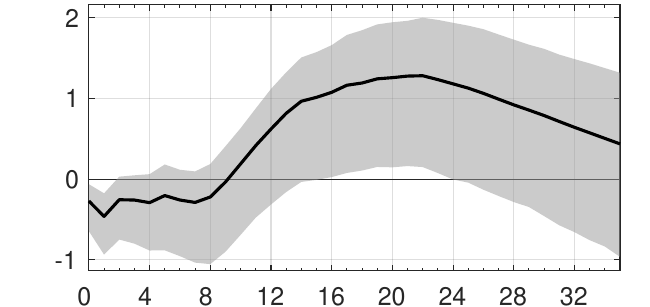} &
\includegraphics[trim=5cm 12cm 5cm 12.5cm, clip, width=0.19\textwidth]{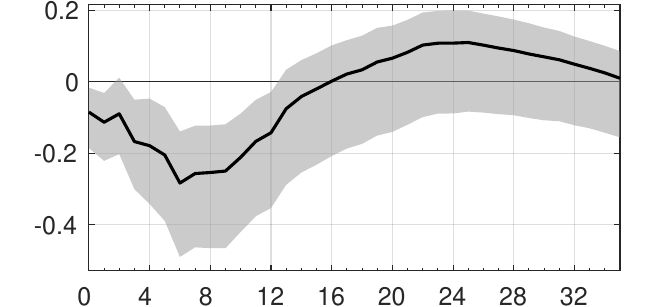} &
\includegraphics[trim=5cm 12cm 5cm 12.5cm, clip, width=0.19\textwidth]{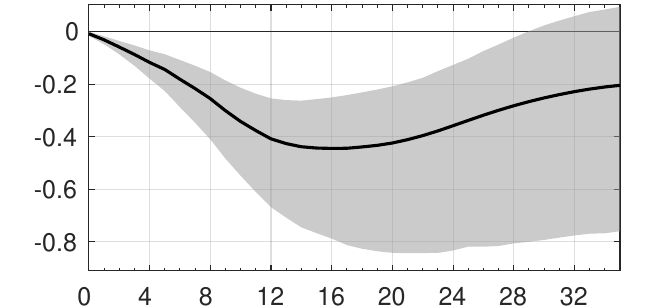} \\

\raisebox{1.5\height}{\rotatebox{90}{\scriptsize FR}} &
\includegraphics[trim=5cm 12cm 5cm 12.5cm, clip, width=0.19\textwidth]{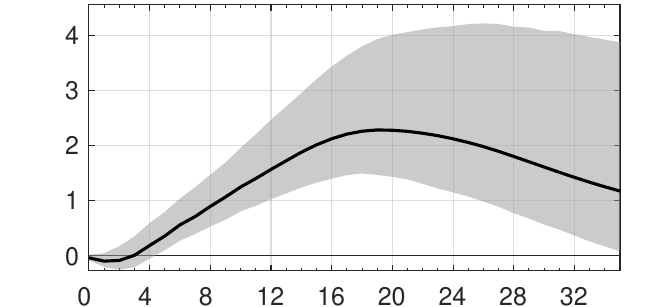} &
\includegraphics[trim=5cm 12cm 5cm 12.5cm, clip, width=0.19\textwidth]{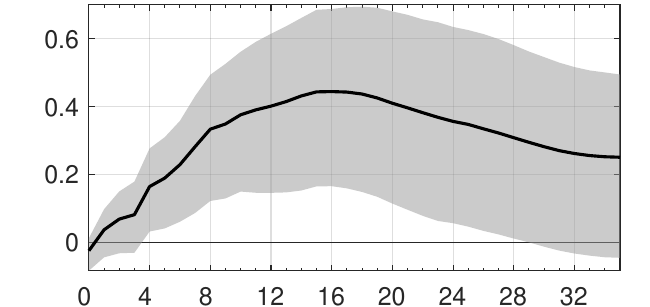} &
\includegraphics[trim=5cm 12cm 5cm 12.5cm, clip, width=0.19\textwidth]{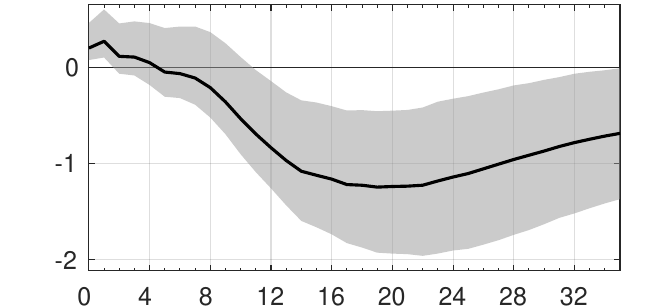} &
\includegraphics[trim=5cm 12cm 5cm 12.5cm, clip, width=0.19\textwidth]{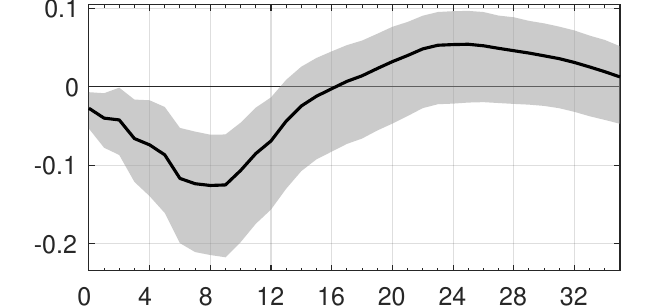} &
\includegraphics[trim=5cm 12cm 5cm 12.5cm, clip, width=0.19\textwidth]{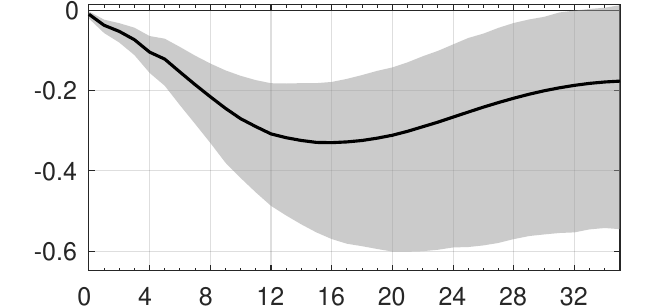} \\

\raisebox{1.5\height}{\rotatebox{90}{\scriptsize NL}} &
\includegraphics[trim=5cm 12cm 5cm 12.5cm, clip, width=0.19\textwidth]{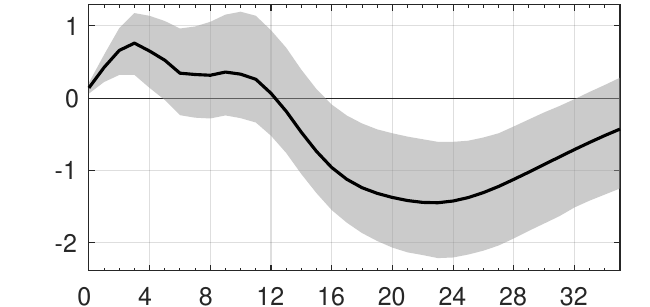} &
\includegraphics[trim=5cm 12cm 5cm 12.5cm, clip, width=0.19\textwidth]{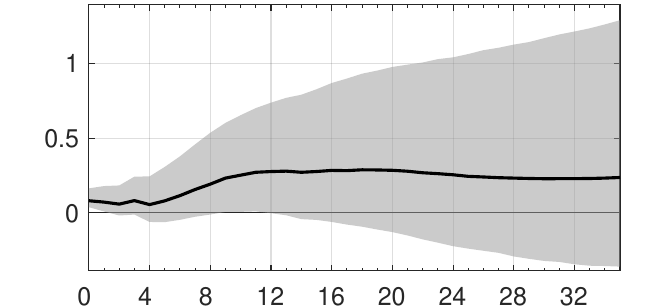} &
\includegraphics[trim=5cm 12cm 5cm 12.5cm, clip, width=0.19\textwidth]{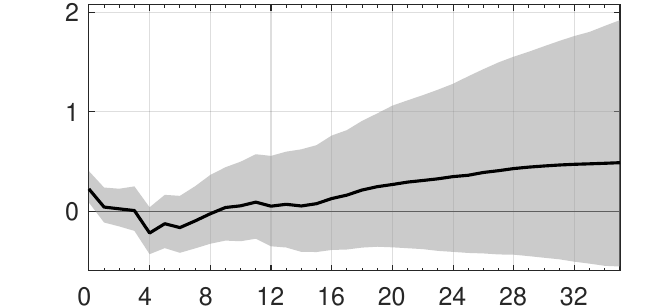} &
\includegraphics[trim=5cm 12cm 5cm 12.5cm, clip, width=0.19\textwidth]{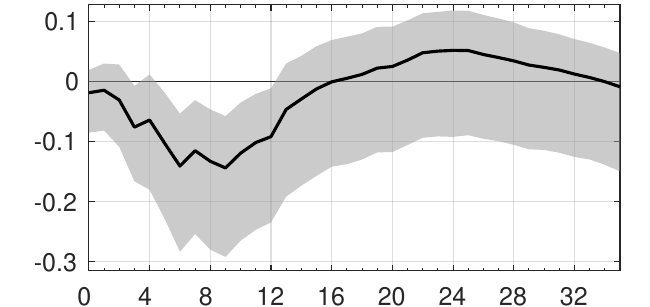} &
\includegraphics[trim=5cm 12cm 5cm 12.5cm, clip, width=0.19\textwidth]{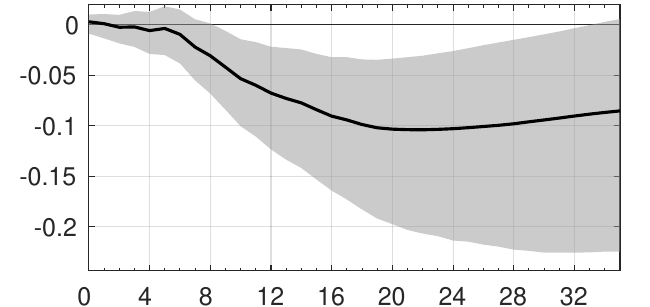} \\

\raisebox{1.7\height}{\rotatebox{90}{\scriptsize EL}} &
\includegraphics[trim=5cm 12cm 5cm 12.5cm, clip, width=0.19\textwidth]{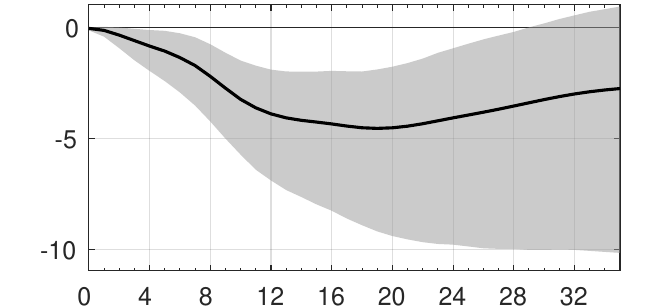} &
\includegraphics[trim=5cm 12cm 5cm 12.5cm, clip, width=0.19\textwidth]{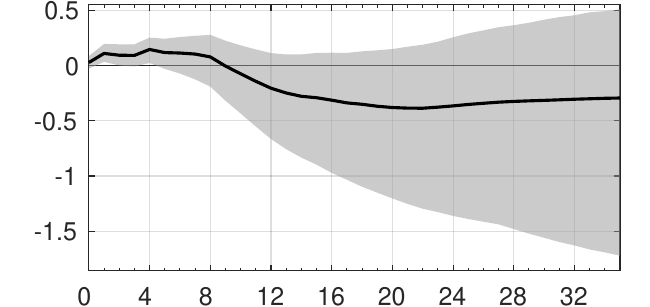} &
\includegraphics[trim=5cm 12cm 5cm 12.5cm, clip, width=0.19\textwidth]{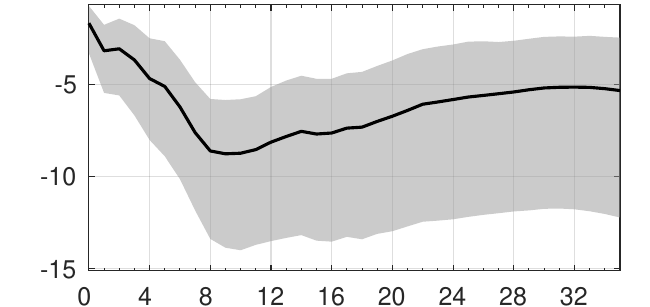} &
\includegraphics[trim=5cm 12cm 5cm 12.5cm, clip, width=0.19\textwidth]{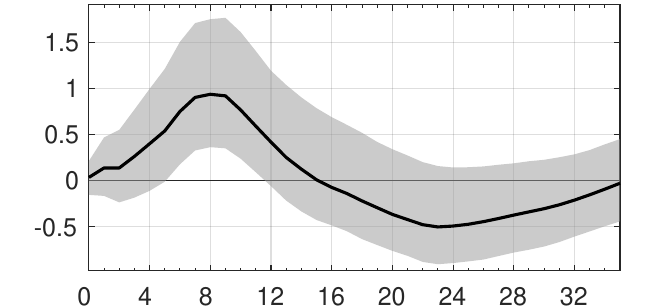} &
\includegraphics[trim=5cm 12cm 5cm 12.5cm, clip, width=0.19\textwidth]{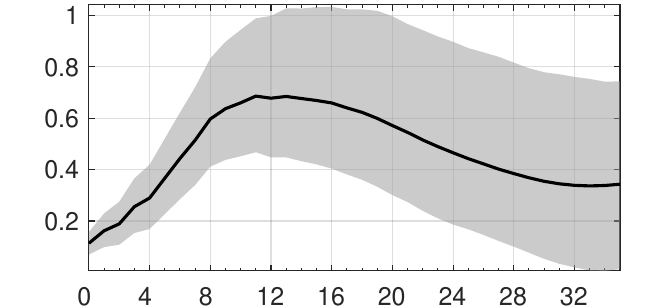} \\

\raisebox{1.7\height}{\rotatebox{90}{\scriptsize ES}} &
\includegraphics[trim=5cm 12cm 5cm 12.5cm, clip, width=0.19\textwidth]{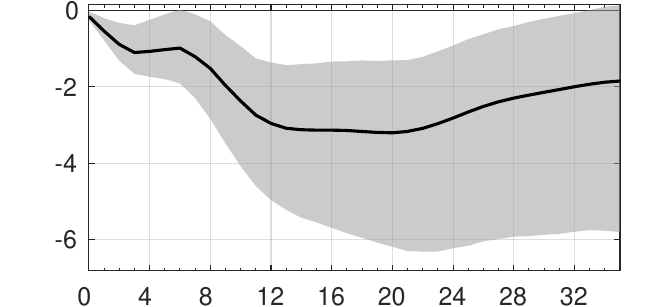} &
\includegraphics[trim=5cm 12cm 5cm 12.5cm, clip, width=0.19\textwidth]{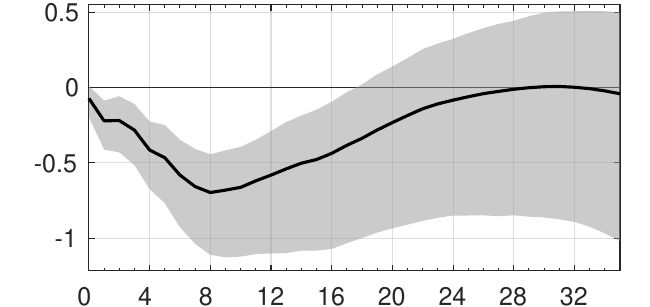} &
\includegraphics[trim=5cm 12cm 5cm 12.5cm, clip, width=0.19\textwidth]{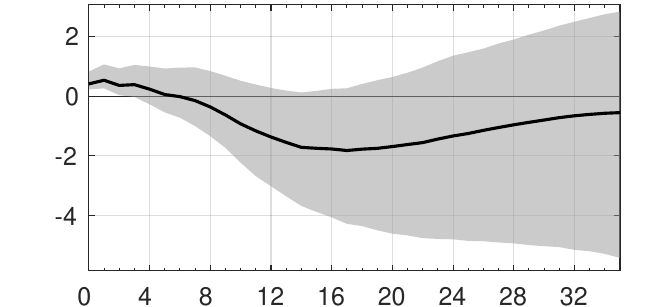} &
\includegraphics[trim=5cm 12cm 5cm 12.5cm, clip, width=0.19\textwidth]{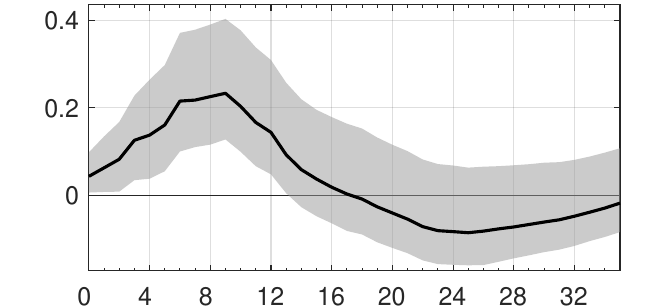} &
\includegraphics[trim=5cm 12cm 5cm 12.5cm, clip, width=0.19\textwidth]{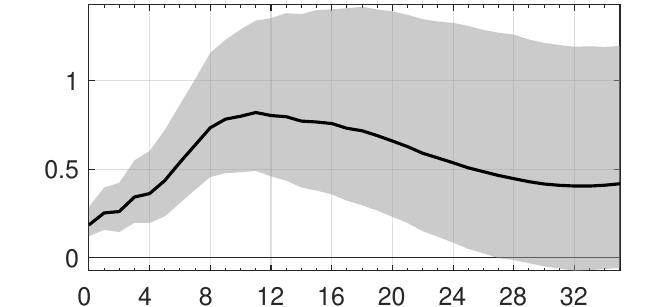} \\

\raisebox{2\height}{\rotatebox{90}{\scriptsize IE}} &
\includegraphics[trim=5cm 12cm 5cm 12.5cm, clip, width=0.19\textwidth]{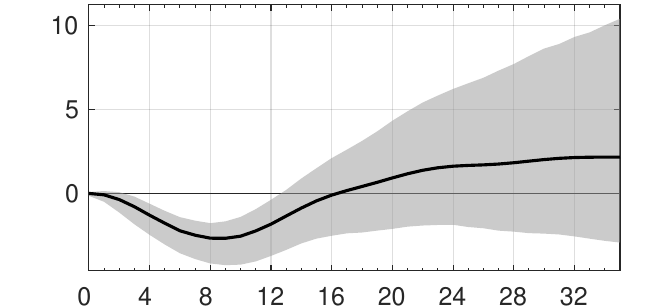} &
\includegraphics[trim=5cm 12cm 5cm 12.5cm, clip, width=0.19\textwidth]{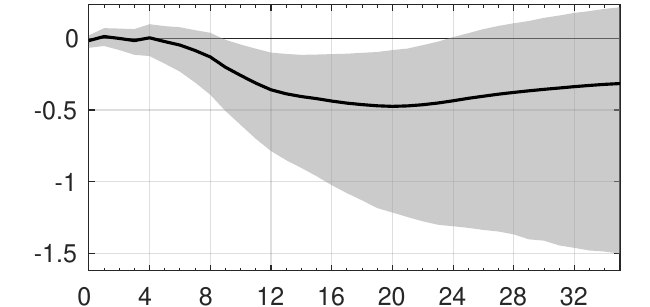} &
\includegraphics[trim=5cm 12cm 5cm 12.5cm, clip, width=0.19\textwidth]{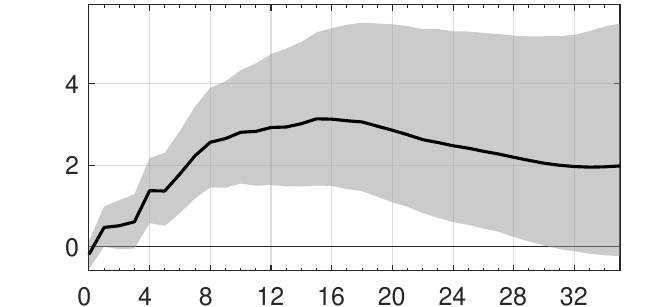} &
\includegraphics[trim=5cm 12cm 5cm 12.5cm, clip, width=0.19\textwidth]{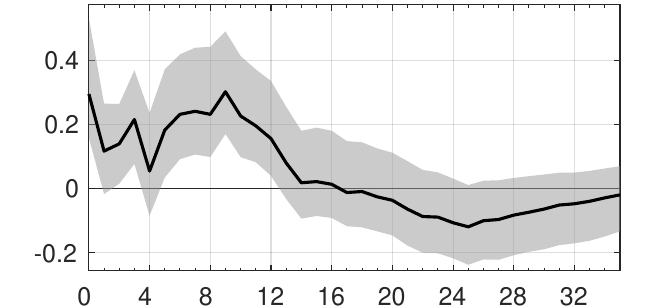} &
\includegraphics[trim=5cm 12cm 5cm 12.5cm, clip, width=0.19\textwidth]{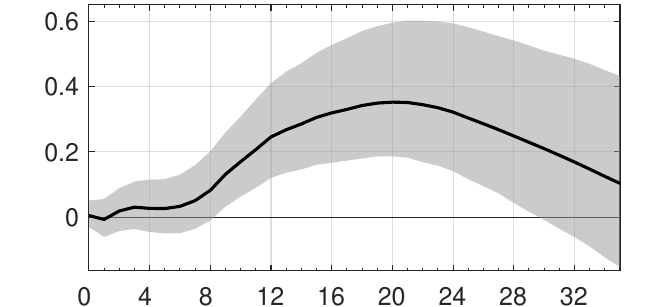} \\

\raisebox{2\height}{\rotatebox{90}{\scriptsize IT}} &
\includegraphics[trim=5cm 12cm 5cm 12.5cm, clip, width=0.19\textwidth]{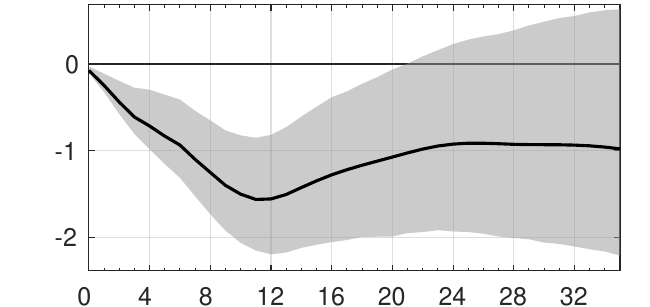} &
\includegraphics[trim=5cm 12cm 5cm 12.5cm, clip, width=0.19\textwidth]{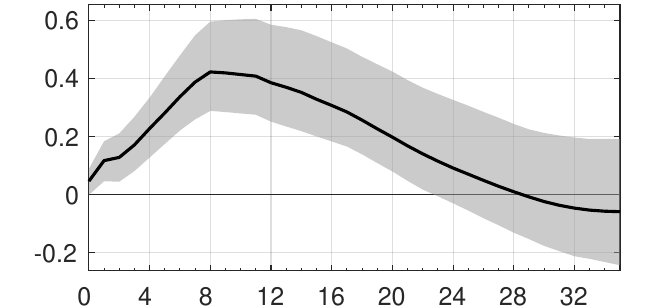} &
\includegraphics[trim=5cm 12cm 5cm 12.5cm, clip, width=0.19\textwidth]{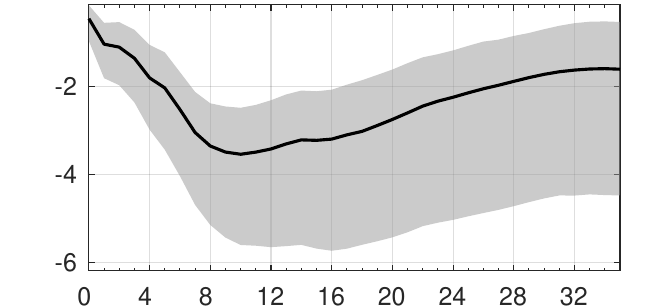} &
\includegraphics[trim=5cm 12cm 5cm 12.5cm, clip, width=0.19\textwidth]{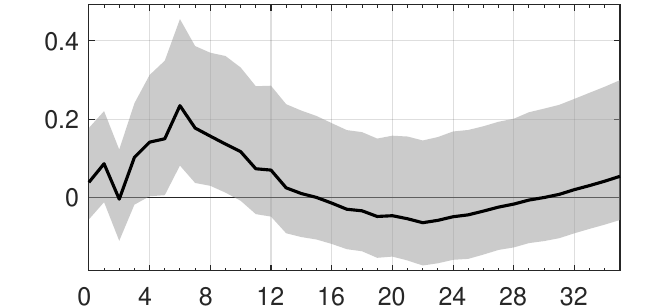} &
\includegraphics[trim=5cm 12cm 5cm 12.5cm, clip, width=0.19\textwidth]{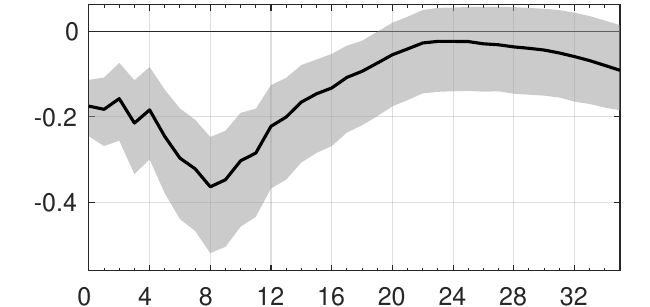} \\

\raisebox{1.5\height}{\rotatebox{90}{\scriptsize PT}} &
\includegraphics[trim=5cm 12cm 5cm 12.5cm, clip, width=0.19\textwidth]{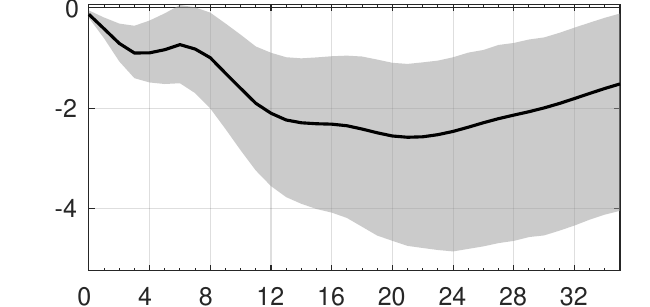} &
\includegraphics[trim=5cm 12cm 5cm 12.5cm, clip, width=0.19\textwidth]{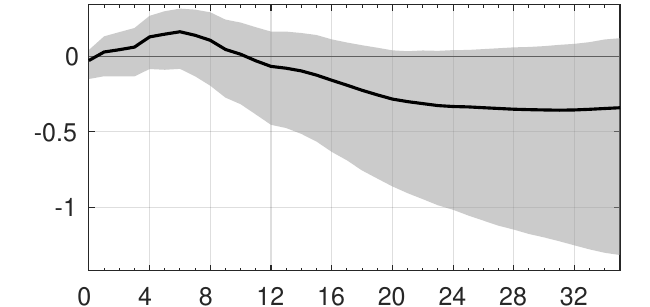} &
\includegraphics[trim=5cm 12cm 5cm 12.5cm, clip, width=0.19\textwidth]{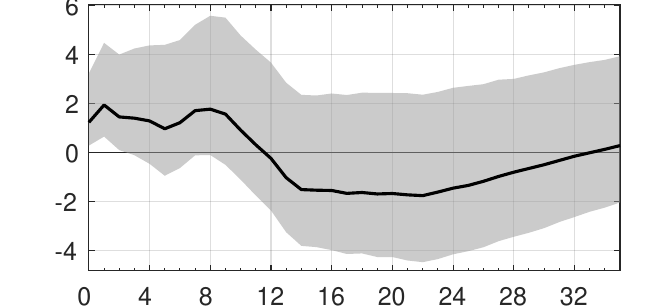} &
\includegraphics[trim=5cm 12cm 5cm 12.5cm, clip, width=0.19\textwidth]{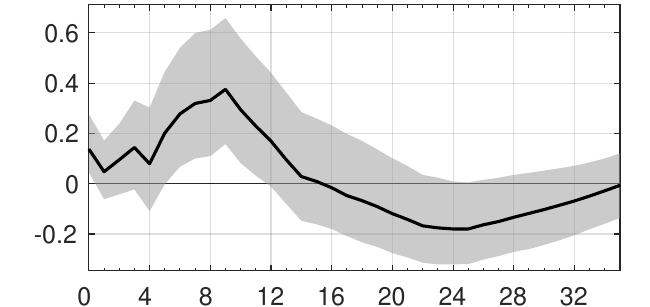} &
\includegraphics[trim=5cm 12cm 5cm 12.5cm, clip, width=0.19\textwidth]{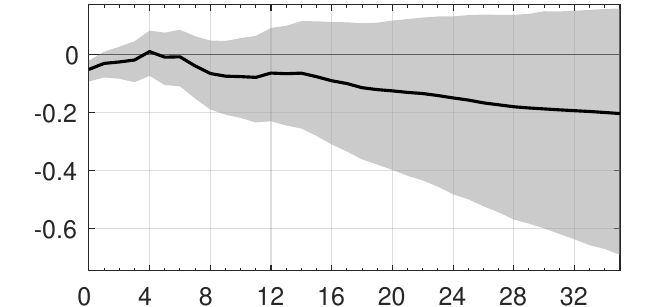} \\
\end{tabular}
\begin{tabular}{p{1\textwidth}} \scriptsize Notes: \rm Each sub-figure plots the difference between the country-level IRF and the corresponding EA counterpart for one variable and one country. Each column of the graph represents a variable, while each row represents a country. The variables considered are: GDP, HICP: Overall (HICPOV), Stock Price Index (SHIX), 10-years Interest Rates (LTIRT) and Unemployment Rate (UNETOT). The black solid line is the point estimate in our baseline setting, while the gray shaded area is the corresponding 68\% confidence interval. The scale in the vertical axis differs across variables and countries.
\end{tabular}
\label{fig::diff_MF}
\end{figure}
\end{document}